\documentclass[11pt
]{article}
\usepackage{amsmath, amsthm,mathtools, amssymb,upgreek,slashed,hyperref}
\usepackage[utf8]{inputenc}
\usepackage{ifpdf}
\ifpdf
  \usepackage[pdftex]{graphicx}
  \usepackage{epstopdf}
\else
  \usepackage[dvips]{graphicx}
\fi
\usepackage[paper=letterpaper,margin=1in]{geometry}
\def\Bbb{\mathbb}
\def\Tr{{\rm Tr}}
\def\cD{{\mathcal D}}
\def\cT{{\mathcal T}}

\def\sK{{\sf K}}
\def\sR{{\sf R}}
\def\SU{{\mathrm{SU}}}

\def\orr{{\mathrm{or}}}

\def\sF{{\sf F}}
\def\U{{\mathrm U}}
\def\RP{{\Bbb{RP}}}
\def\2{{\bf 2}}
\def\3{{\bf 3}}
\def\KB{{\mathrm{KB}}}
\def\4{{\bf 4}}
\def\pin{{\mathrm{pin}}}

\def\i{{\mathrm i}}
\def\h{\widehat}
\def\u{u}

\def\cO{{\mathcal O}}

\def\be{\begin{equation}}
\def\ee{\end{equation}}
 \def\Sp{{\mathrm{Sp}}}
 
 \def\SU{{\mathrm{SU}}}
 \def\SO{{\mathrm{SO}}}

\def\Sp{\mathrm{Sp}}
\def\T{{\mathcal T}}

\def\bar{\overline}

\def\tilde{\widetilde}
\def\t{\widetilde}
\def\R{{\Bbb{R}}}\def\Z{{\Bbb{Z}}}
\def\N{{\mathcal N}}
\def\B{{\mathcal B}}
\def\GL{{\mathrm{GL}}}
\def\hat{\widehat}
\def\P{{\mathcal P}}
\def\Pf{{\mathrm{Pf}}}
\def\D{{\mathcal D}}
\def\DD{{\sf D}}

\def\TT{{\sf  T}}
\def\UU{{\sf U}}
\font\teneurm=eurm10 \font\seveneurm=eurm7 \font\fiveeurm=eurm5
\newfam\eurmfam
\textfont\eurmfam=\teneurm \scriptfont\eurmfam=\seveneurm
\scriptscriptfont\eurmfam=\fiveeurm

\font\teneusm=eusm10 \font\seveneusm=eusm7 \font\fiveeusm=eusm5
\newfam\eusmfam
\textfont\eusmfam=\teneusm \scriptfont\eusmfam=\seveneusm
\scriptscriptfont\eusmfam=\fiveeusm

\font\tencmmib=cmmib10 \skewchar\tencmmib='177
\font\sevencmmib=cmmib7 \skewchar\sevencmmib='177
\font\fivecmmib=cmmib5 \skewchar\fivecmmib='177
\newfam\cmmibfam
\textfont\cmmibfam=\tencmmib \scriptfont\cmmibfam=\sevencmmib
\scriptscriptfont\cmmibfam=\fivecmmib

\numberwithin{equation}{section}

\def\d{\mathrm d}
\def\C{{\Bbb C}}
\def\NS{{\text{NS}}}
\def\Ra{{\text{R}}}
\def\sV{{\mathcal V}}
\def\Z{{\Bbb Z}}
\def\vnu{\updelta}
\def\A{{\mathcal A}}
\def\CC{{\mathcal C}}
\def\S{{\mathcal S}}
\def\H{{\mathcal H}}
\def\bar{\overline}

\def\g{{\mathfrak g}}

\def\M{{\mathcal M}}

\def\sT{{\sf T}}

\def\SL{{\mathrm{SL}}}
\def\OSp{{\mathrm{OSp}}}
\def\detp{{\mathrm{det}}'} 
\def\sC{{\mathcal C}}
\def\ZZ{{\mathcal Z}}
\def\sK{{\sf K}}
\def\Rr{{\sf R}}
\def\diag{{\mathrm{diag}}}
\def\mR{{\mathcal R}}
\def\vol{{\mathrm{vol}}}
\def\cC{C}
\def\AdS{{\mathrm{AdS}}}
\def\NAdS{{\mathrm{NAdS}}}
\def\JT{{\mathrm{JT}}}
\def\osp{{\mathrm{osp}}}
\def\rR{{\mathcal R}}
\newcommand*{\horzbar}{\rule[.5ex]{2.5ex}{0.5pt}}
\def\Ber{{\mathrm{Ber}}}
\def\SJT{{\mathrm{SJT}}}
\def\Str{{\mathrm{Str}}}
\def\sl{{\mathfrak{sl}}}
\def\Vol{{\mathrm{Vol}}}
\def\hM{{\widehat M}}
\def\Pf{{\mathrm{Pf}}}
\def\hw{{\hat\omega}}
\def\O{{\mathrm{O}}}
\def\OO{{\mathrm{O}}}
\def\veps{\varepsilon}

\begin{document}

\begin{titlepage}

\begin{center}

\phantom{ }
\vspace{3cm}

{\bf \Large{JT Gravity and the Ensembles  of Random Matrix Theory}}
\vskip 0.5cm
Douglas Stanford${}^{a,b}$ and Edward Witten${}^a$
\vskip 0.05in
\small{ $^a$ \textit{Institute for Advanced Study}}
\vskip -.4cm
\small{\textit{Einstein Drive, Princeton, NJ 08540 USA}}

\small{$^b$ \textit{Stanford Institute for Theoretical Physics}}
\vskip -.4cm
\small{\textit{Stanford, CA, 94305 USA}}
\begin{abstract}
We generalize the recently  discovered relationship between  JT gravity and double-scaled random matrix theory to the case that the boundary theory
may have time-reversal symmetry  and may  have fermions with or without supersymmetry. 
The matching between variants of  JT gravity and matrix ensembles depends on the  assumed  symmetries.   Time-reversal symmetry in the boundary theory  means that  unorientable
spacetimes must be considered in the bulk.  In such a case, the partition function of JT gravity  is  still  related to the volume  of the moduli  space of conformal structures,
but this volume has a quantum correction and has to  be computed  using Reidemeister-Ray-Singer ``torsion.''
Presence of fermions in the boundary theory (and thus  a symmetry $(-1)^\sF$) means that the bulk has a  spin or pin 
structure. Supersymmetry in the boundary means that the bulk theory 
is  associated to JT supergravity and is
 related  to  the volume of the moduli space of super Riemann surfaces rather than of ordinary Riemann surfaces.   In all cases we match JT gravity or supergravity
with an  appropriate random matrix ensemble.    All ten  standard random matrix ensembles make an appearance -- the three Dyson ensembles  and the seven 
Altland-Zirnbauer ensembles.    To  facilitate the analysis, we extend  to the other ensembles techniques that are most familiar in the case of the original Wigner-Dyson ensemble of hermitian
matrices. We also generalize Mirzakhani's recursion for the volumes of ordinary moduli space to the case of super Riemann surfaces.
\end{abstract}
\end{center}
\end{titlepage}

\setcounter{tocdepth}{2}

{\parskip = .2\baselineskip \tableofcontents}

\section{Introduction}
Jackiw-Teitelboim (JT) gravity \cite{Teitelboim:1983ux,Jackiw:1984je} is a simple theory of two-dimensional quantum gravity that describes rigid hyperbolic spaces. It was used as a model for AdS${}_2$/CFT${}_1$ in \cite{Almheiri:2014cka}, and more generally it describes the low-energy dynamics of any near-extremal black hole.

The theory is so simple that it is almost trivial. On a closed Euclidean manifold $Y$, the path integral reduces to the volume of a finite-dimensional space -- the moduli space of Riemann surfaces with the topology of $Y$. However, in most applications, it is interesting to consider JT gravity on a space $Y$ with at least one asymptotic boundary. In this setting, one also has to do a path-integral over ``wiggles'' associated to each boundary \cite{Jensen:2016pah,Maldacena:2016upp,Engelsoy:2016xyb}.

These wiggles are governed by a solvable theory known as the ``Schwarzian theory.'' This provides a link between JT gravity and the Sachdev-Ye-Kitaev (SYK) model \cite{Sachdev:1992fk,KitaevTalks,Kitaev:2017awl}, which reduces at low energies to the same Schwarzian theory \cite{KitaevTalks,Kitaev:2017awl,Maldacena:2016hyu}. For a review, see \cite{Sarosi:2017ykf}. This correspondence has motivated much recent work on JT gravity.\footnote{This has included many different derivations of the exact path integral \cite{Bagrets:2016cdf,Cotler:2016fpe,Bagrets:2017pwq,Stanford:2017thb,Belokurov:2017eit,Mertens:2017mtv,Kitaev:2018wpr,Yang:2018gdb,Iliesiu:2019xuh}, studies of the Hilbert space structure \cite{Harlow:2018tqv,Kitaev:2018wpr,Lin:2018xkj,Blommaert:2018iqz}, correlation functions \cite{Mertens:2018fds,Blommaert:2018oro,Blommaert:2019hjr}, the physical symmetry charges \cite{Lin:2019qwu}, and the on-shell action \cite{Brown:2018bms}. Further work has included generalizations to include higher spin fields \cite{Gonzalez:2018enk}, supersymmetry \cite{Forste:2017kwy}, different topologies \cite{Maldacena:2018lmt,Saad:2018bqo,Blommaert:2018iqz,Blommaert:2019hjr},  and de Sitter space \cite{Maldacena:2019cbz,Cotler:2019nbi}.}

In \cite{Saad:2019lba} it was shown that the path integral of JT gravity on arbitrary orientable surfaces is computed by a Hermitian matrix integral, with the dictionary 
\be\label{Jru}
Z_{\text{JT}}(\beta_1,\dots \beta_n) \leftrightarrow \left\langle \Tr\,e^{-\beta_1 H}\dots \Tr\,e^{-\beta_n H}\right\rangle.
\ee
On the LHS, $Z_{\text{JT}}(\beta_1,\dots,\beta_n)$ is the JT gravity partition function with $n$ asymptotic boundaries with regularized lengths $\beta_1,\dots \beta_n$. On the RHS, the angle braces imply an average of $H$ over an appropriate random matrix ensemble. The contribution of a particular topology to the path integral on the LHS coincides with the contribution at a given order in the ``genus expansion'' of the matrix integral on the RHS. The crucial fact underlying (\ref{Jru}) is the correspondence \cite{eynard2007weil} between Mirzakhani's recursion relation for the volumes of moduli space \cite{mirzakhani2007simple} and Eynard-Orantin ``topological recursion'' \cite{eynard2007invariants} which gives the genus expansion of a Hermitian matrix integral \cite{eynard2004all}.

In this paper, we will generalize this in multiple directions by studying JT gravity on unorientable surfaces, by including a sum over spin structures, and by studying $\mathcal{N} = 1$ supersymmetric JT gravity \cite{Montano:1990ru,Chamseddine:1991fg,Cangemi:1993mj,Forste:2017kwy}. We also allow for the possibility of including a topological field theory that weights crosscaps and/or spin structures in a particular fashion. All together, we will consider three bosonic theories, twelve
theories with fermions  and no  supersymmetry,  and ten supersymmetric theories. We will find evidence that all of these are related to  matrix integrals.

The relevant matrix integrals are characterized by two pieces of data: first, one has to specify the ``spectral curve,'' or equivalently the leading approximation to the density of eigenvalues (generalizing the Wigner semicircle). Second, one has to specify a discrete choice of one of ten different symmetry classes. These classes consist of the three Dyson $\upbeta$-ensembles \cite{Dyson:1962es} and the seven $(\upalpha,\upbeta)$-ensembles of Altland and Zirnbauer \cite{AZ} (three of these were previously discussed by Verbaarschot and Zahed in \cite{Verbaarschot:1993pm,Verbaarschot:1994qf}).\footnote{We will use the notation $\upbeta$ for the Dyson index, reserving $\beta$ for the inverse temperature.} Which symmetry class is appropriate depends on the algebra of the random matrix, the time-reversal operator $\sT$ (if present) and the symmetry $(-1)^\sF$ (if present). The genus expansion for each of these classes of matrix integrals can be analyzed efficiently using the loop equations \cite{migdal1983loop} in their modern form, following \cite{eynard2004all}.

The basic task of this paper is as follows. For each of the various bosonic, fermionic and supersymmetric bulk theories, we determine the two pieces of data described above, thus specifying a candidate matrix integral dual. Then we check that the bulk theory and the matrix integral agree at higher orders (in some cases these checks are limited, and in some cases they are to all orders). In more detail, to match the leading approximation to the density of eigenvalues, one has to compute the bulk partition function on the disk topology. To match the discrete choice of symmetry class, one has to match anomalies in the discrete symmetries $\sT$ and $(-1)^\sF$. Having fixed this data, to check agreement at higher orders, one can compare the JT path integral on some higher topology space with the prediction of the matrix integral derived using the loop equations.

We will find evidence that the three bosonic JT theories are dual to matrix integrals where the Hamiltonian $H$ is drawn from one of the three Dyson $\upbeta$-ensembles. Similarly, the ten supersymmetric JT theories are dual to matrix integrals where $H = Q^2$ and the supercharge $Q$ is drawn from one of the three Dyson ensembles, or from one of the seven Altland-Zirnbauer ensembles. For the twelve theories involving spin structures but no supersymmetry, we show how to reduce the problem to the three bosonic cases. A summary of the different cases is as follows.

{\it Bosonic, orientable only:} This was the case considered in \cite{Saad:2019lba}.  The volumes of moduli space, and thus the JT path integral, are related to a $\upbeta = 2$ (GUE-like) matrix integral.

{\it Bosonic, orientable + unorientable (two subcases):}  In bosonic  JT  gravity,  it is natural to consider a sum over not necessarily orientable manifolds.   In the  context
of holographic duality, this is appropriate if the boundary  theory  has time-reversal  symmetry. 
There are two versions of the sum, depending on whether or not we include a factor of $(-1)^{n_c}$ where $n_c$ is the number of crosscaps. These two choices amount to two different bulk theories, and they correspond to $\upbeta = 1$ (GOE-like) and $\upbeta = 4$ (GSE-like) matrix integrals. As it turns out, the volume of moduli space of a unorientable surface has a logarithmic divergence due to the contribution of small crosscaps. The corresponding matrix integrals also have a divergence in the relevant double-scaled limit, and they predict the correct formula for the moduli space measure associated to a crosscap. However, because of the divergence, we are not able to study arbitrary-genus correlators in these cases.

{\it Fermionic but nonsupersymmetric (twelve subcases)}: In holographic duality, a bulk geometry should have a  spin structure or a less familiar analog of  this known
as a   $\pin^+$ or $\pin^-$ structure if the boundary theory has fermions and, respectively, no time-reversal  symmetry;  time-reversal
symmetry with $\sT^2=(-1)^\sF$ (in other words, $\sT^2$ anticommutes with elementary fermions); or time-reversal symmetry with $\sT^2=1$.
Hence it is natural to  consider JT gravity on an orientable manifold with  a spin structure  or on a not necessarily orientable manifold with  a $\pin^+$ or $\pin^-$ structure.    
  The sum over spin,  $\pin^+$, or $\pin^-$ structures can be enriched with a topological
field theory, for which the number of possible choices is 2, 2, or 8.   We match these $2+2+8=12$ theories  with random matrix  constructions.   New random matrix ensembles
are not needed; it suffices to consider the three Dyson ensembles  together with the assumed global symmetries.

{\it Supersymmetric, orientable only (two subcases):} New matrix ensembles  are needed if we incorporate not just spin structures but $\N=1$ supersymmetry.
In this  case, JT gravity is replaced with JT supergravity, whose partition function  is an integral over the moduli space of super Riemann surfaces.  Restricting to orientable surfaces, there
are two subcases because of the freedom to include a factor of $(-1)^\zeta$, where $\zeta$ is the mod 2 index, weighting odd spin structures with a minus sign. These two choices of bulk theory correspond to two particular symmetry classes of matrix ensemble. In the case where we weight all spin structures equally, the supercharge $Q$ is drawn from a GUE-like ensemble with a leading distribution of eigenvalues that is nonzero everywhere on the real line. The loop equations for such a matrix integral imply that essentially all correlations vanish.  Correspondence with JT supergravity then predicts that the volume of the moduli space of genus $g$ super Riemann surfaces (summed over even and odd
spin structures) vanishes.. For the second bulk theory in which we weight by $(-1)^\zeta$, the right matrix ensemble is one in which $Q$ is drawn from an Altland-Zirnbauer ensemble. The loop equations for this matrix integral predict a nontrivial recursion for the volumes of super moduli space weighted by $(-1)^\zeta$. We prove this directly using a generalization of Mirzakhani's approach in the bosonic case.

{\it Supersymmetric, orientable + unorientable (eight subcases):} In $\N=1$ JT supergravity with time-reversal symmetry  (which necessarily is of $\pin^-$ type), we have to include a sum over pin$^-$ structures. This sum can be weighted by one of eight possible topological field theories, corresponding to the factor $e^{-\mathrm{i}\pi N'\eta/2}$ where $\eta$ is the $\eta$-invariant of Atiyah-Patodi-Singer \cite{Atiyah:1975jf}, and $N'$ is an integer mod 8. These cases can be matched to matrix integrals by comparing the anomalies in time-reversal $\sT$ and $(-1)^\sF$. In six of these eight cases, there is a divergence from the integral over small crosscaps, as in the bosonic theory. Again, the crosscap measure is precisely matched by the corresponding matrix integral, but the divergence prevents us from going to higher orders. In the other two cases, the contribution of the crosscap is finite. Surprisingly, for the two cases where the crosscap is finite, the matrix integral predicts that all higher-order contributions vanish, indicating a cancellation between volumes of moduli spaces
of orientable and unorientable super-surfaces.

We will now discuss the plan of the paper. In \hyperref[sectionTwo]{{\bf section two}} we review the ten standard ensembles of random matrix theory, and match anomalies in $(-1)^\sF$ and $\sT$ symmetry in order to line them up with JT gravity and supergravity theories. A useful tool in this section is the SYK model. Variants of this theory have an approximate random matrix classification, as well as an approximate relation to JT gravity and JT supergravity. Since both of these relationships are approximate, SYK by itself does not imply an exact correspondence between JT gravity and random matrix theory. However, assuming that such a correspondence does exist, the SYK model can be used to match discrete choices on the two sides.

In this section, in addition to matching the anomalies, we show  that for the twelve fermionic but not supersymmetric theories, agreement with random matrix theory follows from agreement in the three bosonic cases. To demonstrate this, we evaluate the sum over spin, pin$^+$ and pin$^-$ structures weighted by appropriate topological field theories.

In \hyperref[sectionThree]{{\bf section three}} we show how to compute the measure on the moduli space of Riemann surfaces or super Riemann surfaces, starting with the path integral of JT gravity or JT supergravity. These theories can be formulated as $BF$ theories with gauge group $\SL(2,\R)$ or $\OSp(1|2)$ \cite{Fukuyama:1985gg,Chamseddine:1989yz,Montano:1990ru}.   In general, the partition function of a $BF$ theory reduces to an integral over a moduli space of flat connections.   On an orientable
two-manifold, the appropriate measure on the moduli space is defined by classical formulas (for example, the Weil-Petersson measure on moduli space in the case $\SL(2,\R)$,
as discussed in \cite{Saad:2019lba}),
but on an unorientable surface, there is an important one-loop correction, which is given by a ratio of determinants known as the ``torsion.'' 
The role of torsion in what is now known as $BF$ theory was originally shown in \cite{Schwarz}.
We discuss general features of the torsion and then show how to compute it for $\SL(2,\R)$ and $\OSp(1|2)$. As an important special case, we compute explicitly the measure factor associated to the size modulus of a crosscap.   It turns out that the resulting measure has been defined previously from  a different point of view
 \cite{Norbury}.

In \hyperref[sec:loopEquations]{{\bf section four}} we analyze the loop equations \cite{migdal1983loop} of random matrix theory. The loop equations are  a general tool for computing the genus expansion of correlation functions in random matrix ensembles. The method was significantly streamlined in \cite{eynard2004all} for $\upbeta = 2$ ensembles. We explain this streamlining and imitate it for the general $\upbeta$-ensembles and the Altland-Zirnbauer $(\upalpha,\upbeta)$ ensembles. Previous work on the $\upbeta$-ensembles includes \cite{Chekhov:2006rq,Chekhov:2010zg,Marchal:2011iu}.

Finally, in \hyperref[sec:JT]{{\bf section five}} we compare the predictions of the loop equations to JT gravity and JT supergravity. In all cases, we are able in lowest
order to match the crosscap measure computed from the torsion plus topological field theory with  a calculation in random matrix theory. In higher orders,
 most of the cases present difficulties,  either due to a divergent volume of moduli space, or due to the difficulty in evaluating 
 the volume of super-moduli space by independent means. However, for the one supersymmetric case where the matrix 
 integral predicts that the volumes are both finite and nonzero (orientable only, $(-1)^\zeta$ weighting), we use the loop equations 
 to predict a simple recursion relation for the volumes of super moduli space, which we prove directly in appendix \ref{supermirz} by adapting 
 Mirzakhani's approach to super Riemann surfaces.  This recursion relation has also been obtained in unpublished work by  Norbury,  who has studied the  
  spectral curve relevant to JT supergravity  from a different point of view  \cite{Norbury2}, following  earlier work on the related ``Bessel''  spectral
curve \cite{Do:2016odu,NorburyDo}.   The derivation in appendix \ref{supermirz} is based on a super McShane identity, which in the prototypical case of a genus 1 surface
with one puncture has been discovered independently by
Y.  Huang, R. Penner, and A. Zeitlin \cite{HPZ}.

  In appendix \ref{volsymp}, we describe a general
formula for volumes of symplectic supermanifolds that makes contact between our statements and some of Norbury's results.    In  appendix \ref{Euler},  we explain
a relationship between certain pairs of matrix ensembles, generalizing  a known relation between matrix ensembles with  orthogonal and symplectic symmetry
\cite{MulaseWaldron}.  
 In appendix \ref{app:Sch}, we compute the Schwarzian  and super Schwarzian path integral on a  disc or trumpet;
these computations are important inputs to the random matrix analysis of JT gravity and supergravity. In appendix \ref{supermirz}, we review Mirzakhani's recursion relation for
volumes of moduli spaces and generalize it to super Riemann surfaces.   We  arrive at the same recursion relation found in  \hyperref[sec:JT]{{ section five}}, thus
confirming the relation between super JT gravity  and the matrix model.  In appendix \ref{app:nonperturbative} we make preliminary comments about nonperturbative effects in  the matrix models, and in appendix \ref{app:minimal}, we compare
to results about the minimal string.

\section{Random Matrix Ensembles And Bulk Topological Theories}\label{sectionTwo}
For the purposes of this paper, a matrix integral means an integral of the form
\be
\int \mathrm{d}M e^{-L\, \Tr\, V(M)},
\ee
where $M$ is an $L\times L$ matrix. Two types of data are needed to specify this integral.\footnote{In our application, there is actually another type of discrete choice: in the application to JT gravity, we will regard the random matrix $M$ as the Hamiltonian $H$ of the boundary theory. In the application to JT supergravity, we will regard the matrix $M$ as the supercharge $Q$ such that $Q^2 = H$.} One is the potential function $V(M)$. In practice, it is more convenient to specify this function implicitly by giving the ``spectral curve'' or equivalently, the leading large $L$ approximation to the density of eigenvalues. In our application, this piece of data will be determined by comparing to the gravity path integral on the disk topology, as we discuss in section \ref{sec:JT} below. 

The second piece of data is the symmetry class of matrices $M$ over which we integrate. In random matrix theory, there are ten standard classes. The purpose of this section is to explain how the choice of symmetry class is related to a discrete choice of topological field theory that one can include in the bulk 2d gravity theory, and to explore some of the consequences.

We will start in section \ref{sec:ensembles} by reviewing the symmetry classes of random matrix ensembles. In section \ref{sec:bosonic} we then discuss the correspondence between topological field theories and random matrix classes for the simplest purely bosonic cases. In section \ref{sec:strategy} we discuss the strategy for the remaining more complicated cases. In section \ref{sec:spinButNoT}, we include $(-1)^\sF$ symmetry but no time-reversal. In section \ref{sec:bothSpinAndT} we include both $(-1)^\sF$ symmetry and time-reversal, and in section \ref{SUSY} we include $\mathcal{N} = 1$ supersymmetry. Much of the complication in sections \ref{sec:spinButNoT} and \ref{sec:bothSpinAndT} has to do with demonstrating a type of reduction: correspondence with JT gravity in these cases follows from a correspondence in the three purely bosonic cases, together with spin and pin structure identities that we derive. No such reduction is possible in the supersymmetric cases of section \ref{SUSY}.

\subsection{Random Matrix Ensembles}\label{sec:ensembles}
\subsubsection{The Ensembles}\label{ensembles}
There are 10 standard ensembles in random matrix theory, and each of them
will play a role in the present paper.  In each ensemble, one considers a class of $L\times L$ matrices $M$, with a symmetry group $G$ that is either $\U(L)$, $\O(L)$, or  $\Sp(L)$, or a product of two
 groups of one of those types. 
 
(i)   In the original application of random matrix theory to nuclear physics,
the random matrix $M$ was interpreted as the Hamiltonian.   In the absence of time-reversal symmetry,\footnote{Unitary symmetries (as opposed to the antiunitary symmetry
of time-reversal) do not play an important role in classifying matrix ensembles. For example, if a $\Bbb Z_2$ symmetry $g$ is assumed, one simply diagonalizes $g$
and describes the Hamiltonian by one of the Wigner-Dyson ensembles in each eigenspace of $g$. We will encounter this situation shortly with $g=(-1)^\sF$.}
 $M$ is simply a random hermitian matrix $M ^i{}_j,\,i,j=1,\cdots,L,$ and the symmetry group that acts on the ensemble of such matrices is $G=\U(L)$, acting by conjugation.   If time-reversal symmetry $\sT$ is assumed, then $\U(L)$ is reduced to the subgroup
that commutes with $\sT$.   This is $\O(L)$ if $\sT^2=1$ and $\Sp(L)$ if $\sT^2=-1$.\footnote{In nuclear physics, $\sT^2=1$ for a nucleus with an even number of nucleons and
$\sT^2=-1$ for an odd number.}  Hermitian matrices that commute with $\sT$ correspond in the case of $\O(L)$ to
real symmetric matrices $M_{ij}=M_{ji}$.   For $\Sp(L)$, a Hamiltonian that commutes with $\sT$ takes the form $M^i{}_j=\veps^{ik}M_{kj}$, where $\veps^{ik}$
is the invariant antisymmetric tensor of $\Sp(L)$ and $M_{kj}=-M_{jk}$ is antisymmetric.    Energy level statistics associated to these three classes
are usually said to be of type  GUE, GOE, or GSE  (where U, O, or S stand for unitary, orthogonal,  or symplectic, respectively, and GE stands for Gaussian ensemble,
though we are interested in a generalization in which the measure is not really Gaussian).

These three ensembles are sometimes referred to as the Dyson ensembles \cite{Dyson:1962es}. Altland and Zirnbauer \cite{AZ,Zirnbauer:1996zz} described 
seven more classes of random matrix ensemble:

(ii) In four cases, the symmetry group is a simple group and $M$ is a second rank tensor of some kind.   If $G=\U(L)$, then $M$ can be either a symmetric second rank
tensor $M_{ij}=M_{ji}$  or an antisymmetric second rank tensor $M_{ij}=-M_{ji}$.   Note that it does not matter if we consider a covariant or contravariant second rank tensor, since if $M$ is of one type, then its adjoint $M^\dagger$ is of the opposite type.   A tensor $M^i{}_j$ of mixed type  is the hermitian
matrix already considered  in (i).     If $G=\O(L)$, $M$ can be an antisymmetric tensor $M_{ij}=-M_{ji}$, and if $G=\Sp(L)$, $M$ can be a symmetric tensor $M_{ij}=M_{ji}$.
Again, the tensors with the opposite symmetry properties were already considered in (i).

(iii)   Finally, there are three cases\footnote{These three ensembles had been discussed earlier
in  \cite{Verbaarschot:1993pm,Verbaarschot:1994qf} in studying the low energy spectrum of the massless Dirac operator in Euclidean
signature coupled to a generic gauge field. Properties of this spectrum have implications for chiral symmetry breaking. In that application,
the two factors of the symmetry group act on fermion modes of positive or negative chirality, respectively, and the random matrix is the 
chiral Dirac operator, which  is a ``bifundamental field'' that
exchanges the two types of mode.   The symmetry group is $\U(L)\times \U(L)$, $\O(L)\times \O(L)$, or $\Sp(L)\times \Sp(L)$ depending on the nature of the fermion
representation (complex, pseudoreal, or real).}
 in which $G$ is a product $\U(L)\times \U(L)$, $\O(L)\times \O(L)$, or $\Sp(L)\times \Sp(L)$, and $M$ is a bifundamental $M_{ij}$,
with one index transforming under the first factor and one under the second.\footnote{The reader might wonder  why it is not possible to mix and match 
groups of different kinds, and to consider, for example, a bifundamental of $\O(L)\times \Sp(L)$.   The problem is that the bifundamental of $\O(L)\times \Sp(L)$ is pseudoreal
rather than real; as a real vector space its dimension is $2L^2$, not $L^2$.   By contrast, the dimension of the group $\O(L)\times \Sp(L)$ grows as $L^2$ for large  $L$,
not $2L^2$.  So a bifundamental of $\O(L)\times \Sp(L)$ has of order $L^2$ real degrees of freedom that cannot  be eliminated by the symmetry.   By contrast,
in the Wigner-Dyson and Altland-Zirnbauer ensembles, 
the number of independent degrees of freedom, modulo  the symmetry, is always of order $L$.}      These  three examples can be generalized to add another integer $\nu$,
so that  
$G$ is a product $\U(L)\times \U(L+\nu)$, $\O(L)\times  \O(L+\nu)$,  or  $\Sp(L)\times \Sp(L+\nu)$.   $M$ is again taken to be a  bifundamental.

\subsubsection{The Integration Measure}\label{measure}

For each of these random matrix ensembles, by the action of the symmetry group, it is possible to put $M$ in a canonical form in terms of 
real ``eigenvalues'' $\lambda_i$.     For some ensembles, the $\lambda_i$ can vary independently, and for others,
they have a two-fold or four-fold degeneracy.    In the three original Wigner-Dyson ensembles, the independent  $\lambda_i$ are real-valued, but
in the seven Altland-Zirnbauer ensembles, they can be chosen to be positive.

In each case, a $G$-invariant integral over $M$ (by which one means an integral over each independent matrix element of $M$) reduces, after putting $M$ in its canonical
form, to an integral over the $\lambda_i$ or more precisely over those $\lambda_i$ that are independent.      For the three Wigner-Dyson ensembles, the $\lambda_i$
are simply the independent eigenvalues of a hermitian matrix (which may obey a time-reversal constraint, as explained earlier) and 
the measure for integration over the $\lambda_i$ is of the form
\be\label{moggo}\mathrm{d}M \rightarrow \prod_{i<j}|\lambda_i -\lambda_j|^\upbeta \prod_k \d \lambda_k \ee
with $\upbeta = 2, 1, $ or 4 for GUE, GOE, or GSE.  We will sometimes refer to these Dyson ensembles as $\upbeta$-ensembles.    The $\upbeta$-dependent factor comes from the volume of the $G$ orbits,
as we describe presently. For  the seven Altland-Zirnbauer ensembles, the measure is 
\be\label{loggoApp}\mathrm{d}M \rightarrow \prod_{i<j}|\lambda_i^2 -\lambda_j^2|^\upbeta \prod_k |\lambda_k|^\upalpha \d \lambda_k, \ee
with various pairs $\upalpha,$ $\upbeta$. We will sometimes refer to these as $(\upalpha,\upbeta)$-ensembles.

Let us first consider the basic GUE ensemble.   For $L=1$, a hermitian matrix $M$ is just a real number $M=\lambda$, and integration over  $M$ is the same as integration
over $\lambda$.   Now suppose that $L=2$ and the canonical form of $M$ is $\diag(\lambda_1,\lambda_2)$.   Such an $M$ commutes with the group $G_0=\U(1)\times \U(1)$
of diagonal matrices.   Let $\g_0$ be the Lie algebra of this subgroup, and let $\g_\perp$ be its orthocomplement, consisting of $2\times 2$ hermitian matrices that
are strictly off-diagonal. These matrices, when commuted with $M$, generate the tangent space to the orbit of $M$ in the space of all hermitan matrices.  Concretely, $\g_\perp$ is 2-dimensional, and the commutator of any $b\in \g_\perp$ with $M$ is proportional to $\lambda_1-\lambda_2$.
So the volume of the orbit is proportional to $(\lambda_1-\lambda_2)^2$.  More generally, for any $L$, let $\g_0$ be the Lie algebra of diagonal matrices and $\g_\perp$
its orthocomplement, consisting of strictly off-diagonal matrices.  Each pair  of eigenvalues $\lambda_i$, $\lambda_j$ is associated to a two-dimensional subspace
of the full Hilbert space $\H$.   Two generators of $\g_\perp$ act in each such subspace and their commutator with $M$ is proportional to $\lambda_i-\lambda_j$.
So the volume of the group orbit is a multiple of $\prod_{i<j}(\lambda_i-\lambda_j)^2$, leading to the measure (\ref{moggo}) with $\upbeta=2$.  

The GOE and GSE cases are similar.  For GOE, the canonical form of the real symmetric matrix $M$ under the action of $G=\O(L)$  is again $\mathrm{diag}(\lambda_1,\cdots,
\lambda_L)$.  
$G$ has  one broken or off-diagonal generator for each  eigenvalue pair $\lambda_i,\lambda_j$.   The commutator of that generator 
with $M$ is proportional to $\lambda_i-\lambda_j$, leading to the measure (\ref{moggo}) with $\upbeta=1$.   For GSE, the eigenvalues are two-fold degenerate because
of Kramers doubling of energy levels for $\sT^2=-1$.  So each independent variable $\lambda_i$ over which one integrates  actually represents
a pair of eigenvalues.  If the independent $\lambda_i$ are all distinct, the subgroup of $G=\Sp(L)$ that commutes with $M$ is $G_0=\Sp(2)^{L/2}$. Let $\g_0$ be the Lie algebra
of $G_0$ and $\g_\perp$ its orthocomplement.   If a pair $\lambda_i,\lambda_j$ becomes
equal,  then a subgroup $\Sp(2)\times \Sp(2)$ of $G_0$ is enhanced to $\Sp(4)$.    As the difference of dimension between $\Sp(4)$ and $\Sp(2)\times \Sp(2)$ is 4,
for each pair $i\not=j$, there are 4 generators of $\g_0$ whose commutator with $M$ is proportional to $\lambda_i-\lambda_j$.   This leads to the measure
(\ref{loggoApp}) with $\upbeta=4$.

The seven Altland-Zirnbauer ensembles can be treated similarly, but there are  two differences.   First, for some of these ensembles there is symmetry enhancement
when a single eigenvalue $\lambda$ vanishes.   If $\upalpha$ is the increase in the dimension of the unbroken symmetry group when $\lambda\to 0$, then 
$\g_0$ has that number of generators whose commutator with $M$ is proportional to $\lambda$, and this leads to the factor  $\prod_k \lambda_k^\upalpha$ in the measure
(\ref{loggoApp}).  
 Second, for the Altland-Zirnbauer
ensembles, the group $G$ can be used to flip separately the signs of the $\lambda_i$, which is why the $\lambda_i$ can be chosen to be all positive.   But
this means that the same symmetry enhancement that occurs for $\lambda_i=\lambda_j$ must also occur for $\lambda_i=-\lambda_j$.   If, therefore, 
$\upbeta$ is the amount by which the dimension of the symmetry group is enhenced if $\lambda_i=\lambda_j$, the measure
will contain a factor $(\lambda_i-\lambda_j)^\upbeta(\lambda_i+\lambda_j)^\upbeta=(\lambda_i^2-\lambda_j^2)^\upbeta$.    So in short $\upalpha$ and $\upbeta$ can
be determined in all cases just by computing the dimensions of symmetry enhancements.  (In all of these ensembles, symmetry enhancement occurs only if $\lambda_i=\pm
\lambda_j$ or $\lambda_i=0$.) 

For  illustration, we will compute $\upalpha$ and $\upbeta$ for one of the Altland-Zirnbauer ensembles  -- the case that $G=\U(L)\times \U(L)$ and $M$
 is an element of the bifundamental representation, which can be viewed as an $L\times L$ matrix with one $\U(L)$ factor acting on each side.  ($M$ corresponds to $C$
 in eqn.~(\ref{blocks}) below, not to the supercharge $Q$.)
 The canonical form of such a matrix under the action of $\U(L)\times \U(L)$  is $\mathrm{diag}(\lambda_1,\lambda_2,\cdots,\lambda_L)$ where
the $\lambda_i$ can be assumed real and nonnegative.    If the $\lambda_i$ are all distinct, the unbroken symmetry is $G_0=\U(1)^L$.   If a single $\lambda_i$
vanishes, an additional $\U(1)$ symmetry is restored, so $\upalpha=1$.   If $\lambda_i=\lambda_j$ for some $i,j$, there is a symmetry enhancement from $\U(1)\times \U(1)$
to $\U(2)$, so $\upbeta=2$.  The other six Altland-Zirnbauer ensembles can be treated similarly.   The results can be found in Tables 3 and 4 of section \ref{revpairs}.

\subsection{The Purely Bosonic Cases}\label{sec:bosonic}
One can define three basic versions of bosonic JT gravity, and these are dual to matrix integrals of the three Dyson types: GUE, GOE, and GSE. The first version is JT gravity on orientable surfaces only, which is dual to a GUE-type matrix integral \cite{Saad:2019lba}.

The remaining two Dyson ensembles (GOE and GSE) describe matrices that commute with a time-reversal symmetry. In AdS/CFT and related dualities, global symmetries of a boundary theory become gauge symmetries in the bulk description, with the restriction that a gauge  transformation should act trivially on the boundary. For the case that the global symmetry is time-reversal $\sT$, the corresponding bulk theory should include $\sT$ as a gauge symmetry, which means that surfaces can be glued together with a reversal of orientation and unorientable  manifolds are allowed. It is natural to guess that GOE and GSE ensembles should be related to bulk theories of this type.

Once the measure for summing over orientable manifolds  has been fixed,\footnote{This involves an arbitrary parameter that controls the genus expansion,
 $e^{-S_0}$ in random matrix theory or
the string coupling constant in string theory.}  there are two versions of the  sum over possibly unorientable manifolds that are
 consistent with the general principles of topological field theory -- these correspond to including or not including a factor of $(-1)^\chi$, where $\chi$ is the Euler characteristic. On a closed orientable manifold, $\chi$ is even and such a factor would have no effect. But on an unorientable manifold, $\chi$ can be odd and the factor is nontrivial.

The GOE and GSE matrix ensembles were shown to differ by precisely such a factor in \cite{MulaseWaldron}, where $\chi$ labels the Euler characteristic of the 't Hooft double-line diagram in the perturbative expansion of the matrix integral. (For more on this point and its analog for Altland-Zirnbauer
ensembles, see appendix \ref{Euler}.)  So it is natural to conjecture that GOE and GSE-like versions of the matrix integral studied in \cite{Saad:2019lba} will be dual to JT gravity on unorientable surfaces, with the $(-1)^\chi$ factor in the GSE case. Of course, these considerations are not limited to JT gravity. 
 In the context of the $c<1$ minimal string, the conjecture analogous to ours was made in \cite{Harris:1990kc,Brezin:1990xr,Brezin:1990dk}.
   In general the amplitudes for
unoriented open strings  with orthogonal
or symplectic groups differ by a factor of $(-1)^\chi$.\footnote{Unoriented open strings give orthogonal
or symplectic  symmetry depending on the sign of the operator $\Omega$ that exchanges the two ends of the string.  Thus the two cases differ by the sign of $\Tr\,\Omega\exp(-\beta H)$, which is the amplitude for a Mobius strip.   On the other hand, the annulus partition function $\Tr\,\exp(-\beta H)$ does not depend on the sign of $\Omega$.   Since
the annulus and Mobius  strip have $\chi$ differing by 1, a factor $(-1)^\chi$ in the path integral measure gives a relative sign between these  two amplitudes.  
 As we explain momentarily, on surfaces with boundary, it is useful to replace $(-1)^\chi$
with $(-1)^{n_c}$, where $n_c$ is the number of crosscaps.
Since  a Mobius strip can be viewed as a disc with a crosscap attached, we have $(-1)^{n_c}=+1$ for an annulus and $(-1)^{n_c}=-1$ for a Mobius strip.} 

In studying the correspondence of matrix integrals and JT gravity, we will be interested in observables that correspond to path integrals over surfaces with boundaries. In this case there are two further points to consider. First, with an odd number of boundaries, $\chi$ would be an odd integer for an orientable surface, so in order to have a topological field theory that is trivial on orientable surfaces, it is convenient to replace the factor $(-1)^\chi$  by $(-1)^{n_c}$ where $n_c$ is the number of crosscaps in the topological decomposition of the surface.   (We will give an introduction to crosscaps below.) 
The  factors $(-1)^\chi$ and $(-1)^{n_c}$  just differ by a minus sign for each boundary component; using $(-1)^{n_c}$ rather than $(-1)^\chi$ will let us avoid
  minus signs in the map between JT gravity and random matrix resolvents.

Second, in the sum over bulk geometries, the orientations of the boundaries should be regarded as gauge-invariant, because bulk gauge transformations are required to act trivially at the boundaries. So one can define gauge-invariant ``orientation Wilson lines'' that measure the change in orientation along a curve connecting the boundaries. These are $\mathbb{Z}_2$-valued quantities, and with $n$ boundaries there are $n-1$ independent Wilson lines, so there are $2^{n-1}$ topologically distinct contributions. (The simplest case, with two boundaries, will be described in detail below, where we refer to the two possible geometries as the ``double trumpet'' and ``twisted double trumpet.'') The implication is that even on an orientable manifold, the partition function of JT gravity with orientation-reversal gauged will differ from that of JT gravity without orientation-reversal gauged. It will be larger by a factor $2^{n-1}$.

In later sections of this paper, we will compare the predictions of JT gravity on unorientable surfaces to GOE and GSE-like matrix integrals. But for the remainder of this section, we will focus on lining up other possible bulk theories with random matrix descriptions.

\subsection{Strategy For The Remaining Cases}\label{sec:strategy}
So far we have discussed the consequences of a symmetry $\sT$. We would also like to explore the consequence of a symmetry $(-1)^\sF$ (the operator that assigns $1$ to bosonic states and $-1$ to fermionic states), and eventually $\mathcal{N} = 1$ supersymmetry. In discussing the relevant matrix ensembles and their bulk duals, an essential role will be played by anomalies in the realization of the global symmetries $(-1)^\sF$ and $\sT$. These anomalies determine both the topological field theory in the bulk and the random matrix theory class on the boundary.

In principle, the connection between random matrix classes and bulk topological field theories can be made abstractly, using these anomalies. However, we will find it convenient to use the SYK model as a concrete system that connects the concepts together. The SYK model is a quantum mechanical system of $N$ Majorana fermions, and it will be useful for three reasons. First, although is not a random matrix theory in the sense we mean it in this paper, it is ``close enough,'' in that the correlations of nearby energy eigenvalues have a well-defined random matrix classification \cite{You:2016ldz,Li:2017hdt,Kanazawa:2017dpd,Sun:2019yqp}. Second, at low energies it has an approximate bulk description that includes JT gravity or super JT gravity. And finally, it is flexible enough that its variants display all of the needed anomalies and exhaust the Dyson and Altland-Zirnbauer classification of random matrix theory.


The necessary anomalies have to do with the global symmetries $(-1)^\sF$ and time-reversal $\sT$ (if present), and they depend on the number of Majorana fermions $N$. If $N$ is odd, there is an anomaly in a narrow sense -- a violation of a classical symmetry --  while if $N$
is even, the expected symmetries are present, but they do not satisfy  algebraic relations (such as $\sT^2=1$) that would be expected classically.

The same anomalies are relevant in many different problems; examples  without time-reversal symmetry  include 
 the Kitaev chain of Majorana fermions
\cite{Kitaev:2001},  the Ising model \cite{Kapustin:2014gua}, and intersection theory on the  moduli space of Riemann surfaces with boundary 
\cite{Pandharipande:2014qya,Dijkgraaf:2018vnm}.
The extension to include time-reversal also has various applications; for instance, see 
 \cite{Kitaevinteractions} for a time-reversal invariant version of the Kitaev chain with $\sT^2=1$ and \cite{PWW} for a version with $\sT^2=(-1)^\sF$.  At a more   abstract
level, the  mod 2  anomaly
when $(-1)^\sF$ is the only symmetry and the mod 8 anomaly that arises when one includes time-reversal with $\sT^2=1$ are  related to the properties of Dirac operators
in different dimensions and to  the mod 2 periodicity
of complex K-theory  and the mod 8 periodicity of real K-theory \cite{ABS}.  In that context, the applications are too far-flung  to be summarized here.    

These symmetries and their anomalies have implications for the bulk theory. As described above, $\sT$ as a global symmetry of the boundary theory (and gauge symmetry of the bulk) means that one must sum over 
unorientable as well as orientable manifolds; $(-1)^\sF$ means that one must sum over spin structures.   An anomaly in symmetries of a boundary theory must somehow be reflected in the couplings of the bulk dual.    This in general happens
as follows.   Suppose that the boundary theory is formulated on a manifold $X$.   In the bulk dual description, one sums over manifolds $Y$
of one dimension higher.   The theory on $Y$ will then have couplings that are well-defined if $Y$ has no boundary but that, if $Y$ has a boundary
$X$, are anomalous in a way that matches the anomaly of the original theory on $X$.  For example, if the theory on $X$ has continuous global
symmetries that have 't Hooft anomalies (which would obstruct gauging those symmetries), then the bulk theory on $Y$ has gauge fields
with Chern-Simons couplings that are not gauge-invariant on a manifold with boundary.   The analog of this for 
discrete symmetries -- such as we consider here -- is that the bulk description must include a topological field theory that captures the anomaly.
This is a topological field theory that is well-defined on a manifold $Y$ without boundary, but anomalous when $Y$ has a boundary.\footnote{The 
bulk factor $(-1)^{n_c}$ considered in 
section \ref{sec:bosonic} is part of the definition of a purely bosonic topological field theory and could be treated in this framework,
though it appears to be hard to do this in an enlightening way,.}

In what follows, we explain how this works for the discrete symmetries $(-1)^\sF$ and $\sT$ of the SYK model. We will then use the random matrix classification of SYK to relate bulk topological field theories to random matrix symmetry classes.

In discussing anomalies of $(-1)^\sF$ and $\sT$ without supersymmetry, we will encounter twelve cases, but we will not obtain essentially new random matrix classes. In the boundary theory, one finds combinations of the GOE, GUE, GSE ensembles \cite{You:2016ldz,Li:2017hdt,Kanazawa:2017dpd}. In the bulk, we will have to analyze sums over spin and pin structures with weighting given by the topological field theory.   However, the JT path integral does not depend on the spin structure, so we get a result
that is simply the product of a bosonic JT path integral with a factor that comes from summing over spin or pin structures.
By evaluating this factor, one reduces the duality to the three purely bosonic cases described above. 

But  in the ten distinct cases with $\mathcal{N} = 1$ supersymmetry, one finds new random matrix classes \cite{Li:2017hdt,Kanazawa:2017dpd,Sun:2019yqp}, which exhaust the full Altland-Zirnbauer classification.   Moreover, JT supergravity has fermionic fields, so its path integral depends on the spin or pin structure.
Hence,  in the bulk JT supergravity, the sum over spin or pin structures does not just give a simple overall factor, but  an essential and nontrivial part of the bulk theory.

\subsection{Including $(-1)^\sF$ But Not $\sT$}\label{sec:spinButNoT}

\subsubsection{The SYK Model For Even And Odd $N$}\label{evenodd}

To begin, we assume no symmetry except $(-1)^\sF$.   
Let us start with  the SYK model with $N$ Majorana fermions $\psi_1,\cdots,\psi_N$.
The action is
\be\label{uvu}I =\int \d t\left(\frac{\i}{2}\sum_k \psi_k  \frac{\d\psi_k}{\d t}-\i^{q/2}\sum_{i_1\dots i_q}j_{i_1i_2\cdots i_q}\psi_{i_1}\psi_{i_2}\cdots \psi_{i_q} \right)\ee
and the Hamiltonian is $H =\i^{q/2}j_{i_1i_2\cdots i_q}\psi_{i_1}\psi_{i_2}\cdots \psi_{i_q}$.
Classically, there is, for all $N$, a symmetry $(-1)^\sF$ that acts by $\psi_k\to -\psi_k$.  If $q$ is not a multiple of 4, that is generically the only symmetry.

\noindent {\bf If $N$ is even}, then upon quantization we get a Clifford
algebra of rank $N$.
The Clifford algebra  has an irreducible representation in a Hilbert space $\H$ of dimension $2^{N/2}$.     
The symmetry operator $(-1)^\sF$ that anticommutes with the elementary fermions is a multiple of the product
\be\label{ponn}\psi_1\psi_2\cdots \psi_N \ee of all $N$ elementary fermion fields; this operator anticommutes with the elementary fermions.
The SYK path integral,\footnote{We generally consider path integrals without operator insertions.} on a 
circle of circumference $\beta$, computes the partition function $\Tr_\H\, e^{-\beta H}$
if the fermions are antiperiodic in going around the circle, or $\Tr_\H\,(-1)^\sF e^{-\beta H}$ if they are periodic.     Antiperiodic or periodic
fermions correspond to what we will call the Neveu-Schwarz (NS) or Ramond (R) spin structure.  In general, both are nonzero when $N$ is even.  

A special case is that if $H=0$, the path integral in the NS sector computes the dimension of Hilbert space:
\be\label{udfu} \Tr_\H \, 1 = 2^{N/2}. \ee
Since this is the result with $N$ Majorana fermions, the corresponding path integral with a single Majorana fermion equals $\sqrt 2$.

\noindent {\bf If $N$ is odd},  the product in eqn.~(\ref{ponn}) commutes (rather than anticommuting) with the $\psi_k$,
so it is a $c$-number in an irreducible representation of the algebra.   The operation $\psi_k\to -\psi_k$ changes the sign of this
$c$-number, so for odd $N$ the Clifford algebra has two inequivalent irreducible representations, differing by $\psi_k\to-\psi_k$.
Each of them has dimension $2^{(N-1)/2}$.   Pick one of these representations and call it $\H$.   It does not matter which one we pick
since the Hamiltonian $H$, being an even function of the $\psi_k$, is invariant under $\psi_k\to -\psi_k$.

In the NS sector, the path integral of the SYK model on a circle of circumference $\beta$ computes
\be\label{woggoSYK}\sqrt 2 \Tr_\H \,e^{-\beta H}. \ee  To see that the factor of $\sqrt 2$ is necessary, consider the special case
that $H=0$.  The path integral, with a factor $\sqrt 2$ for each of $N$ Majorana fermions, is then $2^{N/2}$.   On the other
hand, since the dimension of $\H$ is $2^{(N-1)/2}$, we have $\Tr_\H 1 =2^{(N-1)/2}$, and therefore an extra factor of $\sqrt 2$
is needed to match the SYK path integral. This factor means that for odd $N$, the SYK path integral does not have a natural interpretation
as a trace in a Hilbert space.  But it is still a well-defined path integral and it makes sense to ask what sort of bulk dual description it would have.

Now let us consider the Ramond sector.    For odd $N$, the path integral of the SYK model in the Ramond sector (with no operator insertions)
 is actually identically zero.
To see this, first note that if $H=0$, so that the action (\ref{uvu}) consists only of the kinetic energy,  then each of the $\psi_k$ has  a zero-mode
in the Ramond sector, so in all there are an odd number of zero-modes.      Now including the Hamiltonian, a term that is proportional to $H^r$
for some $r$ has an insertion of $qr$ fermions, which is an even number since $q$ is even.   An even number of these fermions can be
paired up by propagators, and the remaining ones -- also an even number -- can be used to soak up zero-modes.   Since we started with an
odd number of fermion zero-modes, there is always an odd number left over.   In particular, we can never soak up all of the zero-modes, and 
therefore the Ramond sector path integral of the odd $N$ SYK model is identically zero.

By contrast, the Ramond sector path integral of the model with an insertion of an odd number of elementary fermion operators is generically
nonzero, since with the help of such an insertion (along with the Hamiltonian) all of the zero-modes can be soaked up.   Thus formally
the Ramond sector path integral computes a generically nonzero expectation value 
 $\langle \psi_k\rangle$ (where $\psi_k$ is  any   one of the elementary
fermion fields).   This certainly qualifies as an anomaly, since it explicitly violates a hypothetical symmetry $(-1)^\sF$ that is supposed to
act by $\psi_k\to -\psi_k$.  This is related to the fact that  for odd $N$, the $(-1)^\sF$ symmetry is lost at the quantum level, since $\psi_k\to-\psi_k$
exchanges two different representations of the Clifford algebra. 

\noindent {\bf A summary} of the above discussion is as follows. If we label the path integral of SYK with NS and R boundary conditions as $Z_{\NS}(\beta)$ and $Z_{\Ra}(\beta)$, then the path integrals are related to traces in the Hilbert space as
\be
Z_{\NS}(\beta) = \begin{cases} \Tr\, e^{-\beta H} & N \text{ even}\\ \sqrt{2}\,\Tr\, e^{-\beta H}& N\text{ odd}\end{cases} \hspace{40pt} Z_\Ra(\beta) = \begin{cases} \Tr\, (-1)^\sF e^{-\beta H} & N \text{ even}\\ 0& N\text{ odd}.\end{cases}\label{ZHilbert}
\ee
The trace is in a Hilbert space of dimension $2^{N/2}$ for even $N$, and dimension $2^{(N-1)/2}$ for odd $N$.

\subsubsection{Initial Random Matrix Considerations With $(-1)^\sF$ Symmetry}
Let us now work out the relevant random matrix symmetry classes. For even $N$, we can choose a basis for the Hilbert space so that $(-1)^\sF = \begin{pmatrix} I & 0 \cr 0 & -I\end{pmatrix}$. Then the only symmetry constraint is that $H$ should commute with this matrix. A maximally random matrix consistent with the symmetry would be of type\footnote{A subtlety here is that if $q = 2$ mod 4, then the SYK Hamiltonian actually anticommutes with a time-reversal operator. This affects the random matrix classification, but not at low energy (since a symmetry that anticommutes with $H$ exchanges low and high energies), and it will not have a dual statement in JT gravity. However, to make the discussion above more precise for all energies, one can let $H$ be a linear combination of $q = 0$ mod 4 and $q = 2$ mod 4 terms, so that $H$ does not
commute or anticommute with any version of $\sT$.}
\be
H = \left(\begin{array}{cc}\text{GUE}_1 & 0 \\ 0 & \text{GUE}_2\end{array}\right).\label{HGUE2}
\ee
Here the subscripts on the two blocks indicate two independent GUE matrices. For odd $N$, there is no $(-1)^\sF$ symmetry, so the random matrix class is simply an unconstrained Hermitian matrix,
\be
H = 
\text{GUE}.\label{HGUE}
\ee
We emphasize that although these random matrix ensembles have the correct symmetry properties, they do not describe the SYK model exactly, since SYK is not quite a random matrix theory. However, SYK was useful in motivating (\ref{ZHilbert}) (\ref{HGUE2}) and (\ref{HGUE}), and we will see that random matrix ensembles of these types are precisely dual to appropriate refinements of JT gravity.

\begin{figure}
 \begin{center}
   \includegraphics[width=3in]{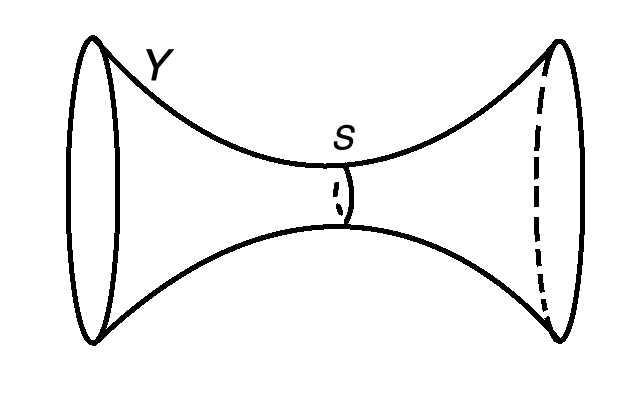}
 \end{center}
\caption{\small A double trumpet $Y$.  At its ``center'' is a closed geodesic $S$. \label{DT}}
\end{figure}
  We will approach the
relation to JT gravity in the following way. Correlation functions of observables $Z_{\NS}(\beta)$ and $Z_{\Ra}(\beta)$ in random matrix theory will be related to JT gravity path integrals on manifolds with boundaries with NS and R spin structures, and regularized lengths proportional to $\beta$. As a first case, consider the expectation value of the quantity $Z_{\NS}(\beta)$. This is large in random matrix theory, 
in the sense that it is of order $L$, the rank of the matrix.\footnote{The types of matrix integral that are actually dual to JT gravity are ``double-scaled'' matrix integrals, in which the analog of $L$ is the parameter $e^{S_0}$ that specifies the density of eigenvalues.}  The leading contribution to $\langle Z_{\NS}(\beta)\rangle$ is dual to a JT gravity path integral on  a spacetime with disc topology.   But $Z_{\NS}(\beta)$ also fluctuates in random matrix
theory.   The fluctuations can be measured by the connected two-point function
\be\label{wondo} \left\langle Z_{\NS}(\beta)^2\right\rangle_c=\left\langle Z_{\NS}(\beta)^2\right\rangle-\left\langle Z_{\NS}(\beta)\right\rangle^2.\ee
In JT gravity, to compute this two-point function, we need a spacetime whose boundary consists of two circles, on each of which
the holographic description would involve a path integral that generates a factor of $\Tr\,\exp(-\beta H)$.   But to get a contribution
to the connected two-point function, the spacetime should be connected.   The simplest possible topology is
the ``double trumpet,'' topologically an annulus (fig.~\ref{DT}), and this makes the dominant contribution to the connected correlation function when $L$ is large.

Now let us consider $Z_{\Ra}(\beta)$.  Its expectation value in JT gravity (without time-reversal symmetry) vanishes, since there is no
oriented two-manifold whose boundary consists of a single circle with R spin structure.    In random matrix theory, one interprets this as follows.
For odd $N$, the R sector path integral is identically zero, as we have seen.
For even $N$, this path integral is not identically zero, but it is zero on the average in random matrix theory: the random matrix class for $H$ in (\ref{HGUE2}) consists of two independent GUE matrices acting on the states of $(-1)^\sF=1$ or $(-1)^\sF=-1$, and there is on average a cancellation between their contributions to $\Tr \,(-1)^\sF e^{-\beta H}$.

Now let us consider in JT gravity the fluctuations of $Z_{\Ra}(\beta)$ around its mean value of zero.
These fluctuations can be measured by the two-point function $\langle Z_{\Ra}(\beta)^2\rangle$.   This two-point function
can receive contributions from connected oriented two-manifolds whose boundary consists of two circles and whose spin structure is of R type on
each boundary.   The simplest possible choice is again the double trumpet (with a different spin structure).    The random matrix 
theory prediction is actually that for even $N$
\be\label{wobbo}\left \langle Z_{\Ra}(\beta)^2\right\rangle= \left\langle Z_{\NS}(\beta)^2\right\rangle_c \hspace{40pt} (N \text{ even}).\ee
To compute in random matrix theory the product of traces on the right or left of eqn.~(\ref{wobbo}), we have to sum over
contributions of pairs of states which may have either eigenvalue 1 or $-1$ of $(-1)^\sF$.   But pairs of states with opposite eigenvalues
make no contribution to the connected correlation functions on either the left or the right of this equation, 
because in (\ref{HGUE2}), the two blocks are considered statistically independent.   On the other hand, pairs of states with the same eigenvalue of $(-1)^\sF$ make equal contributions on the left and right hand side of eqn.~(\ref{wobbo}). For odd $N$, however, $\Tr\,(-1)^\sF e^{-\beta H}$ is identically zero, so the expectation value of its square is also zero, 
\be\label{nobbob}  \left\langle Z_{\Ra}(\beta)^2\right\rangle=0\hspace{40pt} (N\text{ odd}).\ee

By similar reasoning, for any $N$, the cross correlator
\be\label{obbob}\left\langle Z_{\NS}(\beta)Z_{\Ra}(\beta)\right\rangle\ee
vanishes.   It vanishes in random matrix for even $N$   because the states with $(-1)^\sF$ equal to $1$ or $-1$ are statistically independent and contribute
with opposite signs, and for odd $N$ because $Z_\Ra(\beta)$ is identically 0, without fluctuations.   
It vanishes in JT gravity because any oriented two-manifold with boundary has an even number of boundary components of R type.

Our initial goal will be to understand what topological refinement should be added to JT gravity to reproduce the predictions of eqns.~(\ref{wobbo}) and (\ref{nobbob}).

\subsubsection{Bulk Description Of $(-1)^\sF$ Symmetry And Its Anomaly}\label{bulk}

Let $X$ be a compact 1-manifold (a circle or a disjoint union of circles) on which we want to study an SYK model
or a related random matrix theory.    $X$ is endowed with a spin structure (of either NS or R type on each connected component
of $X$) since the SYK model has fermions.    In a bulk dual description (assuming no $\sT$ symmetry), we will sum in a JT-like model over oriented spin manifolds
$Y$ of boundary $X$, such that the spin structure of $Y$ restricts on $X$ to the spin structure of $X$.    Given such a $Y$, we sum
over equivalence classes of spin structures on $Y$, keeping fixed that of $X$.  (The  equivalence relation in this sum involves a subtlety
that is explained later.)   Such a sum represents the bulk dual of an SYK or random matrix
calculation with a particular spin structure on $X$.

We want to find two versions of the sum over spin structures on $X$ that will match the two cases of even and odd $N$.
It will be useful to know the following.   If $Y$ is a two-dimensional compact oriented spin manifold without boundary, then one can consider
the Dirac equation $\slashed{D} \lambda=0$, for a spinor field $\lambda$ on $Y$ of (say) positive chirality.    Let $\zeta$ be the number of
zero-modes of this equation, mod 2.  It is a topological invariant, in the sense that it depends on the spin structure of $Y$ but not on the
choice of a Riemannian metric on $Y$.    In the terminology of Atiyah and Singer, $\zeta$ is  the mod 2 index\footnote{An important
point is that $\zeta$ is not the reduction mod 2 of an ordinary index or of any integer-valued topological invariant.}  of the Dirac operator on $Y$.
For an introduction to the mod 2 index and in particular an explanation of why it is a topological invariant, see for instance section 3.2 of
\cite{Witten:2015aba}.  

On an orientable two-manifold without boundary, there are two ways to sum over spin structures satisfying the general conditions of topological field
theory.    One can sum over spin structures assigning the same weight to each spin structure, or one can sum over spin structures
with relative weights $(-1)^\zeta$.    The reason that it makes sense to include a factor $(-1)^\zeta$ is that this factor is local in the relevant
sense, though to understand that, one has to take a somewhat abstract view of what locality means.   Usually in field theory, one has
some fields $\Phi$ and some Lagrangian density $L(\Phi)$; the integrand of the path integral contains a factor 
$\Upsilon= \exp(-\int_Y \d^Dx \sqrt g L(\Phi))$.   This is local in the sense that if one varies $\Phi$ 
only in a small region $V\subset Y$, then $\Upsilon$ changes by a factor that 
only depends on what is happening inside $V$.   Though this is far from obvious,  $(-1)^\zeta$ is local in the same sense.   The way we will use that is as follows.
Suppose that we are given two spin structures on $Y$ that coincide outside of $V$.   Let $\zeta $ and $\zeta'$ be the mod 2 indices for
these spin structures.   Then the ratio $(-1)^\zeta/(-1)^{\zeta'}$ depends only on $V$ (and the restriction of the two spin structures to $V$).
It does not depend on anything that is happening outside $V$.   This enables one to modify $Y$ away from $V$ in a way that
makes computations simpler.

So on a two-manifold $Y$ without boundary, a sum over spin structures with or without a factor of $(-1)^\zeta$  gives a topological field theory.
However, our definition of $\zeta$ only made sense if $Y$ has no boundary.   (If $Y$ has a boundary, then to make sense of the equation for a fermion zero-mode,
one needs a boundary condition that mixes the two chiralities, so it is not possible to consider fermions of just one chirality.)   If one tries to define $(-1)^\zeta$ on a two-manifold with boundary, then one runs into
an anomaly.  As is explained,
for example, in section 3 of \cite{Dijkgraaf:2018vnm}, this anomaly  is precisely the anomaly of a system with an odd number of Majorana fermions on the boundary.     Therefore, we expect that a simple sum over spin structures can be dual to an SYK-like model
with even $N$, but that such a model with odd $N$ requires including the factor $(-1)^\zeta$.

The only example in which we will actually need to calculate $\zeta$ is the following.    Let $Y$ be a two-torus, obtained
by dividing the real $xy$ plane by the identifications $x\cong x+1$, $y\cong y+1$.    On $Y$, there are four 
spin structures, conveniently represented
as $(\pm,\pm)$ where the two signs appear in the periodicity or antiperiodicity relations \be\label{moglo}\lambda(x+1)=\pm \lambda(x),~~~\lambda(y+1)=\pm \lambda(y)\ee that a spinor field on $Y$ should
obey.    Thus the $(+,+)$ spin structure means that $\lambda$ is periodic in both directions, and in the other cases, $\lambda$ is antiperiodic in one or
both directions.    Setting $z=x+\i y$, the chiral Dirac equation is
\be\label{zondo} \frac{\partial}{\partial\bar z}\lambda=0. \ee
In the case of the $(+,+)$ spin structure, there is a 1-dimensional space of solutions, with constant $\lambda$; for the other cases, there are no
solutions.   So $\zeta=1$ for the $(+,+)$ spin structure, and $\zeta=0$ in the other cases.   

We are now ready to analyze the contribution of the double trumpet in eqns. (\ref{wobbo}) and (\ref{nobbob}).   Once one understands the double trumpet,
it is not difficult to analyze any $Y$ in a similar way.     First of all, the spin structures on the two boundary components of the double trumpet are always of the same
type.    This is because the closed geodesic $S$ at the center of the double trumpet (fig.~\ref{DT}) can be smoothly moved to the left boundary or the right boundary, so the spin
structure is of the same type on each boundary as it is on $S$.   More generally, it is true because in general, for any $Y$, the number of R type boundary components of $Y$
is always even.    

Now having fixed the spin structure on the two boundary components of $Y$, how many spin structures are there on $Y$, up to equivalence?   This is a subtle question
and we have to specify that we consider two spin structures on $Y$ to be ``equivalent'' if they are equivalent under a redefinition of the fermion field that is trivial on the boundary
of $Y$.   The reason for the redefinition to be trivial on the boundary
 is that in general, in holographic duality, a ``gauge equvalence'' in a bulk description is supposed to involve a gauge transformation that is trivial
along the boundary; the restriction of a gauge transformation to the boundary behaves as a global symmetry.

\begin{figure}
 \begin{center}
   \includegraphics[width=3in]{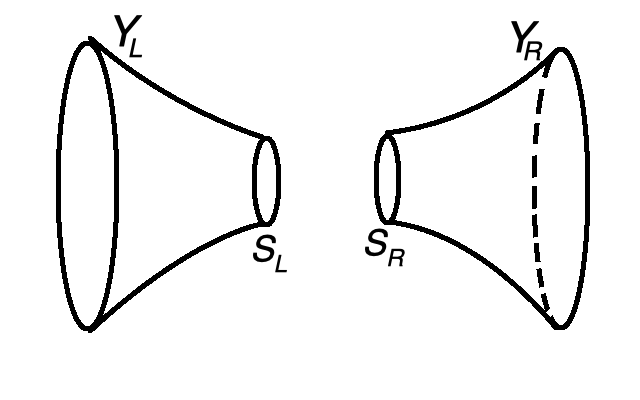}
 \end{center}
\caption{\small Two trumpets $Y_L$ and $Y_R$.   By gluing their ``inner boundaries'' $S_L$ and $S_R$, one can build a ``double trumpet.''   \label{SDT}}
\end{figure}
Given this, we can count the spin structures on $Y$ that have a given restriction to the boundary as follows.   Let $Y_L$ and $Y_R$ be ``trumpets''  (fig.~\ref{SDT}) that can be
glued together along ``inner boundaries'' $S_L$ and $S_R$ to build the ``double trumpet'' $Y$.     The spin structure on $Y_L$ or $Y_R$, once it is specified on
the outer boundary,  is unique 
up to a unique equivalence,\footnote{Here an equivalence is allowed to involve a field redefinition that is nontrivial along $S_L$ or $S_R$, but not on the outer boundary.} since any closed path
in $Y_L$ or in $Y_R$ can be deformed to the outer boundary.    However, when we glue $S_L$ to $S_R$ to build $Y$, a sign choice comes in; if $\lambda_L$
and $\lambda_R$ are fermion fields on $S_L$ and $S_R$, we could identify $ \lambda_L$ with $ \lambda_R$ or with $-\lambda_R$.    This sign can be eliminated if we are allowed to change the
sign of the fermion field on all of $Y_R$ (or $Y_L$), but this is not allowed since a gauge equivalence is supposed to be trivial on the outer boundaries.

Now we can explain eqns. (\ref{wobbo}) and (\ref{nobbob}) in the context of JT gravity.   First of all, if $N$ is even, then we are simply summing over spin structures on $Y$
(restricted to agree with a given one on the boundary) without any signs.   Each spin structure makes the same contribution, given by the double trumpet path integral of JT
gravity, which is not sensitive to the spin structure.   The spin structures that contribute to the left or right hand side of  eqn.~(\ref{wobbo}) are different, 
but there are 2 of them in each case.   So the left or right hand side of eqn.~(\ref{wobbo}) is equal to 2 times the double trumpet path integral of JT gravity.  

\begin{figure}
 \begin{center}
   \includegraphics[width=3in]{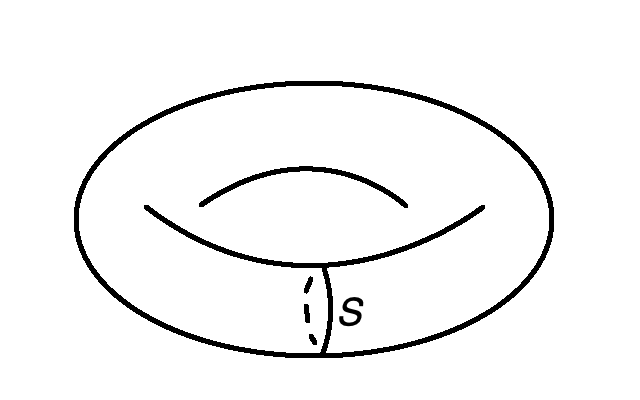}
 \end{center}
\caption{\small A small neighborhood of $S$ has been cut out of the double trumpet and glued into a two-torus.  \label{Regluing}}
\end{figure}
For odd $N$, we have to deal with the factor of $(-1)^\zeta$.   For this, first note that the two spin structures on $Y$ (once the spin structures on the boundaries are given)
are equivalent away from a small neighborhood of $S$, the geodesic at the center of the double trumpet (fig.~\ref{DT}).   They differ by a minus sign that a fermion gets
when it is parallel transported across $S$.   Using the locality of $(-1)^\zeta$, as explained above, we can cut a small neighborhood of $S$ out of the double trumpet
and glue it into, for example, a two-torus (fig.~\ref{Regluing}).   (Alternatively, and by a similar reasoning, one could glue together the two outer boundaries of the double
trumpet, again arriving at a two-torus. By locality, that operation does not affect the comparison of the values
of $(-1)^\zeta$ for a pair of spin structures that differ only near $S$.)

Now recall that spin structures on a two-torus are labeled as $(\pm,\pm)$, where here we can think of the first sign as the sign that a fermion gets when parallel transported
around $S$, and the second sign as the sign that a fermion gets when it is parallel transported in a complementary direction around the torus, on a closed path that intersects
$S$ once.  Moreover, $\zeta=1$ for the $(+,+)$ spin structure and otherwise $\zeta=0$.   The first sign is  $-$ or $+$ depending on whether the spin structures on the boundary
components are of NS or R type (as noted above, the spin structure on $S$ in the original double trumpet is of the same type as the spin structure on either of the two boundaries).  
Suppose that the boundary spin structure is of NS type.    Then the spin structure on the torus after the regluing of fig.~\ref{Regluing} is of type $(-,\pm)$, with some choice of the
second sign.   Regardless of the second sign, these spin structures have $\zeta=0$ and so their contribution to the double trumpet path integral is insensitive to whether $N$
is even or odd.   Hence $\left\langle Z_\NS(\beta)^2\right\rangle$  is not sensitive to whether $N$ is even or odd, in keeping with what one would expect from random matrix theory.

  Suppose on the other hand that the boundary spin structures are of R type.   Then the spin structures on the two-torus after regluing are of type $(+,\pm)$, with some choice of sign.   One of them
  has $(-1)^\zeta=1$ and one has $(-1)^\zeta=-1$, so their contributions cancel.   Thus when $N$ is odd, $\left\langle Z_\Ra(\beta)^2\right\rangle=0$  in JT theory
  plus topological field theory, accounting for the claim in eqn.~(\ref{nobbob}).

If one replaces the double trumpet with any connected two-manifold $Y$ with two or more boundary components including some of R type, 
one can make a similar argument, using for $S$
any closed curve in the interior of $Y$ that is homologous to a boundary of R type.  Pairs of spin structures that differ by a minus sign that a fermion
acquires when transported across $S$ have opposite values of $\zeta$ and
make canceling contributions.  Thus for odd $N$, in JT gravity plus topological field theory,  $\Tr\,(-1)^\sF e^{-\beta H}=0$, in the sense that all of its correlation
functions vanish.
 
\subsubsection{Reduction Of Cases With $(-1)^\sF$ Symmetry To GUE}\label{sec:reduction1}
We would now like to consider a general correlation function, of the type
\be\label{unconnected}
\langle Z_{\Ra}(\beta_1)\dots Z_{\Ra}(\beta_{n_{\Ra}})Z_{\NS}(\beta_{n_{\Ra}+1})\dots Z_{\NS}(\beta_n)\rangle.
\ee
It is somewhat simpler to work with a connected version of this correlator, by subtracting off  all lower order correlation functions. As we will discuss  in detail in section \ref{sec:loopEquations}, in a matrix integral such connected correlations have a ``genus'' expansion 
\be\label{genusExpReduction}
\langle Z_{\Ra}(\beta_1)\dots Z_{\Ra}(\beta_{n_{\Ra}})Z_{\NS}(\beta_{n_{\Ra}+1})\dots Z_{\NS}(\beta_n)\rangle_c \simeq \sum_{g = 0}^\infty \frac{Z_{g,n,n_{\Ra}}(\beta_1,\dots,\beta_n)}{(e^{S_0})^{2g+n-2}}.
\ee
For an ordinary matrix integral, the expansion parameter would be $L$, but for the double-scaled matrix integrals that are relevant to JT gravity, it is a parameter that we  refer to as $e^{S_0}$, proportional to the density of eigenvalues. 

In random matrix theory, this expansion follows from 't Hooft's topological analysis \cite{tHooft:1973alw} of matrix perturbation theory \cite{Brezin:1977sv}. In JT gravity, this expansion corresponds to the sum over different topologies of connected manifold with $n$ boundaries. The weighting by powers of $e^{S_0}$ comes from the contribution of a term $-S_0\chi$ in the action of JT gravity.   The ``genus'' $g$ of a two-manifold is defined by $2-2g-n=\chi$,  so that $g$ is a non-negative integer
in the case of an orientable surface, but in general may be a non-negative integer or half-integer.

For the ensembles defined by (\ref{ZHilbert}), (\ref{HGUE2}), (\ref{HGUE}), we would like to relate the expansion coefficients $Z_{g,n,n_{\Ra}}$ to similar expansion coefficients $Z_{g,n}$ in the ordinary GUE ensemble. One can do this as follows. First, in the case of even $N$, $H$ is a direct sum of two independent matrices of GUE type, and connected correlation functions will be a sum of connected correlators for the two independent blocks. Including a factor of $(-1)^{n_{\Ra}}$ for the block corresponding to fermionic states, one expects $Z_{g,n,n_{\Ra}} = (1 + (-1)^{n_{\Ra}})$ times the GUE answer. Similarly, in the case of odd $N$, (\ref{ZHilbert}) and (\ref{HGUE}) imply that $Z_{g,n,n_{\Ra}}$ should be $(\sqrt{2})^n$ times the GUE answer if $n_{\Ra} = 0$, and zero otherwise. 

By rescaling the expansion coefficient $e^{S_0}$, we can multiply $Z_{g,n,n_{\Ra}}$ by a factor $(\text{const.})^{2g+n-2}$. It is convenient to choose the constant so that the ``disk'' amplitude $Z_{0,1,0}$ is equal to the standard GUE answer. This modifies the naive predictions described above to
\be
Z_{g,n,n_{\Ra}} = 2^{2g+n-2}(1 + (-1)^{n_{\Ra}})Z^{\text{GUE}}_{g,n} \hspace{40pt} (N \text{ even})
\ee
\be
Z_{g,n,n_{\Ra}} = \begin{cases}2^{g+n-1}Z^{\text{GUE}}_{g,n} & n_{\Ra} = 0\\0 & \text{else} \end{cases} \hspace{40pt} (N\text{ odd}).
\ee
This is the random matrix theory prediction. We would like to match it in the two variants of JT gravity that were described above, starting with the statement that JT gravity without spin structures matches $Z_{g,n}^{\text{GUE}}$ \cite{Saad:2019lba}.

In fact, these relations follow from two identities for the sum over spin structures:
\begin{align}\label{spinSumsNoT}
\sum_{\text{spin}}1 = 2^{2g+n-2}(1 + (-1)^{n_{\Ra}})\\
\sum_{\text{spin}}(-1)^{\zeta} = \begin{cases}2^{g+n-1} & n_{\Ra} = 0\\0 & \text{else}. \end{cases}\label{ywq}
\end{align}
In these expressions, the sum is over spin structures on an orientable surface with genus $g$ and $n\ge 1$ boundaries, of which $n_{\Ra}$ are of R type and the rest are NS.

For $n=0$, there are actually twice as many spin structures as these formulas would suggest, but one has to divide by a factor of 2 as a discrete Fadde\'{e}v-Popov gauge fixing, because $(-1)^\sF$ has to be treated as a gauge symmetry.   So for example, the formula $2^{2g+n-1}$ for the total number of spin structures is effectively still valid for $n=0$. For $n>0$, the overall $(-1)^\sF$ is not treated as a gauge symmetry because it is not gauged along the boundaries.

To derive these identities, one can start with fact that on a closed orientable surface, the number of even spin structures is $2^{g-1}(2^g+1)$ and the number of odd ones is $2^{g-1}(2^g-1)$. (These statements can be derived using the facts for the torus described above, and gluing together tori to form the genus $g$ surface.) This implies that on a closed surface, $\sum 1 = 2^{2g}$ and $\sum (-1)^\zeta = 2^g$. To include the boundaries, we cut $n$ holes in the surface and glue in $n$ trumpets. The circles $S_j$ along which we glue are treated as in the discussion above. The sum over spin structure ``orthogonal'' to a given  circle gives a factor of two in all cases except when the circle has R spin structure and we are summing with $(-1)^\zeta$. In this case we get zero. Note that the number of circles with R spin structure is always even.

To get the final results (\ref{spinSumsNoT}) and (\ref{ywq}), one takes the product of the factor for the closed surface times all of the factors for the circles $S_j$ associated to the $n$ trumpets. A final factor of $\frac{1}{2}$ is needed due to the fact that this sum over spin structures counts as distinct situations that differ only by the sign of the spinor in the ``interior'' portion of the surface inside all of the circles $S$. This sign is pure gauge, so we divide by two to remove the overcounting. In the case of the double trumpet, the counting
of spin structures, including this last factor, was explained in a slightly different way in section  \ref{bulk}.

\subsection{Including Both $(-1)^\sF$ And $\sT$}\label{sec:bothSpinAndT}

\subsubsection{Classical Time-Reversal Symmetries In SYK-Like Models}\label{classlike}

The most obvious way to get time-reversal symmetry in the SYK model is to take $q$ to be a multiple of 4.   In this case, the classical SYK action (\ref{uvu}) is invariant
under a time-reversal symmetry that maps $t$ to $-t$, and at $t=0$ leaves the elementary fermions invariant,\footnote{Classically, we view $\sT$ as a transformation
of observables and write its action on an observable $\psi$ as $\sT(\psi)$.    Quantum mechanically, we will view $\sT$ as a linear operator acting on quantum states,
and write the action on operators as $\psi\to \sT\psi\sT^{-1}$.}
\be\label{xono}\sT(\psi_k(0))= \psi_k(0). \ee
Classically, the time-reversal symmetry defined this way satisfies some obvious relations
\be\label{tono} \sT^2=1,~~~~ \sT (-1)^\sF =(-1)^\sF \sT, \ee
where $(-1)^\sF$ is understood to satisfy 
\be\label{wono}\left((-1)^\sF\right)^2 = 1. \ee
As we will discuss, there are anomalies in these statements at the quantum level.

If instead $q$ is congruent to 2 mod 4,  then there is no time-reversal symmetry that acts precisely as in (\ref{xono}).   However, even for $q$ of this form, it is possible to constrain
the SYK couplings to respect such a symmetry.   There are actually two ways to do this.    In one approach, one divides the two sets of fermions into two
subsets, say $\psi_i$, $i=1,\cdots, N_+$ and $\t\psi_j$, $j=1,\dots,N_-$, with $N_++N_-=N$, the total number of Majorana fermions. Then one defines $\sT$ to act at $t=0$ as
\be\label{pono} \sT(\psi_i(0))=\psi_i(0), ~~~~ \sT(\t\psi_j(0))=-\t\psi_j(0). \ee
This is not a symmetry of the generic SYK action (\ref{uvu}).   However, we can constrain the SYK coupling parameters $j_{i_1i_2\cdots i_q}$
to respect the symmetry of eqn.~(\ref{pono}) and otherwise to be random variables with specified variance.  It is reasonable to expect that such a model
 can have SYK-like behavior,
in the sense that the long time behavior is dominated by a Schwarzian mode, as in the case of SYK.

This type of time-reversal symmetry obeys at the classical level the same algebraic relations (\ref{tono}) and (\ref{wono}) as before.   It turns out to have the same
anomalies as in the previous case, with $N$ replaced by $N_+-N_-$.   So from a topological field theory point of view, this generalization does not seem to add anything
new, but in section \ref{SUSY}, we will see that it is useful to be familiar with it.

Still with $q$ congruent to 2 mod 4, it is actually also possible to impose an essentially  different sort of time-reversal symmetry.\footnote{We could do the same
for $q$ a multiple of 4, but the import would be different.  The SYK model with $q$ a multiple  of 4  unavoidably has a standard $\sT$ symmetry with $\sT^2=1$,
so any other $\sT$ symmetry would be the standard one times some global symmetry.}   For this, we assume $N$ to be even,
and we divide the fermions into two groups of equal size, say $\psi_1,\cdots, \psi_{N/2}$ and $\t\psi_1,\cdots \t\psi_{N/2}$.   Then we take $\sT$ to act by
\be\label{plono}\sT(\psi_k(0))=\t\psi_k(0),~~~\sT(\t\psi_k(0))=-\psi_k(0). \ee
Now we have not $\sT^2=1$ at the classical level, but
\be\label{fono} \sT^2=(-1)^\sF.\ee 
Again, the usual SYK model does not respect such a symmetry, but if we choose the coupling parameters to respect this symmetry (and otherwise to be random
variables of specified variance), it is reasonable to expect to find SYK-like behavior.

As remarked in section \ref{sec:strategy}, when a boundary theory has time-reversal symmetry,  a bulk dual description will involve a sum over possibly unoriented two-manifolds.
There are two essentially different generalizations of the notion of a spin structure to unorientable manifolds.   In Euclidean signature (where we will work
in our study of bulk duals), these are called $\pin^+$ structures and $\pin^-$ structures,
and they correspond respectively to theories that have time-reversal symmetries that at the classical level satisfy $\sT^2=(-1)^\sF$ or $\sT^2=1$, respectively.

Thus the usual SYK model with $q$ a multiple of 4 may be expected to have a bulk dual that involves a sum over $\pin^-$ structures.   Instead, an SYK-like model
with $\sT^2=(-1)^\sF$ will potentially have a bulk dual that involves a sum over $\pin^+$ structures.

With $\sT^2=1$, the anomalies depend on the value of $N$ (or more generally $N_+-N_-$) mod 8, so there are 8 different cases.  This matches the fact that
there are 8 topological field theories based on a sum over $\pin^-$ structures in two dimensions.   
With $\sT^2=(-1)^\sF$, $N$ has to be even and it turns out that only the value of $N$ mod 4
matters, so there are two cases, with $N$ congruent to 0 or 2 mod 4.    This matches the fact that two topological field theories can be made by summing
over $\pin^+ $ structures.   We will describe the anomalies and the topological field theories, and try to match what we learn with what one would expect from random matrix
theory.

\subsubsection{Anomalies When \texorpdfstring{$\sT^2=1$}{T*T=1} At The Classical Level}\label{pinminus}

We will consider a theory -- such as the SYK model with $q$ a multiple of 4 -- that has a symmetry group $\Z_2\times \Z_2$ at the classical level.  
One $\Z_2$ is generated by $(-1)^\sF$,
and the other by a time-reversal transformation $\sT$ that commutes with the elementary fermions and obeys $\sT^2=1$.   The other nontrivial element of the group is $\sT'=\sT (-1)^\sF$, which anticommutes with elementary fermions and satisfies
\be\label{welgo} (\sT')^2=1, ~~~~(-1)^\sF =\sT \sT'. \ee
We write $\Z_2^\sT\times \Z_2^\sF$ for the group generated by $\sT $ and $(-1)^\sF$.

It will be convenient to work with the fermion fields $\chi_k=\sqrt 2 \psi_k$, which obey a conventionally normalized Clifford algebra 
\be\label{doffo}\{\chi_k,\chi_l\}=2\delta_{kl}. \ee
This will minimize factors of $\sqrt 2$ in the following formulas.   Also, it is convenient to use the conventional $2\times 2$ Pauli matrices $\sigma_1,\sigma_2,\sigma_3$,
where $\sigma_2$ is imaginary and antisymmetric and $\sigma_1,\sigma_3$ are real and symmetric. We will now work out the anomalies in $(-1)^\sF$ and $\sT$ by working out their operator representations in terms of the fundamental SYK fermions.

\paragraph{The Case That $N$ Is Even} For $N=2$, we can represent the Clifford algebra of the $\chi$'s by $2\times 2$ matrices
\be\label{woffoSYK} \chi_1=\sigma_1, \,\chi_2=\sigma_2. \ee
For $N=4$, we need $4\times 4$ matrices.   We can think of these as matrices that act on a pair of qubits, where $\chi_1$, $\chi_2$ act on the first qubit as before
and $\chi_3,\chi_4$ are new:
\be\label{loffo}\chi_1=\sigma_1\otimes 1, \,\chi_2=\sigma_2\otimes 1, \,\chi_3=\sigma_3\otimes \sigma_1, \,\chi_4=\sigma_3\otimes \sigma_2. \ee
Every time we increase $N$ by 2, we add another qubit, replace the existing $\chi_k$ by $\chi_k\otimes 1$, and add two new $\chi$'s
\be\label{woffor}\chi_{N-1}=\sigma_3\otimes \sigma_3\otimes \cdots \sigma_3\otimes \sigma_1,~~~ \chi_N= \sigma_3\otimes \sigma_3\otimes \cdots \otimes \sigma_3\otimes \sigma_2.
\ee
The purpose of the factors of $\sigma_3$ is to make sure that the new $\chi$'s anticommute with the previous ones.  The $\chi_k$ are real for odd $k$ and imaginary
for even $k$.  These formulas give an irreducible representation of the even $N$ Clifford algebra in a Hilbert space of dimension $2^{N/2}$.

An operator $(-1)^\sF$ that anticommutes with all of the $\chi_k$ and satisfies $\left((-1)^\sF\right)^2=1$ is
\be\label{poffo} (-1)^\sF=\i^{N(N-1)/2}\chi_1\chi_2\cdots \chi_N. \ee
Letting $\sK $ denote complex conjugation, an antiunitary time-reversal transformation $\sT$ that commutes with the $\chi_k$
 and therefore (if $q$ is a multiple of 4) with the usual SYK Hamiltonian is 
\be\label{noffo}  \sT=\begin{cases} \sK  \chi_2\chi_4\chi_6\cdots \chi_N & N = 0 \text{ mod }4 \\ \sK \chi_1\chi_3\chi_5\cdots \chi_{N-1} & N  = 2 \text{ mod }4.\end{cases}\ee
The other symmetry is just  $\sT'=\sT (-1)^\sF$. Based on these formulas, we find that
\be\label{zoffo}\sT^2= \begin{cases}  1 ,~~~~\,{\mathrm {if}}~N= 0,2 \mod 8\\ -1,~~{\mathrm{if}}~N= 4,6 \mod 8  \end{cases}\ee
and
\be\label{zoffor}\sT (-1)^\sF= \begin{cases}  (-1)^\sF\sT ,~~~~\,{\mathrm {if}}~N= 0,4 \mod 8\\ -(-1)^\sF\sT,~~{\mathrm{if}}~N= 2,6 \mod 8 . \end{cases}\ee
Thus, for even $N$, there are anomalies in the statements $\sT^2=1$, $\sT(-1)^\sF=(-1)^\sF\sT$, unless $N$ is a multiple of 8, and these anomalies depend
nontrivially on $N$ mod 8. 

We now discuss the random matrix classes associated with these symmetries. There is a unitary symmetry $(-1)^\sF$, and we can treat the Hamiltonian as a random matrix in each block labeled by the eigenvalue of  $(-1)^\sF$.   If $N=0$ or 4 mod 8, then $\sT$ commutes with $(-1)^\sF$ and constrains  each block separately.  In more detail,
for  $N=0$ mod 8,
$\sT^2=1$ so  the statistics in each block are GOE-like.   For $N=4$ mod 8, $\sT^2=-1$ and the statistics in each block are  GSE-like.    On the other hand, for $N=2$ or 6 mod 8,
$\sT$ anticommutes with $(-1)^\sF$ and exchanges the two blocks, so each block has the same energy levels.   However, in these cases, $\sT$ does not constrain the Hamiltonian within any one block, and the statistics in either block are  GUE-like.

\paragraph{The Case That $N$ Is Odd}  The story is similar for odd $N$, though some details are different. As we noted in section \ref{evenodd}, $(-1)^\sF$ does not act in an irreducible representation.  Since $(-1)^\sF=\sT \sT'$, it will not happen that both $\sT$ and $\sT'$ do act.   What happens is that for each odd $N$, one of them acts within an irreducible representation, and its square is again $\pm 1$.

To represent the Clifford algebra,
we can represent $\chi_1,\cdots,\chi_{N-1}$ as before in a Hilbert space of dimension $2^{(N-1)/2}$.   Then we define
\be\label{onox}\chi_N= \pm \i^{(N-1)/2} \chi_1\chi_2\cdots\chi_{N-1}, \ee which is always real.
Either choice  of sign gives a  representation of the rank $N$ Clifford algebra in a Hilbert space of dimension $2^{(N-1)/2}$.   These representations are irreducible,
and they are inequivalent since they have opposite values of the central element
\be\label{wonox}\chi_1\chi_2\cdots \chi_N\ee
of the Clifford algebra. We choose one of them; nothing that follows depends on this choice.  

We  define a time-reversal operator $\h\sT$ as 
\be\label{plonox}\h \sT=\sK \chi_1\chi_3\chi_5\cdots\chi_N. \ee
It obeys
\be\label{monox} \h\sT\chi_k\h\sT^{-1}=\begin{cases}\chi_k & {\mathrm{if}}~N= 1,5 \mod 8\cr -\chi_k& {\mathrm{if}}~ N = 3,7\mod 8.   \end{cases}.\ee
and
\be\label{wonnox}\h\sT^2=\begin{cases} 1& {\mathrm{if}}~N= 1,7 \mod 8 \cr
         -1 & {\mathrm{if}} ~N= 3,5 \mod 8.\end{cases}.\ee
Eqn.~(\ref{monox}) shows that $\h\sT=\sT$ if $N= 1,5 \mod 8$, while $\h\sT=\sT'$ if $N= 3,7\mod 8$.
This and the sign of  $\h\sT^2$ distinguishes the four cases mod 8.

The random matrix classes are somewhat simpler in this case. Since there is no nontrivial $(-1)^\sF$ symmetry, there is only one block.   For $N=1$ or 7 mod 8, $\sT^2=1$, so the symmetry group that commutes with $\sT$ is $\O(L)$, and the statistics are of GOE type.   For $N=3$ or 5 mod 8, $\sT^2=-1$, and the statistics are of GSE type. This classification, together with the even $N$ case, is summarized in table \ref{tableReduction2} of section \ref{topominus}.

\paragraph{Further Comments} We have described anomalies that depend on the value of $N$ mod 8, but the reader might ask if there could be a more subtle anomaly that we have not found
that remains even if $N$ is a multiple of 8.   The most powerful way to show that this is not the case is the following \cite{Fidkowski:2009dba}.  If $\chi_\alpha$,
$\alpha=1,2,\cdots,8$ is a group of 8 Majorana fermions, all commuting with $\sT$, then a generic quartic SYK-like Hamiltonian\footnote{A simple one that
does the job is $\Delta H=-\chi_1\chi_2\chi_3\chi_4-\chi_5\chi_6\chi_7\chi_8-\chi_1\chi_2\chi_5\chi_6-\chi_1\chi_3\chi_5\chi_7.$} 
$\Delta H=\sum_{\alpha\beta\gamma\delta}t_{\alpha\beta\gamma\delta}\chi_\alpha\chi_\beta\chi_\gamma\chi_\delta$ has a unique ground state, which moreover (if
the coefficients $t_{\alpha\beta\gamma\delta}$ obey a suitable inequality)  is
invariant under both $\sT$ and $(-1)^\sF$.  That means that, by taking the coefficients in $\Delta H$ to be large enough, these 8 Majorana
fermions can be removed from an effective low energy description of the system, without breaking any symmetries.   Since anomalies in global symmetries can always be understood
in terms of any description of the system that is valid at low energies, this implies that anomalies in the symmetries of this system can only depend on the value of $N$ mod 8.
(For $N$ a multiple of 8, it is possible to pick a different representation of the Clifford algebra in which the $\chi_k$ are all real and therefore time-reversal can
act more simply as  $\sT=\sK$.)   

Still assuming that the symmetry group is $\Z_2^\sT\times \Z_2^\sF$, we 
now want to consider the case that at the classical level  there are Majorana fermions $\chi_1,\cdots,\chi_{N_+}$ with
\be\label{uncu} \sT\chi_k\sT^{-1}=\chi_k, \ee
and additional Majorana fermions $\t\chi_1,\cdots,\t\chi_{N_-}$ with
\be\label{wuncu}\sT\t\chi_k\sT^{-1}=-\t\chi_k. \ee

This generalization can be analyzed without much effort.
First of all, there is no anomaly if $N_+=N_-=1$.   We can work in a two-dimensional Hilbert space with
 $\chi=\sigma_1$, $\t\chi=\sigma_2$, along with $\sT=\sK$, $(-1)^\sF=\sigma_3$.
All expected commutation relations are satisfied, and in particular $\sT^2=1$ and $\sT(-1)^\sF=(-1)^\sF\sT$.

More generally, if we have any system of $N$ Majorana fermions with a time-reversal symmetry $\sT$, 
and  two of them, say $\chi$ and $\t\chi$, transform under $\sT$ with opposite signs,
then a perturbation to the Hamiltonian 
$\Delta H=\i m \chi\t\chi$ is $\sT$-invariant, and is invariant under $(-1)^\sF$ if $N$ is even (so that such a symmetry exists).   For large $m$, $\chi$ and $\t\chi$
are removed from the low energy description, without breaking any symmetry.   Such an operation cannot affect any anomalies,
and it reduces $N_+$ and $N_-$ by 1.   So anomalies can only depend on the difference $N_+-N_-$. Combining this with the previous argument, we see that anomalies depend precisely on the value of $N_+-N_-$ mod 8.

There is one last important comment.   
A low energy observer who does not have access to  microscopic fermion fields will not be able to distinguish an anomaly of $N$ from an anomaly
of $8-N$. This means, in particular, that JT gravity together with topological field theory are not sensitive to the sign of the anomaly. 

Exchanging $N$ with $8-N$ changes the sign of the anomaly, but one can compensate for this by exchanging $\sT$ with $\sT'$, which also changes
the sign of the anomaly,  as it exchanges $(N_+,N_-)$ with $(N_-,N_+)$.    For even $N$, the low energy observer sees both $\sT$ and $\sT'$ symmetry,
but -- without access to a microscopic description -- has no way to know which is which.    For odd $N$, the low energy observer has access
to only one of $\sT$ and $\sT'$, and has no way to know which it is.     Whether $N$ is even or odd,
one cannot distinguish $N$ from $8-N$ without access to the elementary fermions.

Note that along with elementary fermion fields $\chi_k$, whose transformation under time-reversal we have discussed, SYK-like models have hermitian fermion fields such as
$\i\chi_k\chi_l\chi_m$ which transform under time-reversal with an opposite sign.   To distinguish $\sT$ from $\sT'$, so that one can distinguish $N$ from $8-N$ mod 8,
one needs access to a fermion field  whose time-reversal properties are related to those
of the elementary fermions in a known way.    An example in which this is possible in a macroscopic description
is the supersymmetric SYK model.   A low energy observer has access to the fermion in the
super-Schwarzian multiplet, whose time-reversal properties are related in a simple way to those of the elementary fermions. The super-Schwarzian multiplet is part of
the description by JT supergravity.  Consequently,
JT supergravity plus topological field theory can distinguish $N$ from $8-N$ mod 8, as we will see in detail.

\subsubsection{Topological Field Theory Interpretation For Cases With $\sT^2 = 1$ Classically}\label{topominus}

To define fermions on an unorientable manifold, one needs a generalization of a spin structure that is called a  $\pin^+$ structure or a $\pin^-$ structure. The meaning of $\pin^+$
or $\pin^-$ 
is that if $\sR$ is a spatial reflection, then $\sR^2$ commutes   or anticommutes with fermion fields.
A relativistic theory has a $\sT$ symmetry if and only if it has an $\sR$ symmetry, but there is a perhaps surprising sign reversal:    if $\sT^2$ commutes with elementary
fermions, then $\sR^2$ anticommutes with them, and vice-versa.\footnote{\label{oddone} The usual explanation in terms of the Dirac equation for a fermion field $\psi$ is as follows.   
One starts with a Clifford algebra $\{\gamma_\mu,\gamma_\nu\}=\pm 2\eta_{\mu\nu}$,
where the choice of sign will lead to $\pin^+$ or $\pin^-$.   Suppressing all space coordinates except one, a relativistic theory will have a symmetry $\sT:\psi(t,x)\to \gamma_0
\psi(-t,x)$ if and only if it has a symmetry $\sR:\psi(t,x)\to\gamma_1\psi(t,-x)$.   Since $\gamma_0^2=-\gamma_1^2$, one of these operators squares to 1 if and
only if the other squares to $(-1)^\sF$.}   So $\sT^2=1$ in Lorentz signature is associated after Wick rotation to $\pin^-$ structures, while $\sT^2=(-1)^\sF$ is associated
to $\pin^+$ structures.  For a detailed introduction to $\pin^-$ and $\pin^+$ 
 structures, see appendix A of \cite{Witten:2015aba}.

\begin{figure}
 \begin{center}
   \includegraphics[width=2.5in]{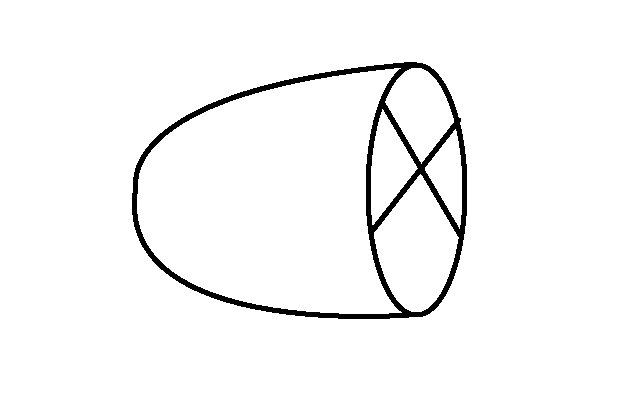}
 \end{center}
\caption{\small This picture is meant to symbolically convey the idea of building $\RP^2$ by closing off a disc with a ``crosscap.''  \label{CrossCap}}
\end{figure}

The topological invariant for a compact two-dimensional $\pin^-$ manifold $Y$ without boundary
 that generalizes $(-1)^\zeta$ for spin manifolds is $\exp(-\i\pi\eta/2)$, where $\eta$ is the Atiyah-Patodi-Singer 
eta-invariant of the self-adjoint operator $\i\slashed{D}$.  For any $Y$, $\exp(-\i\pi\eta/2)$ is an eighth root of unity.\footnote{This can be proved along lines explained in detail
in appendix C of \cite{Witten:2015aba} for an analogous question in four dimensions.    The two-dimensional case is slightly simpler.
In brief, we want to prove that, for any $\pin^-$ bundle $\P\to Y$,  $\left(\exp(-\i\pi\eta/2)\right)^8=1$, or equivalently that $\eta$ is a multiple of $1/2$.  Let $\varepsilon$
be the orientation  bundle of $Y$ (a real line bundle with holonomy $-1$ around  any orientation-reversing  loop)  and let
 $\P'=\P\otimes \varepsilon$ be the complementary $\pin^-$ bundle to $\P$.  
Then $\eta(\P)+\eta(\P')=0$ mod 4, because of general properties of the spectrum of the Dirac operator,
but  also $4\eta(\P)=4\eta(\P')$ mod  4, because the real vector bundle $\varepsilon^{\oplus 4}\to Y$ is trivial for any two-manfold $Y$.   These steps
are explained more fully in \cite{Witten:2015aba}.}   A basic example for which $\eta$ has the minimum value\footnote{Consider an orbifold $T^3/\Z_2$, where $T^3$
is a  three-torus and $\Z_2$ acts as $-1$ on each of the three coordinates.   The $\Z_2$ action has eight fixed points.  Removing a small open neighborhood
of each, one gets a compact three-manifold $W$ whose boundary consists of 8 copies of $\RP^2$.   Applying the Atiyah-Patodi-Singer index theorem to $W$,
one learns that $\eta(\RP^2)=\pm 1/2$, with a sign that depends on the choice of $\pin^-$ structure.   For the details of this argument (in an analogous four-dimensional
case), see appendix C of \cite{Witten:2015aba}.}
$\pm 1/2$ is $\RP^2$.    

There are many possible descriptions of $\RP^2$, but the one that is most useful for us is as follows.   Begin with a closed disc $D$,
whose boundary is a circle $S$, say parametrized by an angle $\theta$, with $\theta\cong \theta+2\pi$.   One can build from $D$ a two-manifold without boundary
by identifying antipodal points on $S$, that is by making the further identification $\theta\cong \theta +\pi$.   This is often described by saying that $D$ is closed off
with a ``crosscap.''  As the result is difficult to draw convincingly, it is often depicted symbolically as in fig.~\ref{CrossCap}.    The resulting compact manifold is
called $\RP^2$ (real projective two-space).   It is unorientable, with Euler characteristic $1$,
so its genus, defined by  $\chi=2-2g$, is $g=1/2$.
  By ``gluing in a crosscap'' to a two-manifold $Y$, one means
that one cuts out of $Y$ an open ball and closes it off with a crosscap in the sense just described. 
Topologically, any compact unorientable two-manifold is an oriented one with a certain number of crosscaps glued in (in fact, one or two crosscaps is enough).   Every time one glues in a crosscap, the Euler characteristic is reduced by 1.   So the surfaces with odd Euler characteristic are the ones with an odd number of crosscaps glued in; they are the ones
with half-integer $g$.

Since $\exp(-\i\pi\eta/2)$ is an eighth root of unity, and has the appropriate locality properties, one can build 8 different topological field theories on a $\pin^-$ manifold $Y$
without  boundary by
summing over $\pin^-$ structures on $Y$ with a factor of $\exp(-\i\pi N\eta/2)$.    When $Y$ has a boundary, $\exp(-\i\pi N\eta/2)$ has an anomaly that precisely matches
the anomaly of $N$ Majorana fermions on the boundary of $Y$.   This is schematically explained in section 5 of \cite{Witten:2015aba}.    For a more abstract explanation
based on cobordism invariance, see \cite{Kap}.

So we expect that in a bulk description of the SYK model with the usual sort of time-reversal symmetry, we should include a sum over $\pin^-$ structures with a factor
of $\exp(-\i \pi N\eta/2)$. We will call the sum over the $\pin^-$ structures on a manifold, weighted by $\exp(-\i \pi N\eta/2)$, the ``$\pin^-$ sum.''   We will denote
this sum on a manifold $Y$ as $F_Y(N)$.  
   If $Y$ is orientable, then a $\pin^-$ structure on $Y$ is the same as a spin structure, and $\exp(-\i\pi \eta/2)$ reduces to $(-1)^\zeta$.   So for orientable
$Y$, the $\pin^-$ sum is just the sum over spin structures weighted by $(-1)^{N\zeta}$.   This is the same sum that we studied in sections \ref{bulk} and \ref{sec:reduction1},
and is only sensitive to $N$ mod 2.   For unorientable $Y$, the $\pin^-$ sum can detect the value of $N$ mod 8.

However, as in section \ref{pinminus},  JT gravity and topological field theory  cannot distinguish $N$ from $8-N$ mod 8.
Here we use the fact that $\pin^-$ bundles over $Y$ come in pairs, in the following way.   If $\P$ is a $\pin^-$ bundle over $Y$, then there is another $\pin^-$ bundle $\P'$ over $Y$,
with the property that parallel transport of a fermion around any closed  loop $\gamma\subset Y$ gives the same result up to sign.  The sign is $+1$ or $-1$ depending on whether
the orientation of $Y$ is invariant or is reversed in going around $\gamma$.   The abstract way to say this is that $\P'=\P\otimes \varepsilon$, where $\varepsilon$ is the orientation
bundle of $Y$.   Then $\eta(\P)=-\eta(\P')$ mod 4, so $\exp(-\pi \i N\eta(\P)/2)=\exp(-\pi\i (8-N)\eta(\P')/2)$.   In other words, exchanging $\P$ with $\P'$ has the same effect
as replacing $N$ by $8-N$ mod 8.   So once one sums over $\pin^-$ structures, the distinction between $N$ and $8-N$ disappears.   Moreover, it disappears by the same
mechanism as in section \ref{pinminus}, since exchanging $\P$ with $\P'$ amounts to exchanging fermions that transform with opposite signs under $\sT$.  

For a variety of reasons, an important special case is the $\pin^-$ sum of $\RP^2$.    Since $\RP^2$ has two $\pin^-$ structures with $\eta=\pm 1/2$, its $\pin^-$ sum is $F_{\RP^2}(N)=2\cos(2\pi N/8)$.   In particular, $F_{\RP^2}(2)=0$ and $F_{\RP^2}(N+4)=-F_{\RP^2}(N)$ for any $N$.

These facts have several interesting implications.
Any two-manifold $Y$ can be built from an oriented one $Y_0$ by gluing in some number $n_c$ of crosscaps.    Here $n_c$ is not uniquely determined, but it is uniquely determined
mod 2; in fact, $n_c$ is congruent mod 2 to the Euler characteristic $\chi(Y)$.
  Using the locality properties of $\exp(-\i\pi\eta/2)$,
one can show that
\be\label{ubic} F_Y(N)=F_{Y_0}(N)F_{\RP^2}(N)^{n_c}. \ee
Since $F_{Y_0}(N)$ only depends on $N$ mod 2, and $F_{\RP^2}(N+4)=-F_{\RP^2}(N),$
we get
\be\label{nubic} F_Y(N+4)=(-1)^{n_c} F_Y(N).
\ee
In other words, shifting $N$ by 4 simply gives a factor of $-1$ for every crosscap.   The random matrix counterpart of this is discussed in appendix \ref{Euler}.

Another interesting consequence is the following.   Since $Y$ is unorientable if and only if $n_c>0$, and since $F_{\RP^2}(2)=F_{\RP^2}(6)=0$, we learn from 
eqn.~(\ref{ubic}) that $F_Y(2)=F_Y(6)=0$ for any unorientable $Y$.   So for those values of $N$ mod 8, only orientable manifolds contribute after the sum over $\pin^-$ structures.

We will now begin to discuss how the connection between JT gravity and random matrix theory is modified by including unorientable surfaces and the $\pin^-$ sum. Suppose first that we want to compute the expectation value of $\langle Z_{\NS}(\beta)\rangle$ or $\langle Z_{\Ra}(\beta)\rangle$. We recall that these are computed in JT gravity by summing over contributions of two-manifolds $Y$ whose boundary is a single circle $X$. In the present context, $Y$ is a possibly unoriented two-manifold with a $\pin^-$ structure. 

\begin{figure}
 \begin{center}
   \includegraphics[width=2.5in]{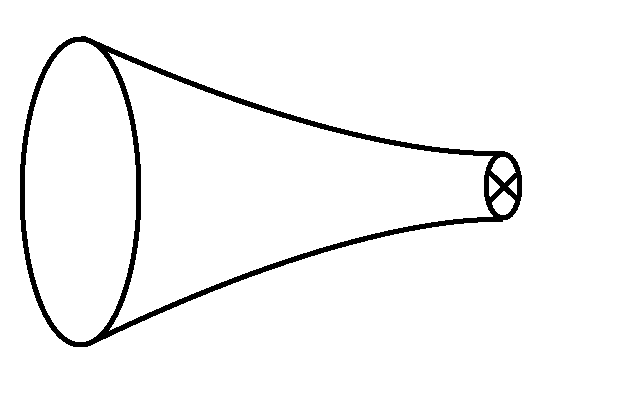}
 \end{center}
\caption{\small A trumpet that ends on a crosscap; we will call this the crosscap spacetime.  \label{tcc}}
\end{figure}

Then $\langle Z_\Ra(\beta)\rangle$ remains zero, because  a $\pin^-$ manifold with boundary always has an even number of boundary components of R type.  If $Y$ has just a single boundary component, it will be of NS type. On the other hand, $\langle Z_\NS(\beta)\rangle$ is nonzero, and can receive contributions from unoriented as well as oriented manifolds. The dominant unoriented contribution comes from the manifold $Y$ with the largest possible Euler characteristic, given that it is supposed to have one boundary component. This is a trumpet that ends on a crosscap (fig.~\ref{tcc}); we will
more briefly call it the crosscap spacetime.

\paragraph{Contribution Of The Crosscap Spacetime} We will now evaluate the contribution of this manifold and compare to the random matrix ensembles identified above. In making this comparison, we will assume the purely bosonic JT gravity / RMT correspondence discussed in section \ref{sec:bosonic}, so that without any minus sign or
sum over pin structures, the JT gravity partition function on the crosscap spacetime agrees with a GOE-like random matrix theory. There are two $\pin^-$ structures on the crosscap, with the same values $\eta=\pm 1/2$ as for $\RP^2$.   Hence the $\pin^-$ sum is just $F_{\RP^2}(N)$.    The dominant crosscap contribution to $\langle Z_{\NS}(\beta)\rangle$ is therefore as follows:

For $N=0$ mod 8, since $F_{\RP^2}(0)=2$,  the crosscap contribution to $Z_{\NS}(\beta)$ is two times the JT path integral for the crosscap spacetime, which we assume is equal to the GOE-like answer. In random matrix theory, for $N = 0$ mod 8, there are two  independent GOE blocks, see table \ref{tableReduction2} below, so the factor of two matches.

For $N=1$ mod 8, the $\pin^-$ sum is $F_{\RP^2}(1)=\sqrt 2$. In random matrix theory, we have a single GOE block, see table \ref{tableReduction2} again, but the path integral and the trace in the Hilbert space are related by the factor of $\sqrt{2}$ in (\ref{ZHilbert}), so we find agreement.

For $N=2$ mod 8, we have $F_{\RP^2}(2) = 0$, so the crosscap does not contribute. This agrees with table \ref{tableReduction2}, since the $N = 2$ case consists of GUE blocks, for which there is no genus 1/2 contribution.

For $N=3$ mod 8, the $\pin^-$ sum is $F_{\RP^2}(3)=-\sqrt 2$.  On the random matrix side, the only relevant difference from $N=1$ mod 8 is that now  $\h\sT^2=-1$, so that we have to use GSE statistics instead of GOE.   Again, this reverses the sign of the crosscap contribution.

For $N=4$ mod 8, the $\pin^-$ sum is $F_{\RP^2}(4)=-2$.   On the random matrix side, the only relevant difference from $N=0$ mod 8 as that now $\sT^2=-1$, so that we have to use GSE statistics instead of GOE.   This reverses the sign of the crosscap contribution.

The remaining cases follow by exchanging $N$ with $8-N$ mod 8, so they do not add much.  

\paragraph{Contribution Of The Double Trumpet} Now we will discuss the leading contributions to the connected correlation functions $\left\langle Z_{\NS}(\beta)^2\right\rangle_c$ and $\left\langle Z_{\Ra}(\beta)^2\right\rangle$.
These come from the same double trumpet studied in section \ref{bulk}, along with another topology that we will describe momentarily. 

In section \ref{bulk}, we constructed the double trumpet $Y$ by gluing together two trumpets $Y_L$ and $Y_R$ along their inner boundaries $S_L$ and $S_R$
(fig.~\ref{SDT}).  
However, in a time-reversal invariant theory, we are free to make a reflection of $S_R$, reversing its orientation, before gluing it onto $S_L$.   This makes a different
manifold that we will call the twisted double trumpet $\t Y$.   $\t Y$ is not equivalent to $Y$ by any diffeomorphism that acts trivially on its boundaries, so in the context
of duality between an SYK-like model (or random matrix theory) and a bulk theory, we should consider $\t Y$ and $Y$ to be inequivalent.\footnote{We could also rotate
$S_R$ (rather than reflecting it, or in addition) before gluing it to $S_L$.    This is important in the path integral of JT gravity, but as it does not affect the topology, it is not
visible in the topological field theory.}

In comparing $\exp(-\i\pi N\eta/2)$ for different $\pin^-$ structures on $Y$ or on $\t Y$, as usual, we can use locality to reduce to a simpler picture.   A convenient way
to use locality is to glue together the outer boundaries of $Y$ or of $\t Y$.    This produces from $Y$ the two-torus of fig.~\ref{Regluing}, while from $\t Y$ it produces
a Klein bottle, which we will call KB.

\begin{figure}
 \begin{center}
   \includegraphics[width=3in]{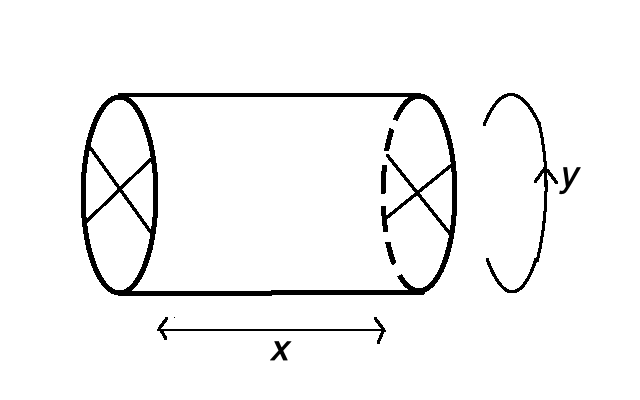}
 \end{center}
\caption{\small A Klein bottle can be constructed by closing off a cylinder at each end with a crosscap.  \label{KB}}
\end{figure}

A standard description of KB is that it can be obtained from the $x-y$ plane by dividing by two symmetries $T_x$ and $T_y$, where
\be\label{twos} T_x(x,y)=(x+1,y), ~~~~~~T_y(x,y)=(-x,y+1). \ee 
It is the $x$ direction here that parametrizes the circle $S_L$ or $S_R$ in the twisted double trumpet (or the external boundaries,
to which $S_L$ and $S_R$ are homologous), since it is the $x$ direction that is reflected
by one of these operations (namely $T_y$).
Another way to look at KB is also useful.   The identifications $(x,y)\cong (x+1,y)$ and $(x,y)\cong (-x,y+1)$ let us restrict to $0\leq x\leq 1/2$.
At generic $x$, the equivalence relation on $y$ is $y\cong y+2$, but at the endpoints there is a further identification $y\cong y+1$.    This means that KB can be
constructed from a cylinder $0\leq x\leq 1/2$, $y\cong y+2$ by closing it off with a crosscap at each end (fig.~\ref{KB}).   This description implies that a complete $\pin^-$ sum of KB is just
$F_{\mathrm{KB}}(N)=F_{\RP^2}(N)^2$.   But that is not quite what we need; to understand contributions to $\langle Z_{\NS}(\beta)^2\rangle_c $
or $\langle Z_{\Ra}(\beta)^2\rangle$, we need to restrict the sum over $\pin^-$ structures on KB to those
for which the $\pin^-$ structure on the boundaries of $\t Y$ are of NS or R type, respectively.   It turns out that these restricted $\pin^-$ sums
are $F_{\mathrm{KB}}^{\NS}(N)=2$,   $F_{\mathrm{KB}}^{\Ra}(N)= 2\cos (2\pi  N/4)$ (this claim is explained at the end of the present
section). Similarly, we can define the restricted  pin$^-$ sums for the  torus. Since this manifold is orientable, pin$^-$ structures are the same as spin structures, and we reduce to the case studied in section \ref{bulk}, which we can summarize as $F_{T}^{\NS}(N) = 2$ and $F_T^{\Ra}(N) = 1 + (-1)^N$.

These results imply that
\be\label{worfo} F_T^\NS(N)+F_\KB^\NS(N)=4 ,\ee
independent of $N$, while
\be\label{zorfo}  F_T^\Ra(N)+F_\KB^\Ra(N) = 1+(-1)^N+2\cos 2\pi N/4  = \begin{cases} 4 &~~ {\mathrm {if}}~N=0~{\mathrm{mod}}~4 \cr
  0&~~{\mathrm{ otherwise}}. \end{cases} \ee

To compare to random matrix theory, we will need one 
standard fact that we will borrow from section \ref{sec:loopEquations}: the 
leading contribution to the connected correlator $\langle Z(\beta)^2\rangle_c$ in  the GOE or GSE-like ensembles is twice 
as big as in the GUE ensemble. More precisely, this is true if we define the GSE case to include both of each pair of degenerate eigenvalues, which is implicit in table \ref{tableReduction2}.

If $N = 0,4$ mod 8,  then   for either $\langle Z_{\NS}(\beta)^2\rangle_c$ or $\langle Z_{\Ra}(\beta)^2\rangle$, the $\pin^-$ sums of the double trumpet  and  the twisted double trumpet
add to $2+2=4$,  according to the above formulas.  This multiplies what we  would get by computing this  correlator
in bosonic JT  gravity. This is consistent with 
random matrix theory, since each of the two GOE or GSE blocks in table \ref{tableReduction2} contribute twice the GUE answer, for a total factor of 4.

If $N$ is odd,  eqns. (\ref{worfo}) and (\ref{zorfo}) imply that
 $\langle Z_{\NS}(\beta)^2\rangle_c$ is four times the GUE answer, and $\langle Z_{\Ra}(\beta)^2\rangle = 0$. Both are consistent with random matrix theory. From (\ref{ZHilbert}), the NS answer is $(\sqrt{2})^2 = 2$ times the GOE or GSE answer, so four times the GUE answer. Also, from (\ref{ZHilbert}), the R answer should be zero because for odd $N$ the path integral in the Ramond sector vanishes.

If $N = 2,6$ mod 8, 
the random matrix ensemble consists of two identical GUE-like blocks. The twofold degeneracy implies a factor of four in  $\langle Z_{\NS}(\beta)^2\rangle_c$, and since the two blocks have identical spectra, the Ramond-sector partition function is identically zero. Both facts are consistent with the pin$^-$ sums (\ref{worfo}) and (\ref{zorfo}).

Finally, to justify the claim about the restricted
 $\pin^-$ sums on KB, we first note that the four $\pin^-$ structures on KB can be  presented by describing a fermion field $\lambda$ on KB 
as a fermion field on $\R^2$ that satisfies
\begin{align}\label{kbspin}   \lambda(x+1,y)& =(-1)^\alpha \lambda(x,y) \cr \lambda(-x,y+1)& =(-1)^\beta \gamma_x \lambda(x,y). \end{align}
Here $\alpha,\beta\in \{0,1\}$ label the four choices of $\pin^-$ structure on KB, and $\gamma_x$ is a gamma matrix (the $\pin^-$ condition means that
$\gamma_x^2=-1$, while $\pin^+$ means $\gamma_x^2=+1$).    Since the $x$ direction parametrizes the external boundaries of the twisted double trumpet
(note the comment after eqn.~(\ref{twos})), this means that parallel transport of a fermion around the external boundary is controlled by $(-1)^\alpha$.
Thus $\alpha=1$ means that the external boundaries are of NS type, and $\alpha=0$ means R type.   

Now we view KB as a cylinder with a crosscap at $x=0$ and another crosscap at $x=1/2$.    Eqn.~(\ref{kbspin}) says that at $x=0$, the
$\pin^-$ structure is described by $\lambda(0,y+1)=(-1)^\beta\gamma_x \lambda(0,y)$.    But at $x=1/2$, the $\pin^-$ structure is instead
described by $\lambda(1/2,y+1)=(-1)^{\alpha+\beta} \gamma_x\lambda(1/2,y)$.     This says that if $\alpha=0$, the $\pin^-$ structures on the two crosscaps
are isomorphic, regardless of $\beta$, but if $\alpha=1$, they are opposite.

This is all we need to compute the restricted $\pin^-$ sums.    If the boundaries are of R type, we set $\alpha=0$.   The two crosscaps have isomorphic
$\pin^-$ structures, each contributing $1/2$ or $-1/2$ to $\eta$, so the sum is $\eta=1$ or $\eta=-1$.     Therefore
 $F_{\mathrm{KB}}^{\Ra}=\exp(-\i \pi N/2)+\exp(\i\pi N/2)=2\cos(2\pi N/4)$, as claimed
earlier.   If the boundaries are of NS type, we set $\alpha=1$.  The two crosscaps have $\pin^-$ structures of opposite types.   If one contributes $\pm1/2$ to $\eta$,
the other contributes $\mp1/2$, so that overall $\eta=0$.   So $F_{\mathrm{KB}}^{\NS}=2$, as also claimed earlier.

\begin{table}[t]
\small
\begin{center}
\begin{tabular}{c|c|c|c}
$N$ mod 8& RMT class & $Z_{g,n,n_{\Ra}}$ & pin$^-$ sum identity ($n_{\text{R}}= 0$)\\
\hline\hline
0 & $\left(\begin{array}{cc}\hspace{-5pt}\text{GOE}_1\hspace{-12pt} & 0 \\ 0 & \text{GOE}_2\hspace{-5pt}\end{array}\right)$ & $2^{2g+n-2}\big(1 + (-1)^{n_{\Ra}}\big)Z_{g,n}^{\text{GOE}}$ & $\sum\limits_{\text{pin}^-} 1= 2^{2g+n-1}$\\
\hline
1 & $\text{GOE}$ & $\begin{cases} 2^{g+n-1}Z_{g,n}^{\text{GOE}}& n_{\Ra}  = 0\\ 0 & \text{else}\end{cases}$ & $\sum\limits_{\text{pin}^-} e^{-\mathrm{i}\pi \eta/2} = 2^{g+n-1}$\\
\hline
2 & $\left(\begin{array}{cc}\hspace{-5pt}\text{GUE}\hspace{-12pt} & 0 \\ 0 & \text{GUE}\hspace{-5pt}\end{array}\right)$ & $\begin{cases} 2^{2g+2n-2}Z_{g,n}^{\text{GUE}}& n_{\Ra}  = 0 \\ 0 & \text{else}\end{cases}$ & $\sum\limits_{\text{pin}^-}e^{-\mathrm{i}\pi \eta}= \begin{cases} 2^{2g+n-1} & n_c = 0 \\ 0 & \text{else}\end{cases}$\\
\hline
3 & GSE & $\begin{cases}  2^{g+n-1}Z_{g,n}^{\text{GSE}}& n_{\Ra}  = 0 \\ 0 & \text{else}\end{cases}$ & $\sum\limits_{\text{pin}^-}e^{-3\mathrm{i}\pi \eta/2} = 2^{g+n-1}(-1)^{n_c}$\\
\hline
4 & $\left(\begin{array}{cc}\hspace{-5pt}\text{GSE}_1\hspace{-12pt} & 0 \\ 0 & \text{GSE}_2\hspace{-5pt}\end{array}\right)$ & $2^{2g+n-2}\big(1 + (-1)^{n_{\Ra}}\big)Z_{g,n}^{\text{GSE}}$ & $\sum\limits_{\text{pin}^-}e^{-2\mathrm{i}\pi \eta} = 2^{2g+n-1}(-1)^{n_c}$
\end{tabular}
\caption{{\small Random matrix classes as a function of $N$ mod 8, for the cases with $\sT^2 = 1$ classically. Column two gives the RMT classes described in section \ref{pinminus}
(for $N=2$, the two GUE blocks  are not independent but are exchanged by $\sT$). 
In the third column we follow the logic of section \ref{sec:reduction1} to reduce the expansion coefficients $Z_{g,n,n_\Ra}$ to expansion coefficients in the ordinary GOE, GUE, GSE ensembles. In the final column, we give the pin$^-$ sum identity (\ref{fullPinSum}) that implies this reduction in the bulk theory. 
The values of $N$ mod 8 not listed here are obtained by $N\rightarrow 8-N$.}}\label{tableReduction2}
\end{center}
\end{table}

\paragraph{Reduction To The GOE, GUE, And GSE Cases}

At this point, we actually have enough information about pin$^-$ structures to give a general reduction that shows 
consistency with random matrix theory, assuming it in the purely bosonic cases from section \ref{sec:bosonic}. Showing this requires several steps. First, one copies the steps in section \ref{sec:reduction1} to reduce the $Z_{g,n,n_\Ra}$ coefficient functions defined in (\ref{genusExpReduction}) to combinations of $Z_{g,n}$ in the GOE, GUE, GSE-like ensembles. The  predictions are given in the third column of table \ref{tableReduction2}.

 Before proceeding further, we note that in  topological field theory, according to eqn.~(\ref{zorfo}), amplitudes
with boundaries of R type  vanish unless $N=0,4$ mod 8.  This agrees with the random matrix theory prediction in the third column of the table.   If $N=0,4$ mod 8,
the restricted $\pin^-$ sums for an NS or R boundary are the  same, so from a topological field theory point of view,
NS and R boundaries  are equivalent except for the constraint that the total
number of R boundaries is even.   This is also a random matrix theory prediction, as stated in the third column in the table.   Therefore, to complete the demonstration
that topological field theory and random matrix theory make compatible predictions, it suffices to consider NS boundaries only.

Next, one uses gluing to derive the pin$^-$ sum on a general surface $Y$. 
 To do so, we start with a closed orientable surface $Y_0$ obtained from $Y$ by removing the holes and crosscaps. $Y_0$ has genus $g_0 = g - \frac{1}{2}n_c$, where $n_c$ is the number of crosscaps in $Y$. Because $Y_0$ is orientable, the pin$^-$ sum is the same as a spin sum, and leads to $2^{2g_0}$ for even $N$ and $2^{g_0}$ for odd $N$. Next, we glue in the $n_c$ crosscaps, using (\ref{ubic}). Finally, we cut $n$ holes with boundaries $S_j$, and glue in $n$ trumpets.  As already
 explained, it suffices to assume the these are of NS type, in which case the sum over the ``orthogonal'' $\pin^-$ structure gives a factor of 2 for each boundary.    The $\pin^-$ sum is therefore
\be\label{fullPinSum}
\frac{1}{2}2^n 2^{2g_0}F_{\RP^2}(N)^{n_c} \ee
for even $N$;  for  odd $N$, it is obtained by replacing $2^{2g_0}$ with  $2^{g_0}$.   
 Using $F_{\RP^2}(N)=2\cos 2\pi  N/8$, this leads to the results stated in the last column of the table for the $\pin^-$ sums.

Comparing the last two columns in the table, we see that in all cases,
the topological field theory prediction is in agreement with  the reduction to GOE, GUE, GSE from the 
first step.\footnote{For $N = 2$ mod 8, one has to remember that the JT gravity path integral in a 
time-reversal-invariant theory is $2^{n-1}$ times the path integral in the GUE-like theory without time-reversal symmetry; see the end of section \ref{sec:bosonic}.}

\subsubsection{Cases With \texorpdfstring{$\sT^2=(-1)^\sF$}{T*T=(-1)F} Classically}\label{pinplus}

Now we want to consider another symmetry group instead of $\Z_2^\sT\times \Z_2^\sF$.   
We still assume a time-reversal transformation $\sT$ and a symmetry $(-1)^\sF$ that distinguishes bosons and fermions. But  now at the classical level we assume
\be\label{expected}\sT^2=(-1)^\sF,~~~~\left((-1)^\sF\right)^2=1.\ee
So $\sT^4=1$ and $\sT$ generates a group that we call $\Z_4^\sT$.
    How to construct an SYK-like model with such a symmetry group was explained in section \ref{classlike}. In particular
 the number $N$ of elementary fermions must be even.   

This case
 is actually simpler than the previous case of $\Z_2^\sT\times \Z_2^\sF$, so we can be more brief.  For even $N$, we described in section \ref{pinminus} an irreducible representation
of the Clifford algebra with $N/2$ real generators and $N/2$ imaginary ones.   Call the even generators $\chi_k$, $k=1,\cdots,N/2$, and the odd ones $\t\chi_k$,
$k=1,\cdots,N/2$.   Define the time-reversal transformation
\be\label{timet} \sT=\frac{\chi_1+\t\chi_1}{\sqrt 2}\frac{\chi_2+\t\chi_2}{\sqrt 2}\cdots \frac{\chi_{N/2}+\t\chi_{N/2}}{\sqrt 2}\sK. \ee
This has been chosen so that
\be\label{limet}\sT \chi_k\sT^{-1}=\t\chi_k,~~~\sT\t\chi_k\sT^{-1}=-\chi_k. \ee
A short computation gives
\be\label{wimet}\sT^2=\i^{N^2/4}(-1)^\sF, \ee
with 
\be\label{gimmick}(-1)^\sF=\i^{N/2}\chi_1\t\chi_1\chi_2\t\chi_2\cdots\chi_{N/2}\t\chi_{N/2}. \ee
Up to an overall sign, the phase of $(-1)^\sF$ has been fixed to ensure $\left((-1)^\sF\right)^2=1$.

If $N= 0$ mod 4, we have the expected relations (\ref{expected}). The random matrix symmetry class of a Hamiltonian with these symmetries consists of two blocks for the $\pm$ eigenvalues of $(-1)^\sF$. Acting on the $+$ subspace, $\sT^2 = 1$, so we have GOE-like statistics. Acting on the $-$ subspace, $\sT^2 = -1$, so we have GSE-like statistics. 

If $N= 2$ mod 4, we have instead
\be\label{wimerick}\sT^2=\i (-1)^\sF.\ee
If we absorb this factor of $\i$ in the definition of $(-1)^\sF$, we would get $\left((-1)^\sF)\right)^2=-1$.   So there is no way to avoid an anomaly, though it can be moved
around. Since $\sT$ commutes with $\sT^2$ and anticommutes with $\i$, eqn.~(\ref{wimerick}) implies the  important fact that if $N=2$ mod 4, $\sT$ anticommutes with $(-1)^\sF$:
\be\label{imerick}\sT(-1)^\sF=-(-1)^\sF\sT. \ee
This implies that the two blocks of the Hamiltonian are exchanged by $\sT$, but are not constrained individually. So we have two GUE-like blocks with identical eigenvalues.

Thus we have found an anomaly that depends on the value of $N$ mod 4.   There is no more subtle anomaly, since by perturbing the Hamiltonian in a suitable fashion,
one can remove two pairs of fermions, say $\chi_1, \t\chi_1$ and $\chi_2,\t\chi_2$, from the low energy spectrum.   One simply adds the ``mass'' term
$\Delta H=-\i m(\chi_1\chi_2 +\t\chi_1\t\chi_2)$.    For large $m$, this set of four fermions is removed from the low energy effective theory, reducing $N$ by 4,
without breaking any symmetry.   

Since $N$ is even and the anomaly only depends on $N$ mod 4, this is a $\Z_2$-valued anomaly.   The bulk dual of a theory with  $\Z_4^\sT$ symmetry
is a theory in which (in Euclidean signature) one sums over $\pin^+$ structures on two-manifolds.    Topological field theories that can be built
by summing over $\pin^+$ structures in suitable ways have a $\Z_2$ classification, which matches what we need.   

The necessary invariant is as  follows.
Let $Y$ be a two-manifold, not necessarily oriented, with a $\pin^+$ structure.   On an unorientable two-manifold $Y$, it is not possible to consider a chiral fermion field,
but we can of course consider a non-chiral, two-component fermion field $\lambda$.  Now we can define a $\Z_2$-valued invariant $\t \zeta$ 
 by imitating the definition of the invariant $\zeta$ in section \ref{bulk}.   We simply define $\t \zeta$ to be the mod 2 index of the nonchiral
Dirac operator on $Y$, that is the dimension mod 2 of the
space of solutions of the Dirac equation $\slashed{D}\lambda=0$.   This is a topological invariant, as discussed in section 3 of \cite{Witten:2015aba}.    On a two-manifold without boundary, one
 can make two topological field theories by summing over $\pin^+$ structures with or without a factor of $(-1)^{\t\zeta}$. As in other cases we have discussed,
 $\t\zeta$ has the necessary locality properties to justify this statement.    On an orientable manifold, the
 two theories are equivalent  because $\t\zeta=0$ if  $Y$ is orientable.\footnote{If
$Y$ is orientable, the dimension of the space of solutions of the Dirac equation is $n_++n_-$, where $n_+$ and $n_-$ are the dimensions of the spaces of solutions
with positive or negative chirality.   But $n_+=n_-$, because complex conjugation reverses the fermion chirality in two dimensions.   So $n_++n_-$ is always even.}
  This is as one would expect, because $\Z_4^\sT$ symmetry forces $N$ to be even.
 
 On a two-manifold with boundary, there is a problem with the definition of $\t\zeta$:   although one can place a local boundary condition on the Dirac equation
 and then count its zero-modes mod 2, such a local boundary condition is not invariant under time-reversal symmetry, that is under a reflection of the boundary.
 So $(-1)^{\t\zeta}$ has an anomaly on a manifold with boundary.   One can show that this anomaly matches the anomaly that we found above for $N$ congruent to 2 mod 4.
 One way to show this is to imitate arguments given in \cite{Dijkgraaf:2018vnm} for the analogous statement about $\zeta$, which was invoked in section \ref{bulk}.
 
Accordingly, we propose that an SYK-like model with the sort  of time-reversal symmetry that we have discussed here should be compared to JT gravity with a sum over
$\pin^+$ structures, with or without a factor of $(-1)^{\t\zeta}$, depending on whether $N$ is congruent to 2 or 0 mod 4.   Equivalently, we can in general include
a factor $(-1)^{\frac{N}{2}\t\zeta}$.

   Since $\t\zeta=0$ if $Y$
is orientable, if follows that the difference between the two theories only comes into play for unorientable $Y$.    This is as we would expect, since $N$ is even in the present
context, and anomalies that can be seen on orientable manifolds only depend on $N$ mod 2.

For illustration and because it will be useful, we will compute $\t\zeta$ for a $\pin^+$ structure on a Klein bottle KB.   Such structures can be described exactly as we described
$\pin^-$ structures in eqn.~(\ref{kbspin}),
\begin{align}\label{kbspintwo}   \lambda(x+1,y)& =(-1)^\alpha \lambda(x,y) \cr \lambda(-x,y+1)& =(-1)^\beta \gamma_x \lambda(x,y), \end{align}
with the only difference being that now $\gamma_x^2=+1$ rather than $-1$.   A zero-mode of the Dirac equation on KB is simply a constant mode of $\lambda$
that obeys the conditions (\ref{kbspintwo}).    If $\alpha=1$, there is no such mode.    But if $\alpha=0$, then since $\gamma_x$ is a $2\times 2$ matrix with
eigenvalues $\pm 1$, there is, regardless of $\beta$, precisely one zero-mode.   So $\t\zeta=0$ if $\alpha=1$ and $\t\zeta=1$ if $\alpha=0$. 

  Now let us view
KB as a cylinder with a crosscap at each end, as in fig.~\ref{KB}.   There are two $\pin^+$ structures on each crosscap.   By the same argument as given for $\pin^-$
structures in section \ref{topominus}, $\alpha=0$ means that the $\pin^+$ structures on the two crosscaps are isomorphic and $\alpha=1$ means that they are opposite.
So changing $\alpha$ from 0 to 1 flips the $\pin^+$ structure on one of the two crosscaps, regardless of $\beta$.    Since changing $\alpha$ also changes $\t\zeta$,
we learn that flipping the $\pin^+$ structure on one crosscap changes the value of $\t\zeta$.  

By locality, this statement is universal and not limited to the particular
case of a Klein bottle.   This leads to a simple statement about the sum over $\pin^+$ structures on any manifold.   Let $G_Y(N)$ be the sum of $\pin^+$ structures
on a manifold $Y$, weighted by $(-1)^{\frac{N}{2}\t\zeta }$.   If $Y$ is unorientable, it can be obtained by gluing a crosscap onto some other manifold $Y'$.   Pairs
of $\pin^+$ structures on $Y$ that differ by flipping the $\pin^+$ structure on this crosscap will, by locality, make canceling contributions to $G_Y(N)$ if $N=2$ mod 4. So for $N=2$ mod 4, only orientable manifolds contribute to the partition function after the sum over $\pin^+$ structures.  This is consistent with the random matrix class identified above, consisting of two identical GUE-like blocks.

On the other hand, for $N = 0$ mod 4, unorientable manifolds do contribute, and lead to something new. A difference between $\pin^+$ structures and $\pin^-$ structures is that if $Y$ is a $\pin^+$ manifold with boundary, it can have an odd number of boundary components of R type.
In fact, the basic crosscap spacetime of fig.~\ref{CrossCap} -- a trumpet that ends on a crosscap -- is an example.  The crosscap spacetime has two $\pin^+$ structures, both of which are of R type on the boundary.   Hence this spacetime contributes to $\left\langle Z_{\Ra}(\beta)\right\rangle$, not to $\left\langle Z_{\NS}(\beta)\right\rangle$.   (This is opposite from the $\pin^-$ case;
a $\pin^-$ structure on the crosscap spacetime is of NS type on the boundary.)  This is consistent with random matrix theory: for $N=0$ mod 4, the Hamiltonian consists of a GOE block with $(-1)^\sF=1$ and a GSE block with $(-1)^\sF=-1$. The crosscap (or genus $1/2$) contributions
in these two random matrix ensembles are equal in magnitude but opposite in sign, so they cancel in $Z_{\NS}(\beta)$ and add in $Z_{\Ra}(\beta)$.

One can go on and look at the connected correlations functions $\langle Z_{\NS}(\beta)^2\rangle_c$ and
$\langle Z_{\Ra}(\beta)^2\rangle_c$.    The lowest order contributions will come from the double trumpet and the twisted double trumpet.

For $N= 0$ mod 4, the double trumpet and twisted double trumpet have two $\pin^+$ structures, each of which contributes $+1$, regardless of whether
the boundaries are of NS or R type.   Thus the contributions to  $\left\langle Z_{\NS}(\beta)^2\right\rangle_c$ and
$\left\langle Z_{\Ra}(\beta)^2\right\rangle_c$ are equal. The cross-correlator $\langle Z_{\Ra}(\beta)Z_{\NS}(\beta)\rangle_c$ is nonzero, but will receive contributions only from spacetimes with odd Euler characteristic. The leading term  would be from $\RP^2$ with two holes removed.

\begin{table}[t]
\small
\begin{center}
\begin{tabular}{c|c|c|c}
$N$ mod 4& RMT class & $Z_{g,n,n_{\Ra}}$ & pin$^+$ structure sum identity\\
\hline
\hline
0 & $\left(\begin{array}{cc}\hspace{-5pt}\text{GOE}\hspace{-12pt} & 0 \\ 0 & \text{GSE}\hspace{-5pt}\end{array}\right)$ & $2^{2g+n-2}\Big( Z_{g,n}^{\text{GOE}} + (-1)^{n_{\Ra}}Z_{g,n}^{\text{GSE}}\Big)$ & $\sum\limits_{\text{pin}^+} 1= 2^{2g+n-2}\big(1 + (-1)^{n_{\Ra}+n_c}\big)$\\
\hline
2 & $\left(\begin{array}{cc}\hspace{-5pt}\text{GUE}\hspace{-12pt} & 0 \\ 0 & \text{GUE}\hspace{-5pt}\end{array}\right)$ & $\begin{cases}  2^{2g+2n-2}Z_{g,n}^{\text{GUE}}& n_{\Ra}  = 0 \\ 0 & \text{else}\end{cases}$ & $\sum\limits_{\text{pin}^+}(-1)^{\widetilde{\zeta}} = \begin{cases}2^{2g+n-1}
& \substack{\hspace{-1pt}n_c\, = \;0 \\ n_{\Ra}\, = \;\text{0}}\\
0 & n_c\neq 0\end{cases}$
\end{tabular}
\caption{{\small Reductions to GOE, GUE, GSE as in table \ref{tableReduction2}, but now for the cases with $\sT^2 = (-1)^\sF$ classically. For $N=2$ mod 4,
$\sT$ exchanges the two GUE blocks.}}\label{tableReduction3}
\end{center}
\end{table}

 For $N=2$ mod 4, the double trumpet and twisted double trumpet contributions
to the correlator
 $\left\langle Z_{\Ra}(\beta)\right\rangle_c$ cancel.   The cancellation happens because, with the boundaries being of R type, one has
$(-1)^{\t\zeta}=1$ for a $\pin^+$ structure on the double trumpet and $(-1)^{\t\zeta}=-1$ for a $\pin^+$ structure on the twisted double trumpet.   To prove this,
using locality, one can glue the two boundaries together, so that the double trumpet and the twisted double trumpet become a torus and a Klein bottle.  
The torus is orientable, so it has $\t\zeta=0$ for any $\pin^+$ structure, and on the Klein bottle, since we are now considering $\pin^+$ structures with $\alpha=0$ in the sense of 
eqn (\ref{kbspintwo}), we have $\t\zeta=1$.  This cancellation is expected in random matrix theory because $Z_{\Ra}(\beta)$ vanishes identically, without fluctuations.  

As in the previous cases, one can make a general argument that pin$^+$ structure sums ensure that random matrix theory answer agrees with the bulk. See table \ref{tableReduction3} for the necessary  $\pin^+$ identities, which follow from the analog of (\ref{fullPinSum}).
 Concretely, the above considerations show that for $N = 0$ mod 4,  the orthogonal $\pin^+$ sum for any  NS or R boundary is 2, so we get eqn.~(\ref{fullPinSum})
 with $F_{\RP^2}(N)$ replaced by  $G_{\RP^2}(0)=2$.    In pin$^+$, crosscaps are in the R sector, so instead of imposing that the number of R boundaries should be even, we impose that the number of R boundaries plus the number of crosscaps should be even.     This  accounts for the  result  claimed in the table for $N=0$ mod 4.
 For $N=2$  mod 4, the orthogonal $\pin^+$ sum  vanishes for an R  boundary, and also $G_{\RP^2}(2)=0$.   So we restrict to $n_\Ra=n_c=0$.
 The orthogonal $\pin^+$ sum still  gives a factor of 2 for every NS boundary. In contrast to the $\pin^-$ case, the $\pin^+$  sum on an orientable
 surface $Y_0$ of genus $g_0$ is $2^{2g_0}$ for any $N$.   These facts lead to the result shown in the table for $N=2$ mod 4.

\subsection{Including $\mathcal{N} = 1$ Supersymmetry}\label{SUSY}

\subsubsection{The Supersymmetric SYK Model}

The supersymmetric SYK model, with minimal or $\N=1$ supersymmetry, is constructed \cite{Fu:2016vas}  by starting 
with a quantum mechanical system of $N$ Majorana fermions
and introducing a random supercharge
\be\label{susy}Q = \i^{\frac{\h q-1}{2}} \sum_{a_1a_2\cdots a_{\h q}}\t j_{a_1a_2\cdots a_{\h q}}\psi_{a_1}\psi_{a_2}\cdots \psi_{a_{\h q}}. \ee
Here $\h q$ is an odd integer $\geq 3$. The Hamiltonian is defined as 
\be\label{polko}H=Q^2. \ee
The assertion that the model is supersymmetric just means that $[H,Q]=0$.

If $Q$ is a homogeneous function of the elementary fermions, as assumed in eqn.~(\ref{susy}), then
 the supersymmetric SYK model inevitably has $\Z_2^\sT\times \Z_2^\sF$ symmetry (possibly with a quantum anomaly).   Indeed, if $\sT$ is a time-reversal symmetry
that commutes with the elementary fermions, then
\be\label{folko}\sT Q= (-1)^{(\h q-1)/2} Q\sT.\ee
Thus $Q$ is either even or odd under $\sT$, and therefore $H$ is even. However, by allowing $Q$ to be a sum of terms with different values of $\h q$, one can break the time-reversal symmetry.

Regardless of how $Q$ is constructed,
 $\sT^2=(-1)^\sF$ is not compatible with $\N=1$ supersymmetry.   If $\sT^2=(-1)^\sF$, then fermionic operators transform under $\sT$ in even-dimensional representations,
 but in the case of $\N=1$ supersymmetry, $Q$ is a fermionic operator that is in a 1-dimensional representation.

As in a nonsupersymmetric SYK model, we can exchange $\sT$ with $\sT (-1)^\sF$, and this will change the sign of the anomaly.   But it also reverses
the sign in eqn.~(\ref{folko}).   This sign is important in the random matrix classification, and also in the bulk supergravity.   So unlike the nonsupersymmetric case,
it is not true that models differing by $N\to 8-N$ will be equivalent.   Rather, a model defined by a pair $(N,\hat q)$ will be equivalent to a model with parameters
$(8-N,\hat q +2)$.   So, for example, it suffices to consider only the case $\hat q =1$ mod 4 if we consider all values of $N$ mod 8, as  in Table 4  of the next section.

\subsubsection{Random Matrix Classes Including Supersymmetry}\label{revpairs}
The random matrix classification of the supersymmetric SYK model was studied in \cite{Li:2017hdt,Kanazawa:2017dpd,Sun:2019yqp}. We will give a self-contained description. First consider the case without time-reversal symmetry, so the only global symmetry is the supercharge $Q$ and a possible $(-1)^\sF$ symmetry. If $N$ is even, there is a $(-1)^\sF$ 
symmetry. The two blocks  of states with $(-1)^\sF=1$ or with $(-1)^\sF=-1$ each have dimension $L=2^{N/2-1}$.    The unitary group acting on either block is $\U(L)$,
so the full group of unitary transformations that commutes with $(-1)^\sF$ is $G=\U(L)\times \U(L)$.   $Q$ anticommutes with $(-1)^\sF$, 
so in a basis in which  $(-1)^\sF$ is block diagonal, $Q$ has only off-diagonal blocks:
\be\label{blocks}Q=\begin{pmatrix}0& C\cr C^\dagger & 0 \end{pmatrix}.\ee
Here $C$ is a complex $L\times L$ matrix obeying no constraint, and $C^\dagger$ is its hermitian adjoint.    $C$ transforms as a bifundamental of $\U(L)\times \U(L)$.
Thus the random matrix statistics will correspond to one of the Altland-Zirnbauer ensembles that were described in section \ref{ensembles}. 
  If $N$  is odd,  then there is no $(-1)^\sF$ symmetry.
The symmetry group $G$ reduces to $\U(L)$ (with now $L=2^{(N-1)/2}$) and $Q$ is simply a hermitian matrix, governed by GUE statistics.

Returning to the  case that there is a $(-1)^\sF$ symmetry, there is a possible generalization that does  not 
occur in the SYK model but that is natural from a random matrix
point of view.   We will interpret it in JT supergravity in section \ref{rpunctures}.
One can assume that the blocks of bosonic and fermionic  states have different dimensions $L+\nu$ and $L$, for some integer  $\nu$.   The symmetry  group  is then
$\U(L+\nu)\times \U(L)$; $C$ is  still a bifundamental.   This corresponds  to  a supersymmetric matrix ensemble in which supersymmetry  is unbroken, the value of
the supersymmetric index being $\Tr\,(-1)^\sF=\nu$. The same generalization is possible in the other examples with bifundamentals.

\begin{table}[t]
\small
\begin{center}
\begin{tabular}{c | c | c | c | c | c | c | c}
$N$ mod 2 & symmetry & $\lambda\rightarrow 0$ enhancement  &  $\lambda_1\rightarrow \lambda_2$ enhancement& $\upalpha$ & $\upbeta$& $\upgamma$ &  bulk TFT\\
\hline
\hline
0 & $\U(L)\times \U(L)$ bif.& $\U(1)\rightarrow \U(1)\times \U(1)$ & $\U(1)\times \U(1)\rightarrow \U(2)$ & $1$ & $2$ & $2$&$ (-1)^\zeta$\\
1 & $\U(L)$ adjoint & $\U(1)\rightarrow \U(1)$ & $\U(1)\times \U(1)\rightarrow \U(2)$&  & $2$  &$\sqrt{2}$&  1
\end{tabular}
\caption{{\small Random matrix symmetry classification for the supercharge $Q$ in the $\mathcal{N} = 1$ super SYK model with $N$ fermions and with no time-reversal symmetry.
The matrix field is a bifundamental of $\U(L)\times \U(L)$ or an adjoint of $\U(L)$, as indicated in the second column.  In the third and fourth columns, we give the symmetry enhancements described in section \ref{measure}. The dimensions of these enhancements directly give $\upalpha$ and $\upbeta$, also shown. The variable $\upgamma$ is the degeneracy of eigenvalues, multiplied by $\sqrt{2}$ for the case of odd $N$, see (\ref{defUpgamma}). In the first row, we have an Altland-Zirnbauer $(\upalpha,\upbeta)$ ensemble, and in the second row, we have an ordinary $\upbeta = 2$ (GUE-like) ensemble (so no value of $\upalpha$ is given).}}\label{table2}
\end{center}
\end{table}
\begin{table}[t]
\small
\begin{center}
\begin{tabular}{c | c | c | c | c | c | c | c|c| c}
$N$& symmetry &$\lambda\rightarrow 0$ enhancement&$\lambda_i\rightarrow \lambda_j$ enhancement & $\mathsf{T}_+^2$ & $\mathsf{T}_-^2$ & $\upalpha$ & $\upbeta$& $\upgamma$&  bulk\\
\hline
\hline
0 & $\O(L)\hspace{-3pt}\times \hspace{-3pt}\O(L)$ bif. &$1\rightarrow 1$&$1\rightarrow \O(2)$& $+1$ & $+1$ & $0$ & 1 &  $2$&$e^{\mathrm{i}\pi \eta/2}$\\
1 & $\O(L)$ symm.  &$1\rightarrow 1$&$1\rightarrow \O(2)$& $+1$ &  & & 1 & $\sqrt{2}$&1\\
 2 & $\U(L)$ symm.  &$1\rightarrow \U(1)$&$1\rightarrow \U(1)$&$+1$ & $-1$ & 1 & 1 & $2$& $e^{-\mathrm{i}\pi \eta/2}$\\
 3 & $\Sp(L)$ symm. &$\U(1)\rightarrow \Sp(2)$&$\U(1)\hspace{-3pt}\times \hspace{-3pt}\U(1)\rightarrow \U(2)$&  & $-1$ & 2 & 2  &$2\sqrt{2}$& $e^{-2\mathrm{i}\pi \eta/2}$\\
 4 & $\Sp(L)\hspace{-3pt}\times \hspace{-3pt}\Sp(L)$  bif. &$\Sp(2)\rightarrow \Sp(2)\hspace{-3pt}\times \hspace{-3pt}\Sp(2)$&$\Sp(2)\hspace{-3pt}\times\hspace{-3pt} \Sp(2)\rightarrow \Sp(4)$&$-1$ & $-1$ & 3 & 4  & $4$&$e^{-3\mathrm{i}\pi \eta/2}$\\
 5 & $\Sp(L)$ anti. &$\Sp(2)\rightarrow \Sp(2)$&$\Sp(2)\hspace{-3pt}\times\hspace{-3pt} \Sp(2)\rightarrow \Sp(4)$&$-1$ &  &  & 4& $2\sqrt{2}$&$e^{-4\mathrm{i}\pi \eta/2}$\\
 6 & $\U(L)$ anti.  &$\SU(2)\rightarrow \U(2)$&$\SU(2)\hspace{-3pt}\times\hspace{-3pt} \SU(2)\rightarrow \Sp(4)$&$-1$ & $+1$ & 1 & 4 & $4$& $e^{-5\mathrm{i}\pi \eta/2}$\\
 7 & $\O(L)$ anti.  &$\O(2)\rightarrow \O(2)$&$\O(2)\hspace{-3pt}\times\hspace{-3pt} \O(2)\rightarrow \U(2)$& & $+1$ & 0 & 2 & $2\sqrt{2}$& $e^{-6\mathrm{i}\pi \eta/2}$
\end{tabular}
\caption{{\small Random matrix symmetry classification for the supercharge $Q$ in the $\mathcal{N} = 1$ super SYK model with $N$ fermions and with $\widehat{q} = 1$ mod 4.  
The first column is $N$ mod 8. The second column indicates the symmetry group and whether the matrix is a bifundamental of a product group or
a  symmetric or antisymmetric  second rank tensor.  The next two columns indicate the symmetry enhancement when an eigenvalue
goes to zero or two eigenvalues become equal (here 1 denotes the trivial group).  The operators $\mathsf{T}_\pm$ are antiunitary operators built out of $\sT$ and $(-1)^\sF$ so that they commute (+) or anticommute ($-$) with the supercharge $Q$; a blank means that there is no such operator. For the cases where $\upalpha$ is blank, the matrix ensemble for $Q$ is a standard Dyson $\upbeta$ ensemble. The others are Altland-Zirnbauer $(\upalpha,\upbeta)$ ensembles. The values of $\upalpha,\upbeta$ follow
from the symmetry enhancements, as explained in section \ref{measure}. $\upgamma$ is the degeneracy, multiplied by $\sqrt{2}$ when $N$ is odd.  The last column  shows a factor that must be included in the sum over $\pin^-$ structures.
(The corresponding table for $\hat q=3$ mod 4 is obtained from this one by $N\to 8-N$ mod 8.)}}\label{table1}
\end{center}
\end{table}

Now we consider the case with time-reversal symmetry, but with $N$ even so that $Q$ has the block structure in eqn.~(\ref{blocks}). Then the role of $\sT$ is to reduce the symmetry group
and place a constraint on $C$.    If $N=0$ mod 8, then $\sT^2=1$ and the symmetry group that commutes with $\sT$ and with $(-1)^\sF$ is $G=\O(L)\times \O(L)$.
$C$ is a bifundamental of this group.  If $N=4$ mod 8, then $\sT^2=-1$,  the symmetry group is $G=\Sp(L)\times \Sp(L)$, and $C$ is again a bifundamental of $G$.
These are two more of the Altland-Zirnbauer ensembles.   If $N=2$ or 6 mod 8, then $\sT$ exchanges the two blocks, and the analysis is slightly more subtle.    
Since $\sT$ exchanges the two blocks, the symmetry group that commutes with $\sT$ (and with $(-1)^\sF$) is just $G=\U(L)$: one can make an arbitrary unitary
transformation of the upper block, accompanied by a $\sT$-conjugate transformation of the lower block.   Let $\H_+$ and $\H_-$ be the subspaces of the super SYK
Hilbert space $\H$ corresponding to the eigenvalue $+1$ or $-1$ of $(-1)^\sF$, respectively.  Since $\H_+$ and $\H_-$ are exchanged by the antiunitary symmetry $\sT$,
they are naturally dual.   Hence instead of thinking of $C$ as a linear map from $\H_+$ to $\H_-$, we can think of it as a bilinear map $\H_+\otimes \H_+\to \C$
or in other words as a second rank tensor.   But we need to decide if $C$ is a symmetric or antisymmetric second rank tensor, and this point is somewhat subtle.  
Denoting as $\mathrm{Spin}(N)$ the group that rotates the $N$ elementary fermions of the SYK model, the elementary fermions transform in the vector
representation $V$ of this group, and  the Hilbert spaces $\H_+$ and $\H_-$ transform, respectively, as the positive and negative
chirality spinor representations, which we will call $S_+$ and $S_-$.   
Suppose that $Q$ is homogeneous of degree $\h q$ in the elementary fermions.   Then $Q$ and $C$ transform in the representation $\wedge^{\h q}V$ (the antisymmetric product of 
$\h q$ copies of $V$).   To decide if $C$ is a symmetric or antisymmetric tensor, we just need to know if $\wedge^{\h q}V$ appears symmetrically or antisymmetrically in
$S_+\otimes S_+$.   For $\h q=1$ mod 4, the group theory answer is that $\wedge^{\h q} V$ appears symmetrically if $N=2$ mod 8 and antisymmetrically if $N=6$ mod 8; these
statements are reversed if $\h q=3$ mod 4.   So for $N=2$ or $6$ mod 8,  we get two more Altland-Zirnbauer ensembles; which one occurs for $N=2$ mod 8 and which
for $N=6$ mod 8 depends  on $\h  q$.

With time-reversal and odd $N$, there is only one block in $Q$.    The
symmetry group is reduced from $\U(L)$ to $\O(L)$ if $\sT^2=1$, which happens for $N=1$ or 7 mod 8,  or to $\Sp(L)$ if $\sT^2=-1$,
as happens for $N=3$ or 5 mod 8 (see section \ref{pinminus} for these statements).   As detailed momentarily, $\sT$ either commutes or anticommutes with $Q$, depending on $N$ and $\h q$.
If $Q$ commutes with $\sT$, then $Q$  is governed by GOE statistics for $\sT^2=1$ or GSE statistics for $\sT^2=-1$.    But if $\sT$ anticommutes with $Q$, then $Q$ is governed
by an ensemble that we have not yet encountered.   If $\sT^2=1$ and $\sT$ anticommutes with $Q$, then $Q=\i M$, where $M$ is a real antisymmetric second
rank tensor of $\O(L)$ (equivalently, an element of the adjoint representation of $\O(L)$).       If $\sT^2=-1$, then $Q$ transforms as a symmetric tensor\footnote{Let $M_{ij}=M_{ji}$
be a symmetric tensor of $\Sp(L)$.   Since the fundamental representation of $\Sp(L)$ is  not real, the components of $M_{ij}$ are not real, but $M$ can satisfy an $\Sp(L)$-invariant
reality condition $\overline M^{ij}=-\varepsilon^{ii'}\varepsilon^{jj'} M_{i'j'}$, where $\varepsilon^{ij}$ is the invariant antisymmetric tensor of $\Sp(L)$ (which we take to  be real).   
Equivalently,  with $\varepsilon$ and $M$ viewed as matrices, the reality condition on $M$ is
$M^\dagger =\varepsilon M \varepsilon$.
Then the matrix $Q^i{}_j=\varepsilon^{ik}M_{kj}$ is hermitian, since $Q^\dagger=(\varepsilon M)^\dagger=
M^\dagger \varepsilon^\dagger= \varepsilon M \varepsilon \varepsilon^\dagger =\varepsilon M = Q$, where we use the fact that $\varepsilon \varepsilon^\dagger=1$.
The hermitian matrix $Q$  is an element of the adjoint representation of $\Sp(L)$.  It is odd under $\sT$, since $\sT(Q)= \varepsilon \overline Q \varepsilon^{-1}
=\varepsilon\overline{\varepsilon M}\varepsilon^{-1}=\overline M \varepsilon =-\varepsilon M =-Q$. We used $\overline \varepsilon=\varepsilon$,
$\varepsilon^2=-1$.}
of $\Sp(L)$ (equivalently, the adjoint representation of $\Sp(L)$).    
  These
are the last two Altland-Zirnbauer ensembles.

We give a summary of this classification in tables \ref{table2} and \ref{table1}. In these tables we also list the values of $\upalpha$ and $\upbeta$, along with a variable $\upgamma$ that characterizes the degeneracy of levels in $Q^2$ and that will be used in section \ref{sec:JT}. In table \ref{table1} for the time-reversal-invariant case, we also give the values of $\sT_+^2$ and $\sT_-^2$. These are defined as the antiunitary operators that either commute (+) or anticommute (-) with $Q$. They are related to $\sT$ and $\sT (-1)^\sF$ in a way that depends on $N$.

\subsubsection{Bulk Description Including Supersymmetry}\label{sect:superBulkDescription}
A bulk description of the supersymmetric SYK model involves JT supergravity at low energies. This motivates a conjecture that pure JT supergravity, together with the appropriate topological field theory, should be dual to matrix integrals of the types just discussed. To determine the topological field theory, we match the anomalies in the SYK model. 

In the case without time-reversal symmetry, the potential anomaly is in $(-1)^\sF$. In the bulk we sum over spin structures on orientable manifolds, with the
weights   $1$ or $(-1)^\zeta$. Naively, one would identify the trivial theory with the non-anomalous random matrix class (even $N$). However, there is an important subtlety. Ordinary JT gravity includes a ``Schwarzian'' mode, which is a bosonic mode that propagates along the boundary. In JT supergravity, the analog is a Schwarzian supermultiplet, a supermultiplet that propagates along the boundary \cite{Fu:2016vas,Forste:2017kwy}.   In particular,
the Schwarzian supermultiplet includes a Majorana fermion. This boundary mode contributes to the anomaly. 

To see how this affects things, suppose we start with a boundary theory without the $(-1)^\sF$ anomaly. We look for a bulk dual description in terms of JT supergravity, including the Schwarzian supermultiplet, plus a bulk topological field theory. The Schwarzian supermultiplet by itself has the $(-1)^\sF$ anomaly, so the bulk topological field theory must cancel it. We conclude that the bulk description of the non-anomalous (even $N$) theory should include $(-1)^\zeta$, and the bulk description of the anomalous (odd $N$) theory should be the trivial theory.

The logic for the case with time-reversal symmetry is similar. In the bulk, we sum over orientable and unorientable manifolds. To match the $\Z_2^\sT\times \Z_2^\sF$ symmetry, we sum over pin$^-$ structures, and the possible topological field theory is $\exp(-\mathrm{i}\pi \eta N'/2)$ with $N'$ an integer mod 8. To relate this $N'$ to the $N$ of SYK, we have to keep track of the contribution to the anomaly of the Schwarzian supermultiplet. The fermion in this multiplet transforms under $\sT$ like the supercharge $Q$.
That means that its contribution to the anomaly is just the sign in eqn.~(\ref{folko}), or $(-1)^{(\h q-1)/2}$.   

Now, suppose we start with a system of $N$ Majorana fermions. We want to match the anomaly in the bulk. The Schwarzian supermultiplet contributes $(-1)^{(\h q-1)/2}$ to the anomaly,
so the bulk topological field theory must contribute what remains, or 
\be\label{nubbo}N'= N-(-1)^{(\h q-1)/2}. \ee

Note that unlike the cases studied in sections \ref{sec:spinButNoT} and \ref{sec:bothSpinAndT}, here we cannot use spin structure sums to reduce the different cases to the GOE, GUE, GSE cases. The bulk reason is simply that JT supergravity contains fermions, so the path integral depends on the spin or pin structure in a nontrivial way.

As a final comment, one  might wonder if JT supergravity plus topological field theory would be unable to distinguish $N'$ from $8-N'$ mod 8, on the grounds that
exchanging $N'$ with $8-N'$ could be compensated by exchanging a $\pin^-$ bundle $\P$ with the complementary one $\P'$.    This is not the
case, because the path integral of JT supergravity is sensitive to the difference between $\P$ and $\P'$, basically because it knows how the gravitino
field transforms under $\sT$.

\section{Torsion}\label{sectionThree}

Our next goal is to understand the  path  integral of JT gravity or supergravity on an unorientable two-manifold.
For this we will have to understand the ``torsion.''   Since JT gravity or supergravity can be formulated as a $BF$ theory, 
we start by explaining how torsion is related to $BF$ theory.  
  
Torsion in topology originally meant the combinatorial torsion of Reidemeister  \cite{Reidemeister}.
Roughly, on a triangulated manifold on which a flat background gauge field is given, one defines the torsion (or R-torsion or combinatorial torsion) 
as a certain product of determinants which is somewhat miraculously
independent of how the manifold was triangulated.   One can calculate the torsion explicitly by triangulating a manifold in a very crude way with only a few
lattice points.   On the other hand, by taking a very fine triangulation, one can approximate a continuum limit.   That last fact, in hindsight, could be a clue that
torsion has a quantum field theory  interpretation.  Our interest in  the torsion is related to the fact that it has a continuum limit, but in our calculations later, we will take
advantage of the fact that it is possible to use a crude triangulation.    

Long after Reidemeister's construction, 
a differential geometric or analytic version of the torsion was defined by Ray and Singer \cite{RS}.   Analytic torsion is defined in terms of a product
of determinants of differential operators.   The differential operators that appear are the Laplace-like operators that arise in quantizing gauge theories.  With hindsight,
this  could
serve as a clue that this form of the torsion is related to a one-loop approximation to gauge theory.   Indeed, it was shown by A. Schwarz that
Ray-Singer analytic torsion can arise as the partition function of what is now called $BF$ theory \cite{Schwarz}.  Later, this result was an ingredient in understanding  the one-loop
approximation to Chern-Simons theory in three dimensions \cite{WittenChernSimons}.

 Ray and Singer introduced $\zeta$-function
regularization in order to define their determinants.  Combinatorial torsion can be viewed as a lattice regularization of the same determinants.   From a physical point of
view, one might expect two regularizations of the same theory to give equivalent results, possibly after adjusting some local counterterms.    In the present
case,  no such counterterms arise in odd dimensions, but there are some in even dimensions.\footnote{In the two-dimensional case relevant to the present paper, there is one
possible counterterm that is consistent with the fact that the analytic and combinatorial torsions are both topological invariants.    This is a factor $\exp(w\chi)$,
where $w$ is a constant and $\chi$ is the Euler characteristic of a manifold.  Such a factor, in general, should be expected whenever one compares two formulations
of  $BF$ theory.   The analytic and combinatorial torsions, however, have been defined in a way that avoids such a factor.}   In hindsight, this might have provided a reason to suspect that combinatorial
and analytic torsion are equivalent.   The equivalence, which was conjectured by Ray and Singer,
 was proved by Cheeger \cite{Cheeger} and M\"{u}ller \cite{Muller}, originally for compact gauge groups.
The proof was later generalized to a larger class of groups that includes noncompact but semi-simple Lie groups such as $SL(2,\R)$ \cite{Muller2,bismut1991metriques,bismut1992extension}.  

When  there are no zero-modes, the torsion is a number.  More generally, under favorable conditions,\footnote{Quantizing a $BF$ theory in higher dimensions requires introducing a hierarchy of ``ghosts for ghosts,'' as explained in \cite{Schwarz}. 
When the higher order ghosts have zero-modes, the torsion does not have a simple interpretation as a measure on a moduli space of classical solutions.}
 such as prevail in the two-dimensional case that
will be of interest here,
 the torsion is instead a measure on the moduli space of flat
connections.   On an even-dimensional orientable manifold (without boundary), the torsion is often said to be ``trivial.''   Triviality  means that the torsion is 1 if there are no zero-modes.  When
the torsion is a measure on moduli space, the ``triviality'' means that this measure can be defined by a simple classical formula, without any quantum correction.
However, on an unorientable two-manifold (or an orientable one with boundary), there is no such ``triviality'' and there is a quantum correction.   For compact
gauge groups, volumes of moduli spaces of flat connections on orientable and unorientable two-manifolds were computed in \cite{WittenGauge} using the combinatorial
definition of the torsion.\footnote{Some important early ideas on this and related matters had been developed in unpublished work by S. Axelrod.}    
Here we will perform an analogous computation for gauge group $\SL(2,\R)$ and its supersymmetric analog $\OSp(1|2)$.   There is a technical difference in the way we will
describe the answer.    In \cite{WittenGauge}, inspired by what was known about two-dimensional Yang-Mills theory \cite{Migdal}, 
the torsion was written in terms of a sum over characters
of the gauge group (which provide a basis for the space of physical states).   While such a representation might be possible for $\SL(2,\R)$ or $\OSp(1|2)$, it turns out to
be very convenient instead to calculate directly with Fenchel-Nielsen or length-twist coordinates.

\subsection{Analytic Torsion}\label{anator}

Let us first review the relationship between $BF$ theory and the torsion.    We consider a theory with a gauge group $G$ whose Lie algebra $\g$ admits an invariant,
nondegenerate quadratic form that we will denote as $\Tr$.   The fields will be a gauge field $A$ with field strength $F=\d A+A\wedge A$,
and an adjoint-valued spin zero field $B$.   The action on a two-manifold $Y$ is
\be\label{thaction}I= -\i\int_Y\Tr \, B F. \ee
This is well-defined even for unorientable $Y$, provided that we take $B$ to be a pseudoscalar.   The path integral of this theory is one-loop exact,
since the integral over $B$ gives a delta function setting $F=0$:
\be\label{waction}\int \D B\exp(- I)=\int \D B \exp(\i\int \Tr \, B F)=\delta(F).  \ee
After gauge-fixing, the  delta function localizes $A$ on the moduli space of flat connections modulo gauge transformations.    The resulting measure on the moduli
space is given by a product of determinants -- a ghost determinant that arises in gauge fixing, and a second determinant that arises in integrating over $A$ after gauge-fixing
with the help of the $\delta(F)$ in eqn.~(\ref{waction}).  

Though it is possible to continue in precisely this way, we will take a small detour so as to be able to present the derivation in
a more familiar language.   Let us add to the action $I$ a $B^2$ term with a small coefficient $\varepsilon$:
\be\label{haction}I_\varepsilon= -\i\int_Y\Tr \, B F-\frac{\varepsilon}{2}\int_Y\d^2x \sqrt g \Tr \,B^2, \ee
where $g$ is a Riemannian metric on $Y$.
Integrating out $B$, we get
\be\label{laction}I'_\varepsilon = -\frac{1}{2\varepsilon} \int\d^2x\sqrt g \Tr F^2. \ee
This is simply weakly coupled Yang-Mills theory.   The one-loop approximation (which is not exact in Yang-Mills theory, but is exact in the limit $\varepsilon\to 0$,
in which we return to the original $BF$ theory) is  given by a standard formula.   Suppose that we  are expanding around a background flat connection $A_0$,
say with $A=A_0+{\sf a}$ where $\sf a$ describes the fluctuations.   Let $\d$ be the exterior derivative  and $D=\d+[A_0,\cdot]$ its gauge-covariant extension, mapping
adjoint-valued $q$-forms to adjoint-valued $(q+1)$-forms for $q=0,1$.   Its adjoint $D^*$ maps $q$-forms to $(q-1)$-forms for $q=2,1$.   The Laplacian acting on adjoint-valued
differential forms is $\Delta=D^*D+DD^*$.    We write $\Delta_q$ for the Laplacian acting on adjoint-valued $q$-forms.   Then a standard computation shows that after gauge-fixing,
the one-loop
path integral of the Yang-Mills theory  gives a ratio of determinants
\be\label{onaction}Z_{1,A_0}= \frac{\det' \Delta_0}{\sqrt {\det' \Delta_1}}. \ee 
Here in a standard approach, the denominator comes from the Gaussian integral over $\sf a$ after gauge-fixing, and the numerator is the ghost determinant.  
The symbol $\det'$ represents a determinant in a subspace orthogonal to the zero-modes.  If $\Delta_1$ has zero-modes, this means that the classical solution $A_0$
around which we are expanding is not unique but represents a point in a moduli space $\M$ of classical solutions; in this case, $Z_{1,A_0}$ must be interpreted as a measure
on $\M$.   If $\Delta_0$ has zero modes, this means that the classical solution $A_0$ leaves unbroken a  positive-dimensional subgroup of $G$ and the Fadde'ev-Popov
gauge-fixing should be discussed more carefully.  This typically does not occur in two dimensions for nonabelian $G$. 

To understand the formula (\ref{onaction}), we need to take into account the following identity:
\be\label{identity}  \detp \Delta_1=\detp \Delta_0\,\, \detp \Delta_2. \ee
The origin of this identity is as follows.  
The Laplacian $\Delta$ commutes with $D$ and $D^*$, so if $\psi\in \Omega^0$ is an eigenstate of $\Delta_0$, then $D\psi\in \Omega^1$ is an eigenstate of
$\Delta_1$ with the same eigenvalue.    Likewise if $\chi\in\Omega^2$ is an eigenstate of $\Delta_2$, then $D^*\chi\in\Omega^1$ is an eigenstate of $\Delta_1$
with the same eigenvalue.   Hodge theory says every nonzero eigenstate of $\Delta_1$ can be identified in a unique way with an eigenstate of $\Delta_0$ or of $\Delta_2$
with the same eigenvalue, leading to the identity (\ref{identity}).    
In fact, Hodge theory gives an orthogonal decomposition of $\Omega^1$ as
\be\label{decompt} \Omega^1=\Omega^1_0 \oplus \Omega^1_2\oplus \Omega^1_{\mathrm{harm}}, \ee
where $\Omega^1_0$ consists of states of the form $D \psi,$ $\psi\in\Omega^0 $, $\Omega^1_2$ consists of the states $D^*\chi$, $\chi\in \Omega^2$,
and $\Omega^1_{\mathrm{harm}}$ is the space of zero-modes, which are known as harmonic forms.  

Using the identity (\ref{identity}), we get 
\be\label{zoneloop}Z_{1,A_0}=\frac{\sqrt{ \det' \Delta_0}}{\sqrt{ \det'\Delta_2}}. \ee
Now we can see why the one-loop correction is trivial if $Y$ is orientable (and without boundary\footnote{If $Y$ has
a boundary, one has to pick a boundary condition, which will typically not be invariant under the Hodge star operator.}), but not otherwise.   On an orientable manifold of dimension $\sf D$, one has the Hodge star operator
mapping $n$-forms to $({\sf D}-n)$-forms and ensuring that $\det' \Delta_n=\det' \Delta_{{\sf D}-n}$.   So in particular on an orientable two-manifold $Y$ without boundary,
$\det'\Delta_2=\det'\Delta_0$,
and the 1-loop correction is trivial.  On an unorientable manifold, $\det'\Delta_2\not=\det'\Delta_0$, and the 1-loop correction is nontrivial.

Now let us discuss what we do with $Z_{1,A_0}$ when it is nontrivial.
To define the path integral of gauge theory, one starts formally with a Riemannian metric on the fluctuation field $\sf a$,
\be\label{wengo}||a||^2=-\int_Y \d^2 x \sqrt g \Tr\, {\sf a}_i{\sf a}_j g^{ij}, \ee
which induces a Riemannian measure.  This measure (and a similar one for the ghosts and antighosts) is used, formally, in defining the functional integral.
The Riemannian metric on the space of all $\sf a$'s induces, in particular, a Riemannian metric on the space of zero-modes.  This in turn determines a Riemannian metric
on the moduli space $\M$ of classical solutions, and therefore this space gets a Riemannian measure $\mu_0$.   
The one-loop determinant corrects this measure to $\mu=\frac{\sqrt{\det'\Delta_0}}{\sqrt{\det'\Delta_2}}\mu_0$, and the 
partition function of $BF$ theory for a compact non-abelian gauge group
is obtained by integrating this measure over the moduli space $\M$ of flat connections:
\be\label{nebbo}Z_{BF}=\int_\M\d\mu =\int_\M\d \mu_0 \frac{\sqrt{\det'\Delta_0}}{\sqrt{\det'\Delta_2}}.\ee
(For a group such as $\SL(2,\R)$, one also wants to divide  by the mapping class group of $Y$ in order to get a finite answer.)   If $Y$ is orientable, the measure that is used here is the elementary one $\mu_0$ whose definition did not require $BF$ theory, but if $Y$ is unorientable, the quantum
correction $\sqrt{\det'\Delta_0/\det'\Delta_2}$ is important.\footnote{A further subtlety arises if the gauge group has a nontrivial center $\ZZ$.   If we are on a manifold
without boundary, because $\ZZ$ acts
trivially on any flat connection, the Fadde'ev-Popov recipe tells us to supplement the ghost determinant with  a   factor of $1/\# \ZZ$,    where $\# \ZZ$ is the number of elements of $\ZZ$.
For $\SL(2,\R)$, this would be a factor of $1/2$.   In our applications, we are usually interested in a two-manifold $Y$ with a nontrivial boundary.    In that
context, we only allow gauge transformations that are trivial along the boundary, so a constant gauge transformation by an element of $\ZZ$ is not allowed and we do not
divide by $\# \ZZ$.   On the contrary, if the boundary of $Y$ has $h$ components, then we have to  sum over $2^{h-1}$ inequivalent classes of gauge bundles on $Y$ that would
become equivalent if we allowed gauge transformations that are nontrivial on the boundary.   This factor was explained in section \ref{pinplus} in the context of the double trumpet,
which is an example with $h=2$.}

\begin{figure}
 \begin{center}
   \includegraphics[width=3in]{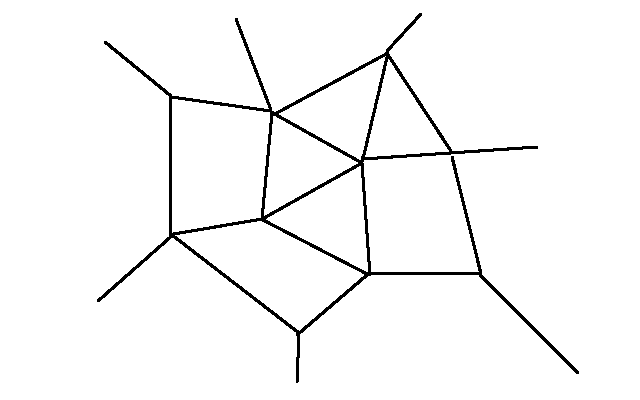}
 \end{center}
\caption{\small A two-manifold build by gluing together polygons (not necessarily triangles) on their boundaries.    \label{triangles}}
\end{figure}

\subsection{Combinatorial Torsion}\label{combitor}

It is possible to imitate  some of these formulas for a triangulated  two-manifold $Y$  and more generally for one  that has been built by gluing together polygons (fig.~\ref{triangles}).  This will lead us to the definition of the Reidemeister torsion.     By a $q$-cell we will mean one of the $q$-dimensional
building blocks in a covering by polygons (a vertex if $q=0$, an edge if $q=1$, and a polygon if $q=2$). 
We write $\S_q$ for the set of all these
 $q$-cells.
Such a covering is sometimes called a cell decomposition.   A somewhat more thorough explanation of the following can
be found, for example,  in \cite{WittenGauge}, section 4.\footnote{There was a much more thorough explanation in  a 1988 preprint by D. Johnson, ``A Geometric Form of Casson's Invariant and Its Connection
With Reidemeister Torsion.''  Unfortunately this article was unpublished and appears to be unavailable online.   Any reader with a copy is invited to share it.}

In the combinatorial approach to torsion, it is convenient to use a language that is dual  to some of what we have said so far. Instead of $q$-forms, one considers functions on $q$-cells
(every $q$-form defines such a function, namely its integral over a given $q$-cell)
and instead of the exterior derivative $D$ mapping $q$-forms to $(q+1)$-forms, one has the boundary operator $\partial$ mapping $q$-cells to $(q-1)$-cells.  The $q$-cells
are considered to be oriented, and if $W$ is a $q$-cell, then its boundary $\partial W$ is an oriented sum of $(q-1)$-cells.  

Suppose that on a two-manifold $Y$, we are given a flat $G$-bundle.  Let $E$ be the associated bundle in the adjoint representation.  
Any particular oriented $q$-cell $w$ is contractible, so when restricted to $w$, the bundle  $E$  has a trivialization by a vector space $\sV_w$ consisting of covariantly constant
sections\footnote{It can happen in a given covering by polygons that two or more vertices (if $q=1,2$) or edges (if $q=2$) of a given $q$-cell are glued together.   We ignore any such
boundary identifications in defining $\sV_w$.   One may say that $\sV_w$ consists of covariantly constant sections of $E$ over the interior of $w$, which is always
contractible, regardless of what gluing occurs on the boundary of $w$. 
If a given edge or vertex  appears more than once in $\partial w$, then we add the different contributions in defining $\partial s$ for $s\in \sV_w$.  }
of $E$ over $w$.  Then we define for $q=0,1,2$ a vector space that is the direct sum of the $\sV_w$ for all $w\in \S_q$:
\be\label{norko}\sC_q=\oplus_{w\in \S_q}\sV_w.\ee

Now we define maps $\partial:\sC_q\to \sC_{q-1}$ as follows.   Let $w$ be a $q$-cell and suppose that geometrically the boundary of $w$ is
$\partial w=\cup_i y_i$, with $y_i$ being $(q-1)$-cells.   Then  one defines $\partial: \sV_w\to \oplus_i \sV_{y_i}$ by $\partial s=\oplus_i s|_{y_i}$. 
More generally, remembering that all the cells are oriented and their orientations may not be compatible, we
write the boundary of $w$ as  $\partial w=\cup_i (-1)^{\lambda_i} y_i$, where $\lambda_i=0,1$ and a minus sign means that the orientation of $y_i$ is not the one that is induced
by the orientation of $w$.   Then we define $\partial s=\oplus_i (-1)^{\lambda_i} s|_{y_i}$.
Having defined maps $\partial: \sV_w\to \sC_{q-1}$ for all $w\in \S_q$, we just sum over $w$ to get $\partial: \sC_q\to \sC_{q-1}$.

The  Lie algebra $\g$ of $G$ can be given a $G$-invariant and translation-invariant measure, which is unique up to a constant multiple. 
For example, such a measure could be deduced from a $G$-invariant inner product on $\g$, which determines a flat $G$-invariant metric on $\g$ and hence a $G$-invariant
Riemannian measure.
  As each $\sV_w$ is a copy of $\g$, each $\sV_w$ gets a measure, and therefore each $\sC_q=\oplus_{w\in \S_q }\sV_w$ gets a measure,
which we will call $\alpha_q$.     These measures will be important in what follows.    Though there was an arbitrary choice of the underlying measure on
$\g$, it can be shown that a change in this measure will just multiply the combinatorial torsion by $\exp(w \chi)$, where $w$ is a constant and $\chi$ is the Euler characteristic
of $Y$.   The proof of this assertion uses the fact that $\chi=n_0-n_1+n_2$, where $n_q$ is the number of $q$-cells in the cell decomposition of $Y$, together with a simple
scaling argument using the  definition (\ref{mumbo}) of the torsion.   

Up to now, we have defined a linear transformation $\partial:\sC_q\to \sC_{q-1}$ for $q=1,2$; when we want to be more precise, we distinguish $\partial_2:\sC_2\to \sC_1$ and $\partial_1:\sC_1\to \sC_0$.    Because the boundary of a boundary vanishes,
these maps satisfy  $\partial_1\partial_2=0$.   That lets one define homology groups $H_q(Y,E)$, which will play a role shortly.  The homology groups
are defined as follows:   $H_2(Y,E)$ is the kernel of $\partial_2:\sC_2\to \sC_1$; $H_1(Y,E)$ is $\mathrm{ker}\,\partial_1/\mathrm{im}\,\partial_2$,
that is it is the quotient of the kernel of $\partial_1$ by the image of $\partial_2$; and $H_0(Y,E)$ is $\sC_0/\partial_1 \sC_1$, that is it is the quotient of $\sC_0$
by the image of $\partial_1$.   A standard fact in topology is that the vector spaces $H_q(Y,E)$ do not depend on the specific cell decomposition of $Y$ that was used
to define them.   

At this point, a quick way to define the Reidemeister or combinatorial torsion is to set $\hat\Delta_2=\partial_2^*\partial_2$, $\hat\Delta_0= \partial_1\partial_1^*$, and then we
can define the Reidemeister torsion:
\be\label{nubb} \tau^\Rr=\frac{\sqrt{\det' \hat\Delta_2}}{\sqrt{\det' \hat\Delta_0}}.\ee
This is exactly in parallel with eqn.~(\ref{zoneloop}) except that the roles of $q$ and $2-q$ have been exchanged.   This has happened because we have employed a dual description based on the operator $\partial$ which reduces $q$ by 1 rather than the operator $\d$ which increases $q$ by 1, used in
the analytic approach.
The remarkable property of $\tau^\Rr$ is that it does not depend on the  combinatorial  description, so we can use a crude triangulation, leading to simple
formulas, or a fine one, approaching a continuum limit.  The main step in showing that the choice of a cell decomposition  does not matter is to show that subdividing 
one of the polygons does not change $\tau^\Rr$.   See, for example, section 4 of  \cite{WittenGauge}.

 Both conceptually and for computational purposes, a slightly different formula for $\tau^\Rr$ is convenient.   In describing this, we will first assume that the 
homology groups $H_q(Y,E)$ vanish.   

We will use the measures $\alpha_q$ on the vector spaces $\sC_q$.   In general, if $V$ is a vector space and $\alpha$ is a measure on $V$, then to any basis
$v_1,v_2,\cdots, v_k$ of $V$, $\alpha$ assigns a number $\alpha(v_1,v_2,\cdots,v_k)$, which is a linear function of each of the $v_i$.    One can think of
 $\alpha(v_1,v_2,\cdots,v_k)$ as the volume of a parallelepiped that  has vertices at the origin in $V$ as well as the points $v_1,v_2,\cdots v_k\in V$.
If $M:V\to V$  is a linear transformation, then
\be\label{transrule} \alpha(M v_1, Mv_2,\cdots, Mv_k)=|\det \,M|\,\alpha(v_1,v_2,\cdots, v_k). \ee
This formula involves  $|\det\,M|$,  not $\det\,M$, because $\alpha$  is a measure, not a differential form of top degree.

Now if $H_2(Y,E)=0$, and $s_1,s_2,\cdots, s_{n_2}$ is any basis  $\sC_2$, then the vectors $\partial s_1,\partial s_2,\cdots,\partial s_{n_2}$ are linearly
independent in $\sC_1$, so they can be completed to a basis $\partial s_1,\partial s_2,\cdots, \partial s_{n_2},t_1,t_2,\cdots t_{n_1-n_2}$ of $\sC_1$.
The conditions $H_1(Y,E)=H_0(Y,E)=0$ mean that $\partial t_1,\partial t_2,\cdots, \partial t_{n_1-n_2}$ provide a basis of $\sC_0$.   Therefore, we have bases
of $\sC_2$, $\sC_1$, and $\sC_0$, and using
the measures $\alpha_q$ on these vector spaces, we can define
\be\label{numbo}\tau^\Rr=\frac{\alpha_2(s_1,s_2,\cdots, s_{n_2})\alpha_0(\partial t_1,\partial t_2,\cdots, \partial t_{n_1-n_2})}{\alpha_1(\partial s_1,\partial s_2,
\cdots \partial s_{n_2}, t_1, t_2,\cdots, t_{n_1-n_2})}. \ee
This formula does not depend on the choices of the $s_i$ or the $t_j$, basically because each $s_i$ and each $t_j$ appears in both the numerator and the denominator
in eqn.~(\ref{numbo}), so a rescaling of the $s_i$ or the $t_j$, or a more general linear transformation of the $s$'s or of the $t$'s, does not affect $\tau^\Rr$.
As we have already remarked, $\tau^\Rr$ also does not depend on the specific cell decomposition that was used.

However, for our application, we need to relax the assumption that the homology is trivial.   For some purposes, we can maintain the assumption $H_0(Y,E)=H_2(Y,E)=0$,
but we should not assume that $H_1(Y,E)=0$.   On the contrary, $H_1(Y,E)$ is the cotangent bundle to the moduli space $\M$ of flat bundles on $Y$,
at the point corresponding to a given flat bundle $E$.    Suppose that $\dim H_1(Y,E)=\dim\M = r$.   Then we should modify the above description as follows.
 To the vectors $\partial s_1,\partial s_2,\cdots, \partial s_{n_2}$, all of which are annihilated by $\partial_1$ since $\partial_1\partial_2=0$, we should adjoin
$r$ more basis vectors $u_1,u_2,\cdots, u_r$ that are annihilated by $\partial_1 $ and provide a basis of $H_1(Y,E)$.   Then  $\partial s_1,\partial s_2,\cdots, \partial s_{n_2},u_1,u_2,\cdots, u_r$ are a basis of the kernel of $\partial_1$.
We complete this
to a basis of $\sC_1$ by adding vectors $t_1,\cdots, t_{n_1-n_2-r}$ which are not annihilated by $\partial_1$, 
after which as before (since we assume $H_0(Y,E)=0$), $\partial t_1,\partial t_2,
\cdots ,\partial t_{n_1-n_2-r}$ provide a basis of $\sC_0$.   So we simply modify the definition of $\tau^\Rr$ by including the $u_i$ among the basis vectors of $\sC_1$:
\be\label{mumbo}\tau^\Rr=\frac{\alpha_2(s_1,s_2,\cdots, s_{n_2})\alpha_0(\partial t_1,\partial t_2,\cdots, \partial t_{n_1-n_2-r})}{\alpha_1(\partial s_1,\partial s_2,
\cdots \partial s_{n_2},u_1,u_2,\cdots, u_r,    t_1, t_2,\cdots, t_{n_1-n_2-r})}. \ee

As before, $\tau^\Rr$ is independent of the $s_i$ and the $t_j$, but it definitely does depend on the $u_k$, since they only appear in the denominator.  In fact,
if we rescale one of the $u$'s by a factor $\lambda$,  $\tau^\Rr$ is multiplied by a factor of $\lambda^{-1}$.   More generally, since the $u_k$ appear in the denominator,
$\tau^\Rr$ transforms in a change of basis as the {\it inverse} of a measure on $H_1(Y,E)$,   The inverse of a measure on a vector space can be regarded as a measure
on its dual space.   In the present context, $H_1(Y,E)$ is the cotangent bundle of the moduli space $\M$ of flat connections on $Y$, and its dual space\footnote{The distinction
between the tangent and cotangent bundle of $\M$ is only important if $Y$ is unorientable, because if $Y$ is orientable, $\M$ has a symplectic structure that does not
depend on any choice of a metric on $Y$, and this can be used to identify the tangent and cotangent bundles of $\M$.}  is
$H^1(Y,E)$, the tangent bundle to $\M$.   A measure on the tangent space to $\M$ at every point in $\M$ determines a measure on $\M$.   In other words, $\tau^\Rr$ is a measure on 
$\M$.

In a combinatorial approach, the partition function of $BF$ theory is the volume of $\M$ computed with this measure:
\be\label{kilo} Z_{BF} =\int_\M \d \tau^\Rr. \ee

If $H^0(Y,E)\not=0$, then we need to add more basis vectors $v_1,\cdots,v_k$ of $\sC_0$, and the definition becomes 
\be\label{tmumbo}\tau^\Rr=\frac{\alpha_2(s_1,s_2,\cdots, s_{n_2})\alpha_0(\partial t_1,\partial t_2,\cdots, \partial t_{n_1-n_2-r},v_1,v_2,\cdots,v_k)}{\alpha_1(\partial s_1,\partial s_2,
\cdots \partial s_{n_2},u_1,u_2,\cdots, u_r,    t_1, t_2,\cdots, t_{n_1-n_2-r})}. \ee

From the foregoing, it is clear that $\tau^\Rr$ can be effectively calculated, using a simple covering of $Y$ by polygons.
But to get a useful result, it is important to know the gluing law for the torsion.

\subsection{Gluing}\label{gluing}

\begin{figure}
 \begin{center}
   \includegraphics[width=3in]{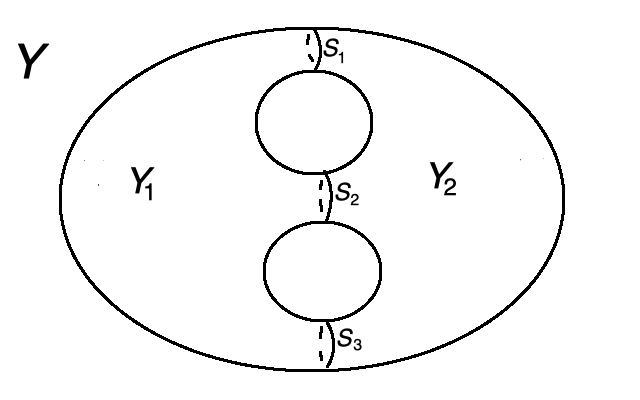}
 \end{center}
\caption{\small A genus 2 surface $Y$ built by gluing two three-holed spheres $Y_1$ and $Y_2$ along boundary circles $S_1,S_2,S_3$.    \label{Glue}}
\end{figure}

In general, a complicated two-manifold can be built by gluing together simple building blocks.   For an example, an oriented surface $Y$ of genus $\geq 2$ can be built
by gluing together three-holed spheres as in 
fig.~\ref{Glue}.

From the point of view of $BF$ theory, if $Y$ is a Riemann surface with boundary $\partial Y$, one can perform the path integral over the fields on $Y$ while keeping
fixed the boundary values of the gauge field $A$ on $\partial Y$.   This path integral will give a function $\Psi(A)$ of the boundary values.   This function defines
an element in a Hilbert space $\H$ of quantum states that live on $\partial Y$.   If $\partial Y$ is a union of several circles $S_i$, $i=1,\dots,n$, then $\H=\otimes_{i=1}^n \H_i$,
where $\H_i$ is a Hilbert space associated to $S_i$.   Suppose $Y_1$ and $Y_2$ are two Riemann surfaces with boundary and we want to glue them along some or all
of their boundary components to make a surface $Y$ which itself may or may not have a boundary.   We can compute the $BF$ path integral on $Y$ by first
computing  $BF$ path integrals on $Y_1$ and $Y_2$ separately, to generate  Hilbert space states, after which, for any circle on which we want to glue
$Y_1$ and $Y_2$, we take an inner product of the corresponding factor in the Hilbert spaces.  Roughly, we ``trace out'' the Hilbert space factor
associated to the circle on which gluing is supposed to occur.

In the example of fig.~\ref{Glue}, we could calculate the $BF$ path integral on $Y$ all at once.  Alternatively, we could calculate the path integral on $Y_1$ and $Y_2$,
keeping fixed the gauge field $A$ on the three circles $S_1,S_2,S_3$ that are the common boundary of $Y_L$ and of $Y_R$.   If we write $A_i$ for the restriction of $A$ to $S_i$,
then the $BF$ path integral on $Y_1$ or on $Y_2$ with fixed boundary values generates a state $\Psi_1(A_1,A_2,A_3)$ or $\Psi_2(A_1,A_2,A_3)$.   If we multiply
the two states together and integrate over $A_1$, $A_2$, $A_3$, we get the inner product $(\Psi_1,\Psi_2)$, which is simply the $BF$ path integral on $Y$.
The fact that the computation can be split up into pieces in this fashion is an important aspect of the locality of quantum field theory.  

A similar cut and paste procedure is possible in the combinatorial approach to the torsion.   However, the most convenient approach involves one difference in detail.
First of all, suppose that $Y$ is a two-manifold with boundary, and that we describe it combinatorially by gluing together polygons.  The procedure described in section
\ref{combitor} is still valid even if $Y$ has a boundary;
 $\tau^\Rr$ as defined in eqn.~(\ref{numbo}) is a measure on the moduli space of flat connections on $Y$, even if $Y$ has a boundary. In particular, let $S$ be
a circle that is part of the boundary of $Y$, and let $a$ be a gauge-invariant function of the holonomy of $A$ around $S$. For example, for $G=\SL(2,\R)$, we can
take $a$ to be (in a sense we make precise later) the logarithm of the holonomy.   Then $a$ is a function on the moduli space
and we can loosely write a measure on $\M$, suppressing other variables, as $f(a)\d a$.   

Now suppose $Y_1$ and $Y_2$ are two-manifolds with a common boundary circle $S$ on which we want to glue them together.   Now we have two 
moduli spaces $\M_1$ and $\M_2$ of flat connections on $Y_1$ and $Y_2$ respectively.   
Working on $Y_1$, the 
 logarithm of the holonomy of a flat connection around $S$ is then a function
on $\M_1$ that we will call $a_1$.   Similarly, working on $Y_2$,  the corresponding logarithm is  a function on $\M_2$ that we will call $a_2$.   The torsion on $Y_1$ or on $Y_2$ gives a measure that we can schematically
call $\tau_{Y_1}=f(a_1)\d a_1$ or $\tau_{Y_2}=g(a_2)\d a_2$, ignoring other variables.  ($Y_1$ and $Y_2$ might be different topologically, so the functions $f$ and $g$ and the other
variables that are being suppressed can be quite different in the two cases.)   

When we glue $Y_1$ and $Y_2$ along $S$ to make a two-manifold $Y$, what sort of gluing law for the torsion might we  one hope for, analogous to what happens in $BF$ theory?   Naively, one might hope that after setting $a_1=a_2$, we would get $\tau^\Rr_Y=\tau^\Rr_{Y_1}\tau^\Rr_{Y_2}$.    But this cannot be right, because after setting
$a_1=a_2$ and writing, say, $a$ for the common value, the product $\tau^\Rr_{Y_1}\tau^\Rr_{Y_2}=f(a)g(a)(\d a)^2$ is not a measure because it has an extra factor of $\d a$.
To get a measure which might equal $\tau^\Rr_Y$, we need a way to remove one factor of $\d a$.   

Intuitively, what is wrong is that  since $S$ is part of $Y_1$ and also part of $Y_2$, when we simply take a product $\tau^\Rr_{Y_1}\tau^\Rr_{Y_2}$,
we are double counting the degrees of freedom that live in $S$.   To get the correct result, we have to remove one copy of $S$, which is accomplished by dividing by
$\tau^\Rr_S$, the torsion of a flat connection on $S$.   $\tau^\Rr_S$ is defined exactly as in section \ref{combitor}, except that as $S$ is a 1-manifold, it can be built from
1-cells and 0-cells, with no 2-cells.  (We will go into more detail when we actually calculate $\tau^\Rr_S$.)   
The correct  gluing law for the Reidemeister or combinatorial torsion, whenever
$Y_1$ and $Y_2$ are glued along $S$ to make $Y$, is\footnote{A mathematical reference on the gluing law for the analytic torsion in its most general form is \cite{bruning2013gluing}.   An  explanation of the gluing law for the combinatorial torsion can be found, for example, in section 4 of \cite{WittenGauge}.}
\be\label{nono}\tau^\Rr_Y=\tau^\Rr_{Y_1}\frac{1}{\tau^\Rr_S} \tau^\Rr_{Y_2}. \ee
We will see the power of this formula when we actually perform computations.  

Dividing by $\tau^\Rr_S$ actually solves two problems.   First, as already explained, it removes a surplus factor of $\d a$.   But it also provides an extra
factor that would be missing in the simple product $\tau^\Rr_{Y_1}\tau^\Rr_{Y_2}$.    A flat bundle on $Y$ can be restricted to $Y_1$ and to $Y_2$, but it
is not uniquely determined by those restrictions.   To build $\M_Y$, the moduli space of flat connections on $Y$, out of the corresponding $\M_{Y_1}$ and $\M_{Y_2}$,
we have to set $a_1=a_2$, as already discussed.  But that is not all.  On $Y$, there can be a new modulus, a relative ``twist parameter'' between fields on $Y_1$ and on $Y_2$.
For $G=\SL(2,\R)$, the twist parameter in question is the ``twist'' variable in Fenchel-Nielsen or length-twist coordinates; see section \ref{hyperbolic}.  
If we denote such a relative twist parameter as $\varrho$, then to build a measure on $\M_Y$ from a product of measures on $\M_{Y_1}$  and on $\M_{Y_2}$,
we need to add a factor $\d\varrho$ as well as removing a factor $\d a$.  As we will see, dividing by $\tau^\Rr_S$ accomplishes both of these tasks.

\subsection{Computations For \texorpdfstring{$G=\SL(2,\R)$}{G=SL(2,R)}}\label{compotwo}

\subsubsection{Flat Connections And Hyperbolic Geometry}\label{hyperbolic}

JT gravity in the absence of time-reversal symmetry
 is related to $BF$ theory with gauge group $\mathrm{PSL}(2,\R)$ in the case of a purely bosonic theory, or $\SL(2,\R)$ in the case of a theory with
fermions and spin structures.\footnote{$\SL(2,\R)$ is the group of $2\times 2$ real matrices of determinant 1, and $\mathrm{PSL}(2,\R)$ is its quotient by $\{\pm 1\}$.}   The difference is that a flat $\SL(2,\R)$ connection encodes the spin structure as well as the geometry, while a flat $\mathrm{PSL}(2,\R)$
connection only encodes the geometry.

We may or may not want to consider a theory with fermions, but in any event, the torsion, with which we are concerned here, is not sensitive
to the difference between $\mathrm{PSL}(2,\R)$ and $\SL(2,\R)$, basically because it is a one-loop effect of fields with values in the adjoint representation.   We will
use the language of $\SL(2,\R)$ because it is more convenient to write $2\times 2$ matrices, not just because we might want to study a theory with fermions.

Actually, the relation of JT gravity to $BF$ theory of $\SL(2,\R)$ refers to just one component\footnote{The group $\SL(2,\R)$ is contractible to $\U(1)$, so an $\SL(2,\R)$ bundle
over a topological space has a first Chern class, just like a $\U(1)$ bundle.  A {\it flat} $\SL(2,\R)$ bundle over a Riemann surface $Y$ of genus $g>1$ can have any value of the
first Chern class from $g-1$ to $-(g-1)$.   The component that is related to hyperbolic geometry is the component of first Chern class $g-1$.} 
of the moduli space of flat $\SL(2,\R)$ connections.  This  is the component we are interested in, though some of our considerations can be applied to other components.  

Three-holed spheres are a basic example, because (with a few exceptions that have non-negative Euler characteristic) any oriented two-manifold $Y$ can be built
by gluing together three-holed spheres.   So let us discuss the hyperbolic geometry of a three-holed sphere.

If $Y$ is a three-holed sphere, then every flat $\SL(2,\R)$ connection on $Y$  (in the appropriate component, but we will not keep stating this qualification in what
follows) is associated to a hyperbolic
metric on $Y$, with constant scalar curvature $R=-2$, and with geodesic boundaries.    The lengths of the three geodesic boundaries are arbitrary positive real numbers, so the relevant moduli space for a three-holed sphere (ignoring the spin structure, about which more in a moment)
is a simple product $\R^3_+$ of three copies of $\R_+$ (the set of positive real numbers).   In particular, the moduli space has dimension 3.

What is the connection between the length or circumference of a geodesic boundary component and the corresponding flat $\SL(2,\R)$ connection?   This
question has a rather simple answer.   The monodromy of the flat connection around a boundary component of length $\ell$ is conjugate to
\be\label{pokko} U_\ell^\pm =\pm \begin{pmatrix}e^{\ell/2} & 0 \cr 0 & e^{-\ell/2}\end{pmatrix}. \ee
The sign, which we do not see if we project to $\mathrm{PSL}(2,\R)$, determines the spin structure; a spin structure of NS or R type around the circle
in question corresponds to a $-$ or $+$ sign, respectively.    We see
that
\be\label{lokk}\Tr\,U_\ell^\pm =\pm 2\cosh \ell/2. \ee
Conversely, any element of $\SL(2,\R)$ whose trace is $\pm 2\cosh\ell/2$ is conjugate to $U_\ell^\pm$.  So the conjugacy class that contains $U_\ell^+$ or $U_\ell^-$ is
(for all $\ell\not=0$) of codimension 1 in $\SL(2,\R)$ and hence of dimension 2.   

Given this, we can do a simple dimension counting and check the claim that the moduli space of flat $\SL(2,\R)$ connections on the three-holed sphere has dimension 3.
Let $U,V,W$ be the monodromies around the three holes; they are subject to one condition $UVW=1$.     The space of all triples $U,V,W\in \SL(2,\R)$ is nine-dimensional;
after imposing the condition $UVW=1$ and dividing by an overall conjugation by an arbitrary element of $\SL(2,\R)$, we are left with a moduli space of dimension
$9-3-3=3$.   One can take the three moduli to be the three boundary lengths or equivalently the three traces $\Tr\,U$, $\Tr\,V$, $\Tr\,W$.

\begin{figure}
 \begin{center} 
   \includegraphics[width=3in]{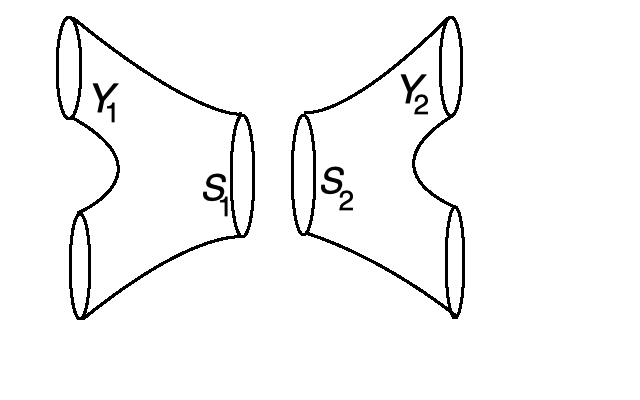}
 \end{center}
\caption{\small $Y_1$ and $Y_2$ are three-holed spheres with geodesic boundaries $S_1$ and $S_2$.   The only necessary condition for being
able to glue $S_1$ onto $S_2$, producing a smooth hyperbolic surface $Y$, is that $S_1$ and $S_2$ should have the same circumference.  If the gluing
is possible, then one can make an arbitrary rotation of $S_1$ relative to $S_2$, by a ``twist'' parameter $\varrho$, prior to the gluing. \label{TwoGliue}}
\end{figure}

It is not difficult to describe explicitly the monodromies of a family of flat connections with  variable boundary lengths on a three-holed sphere $Y$.    We can take, for example,
\be\label{urdu} U=\vnu_a\begin{pmatrix} e^{a/2}& \kappa  \cr 0 & e^{-a/2}\end{pmatrix},~~~~~V=\vnu_b\begin{pmatrix} e^{-b/2} & 0 \cr 1 & e^{b/2}\end{pmatrix}.\ee
depending on three real parameters $a,b,\kappa $.  Here $a$ and $b$ are length  parameters; the geodesics in question have lengths $|a|$ and $|b|$.
 And  $\vnu_a$ and $\vnu_b$ are $1$ or $-1$ for a spin structure of Ramond or NS type.
  To get $UVW=1$, the third monodromy must be
\be\label{wurdu} W=V^{-1}U^{-1}=\vnu_a\vnu_b\begin{pmatrix} e^{-(a-b)/2} & -\kappa  e^{b/2}\cr -e^{-a/2} & e^{(a-b)/2}+\kappa  \end{pmatrix}. \ee
This gives a flat $\SL(2,\R)$ connection for any real $\kappa $, but it is only associated with a hyperbolic metric on $Y$ with geodesic boundaries if
$2\cosh\frac{1}{2}(a-b)+\kappa <-2$.  In that case, the third length is 
 $|c|$ where  $\Tr\,W=2\vnu_c\cosh c/2$ (where $\vnu_c$ controls the spin structure of the third geodesic), and 
\be\label{gurdu}2 \cosh \frac{c}{2} =-\kappa -2\cosh\frac{a-b}{2}.\ee
We have used the fact that $\vnu_a\vnu_b\vnu_c=-1$  for any compatible set of spin structures (since the number of Ramond boundaries is always even).  To
get a hyperbolic three-holed sphere (rather than a flat $\SL(2,\C)$ connection in a different component of the moduli space) we must chose $\kappa$ consistent
with $\cosh\frac{c}{2}>1$.  
Some more detail on this flat connection is described in appendix \ref{uhp}.

Now let $Y_1$ and $Y_2$ be hyperbolic two-manifolds with geodesic boundary components $S_1$ and $S_2$, respectively.  Suppose that  we want to make a hyperbolic
manifold $Y$ by gluing $S_1\subset Y_1$ to $S_2\subset Y_2$.   An obvious necessary condition for this gluing to produce a hyperbolic manifold is that 
$S_1$ and $S_2$ must have the same circumference.   There is actually no further condition, because hyperbolic geometry is rather rigid; the geometry of
a hyperbolic surface near a geodesic boundary component is completely determined by the circumference of that boundary.

Not only are we free to glue $S_1$ to $S_2$ as long as they have the same length, but an arbitrary parameter $\varrho$ enters in this gluing, because we are
free to rotate $S_1$ relative to $S_2$.   This  parameter was mentioned at the end of section \ref{gluing}.  We understand $\varrho$
as the distance (not angle) of the relative rotation between $S_1$ and $S_2$, so the range of $\varrho$ is $0\leq\varrho\leq \ell$, where $\ell$ is the circumference
of $S_1$ or $S_2$. (Alternatively, one can replace $\varrho$ with the angle $\theta=2\pi \varrho/\ell$, which ranges from 0 to $2\pi$.)

The parameter $\varrho$ can be described in the language of gauge theory and flat connections.   When we glue $S_1$ to $S_2$, we are free to make a relative
gauge transformation of the fields on the two sides by any element of $\SL(2,\R)$ that commutes with the monodromy around $S_1$ or equivalently around $S_2$.
If as in eqn.~(\ref{pokko}), the monodromy of the hyperbolic flat connection around $S_1$ or $S_2$ is by the group element $U^\pm_\ell=\pm \diag(e^{\ell/2},e^{-\ell/2})$,
then the element $U^+_\varrho=\diag(e^{\varrho/2},e^{-\varrho/2})$ commutes with $U_\ell$, and a relative gauge transformation by $U^+_\varrho$ corresponds
to the relative rotation between the two sides that was discussed in the last paragraph.   One may wonder about the interpretation of a relative gauge transformation by
the element $U^-_\varrho=-\diag(e^{\varrho/2},e^{-\varrho/2})$, which also commutes with $U_\ell$.   The answer is that the minus sign represents a sign change
of fermions on $S_1$ relative to those on $S_2$, prior to gluing.   This sign change, which produces a different spin structure on the glued manifold, was important
in our discussion of the bulk dual of fermion anomalies.

Incorporating time-reversal symmetry replaces the gauge group by a double cover.   In a purely bosonic theory, incorporating time-reversal symmetry means
replacing $\mathrm{PSL}(2,\R)$ by its double cover $\mathrm{PGL}(2,\R)$ (the group of $2\times 2$ invertible real matrices, modulo multiplication by a nonzero real scalar).   The essential difference is that $\mathrm{PGL}(2,\R)$ contains an element
$\diag(1,-1)$ of determinant $-1$, and $\mathrm{PSL}(2,\R)$ does not.   In the presence of fermions, one has to take a double cover of $\SL(2,\R)$, but there are
two different double covers, corresponding to $\pin^+$ and $\pin^-$.   For $\pin^+$, one wants the group of $2\times 2$ real matrices of determinant $\pm 1$,
meaning that again one includes the element $\diag(1,-1)$.  For $\pin^-$, one wants the group of $2\times 2$ matrices of determinant 1 that are either real or imaginary,
meaning that one extends $\SL(2,\R)$ to include the element $\diag(\i,-\i)=\i\,\diag(1,-1)$.   In any of these cases, one can make an orientation-reversing gluing,
in which $S_1$ is glued to $S_2$ with a relative gauge transformation $\diag(e^{\varrho/2},-e^{-\varrho/2})$ (or $\i \,\diag(e^{\varrho/2},-e^{-\varrho/2}) $ in the $\pin^-$ case);
these commute with $U^\pm_\varrho$.    This more general type of gluing produces, for example, the twisted double trumpet that we studied in section \ref{sectionTwo}.  Note that in either $\pin^+$ or $\pin^-$, one distinguishes the group elements $\diag(-1,-1)$ and $\diag(1,1)$, which are equivalent
in $\mathrm{PGL}(2,\R)$.

From what we have said, every gluing of two three-holed spheres is associated to a pair of moduli $\ell,\varrho$.   They are called length-twist or Fenchel-Nielsen
coordinates.   For example, if we build a genus 2 Riemann surface $Y$ by gluing two three-holed spheres along circles $S_1,S_2,S_3$ (fig.~\ref{Glue}),
one gets three pairs of length-twist coordinates, making 6 parameters in all.   Locally these parametrize the moduli space of Riemann surfaces of genus 2,
whose real dimension is indeed 6.   The subtlety is that there are many ways to build the same $Y$ by gluing together three-holed spheres,
so the same moduli space can be parametrized locally by length-twist coordinates in many different ways.    A given set of length-twist coordinates
gives a good parametrization when the $\ell_i$ are sufficiently small, but when the $\ell_i$ increase beyond a certain point, there exists an alternative description by a different set of length-twist coordinates with smaller lengths.  There is no simple description of the inequalities on the lengths that one should impose to get
a single copy of the moduli space of hyperbolic metrics or conformal structures on $Y$.

\subsubsection{The Torsion Of A Three-Holed Sphere}\label{torsthree}

In computing the torsion of  a three-holed sphere, a circle, and a crosscap, we follow steps that are mostly explained more fully in sections 4.3, 4.4, and 4.6 of 
\cite{WittenGauge}.  There are a few differences in detail because the gauge group is not compact.    

We begin with a three-holed sphere and gauge group $G=\SL(2,\R)$.  We henceforth write just $\tau$ rather than $\tau^\Rr$ for the torsion.   The torsion is a measure; it will be convenient to represent this measure by a differential form.   A measure
is really better understood as the absolute value of a differential form, but we omit the absolute value sign.   For a purely bosonic group such as $\SL(2,\R)$, the torsion is
positive-definite, so we do not need to keep track of overall signs.

In general, let $Y$ be a two-manifold and $P$ a point in $Y$.   Let $\mR$ be the moduli space of flat connections on $Y$, modulo gauge transformations, and $\hat\mR$
the moduli space of ``based'' flat connections, constructed by dividing by gauge transformations that are trivial at $P$.   There is an obvious fibration
$\pi :\h\mR\to \mR$, with fiber $G$, defined by dividing by gauge transformations that are possibly nontrivial at $P$.

 We have already explained how the combinatorial
torsion defines a measure $\tau$ on $\mR$.   By omitting in the definition of $\sC_0$ the 0-cell that corresponds to $P$, but otherwise following exactly the same definitions, one
can use the combinatorial torsion to define a measure $\hat\tau$ on $\hat\mR$.    The definition of the combinatorial torsion made use of a volume form on the Lie algebra $\g$
of $G$.   Such a volume form also determines a volume form on the group manifold.   We will call this volume form $\vol_G$.   (Note that $\vol_G$ is a volume form, not
a number; we are interested in noncompact gauge groups whose volume is infinite.)   The three volume forms $\tau$, $\h\tau$, and $\vol_G$ have a simple multiplicative relationship\footnote{This
statement is equivalent to eqn.~(4.39) in \cite{WittenGauge}, which is formulated a little differently as the volume of $G$ was assumed to be finite.}
\be\label{simpleone} \h\tau =\vol_G\cdot  \pi^*(\tau). \ee

\begin{figure}
 \begin{center} 
   \includegraphics[width=3in]{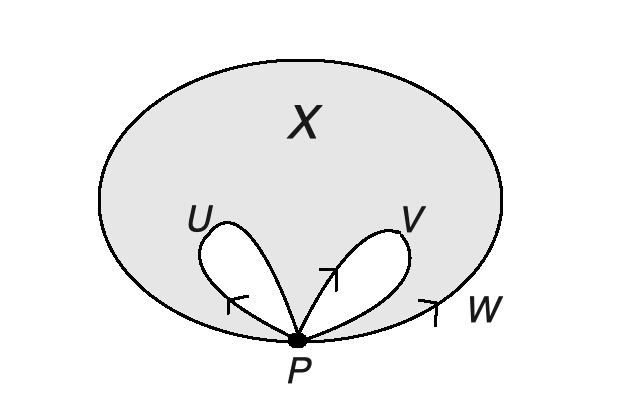}
 \end{center}
\caption{\small  The shaded region is a three-holed sphere $Y$ - or more precisely a topological space that is equivalent to $Y$ up to homotopy.  It has a simple
cell decomposition with 
a single 0-cell $P$, three 1-cells that have been labeled by the holonomies $U,V,W$ of a flat connection, and a single two-cell $X$. \label{simpletri}}
\end{figure}
In general, omitting a single 0-cell in the definition of $\sC_0$ might not lead to a major simplification.   However, for the three-holed sphere (or more exactly for a topological
space that is equivalent to a three-holed sphere up to homotopy, which is good enough in computing the torsion), one can pick a very simple cell decomposition
with only a single 0-cell $P$ (fig.~\ref{simpletri}).    This means that after omitting $P$, we get $\sC_0=0$, and this does lead to a major simplification.    In the cell decomposition of the
figure, there is a single 2-cell $X$, and three 1-cells that have been labeled by the holonomies $U,V,W$ of a flat connection.    Hence $\sC_2=\sV_X$, $\sC_1=\sV_U\oplus \sV_V\oplus \sV_W$.   Since $\sC_0=0$, we can compute the torsion just using the boundary map $\partial:\sC_2\to \sC_1$.   But matters are even simpler than that.  If $\Theta:\sC_1\to \sV_W$
is the obvious projection, then $\Theta\circ \partial: \sC_2\to \sV_W$ is an isomorphism.   It follows then from the definition of the torsion that if we discard $\sC_2$ and $\sV_W$,
the torsion is unchanged.    This means that we can compute the torsion $\hat\tau$ taking $\sC_2=\sC_0=\partial =0$, $\sC_1=\sV_U\oplus \sV_V$.   

On a three-holed sphere, the holonomies of a flat connection are group elements $U,V,W$ satisfying $UVW=1$.    To construct the moduli space $\mR$, we would
identify two such triples that differ by conjugation by an element $R$ of $G$: $(U,V,W)\cong (R U R^{-1}, R V R^{-1}, RWR^{-1})$.    To construct the based moduli space
$\h\mR$, we do not divide by conjugation.   Since $UVW=1$, $\h\mR\cong G\times G$, parametrized by $U$ and $V$.   With $\sC_2=\sC_0=\partial=0$, $\sC_1=\sV_U\oplus \sV_V$,
eqn.~(\ref{mumbo}) just says that $\h\tau$ is the natural volume form on $G\times G$.  Since there are several copies of $G$ in this discussion, we will write
$\vol_G(U)$ for the volume form of a copy of $G$ parametrized by $U$, and similarly for $\vol_G(V)$ and $\vol_G(R)$. So
\be\label{lurdu}\h\tau=\vol_G(U)\cdot \vol_G(V). \ee

We parametrize $U$ and $V$ by the obvious $R$-dependent generalization of eqn.~(\ref{urdu}):  
\be\label{nurdu} U=RU_0R^{-1},~~~~~V=RV_0R^{-1},\ee
with
\be\label{zurdu}U_0=\begin{pmatrix} e^{a/2}& \kappa    \cr 0 & e^{-a/2}\end{pmatrix},~~~~V_0=\begin{pmatrix} e^{-b/2} & 0 \cr 1 & e^{b/2}\end{pmatrix}.\ee
Thus $a,b$, and $\kappa   $ (or $U_0$ and $V_0$)
parametrize the ordinary moduli space $\mR$, while $a,b,\kappa   $, and $R$ parametrize the extended moduli space $\h\mR$.   Eqns. (\ref{simpleone})
and (\ref{lurdu}) tell us that
\be\label{purdu}\vol_G(U)\cdot \vol_G(V)=\vol_G(R) \cdot\tau, \ee
and this condition will determine $\tau$.  

We introduce a standard basis for the Lie algebra $\sl_2$ of $\SL(2,\R)$:
\be\label{conbasis} e=\begin{pmatrix} 0 & 1\cr 0 & 0 \end{pmatrix},~~~f=\begin{pmatrix}0&0\cr 1&0\end{pmatrix},~~~~ h=\begin{pmatrix}1&0\cr 0&-1\end{pmatrix}. \ee
Now we can describe explicitly what measures we will use on the Lie algebra and on the group.    If $x$ is an element of the Lie algebra $\sl_2$, 
write it as 
$x=\begin{pmatrix} x_h & x_e\cr x_f&-x_h\end{pmatrix}=x_e e+x_f f+x_h h$, and then define the measure on $\sl_2$ to be $4\d x_e\d x_f\d x_h$.\footnote{\label{lookback}
The factor of 4 has been chosen with foresight so that our final result for a measure on moduli space will  agree (in the  oriented case) with standard formulas. The measure we are using can be derived from the inner product $(x,x)=2\,\Tr \,x^2= 4x_h^2+4x_e x_f$ on the Lie algebra.
This inner product corresponds to the metric  $\d s^2= 4 \d x_h^2+ 4\d x_e \d x_f$ on $\g$, so that the metric tensor $g$ satisfies $\sqrt{|\det g|}=4$ and the Riemannian measure is
$4 \d x_e \d x_f \d x_h$.    Using a different measure would multiply the final result for the torsion by a factor $(\text{const.})^{\chi}$ where $\chi$ is the Euler characteristic. Such a factor can be absorbed by adjusting the coefficient of $\chi$ in the action of the underlying JT theory. } Similarly, if $U$ is an element of the group
$\SL(2,\R)$, then $U^{-1}\d U$ is a Lie algebra valued one-form, and we expand it as $U^{-1}\d U=(U^{-1}\d U)_e e+(U^{-1}\d U)_f f+ (U^{-1}\d U)_h h. $   Then the volume
form on the group is the three-form
\be\label{woddo}\vol_G(U)= 4(U^{-1}\d U)_e (U^{-1}\d U)_f (U^{-1}\d U)_h.\ee

To use equation (\ref{purdu}) to compute $\tau$, it suffices to work at $R=1$.   Expanding $R=1+r+\cO(r^2)$, we have $\d R=\d r+\mathcal{O}(r)$, $\d R^{-1}=-\d r+\mathcal{O}(r)$.
Thus at $R=1$, $\vol_G(R)=4\d r_e \d r_f \d r_h$.    The general form of $\tau$ is $F(a,b,\kappa   )\d a\,\d b\,\d \kappa   $, for some function $F(a,b,\kappa   )$ that we want to compute.  Thus the right hand
side of eqn.~(\ref{purdu}) reads
\be\label{rightside}4\d r_e\d r_f\d r_h \,F(a,b,\kappa   ) \d a\, \d b\, \d \kappa   ,\ee
and we need to compare this to the left hand side.

To evaluate the left hand side, we first note that at $R=1$,
\be\label{woggle} U^{-1}\d U=U_0^{-1}\d U_0  + U_0^{-1}\d r U_0-\d r,~~~~   V^{-1}\d V=V_0^{-1}\d V_0+V_0^{-1}\d r V_0 - \d r. \ee
Since $U_0$ is upper triangular, there is a simple result for the lower triangular part of $U^{-1}\d U$:
\be\label{zoffoTor} (U^{-1}\d U)_f=(e^a-1) \d r_f. \ee
Similarly, there is a simple result for the upper triangular part of $V^{-1} \d V$:
\be\label{poffoTor} (V^{-1}\d V)_e=(e^{b}-1) \d r_e. \ee
We can now  eliminate $\d r_f $, $(U^{-1}\d U)_f$, $\d r_e$,  and $(V^{-1}\d V)_e$ from eqn.~(\ref{purdu}), reducing to
\be\label{groffo} 4(e^a-1)(e^{b}-1) (U^{-1}\d U)_e(U^{-1}\d U)_h (V^{-1}\d V)_f (V^{-1}\d V)_h=\d r_h\, F(a,b,\kappa   )\d a\,\d b\, \d \kappa   . \ee

In the further evaluation of the left hand side of eqn.~(\ref{groffo}), we can set $\d r_e=\d r_f=0$.  After doing so, we get some simple formulas:
\begin{align}\label{umcro} (U^{-1}\d U)_h& = \frac{1}{2}\d a \cr
        (U^{-1}\d U)_e& =e^{-a/2}\d \kappa   +\frac{\kappa   }{2} e^{-a/2} \d a+2\kappa    e^{-a/2}\d r_h \cr
           (V^{-1}\d V)_h& = -\frac{1}{2}\d b \cr
             (V^{-1}\d V)_f &= \frac{1}{2} e^{-b/2}\d b-2e^{-b/2} \d r_h. \end{align}
The left hand side of eqn.~(\ref{groffo}) is therefore just $8\sinh \frac{a}{2} \sinh \frac{b}{2} \,\d r_h \d a\,\d b\, \d \kappa   $.  
 From eqn.~(\ref{groffo}), we therefore get
\be\label{woffo}\tau=F(a,b,\kappa   )\d a\, \d b \, \d \kappa    = 8\sinh  \frac{a}{2}  \sinh \frac{b}{2} \,\d a\,\d b\,\d \kappa   .\ee
Concretely, starting with eqn.~(\ref{groffo}), we first replace $(U^{-1}\d U)_h\,(V^{-1}\d V)_h$ in (\ref{groffo}) with $-\frac{1}{4}\d a\,\d b$, cancel $\d a$ and $\d b$ from the equation,
and set $\d a=\d b=0$ in the formulas for $(U^{-1}\d U)_e$ and $(V^{-1}\d V)_f$.   Then  $(U^{-1}\d U)_e (V^{-1}\d V)_f$ reduces to $-2 e^{-a/2-b/2} \d \kappa    \d r_h$,
and this leads to eqn.~(\ref{woffo}).

In view of eqn.~(\ref{gurdu}), we can here replace $\d \kappa   $ with $\sinh \frac{c}{2} \,\d c$, where $c$ is the length parameter of the third boundary component.  
 Thus finally
the torsion of a three-holed sphere $Y$ with boundary length parameters $a,b,c$ is
\be\label{koffo} \tau_Y=8\sinh \frac{a}{2} \sinh \frac{b}{2} \sinh \frac{c}{2} \,\d a\,\d b\,\d c.\ee
As a check on the calculation, we note that this is symmetric in $a,b,c$, though the derivation lacked that symmetry.  We recall that the overall sign in eqn.
(\ref{koffo}) is unimportant, as the torsion of the three-holed sphere is really the absolute value of the three-form that appears on the right hand side.   The factor of 8
is meaningful but depended on how we normalized the measure on the Lie algebra.

\subsubsection{The Torsion Of A Circle}\label{torcircle}

\begin{figure}
 \begin{center} 
   \includegraphics[width=2in]{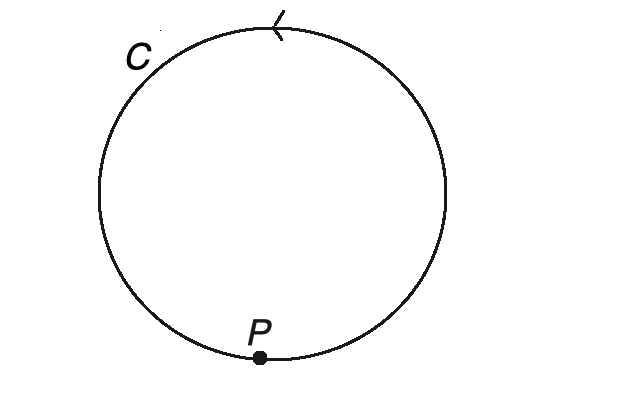}
 \end{center}
\caption{\small A simple cell decomposition of a circle, with one 0-cell $P$, and one 1-cell $C$, with holonomy $U$.  \label{Circle}}
\end{figure}

Now we consider a circle $S$ endowed  with a flat $\SL(2,\R)$ bundle $E\to S$.  We take the holonomy of this flat bundle to be
  $U=\diag(e^{a/2},e^{-a/2})$.  
  
   We can endow $S$ with a very simple cell decomposition,
with a single 0-cell $P$ and a single 1-cell $C$ (fig.~\ref{Circle}).
The computation of the torsion is thus going to be rather simple.   $\sC_1$ and $\sC_0$ are both copies of the Lie algebra $\sl_2$, while $\sC_2=0$.   
 In the definition of the boundary operator
$\partial:{\sV}_1\to {\sV}_0$, we need to  take into account that both ends of $C$ are on $P$.  Let us trivialize the bundle $E$ at and just to the right of the point $P$; we use
this trivialization in identifying ${\sV}_1$  and ${\sV}_0$ with $\sl_2$.   For $s\in {\sV}_1$, the boundary operator acts on $S$ by
\be\label{partgo}\partial s = Us U^{-1} - s. \ee
The two terms are the contributions of the two ends of $C$. The contribution from $C$ ending on $P$ from the left is proportional to $Us U^{-1}$, because to evaluate
it we  have to parallel transport $s$ along $C$ from one end of $C$ to the other.   There is a relative minus sign between the two terms because of a relative orientation
between the two ends of $C$.  

It is convenient to decompose the $\sl_2$ Lie algebra as $\sl_2^0\oplus \sl_2^\perp$, where $\sl_2^0$ consists of diagonal matrices and $\sl_2^\perp$ consists of matrices
whose diagonal matrix elements vanish.   The operator $\partial$ commutes with this decomposition, so
$\partial=\partial^0\oplus \partial^\perp$, where $\partial^0$ and $\partial^\perp$ act on $\sl_2^0$ and on $\sl_2^
\perp$, respectively.    The torsion $\tau_S$ of a flat bundle on $S$ correspondingly factorizes
\be\label{decomp}\tau_S=\tau_S^\perp\cdot \tau_S^0, \ee
where $\tau_S^\perp$ and $\tau_S^0$ are the torsion defined using $\partial^\perp$ and $\partial^0$, respectively.

The operator $\partial^\perp:\sl_2^\perp\to\sl_2^\perp$ is invertible, so we can use the simplest  definition (\ref{numbo}) of the torsion.
In the present case, with ${\sV}_2=0$, and ${\sV}_1,{\sV}_0$ both copies of $\sl_2^\perp$, the definition (\ref{numbo}) reduces to
$\tau^\perp= \alpha_0(\partial^\perp t_1,\partial^\perp t_2)/\alpha_1(t_1,t_2)$, where $t_1,t_2$ is any basis of $\sl_2^\perp$.   But with the two measures $\alpha_0$ and $\alpha_1$ being the same, this ratio is, according to eqn.~(\ref{transrule}),  the absolute value of the  determinant $|\det\partial^\perp|$.      The eigenvalues of $\partial^\perp $ are $e^{\pm a}-1$, so
\be\label{nuvvu}\tau_S^\perp = |\det\partial^\perp|=|(e^a-1)(e^{-a}-1)|=4\sinh^2 \frac{a}{2}. \ee

On the other hand,  $\partial^0$ is actually identically 0, since $UsU^{-1}-s=0$ if $s$ is diagonal.   This being so, the definition (\ref{tmumbo}) reduces 
to $\tau^0=\alpha_0(v)/\alpha_1(u)$, where $v$ and $u$ are basis vectors of the 1-dimensional spaces $\sC_0^0$ and $\sC_1^0$.   This just means that
$\tau^0$ is the ratio of the natural measures on $\sC_0^0$ and $\sC_1^0$.   Since those spaces are both copies of $\sl_2^0$, the natural measures
are the same, and one could claim that $\tau^0=1$.   It is more convenient not to take that step because, although $\sC_0$ and $\sC_1$ are both copies of $\sl_2^0$,
the physical meaning is different.    $\sC_1$ is the tangent space to the moduli space of flat connections on the circle.  $\sC_0$ is the Lie algebra of the subgroup
of $\SL(2,\R)$ that commutes with a given flat connection.    If we parametrize the moduli space by
the parameter $a$ that appears in the monodromy $U=\diag(e^{a/2},e^{-a/2})$, and similarly write $\diag(e^{\varrho/2},e^{-\varrho/2})$ for a group element that commutes
with $U$, then the ratio of the natural measures is
\be\label{ubb}\tau_S^0 =\d a\cdot (\d\varrho)^{-1}. \ee

Combining these statements, we get the torsion of a flat connection on the circle:
\be\label{rubb}  \tau_S=4\sinh^2 \frac{a}{2}\cdot \d a \cdot (\d\varrho)^{-1}. \ee

\subsubsection{The Torsion Of An Oriented Surface}\label{oriented}

Let $Y_1$ and $Y_2$ be two three-holed spheres endowed with hyperbolic metrics, and
with respective boundary components $S_1$ and $S_2$, each with length parameter $a$.   As explained in section
\ref{hyperbolic}, we can build a larger hyperbolic manifold $Y$ by gluing $S_1\subset Y_1$ onto $S_2\subset Y_2$.   According to eqn.~(\ref{nono}),
the torsion of $Y$ is related to the torsions of $Y_1$, $Y_2$, and $S=S_1=S_2$ by $\tau_Y=\tau_{Y_1}\tau_S^{-1}\tau_{Y_2} $.   

Let us isolate in this formula the factors that depend on variables that are defined along $S$.    From both $\tau_{Y_1} $ and $\tau_{Y_2}$, we
get a factor of $2\sinh \frac{a}{2} \d a$.    On the other hand, from $\tau_S^{-1}$, we get $\d \varrho/4\sinh^2 \frac{a}{2} \d a$.    When we multiply these quantities together,
the factors of $\sinh a/2$ cancel, and we are left with just $\d a\d\varrho$.   As promised in section \ref{gluing}, an unwanted extra factor of $\d a$ has disappeared.
We have also gotten a factor of $\d\varrho$, a measure on the subgroup of $\SL(2,\R)$ that commutes with the monodromy around $S$.   We can, in other words,
think of $\d\varrho$ as a measure for integrating over the gluing parameter $\varrho$ that was described in section \ref{gluing}.

Since any oriented two-manifold can be constructed by gluing together three-holed spheres, we can, without further ado, construct the torsion for an arbitrary oriented
two-manifold.    For simplicity, let us consider the case of an oriented two-manifold $Y$ of genus $g$ without boundary.   It can be built by gluing together $2g-2$ three-holed
spheres $Y_\alpha$ along $3g-3$ circles $S_i$ (for $g=2$, this was drawn in fig.~\ref{Glue}).   For each circle, we get a pair of length-twist coordinates $a_i$ and $\varrho_i$,
and the above computation leads to an integration measure $\d a_i\,\d\varrho_i$ for those variables.   We also get a factor $8$ for each three-holed sphere, which cancels the $1/4$
for each circle.   Thus the volume form for the moduli space $\M$ of hyperbolic structures on $Y$ that we get from the torsion
is 
\be\label{zingo}\tau_Y= \prod_{i=1}^{3g-3}\d a_i \d\varrho_i. \ee
For a different normalization of the measure on $\sl_2$, this would be multiplied by $(\text{const.})^\chi$.

A few words about this formula are in order.   
First, length-twist coordinates on $\M$ are far from unique, since there are infinitely many ways to decompose $Y$ into a union of three-holed spheres.   The formula
(\ref{zingo}) is valid for any choice of length-twist coordinates.

Second, this formula for the measure (for the case of an oriented two-manifold $Y$) is actually well-known, but it is not usually
derived using the torsion.   A more common procedure is to use the fact that if $Y$ is oriented, then $\M$ is a symplectic (and in fact Kahler)  manifold.   The
symplectic form of $\M$ in length-twist coordinates is $\sum_i \d a_i\d\varrho_i$, and this leads to the volume form (\ref{zingo}).
The two approaches give the same volume form because,  as we described in the beginning of this section, the torsion is ``trivial'' for orientable $Y$, which just
means that it is equivalent to the volume form that can be defined using the symplectic or Riemannian structure of $\M$.  The symplectic or Riemannian structure of $\M$
is more elementary than the torsion in the sense that these structures are defined just in terms of zero-mode wavefunctions on $Y$, while the torsion involves a quantum
correction coming from  the non-zero modes.    There are two reasons that we have used the torsion to arrive at the formula for the volume form: this is the natural
definition in JT gravity, and (therefore) it generalizes to unorientable two-manifolds.

Finally, though the formula (\ref{zingo}) is simple, it does not lead to any simple results for the volume of $\M$, because the region in length-twist coordinates that
corresponds to $\M$ is not simple to describe.   This is actually the problem that was overcome in the work of Maryam Mirzakhani on the volumes of moduli spaces.

\subsubsection{The Torsion Of a Cross-Cap}\label{crosstor}

To complete the picture, we also need to compute the torsion of a crosscap.

By a crosscap, we mean here a two-manifold $\cC$ that is constructed from a cylinder by making a suitable identification on one boundary.   For example,
we can start with the cylinder $I\times S^1$, where $I$ is the unit interval $0\leq x\leq 1$, and $S^1$ is the circle $\theta\cong \theta+2\pi$.    Then at $x=1$,
we make the identification $\theta\cong \theta+\pi$, while doing nothing at $x=0$.   This gives an unorientable two-manifold $\cC $, whose boundary is a single circle
$S$, the circle at $x=0$.   At $x=1$, there is a second circle $S'$ with $0\leq\theta\leq \pi$, but because of the gluing, this is an ``internal'' circle, not a boundary.
$\cC $ can be viewed as a M\"{o}bius
strip or as a copy of $\RP^2$ with an open ball removed.  

Let $U$ be the monodromy of a flat connection on $\cC $ around $S$,
and let $V$ be the monodromy of the same connection around $S'$.   Clearly they are related by
$U=V^2$.   For hyperbolic monodromies, up to conjugacy we can
 take $U=\diag(e^{a/2},e^{-a/2})$, $V=\diag(e^{a/4},-e^{-a/4})$.    There is an important minus sign in the formula for $V$ 
which reflects the fact that the orientation of $\cC $ is reversed in going around $S'$.   
Spin structures are not important in the discussion of the torsion (because the fields of the $BF$ model are all in the adjoint representation of $\SL(2,\R)$)
so for simplicity we ignore the overall sign of the monodromies  and  subtleties of $\pin^-$ and $\pin^+$.  (We will have to incorporate those subtleties when we get to the supersymmetric case.)

\begin{figure}
 \begin{center} 
   \includegraphics[width=3in]{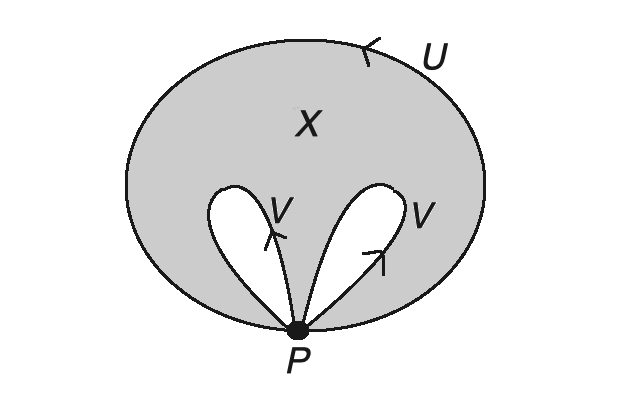}
 \end{center}
\caption{\small  The shaded region in this picture is
the  crosscap manifold $\cC $ or more exactly a topological space equivalent to it in homotopy.  It has a simple cell decomposition
 with one 2-cell $X$, two 1-cells labeled by their holonomies $U$ and $V$,
and one 0-cell $P$.  $V$ appears twice in the drawing, but the two copies are supposed to be identified to create $\cC $.   \label{CrossCapCell}}
\end{figure}

$\cC $ has a simple cell decomposition (fig.~\ref{CrossCapCell}), with one 2-cell $X$, a pair of 1-cells associated to holonomies $U$ and $V$, respectively, and one 0-cell $P$.
Thus $\sC_2\cong \sl_2$, $\sC_1\cong \sl_2\oplus \sl_2$, with the two summands corresponding to the 1-cells labeled $U$ and $V$, respectively, and $\sC_0\cong \sl_2$.
In defining the boundary map, we trivialize the flat bundle $E$ just to the right of the point $P$ in the figure.  
Then the boundary map $\partial_2:\sC_2\to \sC_1$ is
\be\label{bmap} \partial_2(s) = s\oplus (-s-VsV^{-1}) ,\ee
while $\partial_1:\sC_1\to \sC_0$ is
\be\label{cmap}\partial_1(t\oplus u)= UtU^{-1}-t+ VuV^{-1}-u. \ee
One may verify that $\partial_1\circ\partial_2=0$, using $V^2=U$.

Because $V$ and $U$ are diagonal, the torsion of $\cC $ has the same sort of factorization as in section \ref{torcircle}:
\be\label{romap} \tau_\cC =\tau_\cC ^\perp\cdot \tau_\cC ^0.\ee

To compute $\tau_\cC ^\perp$ and $\tau_\cC ^0$, note first that if we compose $\partial_2$ with the projection map on the first summand of $\sC_1=\sl_2\oplus \sl_2$, it is the identity.
This means that $\tau_\cC ^\perp$ is unchanged if we just drop $\sC_2$ and the first summand of $\sC_1$.   This reduces us to the torsion of the map $\partial_1':\sl_2\to\sl_2$ 
defined by $\partial_1'(u)= VuV^{-1}-u$.    

From here on, the discussion is similar to the discussion of the torsion of a circle in section \ref{torcircle}.
Restricted to $\sl_2^\perp$, $\partial_1'$ is invertible, and $\tau_\cC ^\perp$ is just the absolute value of its determinant.  The eigenvalues of $\partial_1'$ acting on $\sl_2^\perp$ are
$-1-e^{\pm a/2}$, so its determinant is $(1+e^{a/2})(1+e^{-a/2})=4\cosh^2 a/4$:
\be\label{pmap} \tau_\cC ^\perp = 4\cosh^2 \frac{a}{4}. \ee
On the other hand, restricted to $\sl_2^0$, $\partial_1'=0$.    So its torsion is just a ratio of two copies of the natural measure on $\sl_2^0$.   As in section \ref{torcircle},
though we could call this ratio 1, it is more useful to think of it as  a ratio of measures on two copies of $\sl_2^0$: one copy is the tangent bundle to the moduli
space of flat  connections on $C$, and the other copy generates the automorphism group of such a flat connection.      Relative to section \ref{torcircle},
we have to be careful with a factor of 2.  The definition of $\partial_1'$ is such that $\partial_1'$ maps  flat sections on $S'$ (elements of
the second summand in $\sC_1$) to flat sections at $p$, 
but when we glue $C$ to some other Riemann surface, the gluing takes place along $S$, not $S'$.   So what was $\d a\cdot\d\varrho^{-1}$ in section \ref{torcircle}
here becomes $\d\bar a\cdot \d\varrho^{-1}$, where $\bar a$ is the length parameter of $S'$, but $\varrho$ generates twists  of  $S$.
    Since $\bar a=a/2$ (the monodromy around $S$
is $\mathrm{diag}(e^{a/2},e^{-a/2}),$ while that around $S'$ is $\mathrm{diag}(e^{\bar a/2},-e^{-\bar a/2})$),  we have 
\be\label{gmapo}\tau_\cC ^0=\d\bar a\cdot (\d\varrho)^{-1}=\frac{1}{2}\d a\cdot (\d\varrho)^{-1}\ee
and 
\be\label{rmap}\tau_\cC =2\cosh^2 \frac{a}{4} \cdot \d a \cdot (\d\varrho)^{-1}. \ee

\subsubsection{The Torsion Of An Unorientable Two-Manifold}\label{torunorientable}

Now let $Y$ be an unorientable two-manifold, without boundary for simplicity.  Such a manifold is not a complex Riemann surface, but -- assuming its Euler characteristic is negative --
it does admit a hyperbolic metric, that is a metric of constant scalar curvature $R=-2$.

\begin{figure}
 \begin{center} 
   \includegraphics[width=3.5in]{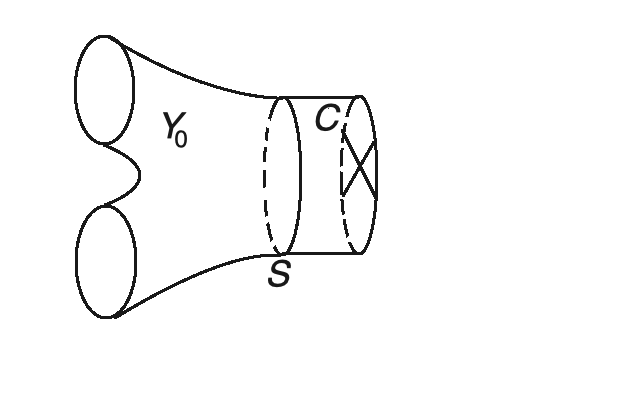}
 \end{center}
\caption{\small A boundary of a two-manifold $Y_0$ can be replaced by a crosscap by gluing  the crosscap manifold $\cC$ to a boundary $S$ of $Y_0$.   \label{NewBuilding}}
\end{figure}

A compact oriented hyperbolic  two-manifold can be built by gluing together three-holed hyperbolic spheres with geodesic boundaries.    
To build a hyperbolic metric on an arbitrary possibly unorientable
$Y$, we need another kind of building block.   This is obtained from a three-holed sphere $Y_0$ by replacing one or more of its boundary circles with crosscaps.
If one of the boundaries of $Y_0$ is a geodesic circle $S$ defined by $\theta\cong \theta+2\pi$, then to replace $S$ with a crosscap, we just make a further
identification $\theta\cong \theta+\pi$, after which $Y_0$ becomes unorientable and $S$ is replaced by a circle $S'$ that is a geodesic of one-half the length of $S$.
$S'$ is a ``one-sided'' geodesic, because in going around $S'$, its two sides are exchanged.     

If we make such a replacement on one or two boundary components of $Y_0$, we get a hyperbolic two-manifold $Y'$ with two or one remaining boundary components
that can be used in further gluing.\footnote{If we make the same replacement on all three boundary
components of $Y_0$, we get a hyperbolic structure on a particular compact two-manifold $Y'$ without boundary. As $Y'$ has no boundary, it is not a building block
for further gluing.   But the discussion in the text applies for computing its torsion.}  In such gluing, we use as ``elementary building blocks'' either a three-holed sphere
or its modification by replacing some boundaries with crosscaps, as just described.
 By gluing such elementary building blocks together along their geodesic boundaries,
one can construct (in general in infinitely many different ways) hyperbolic metrics on an arbitrary possibly unorientable two-manifold $Y$ without boundary.\footnote{The gluing
maps used here can  be orientation-reversing.  If $Y$ is unorientable
but has even (negative)  Euler characteristic, then hyperbolic metrics on $Y$ can  be constructed without the use of crosscaps by gluing together three-holed spheres with gluing maps some of
which are orientation-reversing.}

If $Y$ is presented in such a fashion, then the moduli space of hyperbolic structures on $Y$ can be described locally by a generalization of the length-twist
or Fenchel-Nielsen coordinates that we used in section \ref{oriented}.   Gluing of two elementary building blocks along a boundary circle always produces a pair
of length-twist moduli $a,\varrho$.   But if one of the elementary building blocks has \ a one-sided geodesic as one of its ends, then this one-sided geodesic carries a length
parameter but no corresponding twist parameter, as it is not going to be glued onto anything.   

To get a general formula for the torsion of any two-manifold in length-twist coordinates, we just need to know what happens when a boundary circle is
replaced by a one-sided geodesic.   There is a local universal formula for this.  
Let $Y_0$ be a three-holed sphere and let $\cC$ be the crosscap manifold whose torsion was computed in section \ref{crosstor}.
Thus $\cC$ has a single boundary circle $S$.  By gluing $S$ onto one of the boundary circles of $Y_0$, we make a new manifold $Y'$ in which a boundary
circle has been replaced by a one-sided geodesic (fig.~\ref{NewBuilding}).    Since we already have computed the torsions of $Y_0$, $\cC$, and $S$, it is straightforward
to compute the torsion of $Y'$ using the gluing law:
\be\label{worgo} \tau_{Y'}=\tau_{Y_0}\frac{1}{\tau_S}\tau_{\cC}. \ee
We focus on the factors in the torsion that involve the boundary component $S$ 
of $Y_0$ at which the gluing is occurring.    If the holonomy around $S$ is $\diag(e^{a/2},e^{-a/2})$ (and therefore the length
parameter of the one-sided geodesic in $\cC$ is $\bar a=a/2$), then the factor in $\tau_{Y_0}$ that involves $a$ is $2\sinh \frac {a}{2} \, \d a$.   Including additional $a$-dependent
factors in  eqns. (\ref{rubb}) and
(\ref{rmap}), one finds that the $a$-dependent part of $\tau_{Y_0}\tau_S^{-1}\tau_{\cC}$ is $\frac{1}{2}\coth \frac{a}{4} \,\d a$.

A two-manifold $Y$ of genus $g$ with $n$ crosscaps attached can be built by gluing together $2g-2+n$ building blocks, namely three-holed spheres with a total of $n$
boundaries replaced by crosscaps, along $3g-3+n$ circles.   The Euler characteristic of such a surface is $\chi=2-2g-n$.    Associated with the $i^{th}$ circle
is a pair of length-twist parameters $a_i,\varrho_i$, and associated with the $\alpha^{th}$ crosscap is a length parameter $a_\alpha/2$ (the length of its one-sided geodesic).   
Putting the pieces together,
and keeping track of  factors of 2, we learn that the torsion of $Y$ is
\be\label{yorgo}\tau_Y=\prod_{i=1}^{3g-3} \d a_i\d\varrho_i \prod_{\alpha=1}^n\frac{1}{2} \coth \frac{a_\alpha}{4}\d a_\alpha. \ee

This volume form on the moduli space of hyperbolic structures on an unorientable two-manifold $Y$ was actually first defined by Norbury (see Theorem 5 in \cite{Norbury}; note
that Norbury's $\ell_\alpha$ is our $a_\alpha/2$).   This volume form has been further studied in \cite{Gendulphe}.
The starting point in \cite{Norbury} was not the torsion.
   Rather, the starting point was the hypothesis that there might be a volume form that takes a simple factorized form in length-twist coordinates
and is independent of the particular choice of such coordinates.   As there are infinitely many sets of length-twist coordinates on a given surface,
the condition that a particular factorized expression does not depend on the choice  is very restrictive. It turned out that the expression in eqn.~(\ref{yorgo})
does have this property (and  no other factorized expression does \cite{Gendulphe}).  
   In our derivation here, the starting point was the torsion, which manifestly is a topological invariant, not depending on a choice of length-twist coordinates.   That starting
point did not make it completely obvious that we would get a factorized expression in length-twist coordinates, but the gluing law for the torsion goes a long way towards
explaining why this happens.   

As originally noted in \cite{Norbury} and further elaborated in \cite{Gendulphe}, if one actually tries to integrate the volume form $\tau_Y$ to compute the volume
of the moduli space, one runs into what in string theory would be called an infrared divergence.   Indeed, the measure behaves as $\d a_\alpha/a_\alpha$ for small
$a_\alpha$.

\subsubsection{Application To JT Gravity}\label{zoggo} 

\begin{figure}
 \begin{center} 
   \includegraphics[width=3in]{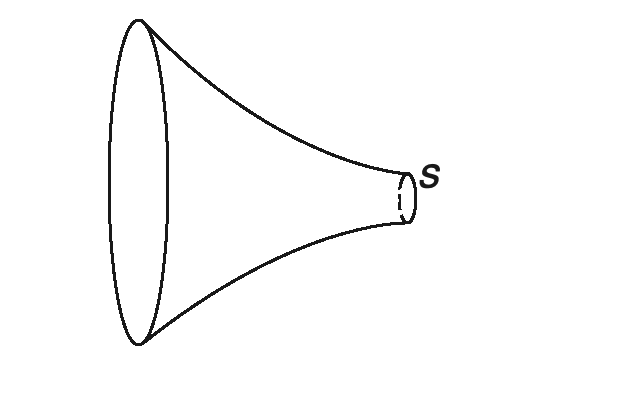}
 \end{center}
\caption{\small  A trumpet.   \label{Trumpet}}
\end{figure}

In applications to  JT gravity, one is usually not interested in a compact two-manifold $Y$, but in a two-manifold with non-compact ``ends'' that are asymptotic
to a cutoff version of $\AdS_2$, which we will call $\NAdS_2$.   A Schwarzian mode propagates along the boundary of $\NAdS_2$. 

The most basic example is $\NAdS_2$ itself.   An important ingredient in describing other possibilities is a ``trumpet'' $T$ (fig.~\ref{Trumpet}), 
which looks like $\NAdS_2$ on the ``outside'' and has
an ``internal'' boundary consisting of a circle $S$ whose boundary is a geodesic with length parameter $a$.   The boundary of $T$ can be glued onto a boundary component 
of a compact two-manifold, or onto a two-manifold with additional noncompact ends.    For example, one can glue together two trumpets along a common boundary
of length $a$ to make a ``double trumpet,'' which we already discussed in section \ref{evenodd}.    
Let $Z^T_\JT(a)$ be the JT path integral on a trumpet $T$ whose internal boundary has length $a$.   An important factor in $Z^T_\JT(a)$ is the path integral of the Schwarzian
mode.   The JT path integral of a double trumpet is the product of two factors of $Z^T_\JT(a)$, one for each trumpet, integrated over $a$ and over the gluing parameter
$\varrho$.   The measure that must be integrated is
\be\label{woggo} Z^T_\JT(a)\, \d a\d \varrho\, Z^T_\JT(a). \ee

It is not immediately clear  how to compute $Z^T_\JT(a)$ in the language of the torsion.   It may be hard to combine the combinatorial approach to the torsion
with the  boundary conditions that are needed to include 
 the Schwarzian mode on the ``outer'' boundary of the trumpet.  Even on the ``inner'' boundary of the trumpet, there is a nontrivial
issue to consider:  the boundary condition that we have used in computing the torsion on a manifold with boundary actually does not coincide with what is usually
used in defining the JT trumpet path integral $Z_\JT(a)$.   That is why in the gluing law for the torsion, which we described in section \ref{gluing}, one has to divide
by $\tau_S$, the torsion of a circle.

Rather than trying to deal with these issues directly, we will take a shortcut.   We will write $\t Z^T_\JT(a)\d a$ for a JT path integral on the trumpet computed with
``Schwarzian'' boundary conditions on the outer boundary, and with ``torsion'' boundary conditions on the inner boundary.   We will not try to compute $\t Z_\JT(a)$ 
directly, but rather we will infer it by comparing to eqn.~(\ref{woggo}).   To get from $\t Z^T_\JT(a)\d a$ to a JT path integral on the double trumpet, we simply glue together
two trumpets on their inner boundaries, 
using the torsion gluing rule since $\t Z^T_\JT(a)\d a$ is defined with ``torsion'' boundary conditions on the inner boundary.   So eqn.~(\ref{woggo}) should
be compared to $\t Z^T_\JT(a) \d a \tau_S^{-1}\t Z^T_\JT(a)\d a$.   Using eqn.~(\ref{rubb}) for  $\tau_S$, we get\footnote{Recall that we are really discussing measures,
so the overall signs in these formulas are not important.}
\be\label{roggo}Z^T_\JT(a) \d a\d\varrho Z^T_\JT(a) = \t Z^T_\JT(a) \frac{\d a\d\varrho }{4\sinh^2 \frac{a}{2}} \t Z^T_\JT(a). \ee
So
\be\label{loggo} \t Z^T_\JT(a) =Z^T_\JT \cdot 2 \sinh \frac{a}{2} . \ee
(Since the bosonic JT path integral is positive and the torsion is a positive measure, we do not need to worry about a sign here in extracting a square root.)

Now that we know the trumpet path integral with  ``torsion''  boundary conditions on the inner boundary, we can predict the path integral for any manifold made by gluing a compact surface $Y$ onto
a trumpet (or a collection of trumpets). We just treat $Y$ and its gluing onto the trumpet using the usual rules for the torsion.
For example, the basic crosscap spacetime is made by replacing the boundary of the trumpet with a crosscap.  
Replacing a boundary with a crosscap  is done by gluing in a copy of the crosscap manifold $\cC$, so it 
gives a factor of $\frac{1}{\tau_S}\tau_\cC $, where $\tau_S$ is given in eqn.~(\ref{rubb}) and $\tau_\cC $ in eqn.~(\ref{rmap}).
Using also eqn.~(\ref{loggo}) for $\t Z^T_\JT(a)$, we find that the path integral measure for the crosscap spacetime is
simply
\be\label{proggo} Z^T_\JT \cdot\frac{1}{2} \coth \frac{a}{4}\,\d a,\ee
where  the circumference of the one-sided geodesic in the crosscap is $\bar a=a/2$.
  
In a theory with fermions, this is the result for a particular $\pin^-$ or $\pin^+$ structure.   It must be multiplied by another factor that arises from the sum over $\pin^\pm$ structures, as described in section \ref{topominus}. 

\subsection{Supersymmetry And The Torsion}\label{osp}

\subsubsection{Supersymmetry and Time-Reversal In Two Dimensions}\label{whypin}
As discussed in section \ref{classlike}, a time-reversal symmetry $\sT$, in a theory with no relevant global symmetry other than $(-1)^\sF$, will satisfy either $\sT^2=1$ or
$\sT^2=(-1)^\sF$.  The two cases correspond to $\pin^-$ and $\pin^+$, respectively.  
In either case, there might be a $c$-number anomaly, so the distinction is really that for $\pin^-$,  $\sT^2$ commutes with fermionic fields,
and for $\pin^+$,  it anticommutes with them.

In a quantum mechanical system with just a single supercharge $Q$, a $\sT$ symmetry is necessarily of $\pin^-$ type,
since $\sT^2=(-1)^\sF$ would force fermionic operators (such as the supercharge $Q$) to come in pairs.

A similar statement holds in a bulk dual theory, when there is one. 
   As we will explain, in a two-dimensional model with $\N=1$ supersymmetry and no relevant global symmetries,
time-reversal symmetry of $\pin^-$ type is the only option.    Explaining this will involve a few details, but the necessary details 
will anyway serve as preparation for analyzing the torsion in $\N=1$ supergravity.

In general, in $D$ dimensions, $\AdS_D$ spacetime has symmetry $\SO(2,D-1)$ in Lorentz signature, or $\SO(1,D)$ in Euclidean signature.   $D=2$ happens to be the only
case in which these groups are isomorphic.   We start in Lorentz signature with $\AdS_2$ defined as the universal cover of the space parametrized by $X,Y,Z$ with a constraint
\be\label{lux} X^2+Y^2-Z^2=R^2,\ee
where $R$ is the radius of curvature, and the metric tensor is $\d s^2=-\d X^2-\d Y^2+\d Z^2.$   The constraint (\ref{lux}) can be solved by introducing a time coordinate $t$
with
\be\label{ux} X=\sqrt{R^2+Z^2}\cos t, ~~~Y=\sqrt{R^2+Z^2}\sin t. \ee
This makes it evident that we can take time-reversal $\sT$ to act by $(X,Y,Z)\to (X,-Y,Z)$, while   $(X,Y,Z)\to (X,Y,-Z)$ is a spatial reflection $\sR$.   In a relativistic
theory, $\sT$ is a symmetry if and only if $\sR$ is, and (in the absence of other symmetries with which $\sT$ or $\sR$ could be combined)
 the properties of $\sT$ determine the properties of $\sR$.   We will consider both $\sT$ and $\sR$ because this will make the passage to Euclidean signature obvious.

To include spin, but not supersymmetry, one replaces $\SO(2,1)$ by its double cover $\SL(2,\R)$.   In the generalization with $\N=1$ supersymmetry,
$\SL(2,\R)$ is extended to
 the supergroup $\OSp(1|2)$, which can be regarded as the group of linear transformations of two bosonic variables $u,v$ and one fermionic variable $\uptheta$
that preserve the symplectic form
\be\label{lymp}\hat\omega=\d u\d v+\frac{1}{2}\d\uptheta^2. \ee   The corresponding Lie superalgebra $\osp(1|2)$ can be described explicitly in terms of
matrices acting on $u,v|\uptheta$.  
It will be helpful to have a concrete description of the Lie  superalgebra $\osp(1|2)$.  This Lie algebra has dimension $3|2$ -- three bosonic generators
and two fermionic ones.   Acting on the triple $\begin{pmatrix}u\cr v\cr \uptheta\end{pmatrix}$, one can pick a basis of bosonic generators 
\begin{equation}\label{bosgen}\newcommand*{\temp}{\multicolumn{1}{r|}{}}
e=\left(\begin{array}{cccccc}
0 &1 \negthinspace\negthinspace\negthinspace &\temp & 0\\ 
0 &0\negthinspace \negthinspace \negthinspace&\temp &0\\
 \cline{1-4}
0 &0\negthinspace \negthinspace \negthinspace&\temp &0\\

\end{array}\right), ~~~~
f=\left(\begin{array}{cccccc}
0 &0 \negthinspace\negthinspace\negthinspace &\temp & 0\\
1 &0\negthinspace \negthinspace \negthinspace&\temp &0\\
 \cline{1-4}
0 &0\negthinspace \negthinspace \negthinspace&\temp &0\\

\end{array}\right),~~~~
h=\left(\begin{array}{cccccc}
1 &0 \negthinspace\negthinspace\negthinspace &\temp & 0\\
0 &-1\negthinspace \negthinspace \negthinspace&\temp &0\\
 \cline{1-4}
0 &0\negthinspace \negthinspace \negthinspace&\temp &0\\

\end{array}\right)
\end{equation}
and fermionic ones
\begin{equation}\label{fermgen}\newcommand*{\temp}{\multicolumn{1}{r|}{}}
q_1=\left(\begin{array}{cccccc}
0 &0 \negthinspace\negthinspace\negthinspace &\temp & 1\\
0 &0\negthinspace \negthinspace \negthinspace&\temp &0\\
 \cline{1-4}
0 &-1\negthinspace \negthinspace \negthinspace&\temp &0\\ \end{array}\right),~~~~~
q_2=\left(\begin{array}{cccccc}
0 &0 \negthinspace\negthinspace\negthinspace &\temp & 0\\
0 &0\negthinspace \negthinspace \negthinspace&\temp &1\\
 \cline{1-4}
1 &0\negthinspace \negthinspace \negthinspace&\temp &0\\
\end{array}\right).     \ee
The Lie superalgebra is characterized by a mixture of commutators and anticommutators, depending on whether an element of $\osp(1|2)$ is ``bosonic'' (supported
in the diagonal blocks of the matrix) or ``fermionic'' (supported in the off-diagonal blocks).    In detail, the anticommutators are
\be\label{antic} q_1^2=-e,~~q_2^2=f,~~\{q_1,q_2\}=h, \ee
and the nonzero commutators are
\be\label{comms}[h,e]=2e, ~[h,f]=-2f, ~[e,f]=h,~ [h,q_1]=q_1,~[h,q_2]=-q_2,~ [e,q_2]=q_1,~[f,q_1]=q_2.\ee

Symmetry groups or supergroups in Lorentz signature are always real, so we consider the real form of $\OSp(1|2)$, with $u,v,\uptheta$ considered real.
Actually, $\OSp(1|2)$ contains a central element that acts as $-1$ on $u,v,\uptheta$.    This element commutes with all of the fields of $BF$ theory (since they transform in the adjoint
representation) so we do not want to regard it as a symmetry.    Thus on an orientable manifold with $\N=1$ supersymmetry, we should consider $BF$ theory of the 
supergroup $\OSp'(1|2)=\OSp(1|2)/\Z_2$.   In $\OSp'(1|2)$, the group elements $\diag(1,1,-1)$ and $\diag(-1,-1,1)$ are equivalent; either one is the operator $(-1)^\sF$ that
distinguishes bosons and fermions.

Now we want to include $\sR$ and $\sT$ and show that it is unavoidable to get $\sR^2=(-1)^\sF$ and $\sT^2=1$, corresponding to a $\pin^-$ structure.\footnote{See  footnote
\ref{oddone} in section \ref{topominus} for an explanation  of why $\sT^2$ and $\sR^2$ have opposite properties.}
As an automorphism of $\sl(2,\R)$, $\sR:(X,Y,Z)\to (X,Y,-Z)$ corresponds to $e\leftrightarrow -f$, $h\to -h$.  The extension of this to the odd generators
is $q_1\to q_2$, $q_2\to -q_1$.     So  as a transformation of the Lie superalgebra, $\sR^2=(-1)^\sF$. Note that the eigenvalues of $\sR$ acting on $q_1,q_2$
are $\pm \i$.    It is a little tricky to analyze $\sT$ because it is anti-unitary
(not unitary) in quantum mechanics.   As a shortcut, we note that $\sT:(X,Y,Z)\to (-X,Y,Z)$ commutes with the subgroup $\SO(1,1)$ of boosts of the $YZ$ plane.   This
subgroup acts on the pair $q_1,q_2$ with distinct real eigenvalues and eigenvectors.   So $\sT$, acting on the fermionic part of the Lie superalgebra,
has 1-dimensional representations, and therefore it must satisfy $\sT^2=1$, not $\sT^2=(-1)^\sF$.

We go to Euclidean signature by continuing $Y\to \i Y$.   The metric is now $\d s^2=-\d X^2+\d Y^2+\d Z^2$.  The reflection $\sR:(X,Y,Z)\to (X,Y,-Z)$
will still have eigenvalues $\pm \i$ in acting on the fermions.     In Euclidean signature, the subgroup of $\SO(2,1)$ that commutes with $\sR$ 
is the group $\SO(1,1)$ of Lorentz boosts of the $X-Z$ plane.   So the eigenvectors of $\sR$, acting on $q_1,q_2$ (or on $u,v$) are the eigenvectors
of a nontrivial element of $\SO(1,1)$.   Such an element corresponds to a hyperbolic element of $\SL(2,\R)$.  

The only unorientable manifold that we will have to study in any detail is the crosscap manifold $\cC$ of section \ref{crosstor}.  We recall that $\cC$ has a boundary
circle $S$ and an ``internal'' circle $S'$ (a one-sided geodesic), such that the monodromies $U$ and $V$ of a flat connection around $S$ and $S'$ satisfy $U=V^2$.   Suppose
that acting on the triple $\begin{pmatrix}u\cr v\cr \uptheta\end{pmatrix}$, $U=\diag(e^{\ell/2},e^{-\ell/2},\pm 1)$ where $\ell$ is a length parameter and a spin structure of R or NS
type  corresponds to a $+$ or $-$ sign.\footnote{In $\SL(2,\R)$, we used the matrices $\pm \diag(e^{\ell/2},e^{-\ell/2})$ to describe R or NS spin
structures  (see eqn.~(\ref{pokko})).   In $\OSp(1|2)/\Z_2$, $\diag(-e^{\ell/2},-e^{-\ell/2},  1)$ is equivalent to $\diag(e^{\ell/2},e^{-\ell/2},-1)$, and the latter form is sometimes
convenient. }     To satisfy $V^2=U$ where $V$ (like $\sR$) should act with imaginary eigenvalues  on $q_1,q_2$, we choose the minus sign in $U$,
and we take $V=\diag(e^{\ell/2},-e^{-\ell/2}, \pm \i)$.  The choice of sign corresponds to a choice of one of the two $\pin^-$ structures on $\cC$.  The fact that we had to take
a minus sign in $U$ explains, in the present context, an assertion that was made in section \ref{topominus}:  in the case of time-reversal symmetry of $\pin^-$ type,
a cylinder can end on a crosscap only if its spin structure is of NS type.    

The fact that $V$ is not real means that as soon as we are on an unorientable manifold, the gravitino field of JT supergravity is not real.\footnote{In terms of $\pin^-$ structures,
one would explain this as follows.   To build a $\pin^-$ structure on a Euclidean signature
two-manifold, one starts with a rank 2 Clifford algebra $\{\gamma_\mu,\gamma_\nu\}=-2\delta_{\mu\nu}$.
This algebra can be represented by $2\times 2$ matrices, but not by real $2\times 2$ matrices, so a fermion field coupled to the $\pin^-$ structure is not real.
In Lorentz signature, one would have $\{\gamma_\mu,\gamma_\nu\}=-2\eta_{\mu\nu}$ (with $\eta=\mathrm{diag}(-1,1)$), and this algebra does have a 
representation by real $2\times 2$ matrices, so the gravitino field in Lorentz signature can be real.}   This is a common occurrence when a Lorentz signature field
theory is continued to Euclidean signature.  For example, in Euclidean signature, the fermions of the Standard Model are not real.

\subsubsection{Generalities About The Torsion}\label{gentor}

The generalities about $BF$ theory and the analytic and combinatorial torsion carry over directly for a semi-simple supergroup such as ${\OSp}'(1|2)$.
In particular, we anticipate that the proof in \cite{Muller2,bismut1991metriques,bismut1992extension} of equivalence of the analytic and the combinatorial torsion generalizes from non-compact semi-simple
groups such as $\SL(2,\R)$ to corresponding supergroups such as ${\OSp'(1|2)}$.  The proof of the topological invariance of the combinatorial
torsion certainly does go through in the same way for a supergroup as for a group, 
as does the proof of the gluing law for the torsion.  If $Y$ is an oriented two-manifold without boundary,
then the moduli space $\rR$ of flat connections on $Y$ valued in a semi-simple supergroup can be given a symplectic structure that is defined by the same formula
as for an ordinary group, and this determines a measure on $\rR$.   On such a manifold, the torsion is ``trivial''  in the sense that it coincides with the symplectic measure.  
For unorientable $Y$, the torsion is not equivalent to anything more elementary and we compute it using the combinatorial definition.  

The definition of the combinatorial torsion in the case of a supergroup can be made by formally imitating the definitions for a bosonic group.
The Lie algebra $\osp(1|2)$ of $\OSp'(1|2)/\Z_2$ has an invariant  nondegenerate quadratic form.
Just as in the bosonic world, a nondegenerate quadratic form  on a vector space such as $\osp(1|2)$ determines a measure on that space.
The meaning of a measure on a super vector space is more subtle than in the bosonic case, but we postpone discussing that point and proceed formally
for the moment.
First of all, the vector spaces $\sC_q$, $q=2,1,0$, are defined in the same way as in section \ref{combitor} as the direct sum of copies of the Lie algebra $\osp(1|2)$,
with one copy for each $q$-cell.  Hence each $\sC_q$ gets a measure $\alpha_q$.
The boundary maps $\partial:\sC_q\to \sC_{q-1}$ are defined precisely as in section \ref{combitor}.   The basic definition (\ref{numbo}) of the torsion and the
generalizations (\ref{mumbo}) and (\ref{tmumbo}) that are needed when the homology is nontrivial are still valid.   However, when one picks the basis vectors
of a vector space, some of these basis vectors are bosonic and some are fermionic.  For example, if $\sC_q$ has dimension $n_q|m_q$ (bosonic dimension $n_q$
and fermionic dimension $m_q$), then a basis of $\sC_2$ consists of $n_2$ bosonic vectors and $m_2$ fermionic ones.   We might write such a basis
schematically as $s_1,\cdots,s_{n_2}|\t s_1,\cdots,\t s_{m_2}$, where the $s_i$ are bosonic and the $\t s_j$ are fermionic.   Similarly, to complete
$\partial s_1\cdots \partial s_{n_2}|\partial \t s_1 \cdots\partial \t s_{m_2}$ to a basis of $\sC_1$ (assuming for simplicity that the homology vanishes) we add
bosonic and fermionic vectors $t_1,\cdots, t_{n_1-n_2}|\t t_1,\cdots, \t t_{m_1-m_2}$.    

All these bosonic and fermionic vectors are then included in the definition of the torsion in eqn.~(\ref{numbo}) and its generalizations, but
some subtlety is hidden in the question of what is meant by a measure on a vector space.   The basic idea is that measures for bosons transform
oppositely to measures for fermions.
 To identify the essential point, suppose
that $V$ is a 1-dimensional bosonic vector space with measure $\alpha$, and let $v\in V$  be a basis vector, and $\lambda$ a complex scalar.  Then
$\alpha(\lambda v)=|\lambda|\alpha(v)$.   But if $V$ is a 1-dimensional fermionic vector space with measure $\alpha$ and basis vector $\psi$,
then $\alpha(\lambda\psi)=\lambda^{-1}\psi$.   More generally, suppose that $V$ is a vector space of dimension $n|m$ with basis $s_1,\cdots, s_n|\t s_1,\cdots,\t s_m$.
Suppose that another basis $r_1,\cdots, r_n|\t r_1,\cdots, \t r_m$ is obtained by acting on the first basis with a supermatrix $M$:
\be\label{bosblo} \begin{pmatrix} r_1\cr \vdots\cr r_n\cr  \horzbar\cr \t r_1\cr \vdots \cr \t r_m\end{pmatrix}=M \begin{pmatrix}s_1 \cr \vdots\cr s_n\cr\horzbar \cr \t s_1 \cr\vdots \cr \t s_m\end{pmatrix} =  \newcommand*{\temp}{\multicolumn{1}{r|}{}}
\left(\begin{array}{cccccc}
A\negthinspace \negthinspace\negthinspace \negthinspace \negthinspace&\temp &B\\
 \cline{1-3}
C\negthinspace \negthinspace\negthinspace\negthinspace \negthinspace&\temp &D\\

\end{array}\right) 
\begin{pmatrix}s_1 \cr \vdots\cr s_n\cr\horzbar \cr \t s_1 \cr\vdots \cr \t s_m\end{pmatrix}, \ee
where $A$ and $D$ are bosonic blocks and $B$ and $C$ are fermionic.
Then the defining property of a measure $\alpha$ (or a density, in the terminology of \cite{DM}, section 3.9; see also \cite{TVoronov,TVoronov2})  is that it is a function on bases such that, if we denote the two bases just as $\vec r$ and $\vec s$ respectively,  then 
\be\label{funbas}\alpha (\vec r)= \Ber'(M) \alpha(\vec s). \ee
This is the superspace generalization of the bosonic formula (\ref{transrule}).

But we need to explain the meaning of $\Ber'$.   First of all,  the superanalog of the determinant is the Berezinian, denoted $\Ber$.   Like the determinant, it obeys $\Ber(M_1M_2)=\Ber(M_1)\Ber(M_2)$ (this is necessary
for the consistency of eqn.~(\ref{funbas})).  There is a particularly simple formula for $\Ber\,M$ 
if $M$ is upper or lower block triangular, in the sense that one of its fermionic blocks  $B$ or $C$ vanishes.   In that case,
\be\label{yggo}\Ber\, M = \det A \cdot \frac{1}{\det D}. \ee  
In evaluating Berezinians, there is also an analog of the usual row and column reduction for determinants.
This is really all we will need to know about the Berezinian.  

However, if $\alpha$ is supposed to be a ${\it measure}$ (rather than a superspace analog of a differential form of top degree), we need a slight refinement of the Berezinian.
$\Ber'$ is defined for a matrix
 $ \newcommand*{\temp}{\multicolumn{1}{r|}{}} M=
\left(\begin{array}{cccccc}
A\negthinspace \negthinspace\negthinspace \negthinspace \negthinspace&\temp &B\\
 \cline{1-3}
C\negthinspace \negthinspace\negthinspace\negthinspace \negthinspace&\temp &D\\
\end{array}\right) $
that has the property  that $A$ is invertible and moreover is real modulo nilpotent  variables.   This being so, $\det A$ is real and nonzero modulo nilpotents
and has a well-defined sign, which  moreover is  invariant  under conjugation  of $M$  (so this sign does not depend on a specific choice of how to decompose $M$
in bosonic and fermionic blocks).
Thus we can define
\be\label{newber} \Ber'(M)={\mathrm{sign}}(\det A)\,\Ber(M). \ee
This is the natural superanalog of $|\det M|$ in the bosonic case.\footnote{The fact that one uses an absolute value for  the block $A$ and not for its fermionic analog $D$
is related to the following fact about  Gaussian integrals.   A bosonic Gaussian integral involves an absolute value; for instance the integral $\frac{1}{2\pi}\int_{-\infty}^\infty \d x\d y
\,\exp(\i\lambda xy)$ is defined for real $\lambda$, and its evaluation  involves an absolute value:
$\frac{1}{2\pi}\int_{-\infty}^\infty \d x\d y
\,\exp(\i\lambda xy)=1/|\lambda|$.
But an analogous fermionic  integral is defined for all $\lambda$ and is holomorphic in $\lambda$: $\int \d \psi \,\d\chi \exp(\lambda\psi\chi)=\lambda$.}

One difference between groups and supergroups is that the torsion for a supergroup has no obvious positivity properties, and therefore, in contrast to the bosonic
case, one does need to be careful with the overall sign.   This is unfortunately easier said than done.
We will see that on an unorientable manifold, the torsion is actually complex-valued.   This should not be too startling,
because in many theories,  fermionic path integrals in Euclidean signature are complex-valued.  The Standard Model is an example.   What is slightly unusual about super
JT gravity is that its path integral is real on an orientable manifold, and becomes complex only on an unorientable manifold.   This, however,
is also not unique; a Majorana fermion in $2+1$ dimensions has the same behavior.  

In section \ref{firstcases}, we compute the torsion of an $\OSp'(1|2)$ flat connection on a circle or a crosscap.  These calculations are an almost immediate
analog of what we have already done in the bosonic case.   With these results, we can generalize the analysis of section \ref{zoggo} and predict the 
contribution in supersymmetric JT gravity  of a spacetime that consists of a trumpet ending on a crosscap.
To be able to compute for a general unorientable spacetime, we also need the torsion of a three-holed sphere, which is evaluated in section \ref{revisited}.   
Concluding remarks are in  section \ref{applications}.

\subsubsection{First Cases and Application}\label{firstcases}

A hyperbolic element $U$ of $\OSp'(1|2)$,  acting on the usual triple $(u,v,\uptheta)$, is of the form $\diag(e^{a/2},e^{-a/2},\vnu)$, where $\vnu=1 $ or  $-1$ for a spin structure
of R or NS type and $a$ is real.      If $U$ is the holonomy of a geodesic in a super Riemann
surface with a hyperbolic metric, then the length of the geodesic is $|a|$.   In computing the torsion of a flat connection on a circle $S$, we can assume that the holonomy is
of this form.   

We can proceed exactly as in section \ref{torcircle}.   $S$ has a cell decomposition with one 0-cell and one 1-cell, so $\sC_0$ and $\sC_1$ are both copies of the Lie superalgebra
$\osp(1|2)$.   As in eqn.~(\ref{partgo}), the boundary map is $\partial s = UsU^{-1}-s$.    Moreover, as in the bosonic case, we  can decompose $\osp(1|2)$ as $\osp(1|2)^0\oplus
\osp(1|2)^\perp$, where $\osp(1|2)^0$ is the Lie algebra  generated by $h$, and $\osp(1|2)^\perp$ is the orthocomplement of this.  (Thus  $\osp(1|2)^\perp$ has dimension $2|2$, and is generated by $e,f,q_1,q_2$.)   
 The boundary operator has a similar decomposition $\partial=\partial^0\oplus \partial^\perp$,
and hence the torsion factorizes:
\be\label{facto}\tau_S=\tau_S^0\cdot \tau_S^\perp. \ee

Just as in the bosonic case, $\partial^0=0$ and hence $\tau_S^0=\d a \cdot (\d\varrho)^{-1}$.   The interpretation of this formula is the same as it was in the bosonic case;
in applications, $\varrho$ will be a gluing parameter.  Moreover, $\partial^\perp$ is invertible, as in the bosonic case, and therefore $\tau_S^\perp=\Ber'(\partial^\perp)$.
This is the analog of the fact that in the bosonic case, we had $\tau_S^\perp=|\det\partial^\perp|$.
The eigenvalues of $\partial^\perp$ acting on bosonic states are $e^a-1$ and $e^{-a}-1$, while its eigenvalues acting on fermions are $\vnu e^{a/2}-1$ and $\vnu e^{-a/2}-1$.
 So $\Ber'(\partial^\perp)= |(e^a-1)(e^{-a}-1)|/(\vnu e^{a/2}-1)(\vnu e^{-a/2}-1)=-\vnu(e^{a/4}+\vnu e^{-a/4})^2$, and
therefore
\begin{align}\label{waus}\tau_S=&-\vnu(e^{a/4}+\vnu e^{-a/4})^2\,\d a\,(\d\varrho)^{-1}\cr =& \begin{cases} 4\sinh^2 \frac{a}{4}\, \d a\,(\d\varrho)^{-1} & \mathrm{NS~ spin~ structure}\cr  -4\cosh^2 \frac{a}{4}\, \d a\,(\d\varrho)^{-1}&\mathrm{R ~spin~ structure}.
 \end{cases}\end{align}

We can treat the crosscap manifold $\cC$ in the same way.   For the holonomy around the boundary circle $S$, we take $U$ as above but with $\vnu=-1$, so
$U=\diag(e^{a/2},e^{-a/2},-1)$.  
For the holonomy $V$ around the ``inner'' circle $S'$, which should satisfy $V^2=U$, we take $V=\mathrm{diag}(e^{a/4},-e^{-a/4},\i)$.    (This corresponds to
one of the two possible $\pin^-$ structures.    The second is obtained by replacing $\i$ with $-\i$, which will have the effect of complex conjugating all of the following
formulas.) 
The boundary operator $\partial$ is defined by the same formulas (\ref{bmap}) and (\ref{cmap}) as in the bosonic case.   It has the familiar decomposition
$\partial=\partial^0\oplus \partial^\perp$, and
  the torsion factorizes again as $\tau_\cC =\tau_\cC^0\cdot
\tau_\cC^\perp$.   The same steps as before show that $\tau_\cC^0=\frac{1}{2}\d a\cdot (\d\varrho)^{-1}$ (with a factor of $1/2$ that has the same origin as before), 
and that $\tau_\cC^\perp=\Ber'(\partial_1')$,   where
$\partial_1':\osp(1|2)^\perp\to \osp(1|2)^\perp$ is defined by $\partial'_1(u)=VuV^{-1}-u$.

The eigenvalues of $\partial'_1$ on bosons are $-e^{a/2}-1$ and $-e^{-a/2}-1$, while its eigenvalues on fermions are $-\i e^{a/4}-1$ and $\i e^{-a/4}-1$.   
So $\Ber'(\partial_1')=2\cosh^2 (\frac{a}{4})/(1+\i\sinh \frac{a}{4})=2(1-\i\sinh \frac{a}{4})$ and the torsion of the crosscap (with the chosen $\pin^-$ structure) is
\be\label{omic}\tau_\cC= \left(1-\i\sinh \frac{a}{4}\right) \d a\cdot (\d \varrho)^{-1}. \ee  

Now we can follow the logic of section \ref{zoggo} and predict, in supersymmetric JT gravity, the contribution of a trumpet that ends on a crosscap.
Let $Z^T_\SJT$ be the path integral in supersymmetric JT gravity for a trumpet $T$ that ends on a circle of circumference $a$.      Then the path
integral measure for a double trumpet  in super JT gravity  is 
\be\label{womic} Z^T_\SJT \,\d a\d \varrho\, Z^T_\SJT. \ee
As in our discussion of ordinary JT gravity, we do not have a convenient way to incorporate the super Schwarzian mode in the framework of combinatorial torsion,
so we proceed indirectly.   We write $\t Z^T_\SJT$ for a path integral of super JT gravity on the trumpet with ``super Schwarzian'' boundary conditions on the outer
boundary and ``torsion'' boundary conditions on the inner one.   Then using the gluing law for the torsion, the path integral on the double trumpet with super Schwarzian
boundary conditions on both boundaries is
\be\label{zomic} \t Z^T_\SJT \, \frac{\d a\,\d\varrho}{4\sinh^2 \frac{a}{4}} \t Z^T_\SJT. \ee
Comparing to (\ref{omic}) gives\footnote{We have to take a square root here.  Because of reflection positivity, we expect that we should take the positive square root.}
\be\label{lomic} \t Z^T_\SJT = 2\sinh\frac{a}{4} \,Z^T_\SJT. \ee

By the gluing law for the torsion,  the supersymmetric JT path integral for a trumpet that ends on a crosscap
 is just $\t Z^T_\SJT \cdot \tau_S^{-1}\tau_\cC$.  Using the above formulas for $\t Z^T_\SJT$, $\tau_S$ and $\tau_\cC$, we get
\be\label{lomioc} Z^T_\SJT\cdot\frac{1}{2} \left(-\i +\frac{1}{\sinh\frac{a}{4}}\right) \d a \ee
for the SJT path integral of a trumpet that ends on a crosscap. 

As explained in section \ref{SUSY}, this has to be multiplied by a factor $\exp(-\i\pi N' \eta/2)$ associated with the anomaly.   With $\eta=1/2$ for a crosscap,
this is $\exp(-\i\pi N'/4)$.    Then we have to sum over $\pin^-$ structures.    The two $\pin^-$ structures give complex conjugate contributions, since
flipping the $\pin^-$ structure changes the sign of $\eta$, and also replaces $V$ in the above derivation by its complex conjugate, as a result of which all the formulas
leading to eqn.~(\ref{lomioc}) get complex conjugated.

So after summing over $\pin^-$ structures, the super JT path integral  measure for a trumpet ending on a crosscap is
 \be\label{komic} Z^T_\SJT \cdot 2\,\mathrm{Re}\left(\exp(-\i\pi N'/4) \cdot\frac{1}{2}   \left(-\i +\frac{1}{\sinh\frac{a}{4}}\right) \right)\,\d a=
 Z^T_\SJT \cdot \left( -\sin\frac{ \pi N'}{4}+ \frac{\cos\frac{ \pi N'}{4}}{\sinh \frac{a}{4}}\right)\,\d a.\ee
Here $a/2$ is the length of the one-sided geodesic on which the crosscap ``ends.''   The integral over $a$ diverges for $a\to 0$, as in the bosonic theory, unless $N'$ is
congruent to 2 or 6 mod 8.  

We should acknowledge a gap in this computation.  Although it is true that the values of $\eta$ for a $\pin^-$ structure on a crosscap are $\pm 1/2$,
and that the two $\pin^-$ structures are characterized by $V=\diag(e^{a/2},-e^{-a/2},\pm \i)$, we did not carefully work out how the two choices of sign are correlated.   Reversing  this  relationship  would  have the same effect as $N'\to -N'$.    Actually, 
since the sign of the anomaly coefficient $N'$ depends on the sign with which time-reversal acts on the elementary fermions in an
underlying description such as an SYK-like model,
a physically meaningful statement about the sign ultimately depends  on comparing to a microscopic definition.

\subsubsection{The Three-Holed Sphere Revisited}\label{revisited}

The torsion of a three-holed sphere can be computed by adapting the considerations of section \ref{torsthree}.    The holonomies $U,V,W$ of a flat connection
around the three holes satisfy $UVW=1$.   This means that we can specify $U$ and $V$ independently (except that we want to restrict $U,V,W$  to be hyperbolic) and then
$W=V^{-1}U^{-1}$.   

The moduli space of flat $\OSp'(1|2)/\Z_2$ connections on a three-holed sphere has dimension $3|2$.   (Indeed, the pair $U,V$ depends on
$3|2+3|2=6|4$ parameters, but $3|2$ parameters are removed when we divide by conjugation, leaving $3|2$ parameters.)   A convenient parametrization is
to take $U=R U_0 R^{-1}$, $V=R V_0 R^{-1}$, where $R$ is an arbitrary element of the gauge group and 
\begin{equation}\label{twom}\newcommand*{\temp}{\multicolumn{1}{r|}{}}
U_0=\vnu_a\left(\begin{array}{cccccc}
e^{a/2} &\kappa    \negthinspace\negthinspace\negthinspace &\temp & 0\\
0 &e^{-a/2}\negthinspace \negthinspace \negthinspace&\temp &0\\
 \cline{1-4}
0 &0\negthinspace \negthinspace \negthinspace&\temp &\vnu_a\\

\end{array}\right)\exp(\xi q_1), ~~~~
V_0=\vnu_b\left(\begin{array}{cccccc}
e^{-b/2} &0 \negthinspace\negthinspace\negthinspace &\temp & 0\\
1 &e^{b/2}\negthinspace \negthinspace \negthinspace&\temp &0\\
 \cline{1-4}
0 &0\negthinspace \negthinspace \negthinspace&\temp &\vnu_b\\

\end{array}\right)\exp(\psi q_2).
\end{equation}
As before, $a$ and $b$ are real length parameters, and  $\vnu_a,\vnu_b=\pm 1$ represent the spin structures
on the corresponding circles.  
 The prefactors $\vnu_a$, $\vnu_b$ are inessential, since we really work in $\OSp'(1|2)=\OSp(1|2)/\Z_2$,
but are included to agree more smoothly with the natural formulas in $\SL(2,\R)$.  
The three bosonic moduli are $a,b,\kappa   $, and the two fermionic moduli are $\xi,\psi$.  
   We have $W=V^{-1}U^{-1}=R W_0 R^{-1}$ with
\begin{equation}\label{wombat}\newcommand*{\temp}{\multicolumn{1}{r|}{}}
W_0=\vnu_a\vnu_b\left(\begin{array}{cccccc}
e^{-(a-b)/2} &-\kappa    e^{b/2} \negthinspace\negthinspace\negthinspace &\temp &- \vnu_a\xi e^{b/2}\\
-e^{-a/2} &\kappa   + e^{(a-b)/2}-\vnu_b \psi\xi e^{a/2}\negthinspace \negthinspace \negthinspace&\temp &\vnu_a\xi-\vnu_a\vnu_b\psi\\
 \cline{1-4}
-\psi e^{-(a-b)/2} &\psi \kappa    e^{b/2}+\vnu_b\xi e^{a/2}\negthinspace \negthinspace \negthinspace&\temp &\vnu_b\vnu_a+\vnu_a\psi\xi e^{b/2}\\
\end{array}\right).\ee
$W$ will be conjugate to $\vnu_c  \diag(e^{c/2},e^{-c/2}|\vnu_c)$ for some $c$ and $\vnu_c$;  the length of the third hole is $|c|$, and its  spin structure
is controlled by $\vnu_c=\pm 1$.  To determine $c$, we use the fact that the supertrace is invariant under conjugation.\footnote{The supertrace of a matrix $ \newcommand*{\temp}{\multicolumn{1}{r|}{}} M=
\left(\begin{array}{cccccc}
A\negthinspace \negthinspace\negthinspace \negthinspace \negthinspace&\temp &B\\
 \cline{1-3}
C\negthinspace \negthinspace\negthinspace\negthinspace \negthinspace&\temp &D\\
\end{array}\right) $ is defined to be $\Str \,M=\Tr \,A-\Tr\,D$.}   So $\Str\, W=\vnu_c(2\cosh c/2-\vnu_c)$.  Setting this equal to $\Str \,W_0$, 
and using the fact that $\vnu_a\vnu_b\vnu_c=-1$, 
we get
\be\label{woffox} -2\cosh\frac{c}{2}=2\cosh \frac{a-b}{2} +\kappa  - \psi\xi (e^{a/2}\vnu_b+e^{b/2}\vnu_a) \ee
or
\be\label{noflox}\kappa =- 2\cosh\frac{c}{2}-2\cosh\frac{a-b}{2} +\psi\xi(e^{a/2}\vnu_b+e^{b/2}\vnu_a). \ee

The same reasoning as in the bosonic case leads to the condition
\be\label{purdux}\vol_G(U)\cdot \vol_G(V)=\vol_G(R) \cdot\tau, \ee
which determines $\tau$.   As in the bosonic case, it suffices to impose this condition at $R=1$.

However, we need a convenient way to describe concretely a measure such as $\vol_G(U)$.
Suppose that $N$ is a super vector space of dimension $n|m$.    Let $\vec f= (f^1,f^2,\cdots f^n |g^1, g^2, \cdots ,g^m)$ be a basis of $N$, and let 
$(f_1,f_2,\cdots, f_n| g_1,g_2,\cdots, g_m)$ be the dual basis of the dual vector space $N^*$.   Then as a tautology there is a measure on $N$
that we denote symbolically as $\alpha= [f_1,f_2,\cdots, f_n|g_1,g_2,\cdots,g_m]$.   It is defined by saying that the value it assigns to the basis $\vec f$ is
$\alpha(\vec f)=1$, while the value for any other basis of $N$ is then determined by eqn.~(\ref{funbas}).  
Under  a change of basis of $N^*$ by
\be\label{nurry}\begin{pmatrix} f_1\cr \vdots\cr  f_n\cr\horzbar \cr g_1 \cr\vdots\cr g_m\end{pmatrix} =M \begin{pmatrix} f'_1\cr \vdots\cr  f'_n\cr\horzbar \cr g'_1 \cr\vdots\cr g'_m\end{pmatrix}\ee 
with some matrix $M$, we have
\be\label{lurry} [f_1,\cdots ,f_n|g_1,\cdots,g_m]=\Ber'(M) [f_1',\cdots, f_n'| g_1,\cdots, g_m']. \ee
As special cases of this, if we rescale a bosonic basis vector of the dual space, say by $f_1\to \lambda f_1$, the measure is rescaled by a factor of $|\lambda|$,
\be\label{murry}[\lambda f_1,f_2,\cdots,f_n|g_1,g_2,\cdots, g_m]=|\lambda|\, [f_1,f_2,\cdots,f_n|g_1,g_2,\cdots,g_m],\ee
but if we similarly rescale a fermionic basis vector of the dual space, say by $g_1\to\lambda g_1$, then the measure is rescaled by a factor of $\lambda^{-1}$:
\be\label{purry}  [f_1,f_2,\cdots,f_n|\lambda g_1,g_2,\cdots,g_m]=\lambda^{-1}[f_1,f_2,\cdots,f_n|g_1,g_2,\cdots,g_m]                          .\ee
This $\lambda^{-1}$ is the reason that, in contrast to  an ordinary manifold, a measure on  a supermanifold cannot be interpreted as a differential form.
Eqn.~(\ref{purry}) generalizes the following property of fermionic integration:  if $\theta$ is a fermionic variable, then $\int \d (\lambda\theta) (A+B\theta) =\lambda^{-1}\int\d\theta(A+B\theta)=\lambda^{-1}B$.

A measure $\vol_G(U)$ is a measure on the tangent bundle of $G$, whose dual space is the cotangent bundle.
In section \ref{torsthree}, to get a basis of the cotangent bundle, we expanded $U^{-1}\d U$  in the basis $e,f,h$ of $\sl_2$ 
as $U^{-1}\d U=(U^{-1}\d U)_e e +(U^{-1}\d U)_f f +(U^{-1}\d U)_h h$.
  Then we defined
the measure $\vol_G(U)$ as the
three-form $4(U^{-1}\d U)_e (U^{-1}\d U)_f (U^{-1}\d U)_h$.       In the present context, to get a basis of the cotangent space,
we make a similar but longer expansion of $U^{-1}\d U$.
As it is valued in $\osp(1|2)$, it can be expanded in the basis described in eqns. (\ref{bosgen}) and (\ref{fermgen}):
\be\label{expbas} U^{-1}\d U =(U^{-1}\d U)_e e+(U^{-1}\d U)_f  f+(U^{-1}\d U)_h h +(U^{-1}\d U)_1 q_1 +(U^{-1}\d U)_2 q_2.\ee
Then we define the measure\footnote{\label{supermeasure} This is the Riemannian measure for the Riemannian metric on the Lie algebra $\osp(1|2)$ associated
with the quadratic form $(x,x)=2\,\Str\,  x^2$.  With $x=x_h h + x_e e+ x_f f +x_1 q_1+x_2q_2$, we have $2\,\Str \,x^2 = 4 x_h^2+4x_ex_f + 8 x_1x_2$.
The Berezinian of the corresponding metric tensor is 1, so the Riemannian measure is $1\cdot [\d x_e\d x_f\d x_h|\d x_1\,d x_2]$.   The metric
tensor that we have used here restricts on $\sl_2\subset \osp(1|2)$ to the metric tensor that we used to define a measure on $\sl_2$ (see footnote \ref{lookback}).
The latter was chosen so that the induced symplectic structure on the moduli space of Riemann surfaces is the Weil-Petersson form with standard normalization.
So the torsion that we will compute using the measure (\ref{wexbas}) will agree with the volume form on the moduli space of super Riemann surfaces that one
would get from a symplectic form on supermoduli space normalized so that on its reduced space (which is the moduli space of Riemann surfaces with spin structure) it coincides with the usual Weil-Petersson form.}
\be\label{wexbas} \vol_G(U) =[(U^{-1}\d U)_e,(U^{-1}\d U)_f,(U^{-1}\d U)_h|(U^{-1}\d U)_1,(U^{-1}\d U)_2] .\ee
The measures $\vol_G(V)$ and $\vol_G(R)$ are defined similarly.    

At $R=1$, we expand $R=1+r+{\mathcal O}(r^2)$, and expand $r=r_e e+r_f f+r_hh+r_1 q_1+r_2 q_2$.
 The general form of the torsion is\footnote{Here and later, when the basis vectors are 1-forms, we omit commas and write 
 $ [\d a\,\d b\,\d \kappa   |\d\xi\,\d\psi] $ rather than the clumsy $[\d a,\d b,\d \kappa   |\d\xi,\d\psi]$.}
\be\label{exbas}\tau=F(a,b,\kappa   |\xi,\psi) [\d a\,\d b\,\d \kappa   |\d\xi\,\d\psi]\, \ee
with an unknown function $F$.  The right hand side of eqn.~(\ref{purdux}) can be written
\be\label{nexbas}[\d r_e\,\d r_f\,\d r_h|\d r_1\,\d r_2] F(a,b,\kappa   |\xi,\psi) [\d a\,\d b\, \d \kappa   |\d\xi\,\d \psi]. \ee

In the bosonic case, what made possible a simple computation of the torsion was that at $R=1$,  $U$ was upper triangular and $V$ was lower triangular; this
led to simple formulas (\ref{zoffoTor}) and (\ref{poffoTor}) for $(U^{-1}\d U)_f$ and $(V^{-1}\d V)_e$ as multiples of $r_f$ and $r_e$.   That made it possible to eliminate $(U^{-1}\d U)_f$, $(V^{-1}\d V)_e$,
and likewise $\d r_e$, $\d r_f$ from eqn.~(\ref{purdux}), and thereby to get a simple result for the torsion.

The ansatz (\ref{twom}) has been chosen to lead to similar simplifications.   First, at $R=1$, we have 
\be\label{loweru} (U^{-1}\d U)_f =(e^a-1)\d r_f,~~~~~   (V^{-1}\d V)_e=(e^{b}-1)\d r_e.\ee
These formulas enable us to eliminate from eqn.~(\ref{purdux}) 
all the 1-forms written on the left or right hand sides of  eqn.~(\ref{loweru}). We may then set $\d r_f = \d r_e = 0$, and compute
\be\label{upperv}(U^{-1}\d U)_2=(\vnu_a e^{a/2}-1)\d  r_2,~~~~~(V^{-1}\d V)_1= (\vnu_b e^{b/2}-1)\d  r_1. \ee
We can use these formulas to remove some more factors from eqn.~(\ref{purdux}), but in doing so,
we have to take into account that rescaling $\d r_1$ or $\d r_2$ by a factor $\lambda$ will multiply the measure by $\lambda^{-1}$, as in eqn.~(\ref{purry}).  After these two steps, the 
reduced version of eqn.~(\ref{purdux}) is \begin{align}\frac{(e^a-1)(e^b-1)}{(\vnu_a e^{a/2}-1)(\vnu_b e^{b/2}-1)}
 & [(U^{-1}\d U)_e\,(U^{-1}\d U)_h|(U^{-1}\d U)_1] \cdot [(V^{-1}\d V)_f \,(V^{-1}\d V)_h|(V^{-1}\d V)_2]\notag \\
&= F(a,b,\kappa   |\xi,\psi)
[\d r_h\, \d a\,\d b\,\d \kappa   |\d\xi\,\d\psi],\label{wupperv}\end{align}
where $R$ is set to 1 and $\d r_e,\d r_f,\d r_1$ and $\d r_2$ are set to 0.
This is the analog of eqn.~(\ref{groffo}).

To complete the calculation, we work out the generalization of eqn.~(\ref{umcro}):
\begin{align}\label{zumcro}  (U^{-1}\d U)_h& = \frac{1}{2}\d a \cr
        (U^{-1}\d U)_e& =e^{-a/2}\d \kappa   +\frac{\kappa   }{2} e^{-a/2} \d a+2\kappa    e^{-a/2}\d r_h +\xi\d\xi \cr
        (U^{-1}\d U)_1 & =  \d\xi+\frac{1}{2}\d a\,\xi +\d r_h\xi\,  \cr
           (V^{-1}\d V)_h& =- \frac{1}{2}\d b \cr
             (V^{-1}\d V)_f &= \frac{1}{2} e^{-b/2}\d b-2e^{-b/2} \d r_h-\psi\d\psi \cr
              (V^{-1}\d V)_2 & =  \d\psi+\frac{1}{2}\d b\,\psi  -\d r_h\psi   .    \end{align}
Using these results in eqn.~(\ref{wupperv}) with the help of the change of variables formula (\ref{lurry}), we finally     arrive at
\be\label{lastone}F(a,b,\kappa   |\xi,\psi)= \frac{2\sinh\frac{a}{2}\sinh\frac{b}{2}}{(\vnu_a e^{a/2}-1)(\vnu_b e^{b/2}-1)} .\ee
  Concretely, the steps that lead to this formula are analogous to the ones that lead to eqn.~(\ref{woffo}).
We first replace $(U^{-1}\d U)_h$ and $(V^{-1}\d V)_h$ in eqn.~(\ref{wupperv}) with $\d a/2$ and $-\d b/2$, cancel $\d a$ and $\d b$ in (\ref{wupperv}),
and set $\d a=\d b=0$ in the remaining equations in (\ref{zumcro}).  At this stage,  eqn.~(\ref{wupperv}) reduces to
\begin{align}\label{goffo}\frac{\frac{1}{4}(e^a-1)(e^{b}-1)}{(\vnu_a e^{a/2}-1)(\vnu_b e^{b/2}-1)} &[(U^{-1}\d U)_e|(U^{-1}\d U)_1] \cdot [(V^{-1}\d V)_f \,|(V^{-1}\d V)_2]\cr = &
F(a,b,\kappa|\xi,\psi)[\d r_h \, \d\kappa|\d\xi\,\d\psi] ,\end{align}
and we have $(U^{-1}\d U)_1=\d \xi +\d r_h \xi$ and $(V^{-1}\d V)_2=\d\psi-\d r_h\psi$.
The Berezinian of a linear transformation from the basis of 1-forms  $\d r_h,\d\kappa|\d\xi,\d\psi$ to the basis $\d r_h,\d\kappa|(\d\xi+\d r_h\xi),( \d\psi-\d r_h \psi)$ is 1,
so on the right hand side of  eqn.~(\ref{goffo}), we can replace $[\d r_h \, \d\kappa|\d\xi\,\d\psi]$ with $[\d r_h \, \d\kappa|(\d\xi+\d r_h\xi)\,(\d\psi-\d r_h\psi)]$. 
Then we can cancel $(U^{-1}\d U)_1$ and $(V^{-1}\d V)_2$ on the left of eqn.~(\ref{goffo}) against $(\d\xi+\d r_h\xi)\,(\d\psi-\d r_h\psi)$ on the right,
and set $\d\xi+\d r_h\xi= \d\psi-\d r_h\psi=0$ in  the remaining equations.  
At this point,  $(U^{-1}\d U)_e (V^{-1}\d V)_f$ reduces to $2 e^{-(a+b)/2} \d \kappa   \d r_h$, and eqn.~(\ref{goffo}) reduces to (\ref{lastone}).   

Instead of writing the torsion as $\tau = F(a,b,\kappa   |\xi,\psi) [\d a\,\d b\,\d \kappa   |\d\xi\,\d\psi]$, we can use eqn.~(\ref{woffox}) and replace $\d \kappa   $ in this
formula with $\sinh \frac{c}{2} \d c$ (there is no  sign here, because  a measure as opposed to a differential form does not change sign in changing variables from $\kappa$
to $c$).   So the torsion is
\be\label{torform}\tau =\frac{2\sinh\frac{a}{2}\sinh \frac{b}{2}\sinh\frac{c}{2} }{(\vnu_a e^{a/2}-1)(\vnu_b e^{b/2}-1)} [\d a \,\d b\,\d c|\d\xi\,\d\psi]. \ee

This formula is not symmetric in $a$, $b$, and $c$.   This happened because we defined the fermionic moduli in a way
that  did not treat the three boundaries symmetrically.  The formula actually is invariant under a symmetry that exchanges $a$ and $b$, because the ansatz that
we  used is invariant up to conjugation under such an operation.   Interchanging $U\leftrightarrow V$ and conjugating by
\be\label{conjmat}
\newcommand*{\temp}{\multicolumn{1}{r|}{}}
\left(\begin{array}{cccccc}
\sqrt{-\kappa} &0    \negthinspace\negthinspace\negthinspace &\temp & 0\\
0 &\frac{1}{\sqrt{-\kappa}}\negthinspace \negthinspace \negthinspace&\temp &0\\
 \cline{1-4}
0 &0\negthinspace \negthinspace \negthinspace&\temp &1  \\

\end{array}\right)
\cdot \left(\begin{array}{cccccc}
0 &-1 \negthinspace\negthinspace\negthinspace &\temp & 0\\
1 &0\negthinspace \negthinspace \negthinspace&\temp &0\\
 \cline{1-4}
0 &0\negthinspace \negthinspace \negthinspace&\temp &1\\

\end{array}\right)
 \ee
(where $-\kappa>0$ according to eqn.~(\ref{noflox})) maps $a\leftrightarrow b$, $\vnu_a\leftrightarrow \vnu_b$ and transforms the fermionic moduli by 
$(\xi,\psi)\to (-\psi\sqrt{-\kappa},\xi/\sqrt{-\kappa})$, or equivalently
\be\label{fermtrans} \xi\pm  \i \psi\sqrt{-\kappa}\to \pm \i ( \xi\pm  \i \psi\sqrt{-\kappa}). \ee
The measure is invariant under this operation, as it should be since  the torsion  is intrinsically defined once we pick  a measure on the $\osp(1|2)$ Lie algebra.  
It may be possible to define the fermionic moduli in a way that respects the full permutation symmetry of the three boundaries, but a convenient way  to
 do this is not immediately obvious.

\subsubsection{Assembling The Pieces}\label{applications}

Since we have computed the torsion of a flat $\OSp'(1|2)/\Z_2$ connection on a three-holed sphere, a circle, or a crosscap, it is straightforward 
now, following the same steps as in sections \ref{oriented} and \ref{torunorientable},  to glue the pieces together and compute
the torsion of such a connection on any two-manifold $Y$.   

For example, we can build a closed oriented  surface $Y$ of genus $g$ without boundary
 by gluing together $2g-2$ three-holed spheres along $3g-3$ circles.   Let $\S$ be the set of three-holed
spheres and $\T$ be the set of circles.   Schematically, the gluing formula says that the torsion of $Y$ is a simple tensor product:
\be\label{omy}\tau_Y=\otimes_{t\in\T} \tau_t \otimes_{s\in \S}\tau_s^{-1} \ee
of the  torsions of the three-holed spheres $t\in\T$ and the inverse torsions of circles $s\in \S$.   The variables that the torsion depends on are a length
parameter $\ell_s$ and a gluing parameter $\rho_s$ for each circle  $s\in \S$, and a pair of fermionic moduli $\xi_t,\psi_t$ for each three-holed  sphere $t\in\T$.

In the product (\ref{omy}),  various  trigonometric  factors appear.   According to eqn.~(\ref{waus}), the inverse torsion of a circle with
circumference $\ell$ and spin structure  $\vnu$  contains a factor   $1/(e^{\ell/4}+\vnu e^{-\ell/4})^2.$
On  the other hand, the $s^{th}$ circle is a  boundary of two  different three-holed spheres $t,t'\in  \T$ (exceptionally it is possible to have $t=t'$).
It is natural to combine one factor $1/(e^{\ell_s/4}+\vnu e^{-\ell_s/4})$ with $\tau_t$ and one with $\tau_{t'}$.   The trigonometric factor that appears in the formula
(\ref{torform}) for the torsion simplifies  when  multiplied by a  product of three factors   $1/(e^{\ell_s/4}+\vnu_s e^{-\ell_s/4})$   in which $\ell_s$ is taken to  be $a,b,$ or $c$.
The  product of all those factors  is  
\be\label{jumbo} \frac{1}{4}\vnu_a \vnu_b  e^{-(a+b)/4} (e^{c/4}-\vnu_c e^{-c/4}). \ee
Multiplying everything together, and multiplying by $\frac{1}{2}$ to take account\footnote{If we want to think of the moduli space of flat $\OSp'(1|2)$ connections
as a supermanifold  (with some orbifold singularities of positive bosonic codimension), then we do not want to divide the parameter space by this $\Z_2$; instead we take the
$\Z_2$ symmetry into account by dividing  $\tau_Y$ by 2.   
} of the $\Z_2$ symmetry generated by  $(-1)^\sF$, the result that we get for the torsion of $Y$ is 
\be\label{torformt}\tau_Y =\frac{1}{2} (-1)^{w_\Ra} \prod_{s\in \S}[ \d\ell_s \,\d\varrho_s] \prod_{t\in\T} \frac{1}{4}\vnu_{a_t}\vnu_{b_t} e^{-(a_t+b_t)/4} (e^{c_t/4}-\vnu_{c_t} e^{-c_t/4}) [\d\xi_t\,\d\psi_t]. \ee
The parameter $w_\Ra$ is the number of circles $s\in \S$ with $\vnu=0$; this factor arises from the sign in the torsion of a circle. 
Eqn.~(\ref{torformt}) is a superanalog  of the usual formula for a measure on the moduli space of ordinary Riemann surfaces, in length-twist coordinates.    By gluing crosscaps
onto some of the three-holed spheres, one can get an analogous formula for unorientable $Y$, generalizing eqn.~(\ref{yorgo}).

In the orientable case, it is possible to obtain the same formula by symplectic methods.  (In a different but somewhat similar coordinate system, the symplectic
structure on the moduli space of super Riemann surfaces has  been  described explicitly in \cite{Pennerone}.)
It is straightforward to generalize  eqn.~(\ref{torformt}) for the case that $Y$ is a  genus
$g$ surface with $n$ boundary components.\footnote{When  $Y$ has
a boundary,  the torsion is not equivalent  to the symplectic volume, since the identity (\ref{identity}) does not hold.}  The only difference is that trigonometric factors associated to the boundary circles do not cancel so neatly; for a boundary circle of length $\ell$, 
one is left with a factor $e^{\ell/4}+\vnu e^{-\ell/4}$.   These factors disappear if the external boundary circles have NS spin structure ($\vnu=-1$) and we glue them onto trumpets,
   In that case, according to eqn.~(\ref{lomic}),  we get a factor $2\sinh \frac{\ell}{4}$ in converting from a ``torsion'' path integral $\t Z_\SJT$ on the trumpet
to a conventional SJT path integral $Z_\SJT$;  gluing via $\tau_S^{-1}$ gives a factor $1/4\sinh^2\frac{\ell}{4}$, according to  eqn.~(\ref{waus}).   The product
of  these factors with $e^{\ell/4}+\vnu e^{-\ell/4}$ is simply $1$.

At first, the formula (\ref{mumbo})  might make one suspect that for  any surface $Y$ obtained by gluing of three-holed spheres (possibly
 with crosscaps), the moduli space of flat connections on $Y$ would have zero volume,
 on the grounds that the volume form in eqn.~(\ref{mumbo}) does not depend on the odd moduli, and so appears to vanish upon integration over those variables.
   The fallacy in this argument is that the region of length-twist coordinates that describes
 the moduli space of super Riemann surfaces is defined by inequalities that involve the fermionic coordinates as well as the bosonic ones.    To see why
 this can lead to a nonzero integral, consider a toy problem with one bosonic variable $\kappa   $, two fermionic variables $\xi,\psi$, and a measure $\d \kappa   \d\xi\d\psi$ (or
 $[\d \kappa   |\d\xi\,\d\psi]$, to be more pedantic).
 The volume of the region $0\leq \kappa   \leq 1$ is 0, as one sees by integrating first over $\kappa   $:
 \be\label{nubboTor}\int_{0\leq \kappa   \leq 1}\d \kappa   \,\d\xi\,\d\psi\cdot 1 =\int\d\xi\,\d\psi\,\cdot 1 = 0.\ee
 But suppose the integration region is $0\leq \kappa   \leq 1+\xi\psi$.   Then integrating first over $\kappa   $,  we get
 \be\label{unbo}\int_{0\leq \kappa   \leq 1+\xi\psi} \d \kappa   \,\d\xi\,\d\psi \cdot 1 =\int \d\xi\,\d \psi (1+\xi\psi) = 1. \ee
 In fact, we will encounter a phenomenon somewhat like that in generalizing Mirzakhani's results to super Riemann surfaces.
 
The volumes do, however, vanish in genus 0, that is if  $Y$ is a hyperbolic sphere with $n$ holes. 
 This  is explained in appendix \ref{volsymp}.
 In the context of JT supergravity, this vanishing implies that a connected correlator  $\left\langle \left(\Tr\,\exp(-\beta H)\right)^n\right\rangle_c$ vanishes in genus 0,
 and receives contributions only in higher orders in the topological expansion.   The analogous statement is certainly not true for ordinary  JT gravity.

\section{Loop Equations}\label{sec:loopEquations}

In this section, we will explain a technique called the ``loop equations'' for analyzing the $1/L$ expansion of matrix integrals. This technology will make it possible to compare matrix integrals to JT gravity and supergravity in section \ref{sec:JT} below. 

The loop equations have been studied since the work of Migdal in \cite{migdal1983loop}, but a significant streamlining was achieved in the work of Eynard for the case of $\upbeta = 2$ ensembles \cite{eynard2004all}. Using a dispersion relation method, it was shown how to turn the loop equations into a recursion relation that closes on correlation functions of resolvents. A nice feature of the resulting equations is that they do not depend explicitly on the potential of the matrix integral, only on the ``spectral curve'' or leading density of eigenvalues. The recursive step in this formalism is a contour integral that reduces to a sum of residues: the particular structure that appears was later abstracted as the formalism of ``topological recursion'' \cite{eynard2007invariants}.

For the case of the general $\upbeta$ ensembles, some of the simplifications of \cite{eynard2004all} are still possible. But the resulting recursion relation is somewhat more complicated: the recursive step involves an integral around a cut, rather than a sum of residues. This has been studied from a somewhat different perspective in works including \cite{Chekhov:2010zg,Marchal:2011iu}. In section \ref{sec:betaensembles}, we will explain the loop equations for general $\upbeta$, pointing out the simplification that occurs for $\upbeta = 2$. We will limit ourselves to so-called ``one-cut'' models, where the large $L$ density of eigenvalues is supported in a single interval of the real axis.

For the Altland-Zirnbauer $(\upalpha,\upbeta)$ ensembles, the basic method of \cite{eynard2004all} is again possible. For a special case $(\upalpha,\upbeta) = (1,2)$, the recursion reduces to a residue computation that is equivalent to a form of ``topological recursion.'' For more general values of $(\upalpha,\upbeta)$, the recursive step involves a nontrivial integral. In section \ref{sec:alphabeta}, we will explain the loop equations for these ensembles.

\subsection{Loop Equations For The \texorpdfstring{$\upbeta$}{beta} Ensembles}\label{sec:betaensembles}
As described in section \ref{ensembles}, the Dyson $\upbeta$ ensembles are characterized by a measure for eigenvalues of the form
\be\label{digDist}
\prod_{1\le i<j\le L}|\lambda_i-\lambda_j|^\upbeta \prod_{i = 1}^L e^{-L\frac{\upbeta}{2}V(\lambda_i)} \mathrm{d}\lambda_i.
\ee
where we included a general potential function $V(\lambda)$ in addition to the measure factor described in section \ref{measure}. $V(x)$ and $\upbeta$ should be thought of as parameters of the distribution. As a probability measure, it makes sense for continuous values of $\upbeta$, but $\upbeta = 1,2,4$, are the values relevant for random matrix theory. The potential $V$ should be a function that grows rapidly enough at infinity that the integral converges. The manipulations below will assume that $V$ is analytic in a neighborhood of the region where the classical (large $L$) eigenvalue distribution is nonzero.

The goal is to set up machinery for computing the $1/L$ expansion of expectation values in the ensemble defined by (\ref{digDist}). As a basis for ``single-trace'' observables, we will use the resolvent. This is a function that depends on a parameter $x$ in addition to all of the eigenvalues $\{\lambda\}$, and it is defined by
\be
R(x,\{\lambda\}) = \sum_{i = 1}^L\frac{1}{x-\lambda_i}.
\ee
In what follows, we will leave the $\lambda$ dependence of the resolvent implicit, and write simply $R(x)$. From this quantity, any other single trace quantity $\sum_{i = 1}^L f(\lambda_i)$ can be obtained. For example, one can start by computing the density of eigenvalues from the discontinuity of $R(x)$ across the real axis:
\be
\rho(x) = \sum_{i = 1}^L \delta(x-\lambda_i) = -\frac{1}{2\pi \mathrm{i}}\left(R(x+i\epsilon) - R(x-i\epsilon)\right).\label{discReal}
\ee
Then one can write $\sum_{i= 1}^L f(\lambda_i) = \int \mathrm{d}x\, \rho(x)f(x)$.

Below, we will use a shorthand notation for products of resolvents
\be\label{resDef}
R(x_1,\dots, x_n) = R(x_1)R(x_2)\dots R(x_n)
\ee
and further abbreviate by writing the LHS as $R(I)$, where 
\be
I = \{x_1,\dots,x_n\}, \hspace{20pt} |I| = n.\label{multiInd}
\ee
Connected correlation functions of resolvents have an asymptotic expansion in powers of $1/L$, which we can write using the above notation:
\be
\langle R(I)\rangle_{\text{c}} \simeq \sum_{g = 0,\frac{1}{2},1,\,\dots}\frac{R_{g}(I)}{L^{2g+|I|-2}}.\label{genusExp}
\ee
This expression defines the quantities $R_{g}(I)$ as the coefficients in the $1/L$ expansion. In general, the ``genus'' summation index $g$ takes both integer and half-integer values, although we will see that only integer values contribute for the special case $\upbeta = 2$.

The loop equations are simply the observation that
\begin{align}
0&=\int_{-\infty}^\infty \mathrm{d}^L\lambda\, \frac{\partial}{\partial\lambda_a}\left[\frac{1}{x-\lambda_a}R(I)\prod_{i<j}|\lambda_i-\lambda_j|^\upbeta \prod_i e^{-L\frac{\upbeta}{2}V(\lambda_i)}\right],
\end{align}
where $a \in \{1,\dots,L\}$ is some fixed index. Distributing the derivative, we find
\be\label{distDer}
0= \left\langle \left[\frac{1}{(x-\lambda_a)^2} + \upbeta\frac{1}{x-\lambda_a}\sum_{j\neq a}\frac{1}{\lambda_a-\lambda_j} - \frac{L\upbeta}{2}\frac{V'(\lambda_a)}{x-\lambda_a}\right]R(I) + \frac{1}{x-\lambda_a}\partial_{\lambda_a}R(I)\right\rangle.
\ee
Here the angle brackets mean an expectation value in the ensemble defined by (\ref{digDist}). Because the eigenvalue distribution is permutation-invariant, we do not learn any new information by setting $a$ to any particular value, and below we will sum over $a$. As much as possible, we would like to reduce (\ref{distDer}) to an expression involving expectation values of products of resolvents. We will do this in three steps.  

(i) The second term inside the brackets in (\ref{distDer}) can be rewritten using
\begin{align}
\frac{1}{x-\lambda_a} \frac{1}{\lambda_a-\lambda_j} &\rightarrow \frac{1}{2}\left(\frac{1}{x-\lambda_a}\frac{1}{\lambda_a-\lambda_j} + \frac{1}{x-\lambda_j}\frac{1}{\lambda_j-\lambda_a}\right)\\
&=\frac{1}{2}\frac{1}{x-\lambda_a}\frac{1}{x-\lambda_j}.
\end{align}
In the first step, we replaced the LHS by its symmetrization under $a\leftrightarrow j$. Summing over $a$, we can then rewrite the first two terms inside brackets in (\ref{distDer}) as
\begin{align}
\sum_{a} \frac{1}{(x-\lambda_a)^2} + \frac{\upbeta}{2}\sum_{a,j\neq a}\frac{1}{x-\lambda_a}\frac{1}{x-\lambda_j} &= \sum_{a} \frac{1-\frac{\upbeta}{2}}{(x-\lambda_a)^2} + \frac{\upbeta}{2}\sum_{a,j}\frac{1}{x-\lambda_a}\frac{1}{x-\lambda_j}\\
&= -(1-\tfrac{\upbeta}{2})\partial_x R(x) + \frac{\upbeta}{2}R(x,x).\label{step1}
\end{align}

(ii) The term involving the potential cannot be completely rewritten in terms of resolvents, but we can improve the situation by writing
\be
\sum_{a}\frac{V'(\lambda_a)}{x-\lambda_a}R(I) = V'(x)R(x,I) - P(x;I)\label{improved}
\ee
where we define
\be
P(x;I) \equiv \sum_a\frac{V'(x)-V'(\lambda_a)}{x-\lambda_a}R(I).
\ee
The first term on the LHS of (\ref{improved}) is now written in terms of resolvents. The remainder term $P(x;I)$ is not, but a key point is that $P(x;I)$ is analytic in $x$ (assuming $V$ is itself analytic). For example, if $V$ is a polynomial, then $P$ is also a polynomial. Below, we will use a dispersion relation that will allow us to ignore the analytic $P(x;I)$ term.

(iii) Finally, the last term in (\ref{distDer}) can be rewritten as follows. We start with
\be
\partial_{\lambda_a}R(I) = \sum_{k = 1}^n \frac{1}{(x_k-\lambda_a)^2}R(I\setminus x_k)
\ee
where $I\setminus x_k$ means the coordinates $\{x_1,\dots x_n\}$ with $x_k$ removed.
Then using
\be
\frac{1}{(x-\lambda_a)(x_k-\lambda_a)^2} = \partial_{x_k}\frac{\frac{1}{x-\lambda_a} - \frac{1}{x_k-\lambda_a}}{x-x_k}
\ee
and summing over $a$, we have
\be
\sum_{a=1}^L \frac{1}{x-\lambda_a}\partial_{\lambda_a}R(I) = \sum_{k = 1}^n \partial_{x_k}\frac{R(x,I\setminus x_k) - R(I)}{x-x_k}.\label{Step3}
\ee
Putting these steps together, and multiplying through by $\frac{2}{\upbeta}$, one can rewrite (\ref{distDer}) as
\begin{align}\notag
&\Bigg\langle (1-\tfrac{2}{\upbeta})\partial_x R(x,I) + R(x,x,I) - LV'(x)R(x,I) + \frac{2}{\upbeta}\sum_{k= 1}^n\partial_{x_k}\frac{R(x,I\setminus x_k) - R(I)}{x-x_k}\Bigg\rangle\\&\hspace{60pt}= -L\langle P(x;I)\rangle.\label{in}
\end{align}

Next, we would like to write (\ref{in}) in terms of connected correlation functions, and use the lower-order equations to simplify. This is a little tricky. It is helpful to focus on a proper subset $J\subsetneq I$ and collect all terms that multiply $\langle R(J)\rangle_{\text{c}}$. One can show that the sum of all terms that multiply this expression is precisely the quantity that vanishes due to (\ref{in}) with $I\rightarrow I\setminus J$. So in writing (\ref{in}) in terms of connected correlators, we can omit any terms that have a factor of $\langle R(J)\rangle_{\text{c}}$ with $J\subsetneq I$. The same argument allows us to omit the terms involving $\langle R(I)\rangle_{\text{c}}$ except for the one that arises from the last term on the first line of (\ref{in}). Writing out what remains, we have
\begin{align}
&(1-\tfrac{2}{\upbeta})\partial_x\langle R(x,I)\rangle_{\text{c}} + \langle R(x,x,I)\rangle_{\text{c}} + \sum_{J\subseteq I}\langle R(x,J)\rangle_{\text{c}}\langle R(x,I\setminus J)\rangle_{\text{c}}\label{connected}\\ &\hspace{40pt}-LV'(x)\langle R(x,I)\rangle_{\text{c}} + \frac{2}{\upbeta}\sum_{k = 1}^n\partial_{x_k}\left(\frac{\langle R(x,I\setminus x_k)\rangle_{\text{c}} - \langle R(I)\rangle_{\text{c}}}{x-x_k} \right) =- L\langle P(x;I)\rangle_{\text{c}}.\notag
\end{align}
We will now discuss the application of this equation to three special cases, before treating the generic case below.

\subsubsection{A First Special Case \texorpdfstring{$R_{0}(x)$}{R0(x)}, And The Spectral Curve}\label{firstspecial}
First, we consider $I = \emptyset$, and further simplify the equation by inserting (\ref{genusExp}) and keeping only the leading terms, proportional to $L^2$. Then (\ref{connected}) reduces to
\be
R_0(x)^2 - V'(x)R_0(x) = (\text{analytic in $x$})
\ee
where the analytic RHS comes from the $P(x)$ term. After adding an analytic quantity $V'(x)^2/4$ to both sides, we can write this as
\be\label{sCurve2}
y^2(x) = (\text{analytic in $x$})
\ee
where $y$ is defined by
\be\label{specCurve}
R_0(x) =  \frac{V'(x)}{2} + y(x).
\ee
Eqn.~(\ref{sCurve2}) defines a hyperelliptic curve, with two sheets differing by the sign of $y$. This is referred to as the spectral curve of the matrix integral.

The quantity $y$ has a physical intepretation as the leading approximation to the density of eigenvalues. To see this, recall that in general, the resolvent is a multivalued function, with a discontinuity across the real axis given by the density of eigenvalues (\ref{discReal}). So, in particular, the discontinuity of the genus-zero resolvent $R_0(x)$ gives the leading approximation to the density of eigenvalues:
\be
 R_0(x+i\epsilon) - R_0(x-i\epsilon) = -2\pi \mathrm{i}\rho_0(x), \hspace{40pt} 
\rho_0(x) = \lim_{L\rightarrow \infty}\frac{1}{L}\langle \rho(x)\rangle.\label{rewriteDisc}
\ee
Here, $\rho_0(x)$ is normalized so that $\int \mathrm{d} x \rho_0(x) = 1$. We will consider the simplest ``one-cut'' matrix integrals, for which $\rho_0(x)$ is supported in a single interval of the real axis $a_-\le x\le a_+$. Resolvents are naturally defined with a branch cut that coincides with this interval. Since $R_0(x)$ and $y(x)$ differ only by an analytic term $\frac{1}{2}V'(x)$, the same is true for $y(x)$, and we can rewrite the LHS of (\ref{rewriteDisc}) as the discontinuity of $y(x)$ across the real axis.  Because $y^2$ is analytic, the only possible discontinuity in $y$ is a change of sign. This implies that if we approach the cut from one side or the other, $y(x)$ has to be one-half of the value of its discontinuity:
 \be\label{yandrho}
 y(x\pm i\epsilon) = \mp\mathrm{i}\pi \rho_0(x).
 \ee
So, along the real axis and inside the region of support of $\rho_0(x)$, $y(x)$ is pure imaginary, with a sign that depends on which half-plane we approach from. Outside the region of support, $y(x)$ is real. At each endpoint of the cut, $y(x)$ has a square-root singularity $\sqrt{a_\pm -x}$.

In principle, one can solve for $y(x)$ starting from the potential $V(x)$ that defines the matrix integral. But, as we will see below, in the recursion relation we will not need to know $V(x)$ directly, only $y$. So in practice it is more convenient to think about things the other way around: we specify the matrix integral by giving the function $y(x)$ rather than the potential.
\begin{figure}[t]
\begin{center}
\includegraphics[width = .8\textwidth]{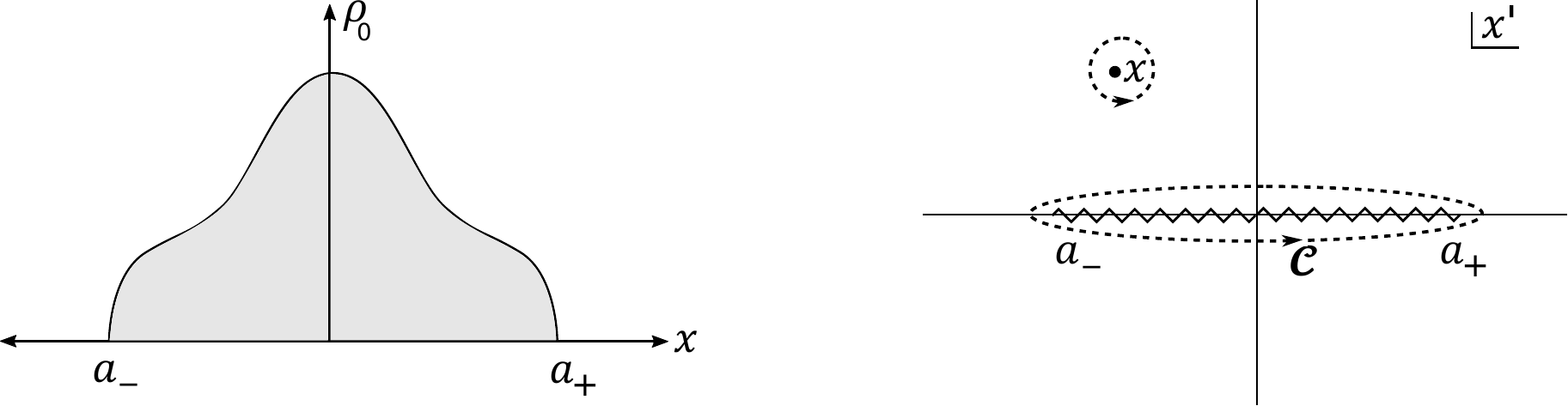}
\caption{{\small At left we show a typical $\rho_0(x)$ function, supported between endpoints $x = a_\pm$. At right we show the $x'$ plane for the manipulation in (\ref{manip}). The original contour surrounds the point $x$, and the final contour surrounds the cut that runs bewteen $x' = a_\pm$. For the special case of an ensemble with $\upbeta = 2$, the cut is replaced by poles at the endpoints.}}\label{figrho0}
\end{center}
\end{figure}

The resolvent can be expressed in terms of the eigenvalue density:
\be\label{expdensity}R_0(x)=\int_{a_-}^{a_+} \d\lambda\,\frac{\rho_0(\lambda)}{x-\lambda}.\ee
This integral is manifestly holomorphic in $x$ on the complement of  the interval $[a_-,a_+]$, where it has a cut.
To be more exact,  the integral defines $R_0(x)$ on what is known as the ``first sheet.''   The formula $R_0(x)=V'(x)/2+y(x)$,
where $y(x)$ is  defined on the spectral curve,   a double cover of the $x$ plane, shows that $R_0(x)$ can be continued through the cut onto
a second sheet.   Multiresolvents can similarly be expressed in terms of a joint eigenvalue density $\rho(\lambda_1,\lambda_2,\cdots,\lambda_k)$.
In lowest order in $1/L$, the joint density is just a product $\rho_0(\lambda_1)\rho_0(\lambda_2)\cdots \rho_0(\lambda_k)$, but in higher orders there 
are corrections.   Thus in general
\be\label{expform} 
R_{g}(I) = \int_{a_-}^{a_+} \frac{\mathrm{d}\lambda_1}{x_1-\lambda_1}\dots \frac{\mathrm{d}\lambda_n}{x_n-\lambda_n}\rho_g(\lambda_1,\lambda_2,\cdots,\lambda_n),
\ee
where $\rho_g(\lambda_1,\lambda_2,\cdots,\lambda_N)$ is the genus $g$ contribution to $\rho(\lambda_1,\lambda_2,\cdots,\lambda_n)$.
Eqn. (\ref{expform}) defines a function that is manifestly holomorphic in $x_1,x_2,\cdots, x_n$ as long as they are on the complement of the interval $[a_-,a_+]$. Moreover,
the formula determines the behavior of this function when any of the $x_i$ become large.
 For any reasonably behaved function $\rho_g(\lambda_1,\lambda_2,\cdots,\lambda_n)$, eqn. (\ref{expform}) implies that $R_g(I)$ vanishes at least as $1/x_i$ for
$x_i\to\infty$.   But since the total number of eigenvalues is a fixed constant $L$, with no fluctuations or higher order corrections, it actually follows that $R_g(I)$ vanishes
as $1/x_i^2$ or faster, except for the special case $g=0$ and $|I|=1$.
In what follows, in writing dispersion relations for the multiresolvents, we begin on the ``first sheet,''  on which $R_g(I)$ has the properties
just stated.   However, from the analysis, we will learn that like $R_0(x)$, the functions $R_g(I)$  can be continued (in each variable separately) to a second sheet, 
on which in general there are additional singularities.   For example, on the second sheet, $R_0(x_1,x_2)$ has a double pole at $x_1=x_2$.

The reader might ask how we know that the functions $\rho_g(\lambda_1,\lambda_2,\cdots,\lambda_n)$ have the same support as the product $\rho_0(\lambda_1)\rho_0(\lambda_2)
\cdots \rho_0(\lambda_n)$.  One answer comes from the fact that the procedure of topological recursion that we will describe leads to a unique answer with this property.
We can also reason as follows.  The fact that the eigenvalues are supported in some interval  is stable against small changes in the effective potential $V$.  But might
$1/L^2$ corrections change the precise interval $[a_-,a_+]$?   In a sense, this does happen, and it is reflected in the fact that $R_g(I)$ as computed by topological recursion has
increasingly strong singularities at the endpoints of the interval as $g$ or $|I|$ is increased.
The Taylor series expansion $\sqrt{\lambda-a_-+c/L^2}=\sqrt{\lambda-a_-}+\frac{c}{2L^2}(\lambda-a_-)^{-1/2}
+\mathcal{O}(1/L^4)$ illustrates the point: a perturbative shift in the endpoint of the cut is equivalent, in the $1/L$ expansion,
to keeping the endpoints of the cut fixed and generating increasingly severe singularities at the endpoints.  

\subsubsection{A Second Special Case \texorpdfstring{$R_{0}(x,x_1)$}{R0(x,x1)}}
Next, we consider the case $I = \{x_1\}$ and again keep only the leading terms in the $1/L$ expansion, this time proportional to $L$. Then (\ref{connected}) gives
\be
2R_0(x)R_{0}(x,x_1) - V'(x) R_0(x,x_1) + \frac{2}{\upbeta}\partial_{x_1}\frac{R_0(x)-R_0(x_1)}{x-x_1} = (\text{analytic in $x$}).
\ee
Using (\ref{specCurve}), and moving some analytic terms to the RHS, this can be rewritten as
\be
2y(x)R_{0}(x,x_1) + \frac{2}{\upbeta}\frac{y(x)}{(x-x_1)^2} = (\text{analytic in $x$ in neighborhood of cut}).\label{usedF}
\ee
Here we are assuming that $x_1$ is away from the cut, so we allowed ourselves to move terms to the RHS that are singular at $x = x_1$. We would like to use this equation to determine $R_{0}(x,x_1)$ using a dispersion relation. In general, $y(x)$ could be growing rapidly at infinity, so before we do this, one would like to simplify the equation by dividing by $y(x)$. However, since $1/y$ is singular at the endpoints of the cut, this will make some of the terms on the RHS singular. Instead, we proceed as follows. We write
\be
\sigma(x) = (x-a_+)(x-a_-)\label{sigmaofx}
\ee
and define $\sqrt{\sigma(x)}$ with a branch cut running between the endpoints of the cut $a_\pm$. Then $\sqrt{\sigma(x)}/y(x)$ is analytic in a neighborhood of the cut, and we can safely multiply (\ref{usedF}) by this quantity, to obtain
\be
R_0(x,x_1)\sqrt{\sigma(x)} + \frac{1}{\upbeta}\frac{\sqrt{\sigma(x)}}{(x-x_1)^2} = (\text{analytic in $x$ in neighborhood of cut}).\label{usedG}
\ee
Now, to determine $R_{0}(x,x_1)$, we write a dispersion relation
\begin{align}
R_0(x,x_1)\sqrt{\sigma(x)} &= \frac{1}{2\pi \mathrm{i}}\oint_x \frac{\mathrm{d}x'}{x'-x}R_0(x',x_1)\sqrt{\sigma(x')}= -\frac{1}{2\pi \mathrm{i}}\int_{\mathcal{C}}\frac{\mathrm{d}x'}{x'-x}R_0(x',x_1)\sqrt{\sigma(x')}\notag\\
&=\frac{1}{2\pi \mathrm{i}}\int_{\mathcal{C}}\frac{\mathrm{d}x'}{x'-x}\frac{1}{\upbeta}\frac{\sqrt{\sigma(x')}}{(x'-x_1)^2}.\label{dispersion}
\end{align}
In the first equality, we used the Cauchy residue formula. In the second equality, we deformed the contour to surround the cut, see figure \ref{figrho0}.   To justify this
deformation, we use
our knowledge that $R_0(x,x_1)$ is holomorphic in $x$ away from the cut.
In the final step we used (\ref{usedG}), taking advantage of the fact that terms that are analytic in a neighborhood of the cut will not contribute. After evaluating the integral, one finds
\be
R_{0}(x_1,x_2) = \frac{1}{\upbeta(x_1-x_2)^2}\left(\frac{x_1x_2-\frac{a_++a_-}{2}(x_1+x_2) + a_+a_-}{\sqrt{\sigma(x_1)}\sqrt{\sigma(x_2)}}-1\right).\label{explicit}
\ee
An important feature of this expression is that it is regular at $x_1 = x_2$ if both coordinates are on the same sheet, but has a double pole if they are on opposite sheets:
\be
R_{0}(x,x) = \frac{(a_+-a_-)^2}{8\upbeta(a_+-x)^2(a_--x)^2}, \hspace{20pt}R_0(\widehat{x}_1,x_2') = -\frac{2}{\upbeta(x_1-x_2)^2} + (\text{regular}).\label{vcb}
\ee
Here we are using the notation $\widehat{x}$ to mean that we continue the $x$ coordinate to position $x$ on the second sheet.

It will be useful below to know eqn.~(\ref{explicit}) in a limit in which $a_+\rightarrow \infty$ with $a_0 = 0$. This is
\be
R_{0}(x_1,x_2) = \frac{1}{2\upbeta \sqrt{-x_1}\sqrt{-x_2}\left(\sqrt{-x_1}+\sqrt{-x_2}\right)^2}.\label{abinf}
\ee

\subsubsection{A Third Special Case \texorpdfstring{$R_0(x,x_1,x_2)$}{R0(x,x1,x2)}}\label{sec:nwm}
As a final special case, we take $I = \{x_1,x_2\}$ and again work at leading order in $1/L$, which is now $L^0$. Then (\ref{connected}) gives
\be
2y(x)R_0(x,x_1,x_2) + 2R_{0}(x,x_1)R_0(x,x_2) + \frac{2}{\upbeta}\left(\frac{R_0(x,x_2)}{(x-x_1)^2} + \frac{R_0(x,x_1)}{(x-x_2)^2}\right) \sim 0.\label{sumof}
\ee
where the $\sim$ means up to terms that are analytic in $x$ in a neighborhood of the cut. After multiplying through by $\sqrt{\sigma(x)}/(2y(x))$, one can write $R_{0}(x,x_1,x_2)$ using a dispersion relation as in (\ref{dispersion}):
\be
R_0(x,x_1,x_2)\sqrt{\sigma(x)} = \frac{1}{2\pi \mathrm{i}}\int_{\mathcal{C}}\frac{\mathrm{d}x'}{x'-x}\frac{\sqrt{\sigma(x')}}{y(x')}\left(R_0(x',x_1)R_0(x',x_2) + \frac{1}{\upbeta}\frac{R_0(x',x_2)}{(x'-x_1)^2} + \frac{1}{\upbeta}\frac{R_0(x',x_1)}{(x'-x_2)^2}\right).\label{disperThreept}
\ee
One can check from the explicit formula (\ref{explicit}) that the integrand is actually meromorphic in a neighborhood of the cut, having a simple pole at each endpoint. So the integral reduces to a sum of two residues.

\subsubsection{The Generic Case}
Now we consider the generic case. To do so, we plug the genus expansion (\ref{genusExp}) into the loop equations (\ref{connected}), and collect the terms that appear at order $L^{2-2g-n}$, where $n$ is the number of elements in the set $I = \{x_1,\dots,x_n\}$. The resulting equation can be written as
\be
2y(x)R_{g}(x,I) + F_g(x,I)\sim 0\label{simin}
\ee
where 
\begin{align}
F_{g}(x,I)&= (1-\tfrac{2}{\upbeta})\partial_xR_{g-\frac{1}{2}}(x,I)+ R_{g-1}(x,x,I)+\sum_{\text{stable}} R_{h}(x,J)R_{g-h}(x,I\setminus J)\label{topfirstlineof}\\
&\hspace{20pt}+ 2\sum_{k = 1}^n\left(R_{0}(x,x_k) +\tfrac{1}{\upbeta} \tfrac{1}{(x-x_k)^2}\right)R_g(x,I\setminus x_k)\label{firstlineof}.
\end{align}
As before, the $\sim$ symbol in (\ref{simin}) means equality up to terms that are analytic in $x$ in a neighborhood of the cut. Working modulo such terms allowed us to drop the expressions involving $P(x;I)$ and $\frac{1}{(x-x_k)^2}R(I)$ in (\ref{connected}).

Naively, the sum on line (\ref{topfirstlineof}) should be over integer and half-integer $h$ satisfying $0\le h\le g$ and over subsets $J\subseteq I$. However, the subscript ``stable'' means that we omit the special cases where one of the factors is $R_{0}(x)$ or $R_{0}(x,x_k)$. These special cases are treated separately: the contribution where one factor is $R_0(x)$ combines with the $-L V'(x)$ term on the second line of (\ref{connected}) to form the combination $2y(x) R_g(x,I)$ present in (\ref{simin}). The contribution where one of the factors is $R_0(x,x_k)$ combines with the $ R(x,I\setminus x_k)$ term in (\ref{connected}) to give the final term on the second line of (\ref{firstlineof}).

Eq.~(\ref{simin}) determines the parts of $y(x)R_g(x,I)$ that are singular at the cut, and knowledge of these terms allows us to recover the full $R_g(x,I)$ using a dispersion relation as in (\ref{dispersion}):
\begin{align}
R_g(x,I)\sqrt{\sigma(x)} &= \frac{1}{2\pi \mathrm{i}}\oint_x \frac{\mathrm{d}x'}{x'-x}R_g(x',I)\sqrt{\sigma(x')}= -\frac{1}{2\pi \mathrm{i}}\int_{\mathcal{C}}\frac{\mathrm{d}x'}{x'-x}R_g(x',I)\sqrt{\sigma(x')}\label{manip}\\
&=\frac{1}{2\pi \mathrm{i}}\int_{\mathcal{C}}\frac{\mathrm{d}x'}{x'-x}F_g(x',I)\frac{\sqrt{\sigma(x')}}{2y(x')}.\label{manip2}
\end{align}
In the second equality, we deformed the contour as in figure \ref{figrho0}. 
The properties needed to justify this contour deformation were argued at the end of section \ref{firstspecial}: $R_g(x,I)$ is holomorphic in $x$ away from the cut and
vanishes  as $1/x^2$ or faster
at $x=\infty$.       It can further be proved inductively using (\ref{manip2})
that on the first sheet, $R_g(x,I)$ is actually holomorphic in $x$ except at the endpoints of the cut (and hence can be continued through the cut).   
This is equivalent to saying that the eigenvalue distribution $\rho_g(I)$
in eqn. (\ref{expform})  is real analytic except at the endpoints.

The relation (\ref{manip}), together with the special cases discussed above, gives a recursion that determines all $R_{g}(x,I)$.

\subsubsection{Why \texorpdfstring{$\upbeta = 2$}{beta = 2} Is Special}\label{sec:beta2special}
When $\upbeta = 2$, the first term in (\ref{firstlineof}) is absent. This leads to two simplifications. First, it is easy to see that this implies that all $R_g(I)$ with half-integer $g$ vanish, so the $1/L$ expansion becomes a $1/L^2$ expansion. Second, the contour integral around the cut in (\ref{manip2}) reduces to a sum of residues at the endpoints. 

The second point is due to the fact that for $\upbeta = 2$, the multivaluedness of the $R_{g}(I)$ becomes very simple. We will use the notation $\widehat{x}$ in the argument of a function to indicate a coordinate that has been continued to location $x$ but on the second sheet. So, for example, $y(\widehat{x}) = -y(x)$ and $\sqrt{\sigma(\widehat{x})} = -\sqrt{\sigma(x)}$. Then with the exception of $R_0(x_1)$ and $R_0(x_1,x_2)$, we claim that
\be
R_g(\widehat{x}_1,x_2,\dots x_n) = -R_g(x_1,x_2,\dots x_n).\label{switchSign}
\ee
This can be proven by induction. First we note that if (\ref{switchSign}) holds, then $F_g(x,I)$ is actually single-valued in $x$. This is obvious for the terms $R_{g-1}(x,x,I)$ and $R_{h}(x,J)R_{g-h}(x,I\setminus J)$ that appear in (\ref{firstlineof}). It would not be true for the $\partial_xR_{g-\frac{1}{2}}(x,I)$ term, but this is absent for $\upbeta = 2$. For the final term in (\ref{firstlineof}), it is true using the explicit formula (\ref{explicit}).

Next, we note that single-valuedness of $F_g$ implies that the integrand in (\ref{manip2}) is itself single-valued, because $\sqrt{\sigma(x)}/y(x)$ is also single-valued. So, assuming (\ref{switchSign}), one finds that the contour integral around the cut just picks up the contribution of poles at the endpoints of the cut. The result of this integral will be analytic in $x$, and therefore $R_{g}(x,I)\sqrt{\sigma(x)}$ is analytic in $x$, so $R_g(x,I)$ satisfies (\ref{switchSign}). To complete the induction, we also need to check explicitly the special base case $R_0(x,x_1,x_2)$, which amounts to showing that the sum of the second and third terms from (\ref{sumof}) are single-valued in $x$. This follows from (\ref{explicit}).

The fact that the integral (\ref{manip2}) reduces to residues at the endpoints of the cut is a major simplification. In practical terms, it makes the loop equations much easier to implement. Conceptually, it makes the loop equations for $\upbeta = 2$ a special case of the Eynard-Orantin ``topological recursion'' \cite{eynard2007invariants}.

\subsubsection{Relation Between \texorpdfstring{$\upbeta = 1$}{beta = 1} And \texorpdfstring{$\upbeta = 4$}{beta = 4}}\label{sec:relationbeta1beta4}
The $\upbeta = 1$ and $\upbeta = 4$ ensembles are very closely related to each other \cite{MulaseWaldron}. The relationship is clearest if, on the $\upbeta = 4$ side, we work in terms of $\widetilde{R}$ defined by (assuming $L$ even)
\be
\widetilde{R}(x;L) = 2R(x;\tfrac{L}{2}), \hspace{40pt} \tilde{R}(I;L) = 2^{|I|}R(I;\tfrac{L}{2}).
\ee
Here we temporarily made the $L$ argument explicit. One can interpret $\widetilde{R}(x)$ as $\text{Tr}\frac{1}{x-H}$ where $H$ is a matrix that has $\frac{L}{2}$ independent eigenvalues from the $\upbeta = 4$ ensemble, each with two-fold degeneracy. This is actually the situation that describes a ensemble of matrices with $\Sp(L)$ symmetry: such matrices have $\upbeta = 4$ statistics and twofold degeneracy. We can write a $1/L$ expansion of connected correlators of $\widetilde{R}(I)$ as
\be
\langle \widetilde{R}(I;L)\rangle_{\text{c}} \simeq \sum_{g}\frac{\widetilde{R}_g(I)}{L^{2g+|I|-2}} = 2^{|I|}\sum_g\frac{R_g(I)}{(\frac{L}{2})^{2g+|I|-2}}
\ee
which implies that
\be\label{givesb}
\widetilde{R}_{g}(I) = 2^{2(g+|I|-1)}R_g(I).
\ee
Substituting (\ref{givesb}) into (\ref{topfirstlineof}), one finds that the $\widetilde{R}_g$ quantities for the $\upbeta = 4$ ensemble satisfy
\be
2y(x) \widetilde{R}_g(x,I) + \widetilde{F}_g(x,I) \sim 0
\ee
where
\begin{align}
\widetilde{F}_{g}(x,I)&= \partial_x\widetilde{R}_{g-\frac{1}{2}}(x,I)+ \widetilde{R}_{g-1}(x,x,I)+\sum_{\text{stable}} \widetilde{R}_{h}(x,J)\widetilde{R}_{g-h}(x,I\setminus J)\label{topfirstlineofTilde}\\
&\hspace{20pt}+ 2\sum_{k = 1}^n\left(\widetilde{R}_{0}(x,x_k) + \tfrac{1}{(x-x_k)^2}\right)\widetilde{R}_g(x,I\setminus x_k)\label{firstlineofTilde}.
\end{align}
Now, the point is that this is the same recursion satisfied by $R_{g}$ for the $\upbeta = 1$ ensemble, except for the sign of the first $\partial_x R_{g-\frac{1}{2}}$ term (which would have been $-1$ for the $\upbeta = 1$ ensemble). In the solution to the recursion, changing this sign has the effect of reversing the sign of all terms with half-integer $g$. In terms of 't Hooft double-line diagrams (or in the JT gravity interpretation below), these correspond to unorientable surfaces with an odd number of crosscaps, so we conclude that the difference between the $\upbeta =1$ and $\upbeta = 4$ ensembles is to insert a relative factor of $(-1)^{n_c}$ where $n_c$ is the number of crosscaps \cite{MulaseWaldron}.   Related
matters are discussed in appendix \ref{Euler}.

\subsection{Loop Equations For The \texorpdfstring{$(\upalpha,\upbeta)$}{(alpha,beta)} Ensembles}\label{sec:alphabeta}
In this section, we will study the loop equations for the $(\upalpha,\upbeta)$ ensembles of Altland and Zirnbauer \cite{AZ}.\footnote{At least one of these
ensembles, namely the bifundamental of $\U(L)\times \U(L)$, was studied in early literature on matrix models and two-dimensional gravity \cite{Morris,Johnson1,Johnson2}.}  
 As discussed in section \ref{measure}, these ensembles are characterized by a measure 
\be
\prod_{1\le i<j\le L}|\omega_i^2-\omega_j^2|^\upbeta\prod_{i=1}^L|\omega_i|^\upalpha e^{-L\frac{\upbeta}{2} V(\omega_i^2)}\mathrm{d}\omega_i\label{probOmega}
\ee
where $V(x),\upalpha,\upbeta$ are parameters. The distribution makes sense for continuous values of $\upalpha,\upbeta$, but for special values of these parameters, it has an interpretation in terms of random matrix theory, as reviewed in section \ref{sec:ensembles}.

As shown in \cite{Li:2017hdt,Kanazawa:2017dpd,Sun:2019yqp} and discussed in section \ref{SUSY} above, the symmetry classes associated to these ensembles are relevant the supercharge $Q$ in the $\mathcal{N} = 1$ supersymmetric SYK model \cite{Fu:2016vas}. An important detail is that the eigenvalues of $Q$ come in pairs $\omega,-\omega$. Also, for $\upbeta = 4$, there is twofold degeneracy. We emphasize that the probability distribution (\ref{probOmega}) is written for a set of independent variables, meaning just one of each pair of distinct eigenvalues $\omega,-\omega$. We will deal with the degeneracy later when we compare to JT supergravity.

In supersymmetric quantum mechanics, the Hamiltonian is the square of the supercharge $H = Q^2$, so eigenvalues of the Hamiltonian are $\lambda = \omega^2$. The measure (\ref{probOmega}) for the independent eigenvalues of $Q$ implies a measure for the distinct eigenvalues of $H$:
\be
\prod_{1\le i<j\le L}|\lambda_i-\lambda_j|^\upbeta\prod_{i=1}^L|\lambda_i|^{\frac{\upalpha-1}{2}} e^{-L\frac{\upbeta}{2} V(\lambda)} \d\lambda_i, \hspace{20pt} \lambda_i\ge 0.\label{appearingIn}
\ee
It will be slightly easier to derive the loop equations using the variables $\omega$. However, because we will ultimately be interested in viewing $\lambda_i = \omega_i^2$ as the physical eigenvalues, we work in terms of resolvents defined as
\begin{align}
R(x)& = \sum_{i = 1}^L\frac{1}{x-\omega_i^2}.
\end{align}
As before, we use the notation $R(x_1,\dots,x_n) = R(x_1)\dots R(x_n)$, and $I = \{x_1,\dots,x_n\}$. Connected correlators of these resolvents still have a $1/L$ expansion of the same form (\ref{genusExp}), and our goal will be to find a recursion for the $R_{g}(I)$ using the loop equations.

The loop equations follow from the statement
\begin{align}
0 = \frac{1}{\upbeta x}\int_{-\infty}^\infty d^L\omega \frac{\partial}{\partial \omega_a}\left[\frac{\omega_a}{x-\omega_a^2}R(I)\prod_{i<j}|\omega_i^2-\omega_j^2|^\upbeta\prod_{i=1}^L|\omega_i|^\upalpha e^{-L\frac{\upbeta}{2} V(\omega_i^2)}\right].
\end{align}
Distributing the derivative and simplifying using analogs of the three steps in (\ref{step1}), (\ref{improved}), (\ref{Step3}), one finds
\begin{align}\notag
&\Bigg\langle \left((1-\tfrac{2}{\upbeta})\partial_x + \frac{\upalpha{-}1}{\upbeta x} - L V'(x)\right)R(x,I)+ R(x,x,I) + \frac{2}{\upbeta x}\sum_{k= 1}^n\partial_{x_k}\frac{xR(x,I\setminus x_k) - x_kR(I)}{x-x_k}\Bigg\rangle\\&\hspace{60pt}= -\frac{L}{x}\langle P(x;I)\rangle,
\end{align}
where now
\be
P(x;I) = \sum_{a = 1}^L \frac{x V'(x) - \omega_a^2V'(\omega_a^2)}{x-\omega_a^2}R(I).
\ee
Again, $P(x;I)$ is analytic in $x$ assuming an analytic potential. This can be rewritten in terms of connected correlators using the same logic as in the $\upbeta$-ensemble case. One finds
\begin{align}\notag
&\left((1-\tfrac{2}{\upbeta})\partial_x + \frac{\upalpha{-}1}{\upbeta x} - L V'(x)\right)\langle R(x,I)\rangle_{\text{c}}+ \langle R(x,x,I)\rangle_{\text{c}} + \sum_{J\subseteq I}\langle R(x,J)\rangle_{\text{c}}\langle R(x,I\setminus J)\rangle_{\text{c}} \\ & \hspace{20pt} + \frac{2}{\upbeta x}\sum_{k= 1}^n\partial_{x_k}\frac{x\langle R(x,I\setminus x_k)\rangle_{\text{c}} - x_k\langle R(I)\rangle_{\text{c}}}{x-x_k}= -\frac{L}{x}\langle P(x;I)\rangle_{\text{c}}.\label{inSuper}
\end{align}
We will use this equation to compute the $R_g(I)$ following our steps for the $\upbeta$-ensemble. First, we will work out three special cases. Then we will write a recursion relation for the other cases.

\subsubsection{A First Special Case \texorpdfstring{$R_{0}(x)$}{R0(x)}}
As in the $\upbeta$-ensemble case, we start with the special case $I = \emptyset$, and further simplify by keeping only the terms proportional to $L^2$. Then (\ref{inSuper}) reduces to
\be
R_0(x)^2 - V'(x)R_0(x) = \frac{1}{x}(\text{analytic in $x$}).
\ee
which can be written as
\be
y(x)^2 = \frac{1}{x}(\text{analytic in $x$}), \hspace{20pt} R_0(x) = \frac{V'(x)}{2} + y(x).\label{intermsofy}
\ee
Again, we find a hyperelliptic curve, and the relationship between $y(x)$ and $\rho_0(x)$ in (\ref{yandrho}) remains true, where $\rho_0(x)$ is interpreted as the density of eigenvalues $\lambda = \omega^2$. Since these are non-negative, the density $\rho_0(x)$ has to be supported for $x\ge 0$. We will specialize to the simplest ``one-cut'' case, where $\rho_0(x)$ is nonzero in a single interval between zero and some value $a_+>0$. This interval will be referred to as ``the cut.''

The important difference between the present case and the $\upbeta$-ensembles is the factor of $1/x$ in (\ref{intermsofy}). What this means is that $y(x)$ will have a $1/\sqrt{x}$ singularity at the origin, and an ordinary $\sqrt{x}$ branchpoint at $a_+$. As an example, we can work out $y(x)$ for the case $V(x) = x$. Then (\ref{intermsofy}), together with the requirement that $R_0(x) \approx 1/x$ for large $x$, implies that 
\be
y(x) = -\frac{1}{2}\sqrt{\frac{x-4}{x}} \hspace{20pt} \implies \hspace{20pt} \rho_0(x) = \frac{1}{2\pi}\sqrt{\frac{4-x}{x}}.
\ee
This is a special case of the Marchenko-Pastur distribution. It is the analog of the Wigner semicircle distribution for Wishart matrices $H = Q^2$. See figure \ref{fig3}.
\begin{figure}
\begin{center}
\includegraphics[width = .65\textwidth]{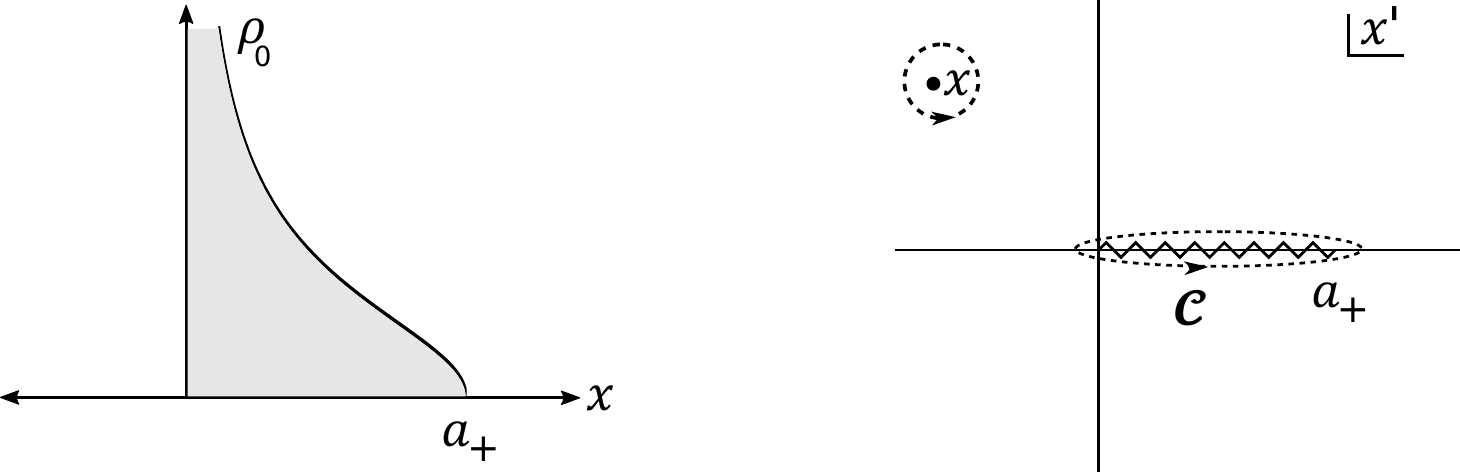}
\caption{At left we plot a typical $\rho_0(x)$, the density of eigenvalues for $H = Q^2$. Note the $1/\sqrt{x}$ divergence at the origin, and the $\sqrt{x}$ vanishing at $x = x_+$. At right we show the cut $x'$ plane and the contour manipulation used in deriving the dispersion relation.}\label{fig3}
\end{center}
\end{figure}

The $1/\sqrt x$ singularity in $y$, compared to the typical behavior  $y\sim (x-a)^{1/2}$   near a branch point in the case of a Dyson ensemble, reflects the
fact that the eigenvalues $\lambda_i=\omega_i^2$ are strictly nonnegative in an Altland-Zirnbauer ensemble.   This constraint leads to a divergence in the eigenvalue
density at $x=0$.    Thus regardless of the values of $\upalpha,\upbeta$, an Altland-Zirnbauer ensemble is never equivalent to a Dyson ensemble, even though
if one ignores the constraint $\lambda_i\geq  0$, the measure (\ref{appearingIn}) in the special case $\alpha=1$ would coincide with that of a Dyson ensemble. 

\subsubsection{A Second Special Case \texorpdfstring{$R_{0}(x,x_1)$}{R0(x,x1)}}
Next, take the case $I = \{x_1\}$ and study (\ref{inSuper}) at leading order $L^1$. This leads to
\be
-V'(x) R_0(x,x_1) + 2R_0(x)R_{0}(x,x_1) + \frac{2}{\upbeta x}\partial_{x_1}\frac{xR_0(x) - x_1R_0(x_1)}{x-x_1} = \frac{1}{x}(\text{analytic in $x$}).
\ee
Using (\ref{intermsofy}), and moving some terms to the RHS, this can be rewritten as
\be
2y(x)R_0(x,x_1) + \frac{2}{\upbeta}\frac{y(x)}{(x-x_1)^2} = \frac{1}{x}(\text{analytic in $x$ in a neighborbood of the cut}).\label{tomultby}
\ee
We assume that $x_1$ is located away from the cut, so expressions like $1/(x-x_1)^p$ are analytic in $x$ in a neighborhood of the cut. After multiplying through by $x$, this is
\be
xy(x)R_0(x,x_1) + \frac{x}{\upbeta}\frac{y(x)}{(x-x_1)^2} \sim 0.
\ee
where as before, the $\sim$ means equal up to terms that are analytic in $x$ in a neighborhood of the cut. To analyze this equation, we specalize $\sigma(x)$ from (\ref{sigmaofx}) to the case with $a_- = 0$,
\be
\sigma(x) = x(x-a_+),
\ee
and then multiply (\ref{tomultby}) by $\sqrt{\sigma(x)}/(xy(x))$. This ratio is analytic in $x$ in a neighborhood of the cut, thanks to the fact that $y(x)$ has a $1/\sqrt{x}$ singularity at $x = 0$, and a $\sqrt{x}$ branch point at $x = a_+$. The resulting equation
\be
R_0(x,x_1)\sqrt{\sigma(x)} + \frac{2}{\upbeta}\frac{\sqrt{\sigma(x)}}{(x-x_1)^2} \sim 0
\ee
is the same as we had for the $\upbeta$-ensemble. So we conclude that $R_{0}(x_1,x_2)$ is still given by the formula (\ref{explicit}), but now with $a_- = 0$. 

\subsubsection{A Third Special Case \texorpdfstring{$R_0(x,x_1,x_2)$}{R0(x,x1,x2)}}\label{sec:uwh}
As a final special case, we take $I = \{x_1,x_2\}$ and again work at leading order in $1/L$, which is now $L^0$. Then (\ref{inSuper}) implies
\be
xy(x)R_0(x,x_1,x_2) + xR_{0}(x,x_1)R_0(x,x_2) + \frac{x}{\upbeta}\left(\frac{R_0(x,x_2)}{(x-x_1)^2} + \frac{R_0(x,x_1)}{(x-x_2)^2}\right) \sim 0.
\ee
where the $\sim$ means up to terms that are analytic in $x$ in a neighborhood of the cut. After multiplying through by $\sqrt{\sigma(x)}/(xy(x))$, one can write exactly the same dispersion relation as (\ref{disperThreept}). A difference is that in the present case, the factor $\sqrt{\sigma(x)}/y(x)$ has a zero at $x = 0$ which cancels the pole from the rest of the integrand in (\ref{disperThreept}). This means that $R_{0}(x,x_1,x_2)$ receives a contribution only from the $a_+$ endpoint. This implies that $R_{0}(x_1,x_2,x_3)$ will vanish in a double-scaled $(\alpha,\upbeta)$ ensemble, where we take $a_+$ to infinity. In turn, this implies that all correlators $R_{0}(x_1,...,x_n)$ with $g = 0$ and $n\ge 3$ vanish in a double-scaled ensemble of this type.

\subsubsection{The Generic Case}
To consider the generic case, one plugs the general form of the genus expansion (\ref{genusExp}) into (\ref{inSuper}) and collects terms at order $L^{2-2g-|I|}$. The result can be written
\be
2xy(x) R_g(x,I) + xF_g(x,I) \sim 0\label{ywu}
\ee
where
\begin{align}
F_{g}(x,I)&= \left((1-\tfrac{2}{\upbeta})\partial_x + \frac{\upalpha{-}1}{\upbeta x}\right)R_{g-\frac{1}{2}}(x,I)+ R_{g-1}(x,x,I)+\sum_{\text{stable}} R_{h}(x,J)R_{g-h}(x,I\setminus J)\notag\\
&\hspace{20pt}+ 2\sum_{k = 1}^n\left(R_{0}(x,x_k) +\tfrac{1}{\upbeta} \tfrac{1}{(x-x_k)^2}\right)R_g(x,I\setminus x_k)\label{firstlineofalpha}.
\end{align}
The resulting dispersion relation is
\be
R_g(x,I)\sqrt{\sigma(x)} = \frac{1}{2\pi \mathrm{i}}\int_{\mathcal{C}}\frac{\mathrm{d}x'}{x'-x}F_g(x',I)\frac{\sqrt{\sigma(x')}}{2y(x')}.\label{disSuper}
\ee

We will have more to say about special cases of these equations below. For now, we note that in the special case $(\upalpha,\upbeta) = (1,2)$, the first term in (\ref{firstlineofalpha}) vanishes. This implies by induction that all half-integer $g$ coefficients vanish, and the dispersion relation reduces to a sum of residues at the endpoints. The argument for this is the same as for ordinary $\upbeta = 2$ ensembles in section \ref{sec:beta2special}. 

Also, one can show that the ensembles with $(\upalpha,\upbeta)$ equal to $(\upalpha,1)$ and $(-2\upalpha+3,4)$ are related in the same way as discussed for the $\upbeta = 1$ and $\upbeta = 4$ ensembles in section \ref{sec:relationbeta1beta4}. After rescaling the $(-2\upalpha+3,4)$ resolvents as in (\ref{givesb}), the results differ  from the $(\upalpha,1)$ case only by the sign of the terms with half-integer $g$. Looking back to Table \ref{table1} of  section  \ref{revpairs}, we deduce in the context of JT supergravity\footnote{We state
the following for $\hat q=1$ mod 4, as assumed in the table.  For $\hat q=3$ mod 4, one must exchange $N\to 8-N$ in all statements.}
 that if $N$ is even, then a shift $N\to N+4$
has the effect only of changing the sign of crosscap contributions.   The same is true for odd $N$.   For $N=3,7$, the   conclusion follows
from the behavior under an exchange
$(\upalpha,2)\leftrightarrow (2-\upalpha,2)$.
  For $N=1,5$, one uses the original statement  for $\upbeta=1$  and $\upbeta=4$  Dyson ensembles.  See also appendix \ref{Euler}.

\subsubsection{The Crosscap \texorpdfstring{$R_{\frac{1}{2}}(x)$}{R1/2(x)}}
For comparision to JT supergravity in  section \ref{srt}, it will be helpful to have a formula for the one-crosscap contribution to the resolvent, $R_{\frac{1}{2}}(x)$. This is determined by setting $g = \frac{1}{2}$ and $I = \emptyset$ in (\ref{firstlineofalpha}). The resulting equation can be written as
\be
2xy(x)R_{\frac{1}{2}}(x) + x \left((1-\tfrac{2}{\upbeta})\partial_x + \frac{\upalpha{-}1}{\upbeta x}\right) y(x)\sim 0. \label{nnwy}
\ee
This leads to
\begin{align}
R_{\frac{1}{2}}(x) &= \frac{1}{2\pi\mathrm{i}}\int_{\mathcal{C}}\frac{\mathrm{d}x'}{x'-x}\frac{\sqrt{\sigma(x')}}{\sqrt{\sigma(x)}}\left(\frac{(1-\tfrac{2}{\upbeta})y'(x')}{2y(x')} + \frac{\upalpha{-}1}{2\upbeta x'}\right)\label{hwy}
\end{align}

\subsubsection{Double-Scaling}
An important detail is that JT gravity or supergravity  is related not to an ordinary matrix integral of the type we have discussed here, but instead a particular type of limit that is referred to as a double-scaled matrix integral. We will briefly describe what this means.

In terms of the data of the spectral curve, to study a double-scaled matrix integral we scale the upper endpoint of the cut to infinity $a_+\rightarrow \infty$, and we relax the constraint that $\rho_0(x)$ should have unit integral. Such a system can be understood as a limit of an ordinary matrix integral in which $L$ is taken to infinity holding fixed the classical approximation to the density of eigenvalues near the lower endpoint. In an ordinary matrix integral, the classical approximation to the density of eigenvalues is $\rho_{0}^\text{total}(x) = L \rho_0(x)$, where $\rho_0$ is normalized so that the integral is one. To define a double-scaled integral, we write
\be\label{ywup}
\rho_0^{\text{total}}(x) = e^{S_0}\rho_0(x).
\ee
Then we take $L$ to infinity and adjust the potential so that $e^{S_0}$ remains fixed and $\rho_0(x)$ approaches the desired function. 

We retain the definition $y(x+\mathrm{i}\epsilon) = -\mathrm{i}\pi\rho_0(x)$. Since the loop equations depend only on $y(x)$, they commute with the double-scaled limit. However, because $L$ has been replaced by $e^{S_0}$ in (\ref{ywup}), the genus expansion is now in powers of $e^{-S_0}$:
\be
\langle R(I)\rangle\simeq\sum_{g = 0,\frac{1}{2},1,\,\dots}\frac{R_g(I)}{\left(e^{S_0}\right)^{2g+|I|-2}}.
\ee

\section{Comparison To JT Gravity And Supergravity}\label{sec:JT}
In this section, we will compare JT gravity and JT supergravity to the predictions of the loop equations for the matrix ensembles described in section \ref{sec:ensembles}. For ordinary JT gravity, there are three basic cases that correspond to the Dyson ensembles. As described in detail in sections \ref{sec:spinButNoT} and \ref{sec:bothSpinAndT}, ordinary JT gravity theories with spin or pin structure sums reduce to these cases and need not be discussed further. On the other hand, there are ten versions of super JT gravity, covering both the Dyson and the Altland-Zirnbauer ensembles.

We will go through the cases and compare what we can between JT (super) gravity and the predictions of the loop equations. For quantities where we can compute the predictions of JT gravity and supergravity, we will find agreement with the loop equations. For other quantities, we will find predictions from the loop equations for the volumes of supermoduli space.

The basic dictionary that we will use to compare matrix integral quantities and JT gravity quantities is
\be\label{dictionary}
\left\langle \Tr\big(e^{-\beta_1 H}\big)\dots \Tr\big(e^{-\beta_nH}\big)\right\rangle_{\text{matrix int.}}\leftrightarrow\;\;\;\text{\parbox{6cm}{ {\footnotesize(super) JT path integral summed over topologies, with $n$ asymptotic ``trumpet'' boundaries of regularized lengths $ \beta_j$}}}
\vspace{8pt}
\ee
On the LHS, $H$ is the random matrix from an appropriate ensemble, including degeneracy if there is any.\footnote{In cases where the relevant ensemble corresponds to a theory with an odd number of fermions, then we should multiply each trace in the LHS by $\sqrt{2}$, as explained in (\ref{woggoSYK}). For the most part we will consider NS boundaries, but we will comment on R boundaries in section \ref{sec:RBoundaries}.} On the RHS, we have a path integral in JT gravity or super gravity, with an asymptotic boundary of regularized length $\beta$ for each insertion of $\Tr\, e^{-\beta H}$ on the LHS.

In (\ref{dictionary}), we are studying the full (connected plus disconnected) correlation function on the LHS. And on the RHS, we have a sum over all bulk topologies, including disconnected ones. But to establish (\ref{dictionary}), it is sufficient to compare the connected correlators (or ``cumulants'') constructed from both sides of (\ref{dictionary}). On the LHS this means that we subtract products of lower-order correlators. On the RHS, it means that we consider only connected geometries. After making this restriction, both sides have a ``genus'' expansion in powers of $e^{-S_0}$
\be
\sum_{g = 0}^\infty \frac{Z_{g}(\beta_1,\dots\beta_n)}{(e^{S_0})^{2g+n-2}},
\ee
and we will compare the coefficients in the expansion $Z_g(\beta_1,\dots,\beta_n)$. In general, the sum will involve both half-integer and integer values of the ``genus'' summation index $g$. 

On the matrix integral side, this genus expansion is determined by the loop equations. More precisely, in our discussion of the loop equations, we saw how to compute the $e^{-S_0}$ expansion of correlation functions of resolvents, but one can go back and forth between resolvents and partition functions using Laplace and inverse Laplace transforms.

On the JT gravity side, $Z_g(\beta_1,\dots,\beta_n)$ is the path integral over two-dimensional connected geometries with fixed genus $g$. The weighting by powers of $e^{S_0}$ comes from an Euler characteristic term we can add to the bulk JT action, as is familiar in string theory. The rest of the bulk action can be written as a topological $BF$ theory, with gauge group $G = \SL(2,\R)$ for ordinary JT gravity and $G = \OSp'(1|2)$ for JT supergravity. As described in section \ref{sectionThree}, the path integral of a $BF$ theory reduces to an integral over the moduli space of flat connections, with measure given by the torsion.\footnote{One also has to restrict to the appropriate topological component of the moduli space, and quotient by the action of the mapping class group.} So, the main part of the computation of $Z_{g}(\beta_1,\dots,\beta_n)$ is a computation of the volume of a moduli space with this measure.

More precisely, the path integral reduces to an integral over such a moduli space, together with a path integral over ``wiggles'' associated to the $n$ asymptotic boundaries:
\be
Z_{g}(\beta_1,\dots,\beta_n) = \int \mathrm{d}(\text{moduli})\int \mathcal{D}(\text{wiggles})e^{-I_{\text{Sch}}}.
\ee
These wiggles are described by the Schwarzian theory in the ordinary JT gravity case \cite{Jensen:2016pah,Maldacena:2016upp,Engelsoy:2016xyb}, and by the super-Schwarzian theory in the super JT gravity case \cite{Forste:2017kwy}. These path integrals can be done in advance, and then glued together with the computation of the moduli space volume. For more on this, see appendix \ref{app:Sch}.

Two different Schwarzian and super-Schwarzian path integrals will be needed, see figure \ref{figdisktrumpet}. The first path integral, $Z^D(\beta)$ is an integral over the wiggles at the boundary of a disk. In this situation, the only parameter is the regularized length of the wiggly boundary, $\beta$. The second path integral, $Z^T(\beta,b)$, is an integral over the wiggles at the ``big end'' of a trumpet. The trumpet is defined so that the ``small end'' is a geodesic of length $b$, and the answer depends on this parameter as well as $\beta$, the regularized length of the big end. For both cases, the path integrals depend on the normalization of the path integral measure, and also on the normalization of $\beta$. Both of these normalizations are arbitrary, but must be chosen consistently. With a particular choice that we will describe in more detail in appendix \ref{app:Sch}, the Schwarzian and super-Schwarzian path integrals for the disk and the trumpet are explicitly
\begin{align}
Z_{\text{JT}}^D(\beta) &= \frac{1}{4\pi^{1/2}\beta^{3/2}}e^{\frac{\pi^2}{\beta}}\hspace{30pt} Z_{\text{JT}}^{T}(\beta,b) = \frac{1}{2\sqrt{\pi\beta}}e^{-\frac{b^2}{4\beta}}\\
Z_{\text{SJT}}^D(\beta) &= \sqrt{\frac{2}{\pi\beta}}e^{\frac{\pi^2}{\beta}}\hspace{45pt} Z_{\text{SJT}}^T(\beta,b) = \frac{1}{\sqrt{2\pi\beta}}e^{-\frac{b^2}{4\beta}}.\label{diskSJT}
\end{align}
The ordinary JT cases were computed in \cite{Bagrets:2016cdf,Cotler:2016fpe,Bagrets:2017pwq,Stanford:2017thb,Belokurov:2017eit,Mertens:2017mtv,Kitaev:2018wpr,Yang:2018gdb,Iliesiu:2019xuh} and the super cases are computed in appendix \ref{app:Sch}. 

For the disk topology, there is no moduli space to integrate over, so the path integral over the Schwarzian ``wiggles'' is actually the whole answer:
\be
Z_0(\beta) = Z^D_{\text{(S)JT}}(\beta).
\ee
In comparing to matrix integrals, the disk plays a special role, because it determines the spectral curve, or equivalently the $\rho_0(x)$ function. We can read this off by interpreting $Z_0(\beta) = \int_0^\infty \mathrm{d}x\, e^{-\beta x}\rho_0(x)$ and solving for $\rho_0(x)$. For the two cases of JT gravity and super JT gravity, one finds the $\rho_0(x)$ functions and corresponding spectral curves
\begin{align}
\rho_{0\,\text{JT}}(x) &= \frac{\sinh(2\pi \sqrt{x})}{4\pi^2} \hspace{51pt} y_{\text{JT}}(x) = \frac{\sin(2\pi\sqrt{-x})}{4\pi}\label{specCurveY}\\
\rho_{0\;\text{SJT}}(x) &= \frac{\sqrt{2}\cosh(2\pi\sqrt{x})}{\pi\sqrt{x}} \hspace{30pt} y_{\text{SJT}}(x) = -\frac{\sqrt{2}\cos(2\pi\sqrt{-x})}{\sqrt{-x}}.\label{rhoSJT}
\end{align}
As we have seen, together with the discrete choice of matrix ensemble, the data of the spectral curve completely determines the genus expansion of a matrix integral. So from the perspective of a correspondence between (super) JT gravity and matrix integrals, the disk topology should be regarded as input that allows us to fit the correct matrix integral. All other topologies constitute nontrivial tests of the correspondence.

\begin{figure}[t]
\begin{center}
\includegraphics[width = .75\textwidth]{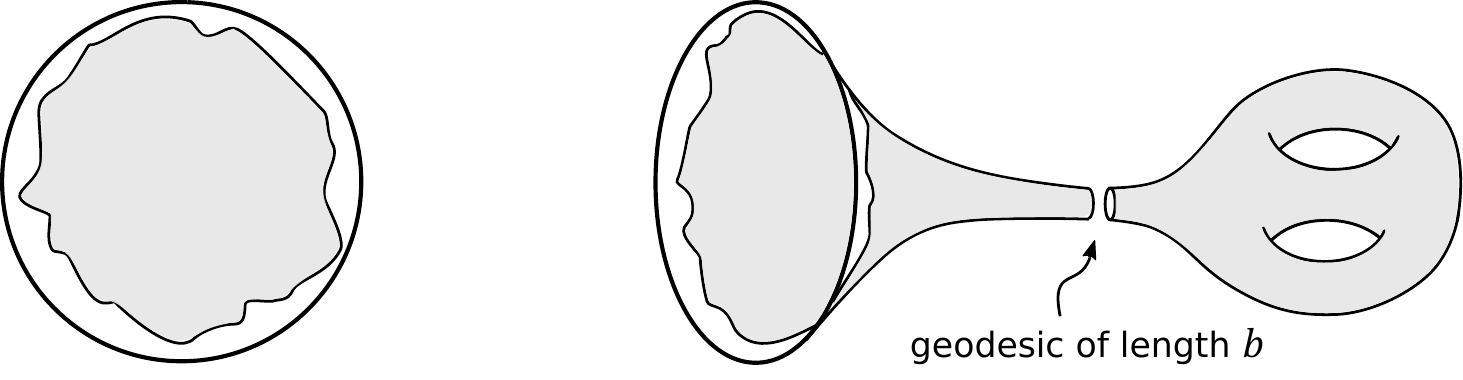}
\caption{\small{The disk (left) and trumpet (right) partition functions are path integrals over a wiggly regularized AdS${}_2$ boundary, together with a fermionic partner mode in the super case. The length of the wiggly boundary is proportional to $\beta$. In the case of the trumpet, the other end of the trumpet is a geodesic of length $b$, across which the geometry can be glued to a bordered (super) Riemann surface.}}\label{figdisktrumpet}
\end{center}
\end{figure} 
In (super) JT gravity, the path integrals on other topologies $Z_g$ are given by a combination of the Schwarizan trumpet path integral and the volumes of moduli space. There are two cases that have to be treated individually. One of them is the ``double trumpet,'' for which the path integral gives
\be
Z_{0}(\beta_1,\beta_2) = c\int_0^\infty b\,\mathrm{d}b\,Z^T_{\text{(S)JT}}(\beta_1,b)Z^T_{\text{(S)JT}}(\beta_2,b).\label{doubletrumpetSpecial}
\ee
In this expression, two trumpets have been glued together by integrating over the length of their shared geodesic $b$ and the relative twist. The integral over the relative twist gives the factor of $b$ in the measure. In a theory where we sum over neither spin structures nor orientation reversal, the constant $c$ should be one. In a theory where we sum over one but not the other, we have $c = 2$. Finally, in a theory where we sum over both spin structures and orientation reversal, we have $c = 4$.

Another special case is the crosscap spacetime:
\be
Z_{\frac{1}{2}}(\beta) = \int_0^\infty \mathrm{d}b\, V_{\frac{1}{2}}(b)Z^T_{\text{(S)JT}}(\beta,b).\label{crosscapG}
\ee
This case is special because when we glue the trumpet to the crosscap, there is only a single modulus involved in the gluing, the size. There is no twist modulus because of the rotational symmetry of the crosscap. The result of this is that the measure factor is simply $\mathrm{d}b$ rather than $b\mathrm{d}b$ as we have in other cases.

All other $Z_g(\beta_1,\dots \beta_n)$ are given by the generic formula\footnote{\label{footnotefactorFirst}In principle, the genus one case with a single boundary is also special: because the torus with one hole has a $\mathbb{Z}_2$ symmetry, we should integrate the twist from zero to $b/2$ instead of $b$. Instead, we simply define $V_1(b)$ to be one-half of the true moduli space volume of the torus with one hole, and integrate the twist as usual. See also footnote \ref{footnotefactoroftwo}.}
\be
Z_{g}(\beta_1,\dots,\beta_n) = c_n\int_0^\infty\prod_{j = 1}^n\left[ b_j\mathrm{d}b_jZ^T_{\text{(S)JT}}(\beta_j,b_j)\right]V_{g}(b_1,\dots b_n).\label{c_n}
\ee
The volumes $V_g(b_1,\dots,b_n)$ are computed using the torsion as described in section \ref{zoggo} and section \ref{firstcases}. We define them to also include the sum over spin structures (if present) holding fixed the NS spin structure on the boundaries.\footnote{We will consider Ramond boundaries briefly in section \ref{sec:RBoundaries}.} The constant $c_n$ should be one in a theory where we do not gauge orientation reversal, and it should be $2^{n-1}$ in a theory where we do. This accounts for the possibility of gluing the $n$ trumpets in with independent orientation-reversals, up to an overall orientation reversal.

In order to compare to the loop equations, we will often work in terms of resolvents instead of partition functions. These can be obtained using $R(x) = -\int_0^\infty \mathrm{d}\beta e^{\beta x}\Tr(e^{-\beta H})$. In order to apply this integral transform to the (super) JT partition functions, we will often use the formulas
\be
\int_0^\infty d\beta e^{-\beta z^2} Z^T_{\text{JT}}(\beta,b) = \frac{1}{2z}e^{-b z}, \hspace{20pt}
\int_0^\infty d\beta e^{-\beta z^2} Z^T_{\text{SJT}}(\beta,b) = \frac{1}{\sqrt{2}z}e^{-b z}.\label{intTrumpet}
\ee

\subsection{Ordinary JT Gravity With Time-Reversal Symmetry}\label{trt}
JT gravity on orientable surfaces was related to a double-scaled $\upbeta = 2$ matrix ensemble in \cite{Saad:2019lba}.  In this section we will relate $\upbeta = 1,4$ versions of the same matrix ensemble to a JT gravity theory in which orientation-reversal is viewed as a gauge symmetry. 

Concretely, gauging orientation-reversal means we can glue surfaces together with an orientation reversal, so in particular we should include unorientable surfaces as well as orientable ones. There are two versions of the bulk theory, depending on whether or not we include a factor of $(-1)^{n_c}$ where $n_c$ is the number of crosscaps. These two versions are related to $\upbeta = 1$ and $\upbeta = 4$ ensembles. Because of the discussion in \ref{sec:relationbeta1beta4}, it will be sufficient to compare just one of these cases, and we choose the $\upbeta = 1$ ensemble, which is dual to a bulk theory in which we do not include $(-1)^{n_c}$.

To compare the matrix and JT gravity predictions, we will start with genus zero. Here we can use as input the fact from \cite{Saad:2019lba} that JT gravity is dual to a $\upbeta = 2$ ensemble. It is easy to check from the recursion that genus zero resolvents in the $\upbeta = 1$ and $\upbeta = 2$ ensemble are related by the simple factor
\be
R_0^{\upbeta = 1}(I) = 2^{n-1}R_0^{\upbeta = 2}(I), \hspace{20pt} n = |I|.\label{genuszerofactor}
\ee
We wrote this equation in terms of resolvents, but partition functions $Z_{0}(\beta_1,...,\beta_n)$ will also be related by the same factor of $2^{n-1}$. So, to show agreement with the matrix integral, we need to find the same factor in the relationship between JT gravity partition functions in the theories with and without orientation-reversal gauged. Indeed, this is just the factor $c_n$ discussed in (\ref{c_n}). Topologically, $Z_{0}(\beta_1,\dots,\beta_n)$ involves $n$ trumpets glued into a sphere. We can glue each trumpet to the sphere with or without an orientation reversal, giving a factor of $2^n$. Gauging the overall orientation gives a single factor of $1/2$, so $c_n = 2^{n-1}$, which matches (\ref{genuszerofactor}).

Next we discuss genus one-half with a single boundary. From the loop equations for the ordinary $\upbeta$-ensembles, the equation (\ref{simin}) for $R_{\frac{1}{2}}(x)$  is 
\be
2y(x)R_{\frac{1}{2}}(x) +\left(1-\frac{2}{\upbeta}\right) \partial_x y(x)\sim 0.
\ee
where we replaced $\partial_xR_0(x)$ by $\partial_xy(x)$ at the cost of introducing an analytic term that we ignore in this expression. The corresponding dispersion relation is
\be
R_{\frac{1}{2}}(x)\sqrt{\sigma(x)} = \frac{1-\frac{2}{\upbeta}}{2\pi \mathrm{i}}\int_{\mathcal{C}}\frac{\mathrm{d}x'}{x'-x}\sqrt{\sigma(x')}\frac{\partial_{x'}y(x')}{2y(x')}.\label{RQ12}
\ee
For a double-scaled matrix integral with $a_- = 0$ and $a_+ = \infty$, this becomes (for $\upbeta = 1$)
\be
\sqrt{-x}R_{\frac{1}{2}}(x) = -\frac{1}{2\pi}\int_0^\infty \frac{\sqrt{x'}\,\mathrm{d}x'}{x'-x}\frac{\partial_{x'}y(x')}{y(x')}.\label{R12}
\ee
For a spectral curve $y(x)$ that is asymptotically a power of $x$, this integral would converge. But for the $y(x)$ in (\ref{specCurveY}), it is logarithmically divergent at large $x'$. What this means is that the limiting procedure needed to define the double-scaled $\upbeta = 1$ theory for JT gravity doesn't produce a finite $e^{-S_0}$ expansion. (The divergence is regulated in the minimal string, see appendix \ref{app:minimal}.)

For our purposes, this is actually a good thing, because as we saw  in section \ref{torunorientable}, the volume of the moduli space of unorientable surfaces is in general divergent. For the quantity $R_{\frac{1}{2}}$, we will be able to compare the integrands of the matrix and JT expressions, and find precise agreement. To proceed, we substitute in (\ref{specCurveY}) and rewrite (\ref{R12}) as
\be
R_{\frac{1}{2}\,\text{M}}(-z^2) = -\frac{1}{2z}\int_0^\infty \frac{\mathrm{d}x'}{x'+z^2}\coth(2\pi \sqrt{x'}).\label{nxn}
\ee
Here we added a subscript ``M'' to indicate that it is the matrix integral answer. One would like to use this formula to predict an answer for the path integral measure for the crosscap, and then check it against the torsion computation (\ref{proggo}). If we didn't know the measure for the crosscap in JT gravity, we would write (\ref{crosscapG}) with $V_{\frac{1}{2}}(b)$ unknown. This can be converted to an expression for the resolvent:
\be
R_{\frac{1}{2}\,\text{JT}}(-z^2) = -\int_0^\infty \mathrm{d}\beta\, e^{-\beta z^2}Z_{\frac{1}{2}\,\text{JT}}(\beta) = -\frac{1}{2z}\int_0^\infty \mathrm{d}b\, e^{-a z}V_{\frac{1}{2}}(b).\label{nxm}
\ee
In the last step we inserted (\ref{crosscapG}) and did the integral over $\beta$ explicitly using (\ref{intTrumpet}). So, matching (\ref{nxn}) and (\ref{nxm}), the equation to be solved for $V_{\frac{1}{2}}(b)$ is
\be
\int_0^\infty \mathrm{d}b\, e^{-b z}V_{\frac{1}{2}}(b) = \int_0^\infty \frac{dx'}{x'+z^2}\coth(2\pi\sqrt{x'}).
\ee
Taking an inverse Laplace transform of both sides of the equation, one finds
\begin{align}
V_{\frac{1}{2}}(b) &= \int \frac{\mathrm{d}z}{2\pi\mathrm{i}}e^{bz}\int_0^\infty \frac{dx'}{x'+z^2}\coth(2\pi\sqrt{x'})\\
&= \int_0^\infty \frac{dx'}{\sqrt{x'}}\sin(b\sqrt{x'})\coth(2\pi\sqrt{x'})\\
&=\tfrac{1}{2}\coth(\tfrac{b}{4}).
\end{align}
This matches the answer we got from the computation of the torsion of the crosscap (\ref{proggo}).

It may be possible to make further comparisons between these two divergent theories, but we will stop at this point and move to super JT gravity.

\subsection{Super JT Without Time-Reversal Symmetry}\label{superjtnot}
We will begin by considering a version of JT supergravity in which we do not gauge orientation-reversal. This means that we are restricted in how surfaces can be glued together. In particular, the twisted double-trumpet and its relatives are not allowed, and nor are unorientable surfaces. This is expected to be dual to a matrix ensemble that does not respect any time-reversal symmetry. As described in section \ref{SUSY} and table \ref{table2}, there are two versions of the matrix ensemble, and two versions of the bulk theory. 

In this section and in the next one,
 we will focus on the case that the boundaries have NS spin structures. We will comment on the case with R boundaries in  section \ref{sec:RBoundaries}.

\subsubsection{Sum Of Even And Odd Spin Structures}\label{sec:evenodd}
We start with the case corresponding to odd $N$. This is a theory with an anomaly in $(-1)^\sF$, which (as described in section \ref{sect:superBulkDescription}) is already accounted for in the contribution of the boundary fermion in the Schwarzian supermultiplet. So in the bulk theory we should not include the $(-1)^\zeta$ TFT. This means that bulk spin structures are all weighted equally, with no minus signs. As argued in section \ref{SUSY}, the random matrix ensemble is specified by saying that the supercharge $Q$ is drawn from an ordinary $\upbeta = 2$ ensemble. 

Since we are in an ensemble with an odd number of fermions, the relationship between the path integral and the trace in the Hilbert space involves a factor of $\sqrt{2}$, see (\ref{woggoSYK}). So we expect to have the relationship
\be\label{sqrt2}
Z_{\text{SJT}}(\beta)  = \sqrt{2}\,\text{Tr}(e^{-\beta Q^2}).
\ee
To determine the spectral curve of the matrix ensemble, we need the leading density of eigenvalues of $Q$. This can be obtained by taking the genus zero approximation of both sides of (\ref{sqrt2}):
\be
Z^D_{\text{SJT}}(\beta) = \sqrt{2}\int_{-\infty}^\infty \mathrm{d}x\, \rho_0(x) e^{-\beta x^2}.
\ee
After plugging in (\ref{diskSJT}) for the LHS, we need to solve this equation for $\rho_0(x)$. It is easy to check the solution
\be
\rho_0(x) = \frac{\cosh(2\pi x)}{\pi}.\label{coshspec}
\ee
This is the leading density of eigenvalues for the matrix $Q$. The corresponding spectral curve is determined by (\ref{yandrho}). It is somewhat unusual, with a branch cut running along the whole real axis. This can be arranged as a limit of a conventional matrix integral where both endpoints $a_{\pm}$ have been scaled away, to plus and minus infinity. 

We can get the expression for $R_0(x_1,x_2)$ for such a matrix integral by taking $a_\pm = \pm \infty$ in (\ref{explicit}). The result for $\upbeta = 2$ is 
\be
R_0(x_1,x_2) = \begin{cases}0 & x_1,x_2\text{ on same side of cut}\\ -\frac{1}{ (x_1-x_2)^2} & x_1,x_2\text{ on opposite sides.}\end{cases}\label{riu}
\ee
Here the cut runs along the whole real axis, so the statement is that $R_0$ is nonzero (on the principal sheet) only if the two arguments are in opposite half-planes. Eq.~(\ref{riu}) gives the two-resolvent correlator for resolvents constructed from the eigenvalues of $Q$. To compare to JT supergravity, it is helpful to first convert to a formula for the resolvents of $H$. Using
\be
\frac{1}{x+\lambda_Q} - \frac{1}{x-\lambda_Q} = \frac{2x}{x^2-\lambda_Q^2} = \frac{2x}{x^2 - \lambda_H}
\ee
we can relate the resolvents (using an obvious notation) as
\be
R^{(Q)}(x) - R^{(Q)}(-x) = 2x R^{(H)}(x^2).
\ee
The pair correlator of the resolvents of $H$ is determined by this formula and the pair correlator of resolvents of $Q$ 
in (\ref{riu}). The result is
\be\label{needstwice}
R_0^{(H)}(-z_1^2,-z_2^2) = \frac{1}{2z_1 z_2(z_1+z_2)^2}.
\ee
Including the factor of $\sqrt{2}$ in (\ref{sqrt2}), the matrix integral result will be consistent with JT supergravity if $R_{0\,\text{SJT}}$ is equal to $(\sqrt{2})^2 =2$ times this value. Let's now check against the prediction from JT supergravity. From (\ref{doubletrumpetSpecial}) and after plugging in (\ref{diskSJT}), we have
\be
Z_{0\;\text{SJT}}(\beta_1,\beta_2) = 2\int_0^\infty b\, \mathrm{d}b\, Z^T_{\text{SJT}}(\beta_1,b)Z^T_{\text{SJT}}(\beta_2,b) = \frac{2\sqrt{\beta_1\beta_2}}{\pi(\beta_1+\beta_2)}.\label{ZpairJT1}
\ee
The factor of two out front corresponds to the value $c = 2$ in (\ref{doubletrumpetSpecial}) which is appropriate for the current case where we sum over spin structures but not over orientation-reversal. Translating to resolvents using $R(x) = -\int_0^\infty \mathrm{d}\beta e^{\beta x}Z(\beta)$ and (\ref{intTrumpet}), we indeed find twice the value in (\ref{needstwice}).

A property of this type of scaled matrix integral is that all higher-order correlators vanish. This is due to the following. In a $\upbeta = 2$ ensemble, as we remarked in section \ref{sec:beta2special}, the recursion reduces to a sum of residues at the endpoints of the cut. Because of the factor $1/(x'-x)$ in the integrand of the dispersion relation (\ref{manip2}), these residues vanish in the limit that the endpoints go to infinity. So for a $\upbeta = 2$ matrix integral that has been double scaled to give a spectral curve like (\ref{coshspec}), all higher order correlators are zero. 

The conjectured correspondence of JT supergravity with a matrix integral then predicts that the volumes of the moduli spaces of bordered super-Riemann surfaces (summing equally over even and odd spin structures) are zero. We confirm this at the end of appendix \ref{supermirz}.

\subsubsection{Difference Of Even And Odd Spin Structures}\label{sec:diffEvenOdd}
Now we consider the random matrix ensemble appropriate to the case with an even number of Majorana fermions. We expect this to be related to JT supergravity including a factor of $(-1)^\zeta$, which means that we take the sum of even spin structures minus the sum of odd ones. As argued in section \ref{SUSY}, the appropriate random matrix ensemble is a $(\upalpha,\upbeta) = (1,2)$ ensemble. Because the Hamiltonian is two-fold degenerate, we expect
\be
Z_{\text{SJT}}(\beta) = \text{Tr}(e^{-\beta H}) = 2\sum_{\substack{\text{distinct}\\ \text{eigs.~of $H$}}}e^{-\beta\lambda_i} = 2Z_{\text{M}}(\beta).
\ee
Here, we are using the subscript M to indicate a matrix integral quantity where we sum only over the distinct eigenvalues of $H$, as we did in our study of the loop equations above. 

Because of the degeneracy, the density of distinct eigenvalues $\rho_{\text{M}}$ is half of the total density of eigenvalues. For JT supergravity, the genus zero expression for the total density of eigenvalues is given in (\ref{rhoSJT}). So we learn that $\rho_{0\,\text{M}}$ and the spectral curve $y$ of the matrix integral are given by
\be
\rho_{0\,\text{M}}(x) = \frac{\rho_{0\,\text{SJT}}(x)}{2} =  \frac{\cosh(2\pi\sqrt{x})}{\pi\sqrt{2x}},\hspace{20pt} y_{\text{M}}(x) = -\frac{\cos(2\pi\sqrt{-x})}{\sqrt{-2x}}.\label{specCurvebeta2}
\ee
Recall that for the $(\upalpha,\upbeta)$-ensembles, we work directly with resolvents for $H$, and not for $Q$. So here we have written the leading density of distinct eigenvalues of $H$. 

We would now like to use the loop equations of the matrix integral to predict answers for JT supergravity. It is convenient to compare the resolvents in the two theories. Because of the degeneracy, multi-resolvents in JT supergravity will be related to multi-resolvents of the distinct eigenvalues as
\be
R_{\text{SJT}}(I) = 2^{|I|}R_{\text{M}}(I)\label{ncu}.
\ee
As a first example, we can compare the leading two-point correlator. From the expression (\ref{abinf}) specialized to $(\upalpha,\upbeta) = (1,2)$, one finds
\be
R_{0\,\text{M}}(-z_1^2,-z_2^2) = \frac{1}{4z_1z_2(z_1+z_2)^2}.
\ee
Because of (\ref{ncu}), the prediction for JT gravity is four times this, which is indeed the answer that follows from (\ref{ZpairJT1}).

To evaluate higher order correlators on the matrix side, we use the recursion relation (\ref{disSuper}). For a double-scaled theory of the type we are studying, it is convenient to write this dispersion relation using a uniformizing coordinate $z = \sqrt{-x}$ as 
\be
R_{g\,\text{M}}(-z^2,I) = \frac{1}{2\pi\mathrm{i}z}\int_{\epsilon - \mathrm{i}\infty}^{\epsilon + \mathrm{i}\infty}\frac{z'^2\mathrm{d}z'}{z'^2-z^2}\frac{F_{g\,\text{M}}(-z'^2,I)}{y_{\text{M}}(-z'^2)}.\label{topRecurF}
\ee
In terms of $z$, the spectral curve (\ref{specCurvebeta2}) is
\be\label{specCURVE}
x = -z^2,\hspace{20pt} y_{\text{M}} = -\frac{\cos(2\pi z)}{\sqrt{2}\,z}.
\ee
The quantity $F_g$ is given in (\ref{firstlineofalpha}). An important feature is that the first term in the definition of $F_g$ vanishes for the ensemble $(\upalpha,\upbeta)=(1,2)$. An argument similar to the one in section \ref{sec:beta2special} then shows that the dispersion relation (\ref{disSuper}) reduces to a sum over residues at the endpoints of the cut in the $x$ plane. (For the double-scaled case (\ref{topRecurF}), this is simply one-half the residue at $z' = 0$.) This implies that the loop equations reduce to ``topological recursion'' for the spectral curve (\ref{specCURVE}).

Now, assuming a correspondence between JT supergravity and matrix integrals, this recursion can be used to predict the volumes of super moduli space. The idea is as follows. In JT supergravity, correlators of partition functions are related to volumes of super moduli space by (\ref{c_n}). For the present case where we do not sum over orientation-reversal, we have $c_n = 1$. Using $R(x) = -\int_0^\infty \mathrm{d}\beta e^{\beta x}Z(\beta)$, and using (\ref{intTrumpet}), we find the JT supergravity formula for the resolvents in terms of the volumes
\be
R_{g\,\text{SJT}}(-z_1^2,\dots,-z_n^2) = (-1)^n\int_0^\infty V_g(b_1,\dots,b_n)\prod_{j = 1}^n b_j\mathrm{d}b_j\frac{e^{-b_j z_j}}{\sqrt{2}\,z_j}.\label{dirExp}
\ee
In this expression, the $V_g(b_1,\dots,b_n)$ are the volumes of moduli space of bordered super Riemann surfaces, with sum over spin structures weighted by $(-1)^\zeta$.\footnote{Alternatively, since we know that the sum without this factor vanishes, we can regard it as twice the sum over even spin structures, or minus two times the sum over odd spin structures.} An inverse expression gives the volumes in terms of the resolvents:
\be
V_{g}(b_1,\dots,b_n) = \left(-1\right)^n \int_{c_0 + \mathrm{i}\mathbb{R}}R_{g\,\text{SJT}}(-z_1^2,\dots,-z_n^2)\prod_{j = 1}^n\frac{\mathrm{d}z_j}{2\pi \mathrm{i}}\frac{\sqrt{2}\,z_j}{b_j}e^{b_j z_j}.\label{invExp}
\ee
Assuming the correspondence between SJT and a matrix integral (\ref{ncu}) and using the matrix intgral recursion (\ref{topRecurF}), one can efficiently compute the $R_{g\,\text{SJT}}$ functions. Then one can compute the inverse Laplace transform (\ref{invExp}) to obtain $V_{g}$. As an example, the following three cases can be computed (in order) using the recursion:
\be
R_{1\,\text{SJT}}(-z_1^2) = \frac{1}{2^{7/2}z_1^3} \hspace{20pt} R_{1\,\text{SJT}}(-z_1^2,-z_2^2) = \frac{1}{2^4z_1^3z_2^3}\hspace{20pt} R_{2\,\text{SJT}}(-z_1^2) = \frac{9}{2^{17/2}z_1^5} + \frac{9\pi^2}{2^{15/2}z_1^3}.
\ee
After using (\ref{invExp}), one finds the corresponding volumes (see footnote \ref{footnotefactorFirst})
\be\label{examples}
V_1(b_1) = -\frac{1}{8}\hspace{20pt} V_1(b_1,b_2) = \frac{1}{8}\hspace{20pt} V_{2}(b_1) = -\frac{3}{2^{9}}(b_1^2 + 12\pi^2).
\ee
In practical terms, this is the most efficient method to compute the volumes. But it is also possible to rewrite the recursion (\ref{topRecurF}) directly in terms of the volumes. This gives a recursion that is structurally identical to the one derived by Mirzakhani for the volumes of ordinary moduli spaces.

To derive this relation, one can substitute the recursion (\ref{topRecurF}) into the RHS of (\ref{invExp}) taking care of the factors of two due to (\ref{ncu}). This gives an expresion for $V_{g}(b_1,\dots,b_n)$ in terms of some integrals of resolvents. Then using (\ref{dirExp}) one rewrites these resolvents in terms of ``lower order'' volumes. The resulting expression contains many integrals, but most of them are Laplace and inverse Laplace transforms that cancel each other. What remains is a small number of $z$ and $b$ integrals. One of the $z$ integrals can be done by trivial contour integration, and the other can be treated using the following function
\be
D(x,y) = \int_{-\mathrm{i}\infty}^{\mathrm{i}\infty} \frac{dz}{2\pi \mathrm{i}} \frac{e^{-xz}}{\cos(2\pi z)}\sinh(y z) = \frac{1}{8\pi}\left(\frac{1}{\cosh\frac{x-y}{4}} -\frac{1}{\cosh\frac{x+y}{4}}\right).
\ee
The final $b$ integrals are explicit in the final form of the recursion relation, which is
\begin{align}
bV_{g}(b,B) &= -\frac{1}{2}\int_0^\infty (b'\mathrm{d}b') (b''\mathrm{d}b'') D(b'+b'',b)\left(V_{g-1}(b',b'',B) + \sum_{\text{stable}}V_{h_1}(b',B_1)V_{h_2}(b'',B_2)\right)\notag\\
&\hspace{20pt} -\sum_{k = 1}^{|B|}\int_0^\infty b'\mathrm{d}b' \Big(D(b'+b_k,b)+D(b'-b_k,b)\Big)V_g(b',B\setminus b_k).\label{recurVols}
\end{align}
In this expression, we are using the notation $B = \{b_1,\dots,b_n\}$. In the second term on the first line, the sum marked ``stable'' is over $h_1+h_2 = g$ and $B_1\cup B_2 = B$, with terms involving the unstable cases $V_0(b_i)$ and $V_0(b_i,b_j)$ excluded.
We can start the recursion with $V_1(b_1)=-1/8$ along with the vanishing of the volumes in genus 0.


The volumes of supermoduli space, like the volumes of moduli space, depend on the normalization of the symplectic form. Rescaling this form will multiply the entire RHS of (\ref{recurVols}) by a constant factor. Formula (\ref{recurVols}) is supposed to be correct with the symplectic form normalized so that the bosonic part of it reduces to the form $\omega = \sum_{i}\mathrm{d}b_i\,\mathrm{d}\tau_i$ in terms of Fenchel-Nielsen (length-twist) coordinates measured in units where the hyperbolic curvature length is one, so the scalar curvature is $R = -2$.

The recursion relation (\ref{recurVols}) has also been obtained in unpublished work by  Norbury, extending  work on the ``cosine'' spectral curve   \cite{Norbury2}   
(and  earlier work on the related ``Bessel''  spectral
curve \cite{Do:2016odu}).    Norbury
has also computed examples such as those of eqn.~(\ref{examples}).

In the case of ordinary moduli space, a recursion relation structurally similar to (\ref{recurVols}) was derived by Mirzakhani \cite{mirzakhani2007simple} using a sum rule for geodesics on hyperbolic surfaces. Shortly afterwards, Eynard and Orantin showed that Laplace transforms of Mirzakhani's volumes satisfy topological recursion  \cite{eynard2007weil}. 
We  have here obtained eqn.~(\ref{recurVols}) purely as a statement about a matrix  ensemble, without reference to super Riemann surfaces.
  However, in appendix \ref{supermirz}, we show how eqn.~(\ref{recurVols}) can be obtained, as a statement
about supermoduli space volumes,  by a supersymmetric analog of
Mirzakhani's approach.   This will establish the connection between super JT gravity and the random matrix ensemble, in the same sense that the relation
between bosonic JT gravity and a matrix ensemble was shown in \cite{Saad:2019lba}.

\subsection{Super JT With Time-Reversal Symmetry}\label{srt}
Next we will study the correspondence between JT supergravity and matrix integrals in the case where we include 
time-reversal symmetry. In JT supergravity, this means that we will gauge orientation-reversal in the bulk. In section \ref{whypin}, 
it was shown that the bulk theory will involve a sum over $\text{pin}^-$ structures, rather than pin${}^+$. In section \ref{SUSY} and 
table \ref{table1}, we described the various matrix ensembles that appear, and the eight corresponding bulk topological field theories 
$e^{-\mathrm{i}\pi \eta N'/2}$. We will now compare what we can between the matrix ensembles and JT theories, starting with genus zero.

\subsubsection{Genus Zero}
\paragraph{One boundary} At genus zero with one boundary, we do not have any prediction from the matrix integral. 
Instead, by matching to super JT gravity, we simply learn what the spectral curve of the matrix integral should be.

To get this right, we have to understand a normalization factor in the relationship between matrix integral quantities and 
JT gravity quantities. One nontrivial factor was noted in (\ref{woggoSYK}) above: in the case with an odd number of fermions, 
the path integral gives $\sqrt{2}$ times the trace in the Hilbert space. The other factor is due to the degeneracy of the levels of the Hamiltonian.
It is convenient to combine these together into a variable $\upgamma$
\be\label{defUpgamma}
\upgamma = \begin{cases} (\text{degeneracy in $H$}) & N \text{ even}\\
\sqrt{2}\cdot(\text{degeneracy in $H$}) & N\text{ odd}\end{cases}
\ee
so that the relationship between the super JT quantities and the matrix quantities is expected to be
\be
R_{\text{SJT}}(I) = \upgamma^{|I|} R_{\text{M}}(\beta).\label{upgammafactor}
\ee
The values of $\upgamma$ for the various cases are given in table \ref{table1}. For the six cases corresponding to $(\upalpha,\upbeta)$ ensembles, it is straightforward to obtain the spectral curve, generalizing the argument that led to (\ref{specCurvebeta2}) for the case $\upgamma = 2$. The density of distinct eigenvalues of $H$ is equal to $1/\upgamma$ times the total density inferred from the JT supergravity path integral, and the spectral curve is therefore
\be
y_{\text{M}}(x) = \frac{1}{\upgamma}y_{\text{SJT}}(x),\hspace{40pt} y_{\text{SJT}}(x) = -\frac{\sqrt{2}\cos(2\pi\sqrt{-x})}{\sqrt{-x}}.\label{matspeccurve}
\ee

For the two $\upbeta$-ensemble cases, the relevant spectral curve is the one related to the density of eigenvalues of $Q$. In (\ref{coshspec}) we determined the density of distinct eigenvalues of $Q$ for the case with $\upgamma = \sqrt{2}$. Repeating the argument for a general value of $\upgamma$, we find
\be
\rho_{0}^{(Q)}(x) = \frac{\sqrt{2}}{\upgamma}\frac{\cosh(2\pi x)}{\pi}.
\ee

\paragraph{Two boundaries} At genus zero with two boundaries, the SJT gravity prediction is
\be
Z_{0\;\text{SJT}}(\beta_1,\beta_2) = 2\cdot 2\int_0^\infty b\, \mathrm{d}b\, Z^T_{\text{SJT}}(\beta_1,b)Z^T_{\text{SJT}}(\beta_2,b) = \frac{4\sqrt{\beta_1\beta_2}}{\pi(\beta_1+\beta_2)}.\label{ZpairJT}
\ee
The factors of two out front are because of the sum over pin${}^-$ structures and the sum over the double trumpet and twisted double trumpet. Using $R(x) = -\int_0^\infty \mathrm{d}\beta e^{\beta x}Z(\beta)$, this implies
\be
R_{0\;\text{SJT}}(-z_1^2,-z_2^2) = \frac{2}{z_1z_2\left(z_1+z_2\right)^2}.\label{r0p}
\ee

We would like to compare this to the random matrix ensembles. For the six $(\upalpha,\upbeta)$ cases in table \ref{table1}, the leading two-resolvent correlator is given in (\ref{abinf}). The $x$ dependence is the same as in (\ref{r0p}), but the prefactor is $\frac{1}{2\upbeta}$ instead of 2. However, to convert from $R_{0\,\text{M}}(x_1,x_2)$ to $R_{0\,\text{SJT}}(x_1,x_2)$, we need to multiply by $\upgamma^2$. For each of the six $(\upalpha,\upbeta)$ cases in the table, one finds that $\gamma^2/(2\upbeta) = 2$, so we find agreement with (\ref{ZpairJT}). 

For the remaining two $\upbeta$-ensemble cases from table \ref{table1}, we need a separate argument. For these cases, $H = Q^2$, where $Q$ is either a $\upbeta = 1$ or $\upbeta = 4$ matrix from a regular $\upbeta$-ensemble. Repeating the steps that led to (\ref{needstwice}) for a general value of $\upbeta$, one finds
\be
R_{0\,\text{M}}(-z_1^2,-z_2^2) = \frac{1}{\upbeta z_1 z_2(z_1+z_2)^2}.
\ee
Multiplying by $\upgamma^2$, we again find agreement with (\ref{r0p}) for the two $\upbeta$-ensemble cases in table \ref{table1}.

\paragraph{More boundaries} As shown in appendix \ref{volsymp}, in super JT gravity, all higher correlators vanish at genus zero, because they are proportional to the volume of a symplectic supermanifold with more fermionic than bosonic coordinates. This is consistent with the matrix integral. For the six $(\upalpha,\upbeta)$ cases, this follows from the discussion in section \ref{sec:uwh}. For the two $\upbeta$-ensemble cases, one can argue as follows. At genus zero, we noted in our discussion of $\upbeta$-ensembles in section \ref{sec:nwm} that the dispersion relation for $R_{0}(x_1,x_2,x_3)$ reduces to a sum of residues at the endpoints of the cut. So we can apply the logic from the end of section \ref{sec:evenodd}: in the double-scaled limit that is applied to $Q$, both endpoints move to $\pm \infty$. Then the residues vanish due to the factor of $\frac{1}{x'-x}$, where $x$ remains finite and $x'$ is evaluated at one of the endpoints. It is easy to check from the recursion that vanishing of $R_{0}(x_1,x_2,x_3)$ implies that all higher $R_0(I)$ vanish.

\subsubsection{Genus One-Half}
At genus one-half, we will get a more detailed check on the correspondence by computing $R_{\frac{1}{2}}(x)$. As before, we start with the six $(\upalpha,\upbeta)$ ensemble cases. The matrix integral formula for $R_{\frac{1}{2}}(x)$ is (\ref{hwy}). Substituting in the spectral curve (\ref{matspeccurve}) and simplifying, one finds
\be
R_{\frac{1}{2}\,\text{M}}(-z^2) = \frac{1}{2z}\int_{0}^\infty \frac{\sqrt{x'}\mathrm{d}x'}{x'+z^2}\left[\left(1-\tfrac{2}{\upbeta}\right)\frac{\tanh(2\pi\sqrt{x'})}{\sqrt{x'}} + \frac{\upalpha - \frac{\upbeta}{2}}{\pi\upbeta x'}\right].\label{toRe}
\ee
As in the bosonic case, let us pretend that we don't know the answer in super JT, and write
\begin{align}
R_{\frac{1}{2}\,\text{SJT}}(x) &= -\int_0^\infty \mathrm{d}\beta e^{\beta x}Z_{\frac{1}{2}\,\text{SJT}}(x)\\
&=-\int_0^\infty \mathrm{d}\beta e^{\beta x}\int_0^\infty \mathrm{d}b\, Z^T_{\text{SJT}}(\beta,b) V_{\frac{1}{2}}(b)
\end{align}
with an undetermined $V_{\frac{1}{2}}(b)$. Evaluating this at $x = -z^2$ and doing the integral over $\beta$ using (\ref{intTrumpet}), we find
\be
R_{\frac{1}{2}\,\text{SJT}}(-z^2) =-\frac{1}{\sqrt{2}\,z}\int_0^\infty \mathrm{d}b\, e^{-bz}V_{\frac{1}{2}}(b).
\ee
Setting $R_{\frac{1}{2}\,\text{SJT}} = \upgamma R_{\frac{1}{2}\,\text{M}}$, and taking the inverse Laplace transform with respect to $z$, we find
\begin{align}
V_{\frac{1}{2}}(b) &= -\frac{\upgamma}{\sqrt{2}}\int\frac{\mathrm{d}z}{2\pi i}e^{bz}\int_0^\infty\frac{\sqrt{x'}\mathrm{d}x'}{x'+z^2}\left[\left(1-\tfrac{2}{\upbeta}\right)\frac{\tanh(2\pi\sqrt{x'})}{\sqrt{x'}} + \frac{\upalpha- \frac{\upbeta}{2}}{\pi\upbeta x'}\right]\\
&=-\frac{\upgamma}{\sqrt{2}}\int_0^\infty \mathrm{d}x'\sin(b\sqrt{x'})\left[\left(1-\tfrac{2}{\upbeta}\right)\frac{\tanh(2\pi\sqrt{x'})}{\sqrt{x'}} + \frac{\upalpha- \frac{\upbeta}{2}}{\pi\upbeta x'}\right]\\
&=-\frac{\upgamma}{\sqrt{2}}\left[\frac{1-\frac{2}{\upbeta}}{2\sinh(\frac{b}{4})}+\frac{\upalpha-\frac{\upbeta}{2}}{\upbeta}\right].\label{comparingTO}
\end{align}
We will compare this to the JT supergravity answer in a moment.

For the two $\upbeta$-ensemble cases, the genus one-half resolvent for $Q$ is (\ref{RQ12}). We can then use $R_{H}(-z^2) = \frac{1}{2iz}(R_Q(iz) - R_Q(-iz))$ to find the resolvent for $H$:\footnote{One has to take care with $\sqrt{\sigma(x)}$ and $\sqrt{\sigma(x')}$. In the limit where both endpoints of the cut go to infinity, the ratio of these factors is one if $x,x'$ are in the same half-plane, and minus one if they are in opposite half-planes.}
\be
R_{\frac{1}{2}\,\text{M}}(-z^2) = \frac{1-\frac{2}{\upbeta}}{z}\int_{-\infty}^\infty \frac{\mathrm{d}x_Q'}{x_Q'^2+z^2}\tanh(2\pi x_Q').
\ee
After changing variables to $x' = x_Q'^2$, we recognize this as two times the expression (\ref{toRe}) with $\upalpha = \frac{\upbeta}{2}$. So, by comparing to (\ref{comparingTO}), the prediction for the volume is
\be
V_{\frac{1}{2}}(b) = -\sqrt{2}\upgamma \frac{1-\frac{2}{\upbeta}}{2\sinh(\frac{b}{4})}.\label{comparintTO2}
\ee

Now, using the values in table \ref{table1}, and applying (\ref{comparingTO}) for the $(\upalpha,\upbeta)$-ensemble cases and (\ref{comparintTO2}) for the $\upbeta$-ensemble cases, one finds that the answer can be written uniformly as
\be
V_{\frac{1}{2}}(b)  = \begin{cases}  \frac{\cos\frac{\pi N'}{4}}{\sinh\frac{b}{4}}-\sin\frac{\pi N'}{4} & \widehat{q} = 1\text{ mod }4\\ \frac{\cos\frac{\pi N'}{4}}{\sinh\frac{b}{4}}+\sin\frac{\pi N'}{4} & \widehat{q} = 3\text{ mod }4.
\end{cases}\label{nwy}
\ee
This agrees with the computation of the torsion in (\ref{komic}), up to the sign that was not fixed in that computation. Note that although a sign was not fixed, it is easy to see from a bulk perspective that the $\widehat{q} = 1$ and $\widehat{q} = 3$ cases should be related by flipping the sign of the second term. This is because the action of $\sT$ on the supercharge (and thus the gravitino) has opposite sign in the two cases, which leads to the sign of the $\eta$ invariant reversing. Since $\eta$ appears in the combination $\exp(-\mathrm{i}\pi\eta N'/2)$, reversing the sign of $\eta$ is equivalent to reversing the sign of $N'$, as in (\ref{nwy}).

Generically, this volume is divergent at $b = 0$ because of the $1/\sinh\frac{b}{4}$ term. This is similar to the divergent $\coth\frac{b}{4}$ term in the purely bosonic case.  So for the same reason one finds that the volume of supermoduli space, weighted by the $\eta$-invariant factor, is divergent. However, for the special cases with $\upbeta = 2$ (which correspond to $(\upalpha,\upbeta) = (0,2)$ and $(2,2)$), the coefficient of the first term vanishes, and we find $V_{\frac{1}{2}}(b) = \pm 1$. For these two cases, the volume is finite and we can proceed to higher orders.

\subsubsection{Higher Orders}\label{sec:higherOrdersZero}
In fact, the matrix integral prediction for these two cases is that all further correlators vanish in the double-scaled limit. In other words, we claim that for $(\upalpha,\upbeta) = (0,2)$ or $(2,2)$, the following is a solution to (\ref{inSuper}):
\begin{align}
\langle R(x)\rangle - \frac{LV'(x)}{2} &= e^{S_0}y(x) - \frac{\upalpha{-}1}{4x} \label{ex1}\\
\langle R(x_1)R(x_2)\rangle_{\text{c}} &= \frac{1}{4\sqrt{-x_1}\sqrt{-x_2}\left(\sqrt{-x_1}+\sqrt{-x_2}\right)^2}\label{ex2}\\
\langle R(I)\rangle_{\text{c}} &= 0 \hspace{20pt} |I|\ge 3.\label{ex3}
\end{align}
Here, we are assuming that the matrix integral is a double-scaled one-cut model, with a single branch point at the origin where $y(x)$ has a $1/\sqrt{-x}$ singularity. We are using a notation appropriate for a double-scaled limit, with both $L$ and $e^{S_0}$ appearing; for the purposes of power-counting, $e^{S_0}$ should be considered of order $L$. 

We can summarize (\ref{ex1}), (\ref{ex2}), (\ref{ex3}) by saying that only terms of order $L$ and order one appear; there are no terms proportional to negative powers of $L$. To check that we have a solution to the loop equations, one can substitute into (\ref{inSuper}) and consider a $1/L$ expansion. The terms at order $L^2$ in (\ref{inSuper}) determine $y(x)$ as a function of the potential (or vice versa). The terms at order $L$ determine the order-one pieces of $\langle R(x)\rangle$ and $\langle R(x_1)R(x_2)\rangle_{\text{c}}$. Since (\ref{ex1}), (\ref{ex2}), (\ref{ex3}) do not contain any negative powers of $L$, all that remains is to check that (\ref{inSuper}) is satisfied at order one. This amounts to checking that the loop equations in the form
\be
2xy(x)R_g(x,I) + xF_g(x,I) \sim 0\label{rewrittenywu}
\ee
are satisfied for the three cases $(g,|I|) \in\{(1,0),(\frac{1}{2},1),(0,2)\}$. We will now check each case.

For $(g,|I|) = (1,0)$, the $F_g(I)$ quantity is
\be
F_1(x) = \frac{\upalpha{-}1}{2 x}R_{\frac{1}{2}}(x) + R_0(x,x) + R_{\frac{1}{2}}(x)^2.
\ee
For $\upalpha \in\{0,2\}$, this vanishes exactly after using $R_{\frac{1}{2}}(x) = -(\upalpha-1)/(4x)$ and also using the formula for $R_0(x,x_1)$ in (\ref{ex2}). This means that (\ref{rewrittenywu}) is indeed satisfied with $R_{1}(x) = 0$. Next, for the case $(g,|I|) = (\frac{1}{2},1)$, we have
\begin{align}
F_{\frac{1}{2}}(x,x_1) &= \frac{\upalpha{-}1}{2x}R_0(x,x_1) + 2\left(R_0(x,x_1) + \frac{1}{2}\frac{1}{(x-x_1)^2}\right)R_{\frac{1}{2}}(x)\\&= -\frac{\upalpha{-}1}{4x(x-x_1)^2}.\label{cnm}
\end{align}
This expression is nonzero, but $x F_g(x,I)$ is analytic in $x$ in a neighborhood of the cut so (\ref{rewrittenywu}) is satisfied. The final case to consider is $(g,|I|) = (0,1)$, which determines $R_{0}(x_1,x_2,x_3)$. This always vanishes in the double-scaled $(\upalpha,\upbeta)$-ensembles, as discussed in section \ref{sec:uwh}.

This matrix integral result, together with the correspondence with JT supergravity, predicts that certain super moduli space volumes are zero. Specifically, the volumes where we sum over pin${}^-$ structures, weighting by $\exp(\mathrm{i}\pi \eta)$ or by $\exp(-\mathrm{i}\pi\eta)$. The first nontrivial case is at genus one with one boundary, where there is apparently a cancellation between the disk with a handle glued in, and the disk with two crosscaps glued in.

In the matrix integral, we expect that although there are no perturbative corrections in powers of $e^{-S_0}$, there will be nonperturbative effects; see appendix \ref{app:nonperturbative}.

\subsection{Ramond Boundaries In The Super Case}\label{sec:RBoundaries}
So far, in our discussion of super JT gravity, we have focused on the case that all boundaries have NS spin structure. From the perspective of random matrix theory, any 
correlator with Ramond spin structures should vanish. For odd $N$, this is because the Ramond path integral vanishes, as in eqn.~(\ref{ZHilbert}) (or is a constant, still with vanishing
correlators,
in the ensembles with $\nu$), and for even $N$ it is because for the models we have discussed, the fermionic and bosonic states have precisely the same energies, 
so $\Tr\;(-1)^\sF e^{-\beta H}$ vanishes.

One would like to use this to learn something about the volumes of moduli spaces of super Riemann surfaces with R boundaries.\footnote{For some anomalies involving these objects,
see appendix \ref{punctureapp}.} However, in super JT gravity, this vanishing can actually be explained in a simpler way, just from the fact that the super-Schwarzian path integral gives zero in the Ramond sector. On the disk this is because no R spin structure is possible. For the trumpet, it is because the Schwarzian path integral vanishes (see eqn.~(\ref{rtrumpet})).

To get around this problem, one can consider instead the quantity $\Tr\left(Q (-1)^\sF e^{-\beta H}\right)$, where $Q$ is the supercharge. Now the trumpet super-Schwarzian path integral is nonzero: to leading order for small fluctuations about a classical solution, the supercharge $Q$ of the super-Schwarzian theory is proportional to $S\eta$, where $S$ is the classical value of the bosonic Schwarzian derivative. $S$ is nonzero for the ``trumpet'' solution, see appendix \ref{app:Sch}, so an insertion of $S\eta$ will soak up the zero mode of $\eta$. Since the trumpet is nonzero, if we can show that $\Tr\left(Q (-1)^\sF e^{-\beta H}\right)$ has to vanish in a given matrix ensemble, this will imply that in the corresponding bulk theory, volumes of super moduli space with Ramond boundaries must vanish.

In fact, for the matrix ensembles with even $N$, it is easy to show that $\Tr\left(Q (-1)^\sF e^{-\beta H}\right)$ vanishes identically. The operator inside the trace takes bosonic states to fermionic ones and vice versa, so it has no diagonal components and zero trace. For odd $N$, the situation is more complicated: in this case by ``$\Tr\left(Q (-1)^\sF e^{-\beta H}\right)$'' we really mean a path integral with periodic boundary conditions and an insertion of $Q$. Such a path integral can be written in operator language in the Hilbert space of dimension $2^{(N-1)/2}$ where we defined the odd $N$ SYK model. One can show that it reduces to $\pm\sqrt{2}\Tr\;Q e^{-\beta H}$, with an ambiguous overall sign. For the cases where $Q$ is drawn from an Altland-Zirnbauer ensemble, this vanishes identically because the eigenvalues of $Q$ come in pairs $\omega,-\omega$. For the case where $Q$ is drawn from a GUE ensemble (odd $N$, no time-reversal), all ``stable'' correlators vanish, because we argued in section \ref{sec:evenodd} that all correlators of resolvents of $Q$ beyond $R_0(x)$ and $R_0(x_1,x_2)$ are zero. This argument extends to correlators of any single-trace quantities (including $\Tr\;Qe^{-\beta H}$), since they can be written in terms of resolvents. This leaves the two cases in which $Q$ is drawn from a GOE-like or GSE-like ensemble. For these cases, all genus-zero contributions will vanish, since GOE and GSE are
equivalent to GUE in genus zero.  But it seems that in general the correlators with Ramond boundaries are nonzero.

Our conclusion is that correspondence with the matrix integral predicts that volumes of super moduli space with 
Ramond boundaries must vanish for all of the bulk theories considered, with the exception of unorientable surfaces in the theories with bulk TFT equal to 1 or $\exp(-4\mathrm{i}\pi\eta/2)$. For these cases, the volumes will be infinite due to the crosscap divergence, so it seems that there are no cases with finite and nonzero volumes with Ramond boundaries.

\subsection{Unbroken Supersymmetry And Ramond Punctures}\label{rpunctures}

Several of the ensembles that we have discussed can be generalized to depend on another parameter.   This happens if the random matrix $C$ is
a bifundamental 
 of a symmetry group $\U(L)\times \U(L)$, $\O(L)\times \O(L)$, or $\Sp(L)\times \Sp(L)$.
  As discussed in section \ref{revpairs}, these ensembles have a generalization in which $C$ is a bifundamental of $\U(L+\nu)\times \U(L)$,  $\O(L+\nu)\times \O(L)$, or $\Sp(L+\nu)\times \Sp(L)$ 
  for some integer $\nu$ (an even integer in the $\Sp$ case).
This corresponds in the ansatz of eqn.~(\ref{blocks}) to a theory with $L+\nu$ bosonic states and $L$ fermionic states.    Hence the value of the supersymmetric index  is $\nu$, and in fact 
\be\label{supindex} \Tr\,(-1)^\sF \exp(-\beta H)=\nu, \ee
independent of $\beta$.
For $\nu\not=0$, supersymmetry is  unbroken; for generic $C$ in this ensemble, there are $|\nu|$ supersymmetric states, of precisely zero energy.

Let us  first consider a bifundamental of $\U(L+\nu)\times  \U(L)$, corresponding to a theory with no $\sT$ symmetry and with even $N$.
By slightly generalizing the derivation that was given at the end of section \ref{measure}, one can show that the eigenvalue measure for such an ensemble takes
the Altland-Zirnbauer form (\ref{loggoApp}) with
\be\label{azmeasure} (\upalpha,\upbeta)=(1+2|\nu|,2). \ee 

How can JT supergravity be modified to accomodate this and to give a model in which supersymmetry is unbroken?    
We claim that the parameter $\nu$ can be incorporated in JT supergravity by including Ramond
punctures.\footnote{A Ramond puncture is  not simply the small $b$ limit of a Ramond hole of circumference $b$, the objects considered in
section \ref{sec:RBoundaries}.  Roughly, a Ramond puncture is obtained by taking $b\to 0$ and discarding a fermionic mode.   See appendix \ref{application}.}    Ideally, we would
include Ramond punctures by adding some sort of spin field to the action of JT supergravity.   As is usual in string theory, it is difficult to find a good framework for doing that,
but there is no problem to incorporate Ramond punctures perturbatively.   The recipe is to sum, in every computation, over the number of Ramond punctures
that are inserted, weighting each such puncture with a factor of $\nu$. Adding a puncture  decreases the Euler characteristic by one, so the total factor associated to such 
an insertion is
\be\label{ramondwt} f=\nu e^{-S_0}. \ee
The factor of $e^{-S_0}$ means that a Ramond puncture has the same effect in the genus expansion as a crosscap.
In each order of the topological expansion, there is a maximum possible number of Ramond punctures.

Let us first see how this prescription reproduces eqn.~(\ref{supindex}).   The lowest order contribution to $\Tr\,(-1)^\sF \exp(-\beta H)$ comes from a disc with a  single
Ramond puncture.  The Schwarzian path integral  in  this  situation is equal simply to 1 (eqn.~(\ref{rpuncture}); note that the factor $e^{-S_0}$ in the Ramond weight
eqn. (\ref{ramondwt}) was included in writing this formula), and the Euler characteristic of the punctured disk is zero, so the contribution to $\Tr\,(-1)^\sF
\exp(-\beta H)$ is simply $\nu$, which is the expected result. 
 In the matrix ensemble, the result $\Tr\,(-1)^\sF \exp(-\beta H)=\nu$ is an exact formula, with no corrections and no fluctuations in the presence of other
operator insertions.  The same is true in JT supergravity because any other topology contributing to $\Tr\,(-1)^\sF \exp(-\beta H)$ can be constructed by gluing a trumpet,
with Ramond spin structure, onto something else.    But the Schwarzian path integral of the trumpet vanishes (eqn.~(\ref{rtrumpet})).

\def\MM{{\mathrm M}}
The NS sector  provides a more elaborate test.   Let us first express the partition function 
$Z_\SJT(\beta)$ of SJT supergravity in terms of the corresponding
partition function $Z_\MM(\beta)$ of the Altland-Zirnbauer ensemble with $(\upalpha,\upbeta)=(1+2|\nu|,2)$:
\be\label{compar} Z_\SJT(\beta)= |\nu| + 2 Z_\MM(\beta). \ee
The additive term $|\nu|$ is the contribution of the zero-energy ground states, and as usual the factor of 2 reflects the degeneracy of the energy levels.
Equivalently, the SJT and matrix model resolvents are related by
\be\label{comres} R_{{\SJT}}(x)=\frac{|\nu|}{x}+2 R_{\MM}(x). \ee
A similar formula holds for multi-resolvents.

$R_{\SJT}(x)$ is supposed to  be computed by  summing over two-manifolds with a single boundary that has NS spin structure.    The number of Ramond
punctures in such a situation will always be even.  Hence only even powers of $\nu$ will appear, and  in particular, there is no contribution linear in $|\nu|$  to $R_{\mathrm{SJT}}(x)$.
At first sight, this appears to contradict eqn.~(\ref{comres}).   But actually, the linear term cancels, as one  can learn by evaluating $R_{\MM}(x)$.   From eqn.~(\ref{nnwy}) or (\ref{hwy}),
one sees that shifting $\alpha$ by $2|\nu|$ will shift $R_{\frac{1}{2},\MM}$ by $-|\nu|/2x$.  Thus
\be\label{rcorr}  \h  R_{\frac{1}{2},\MM}(x) =-\frac{|\nu|}{2x},\ee
where $\h R$ will denote the lowest order $|\nu|$-dependent contribution.   When this is inserted in eqn.~(\ref{comres}), the term linear in $|\nu|$ cancels and a contradiction
with SJT supergravity is avoided.

In SJT supergravity, the lowest order $\nu$-dependent contribution to the resolvent comes from a disc with two Ramond punctures, with $\chi = -1$.   Such a disc can be constructed
by gluing a trumpet to a sphere with one hole and two Ramond punctures.   The moduli space of super Riemann surface structures on such a sphere is a point, of volume 1.
So the contribution of a disc with two punctures to $Z_\SJT(\beta)$ is just
\be\label{lowestcon} e^{-S_0}\frac{\nu^2}{2}\int_0^\infty b \d b \,Z^T_\SJT(\beta,b)= \nu^2 e^{-S_0}\sqrt{\frac{\beta}{2\pi}}  . \ee
We included a factor of $\nu$ for each Ramond puncture, and a factor of $1/2$ because a $\pi$ rotation of the circle on which gluing occurs will exchange the two
Ramond punctures.  We also  used (\ref{nstrumpet}) for $Z^T_\SJT$.   Integrating over  $\beta$ to convert this  to a resolvent, we get\footnote{There is no
factor $e^{-S_0}$ here merely  because the coefficients $R_{g,\SJT}(x)$ in the genus expansion are defined with powers of $e^{S_0}$ removed.}
\be\label{lowestsjt}\h R_{1,\SJT}(x) =-\frac{\nu^2 }{2\sqrt 2 (-x)^{3/2}}. \ee

On the matrix model side, the $\nu^2$ term in $F_1$ is, from eqn.~(\ref{firstlineofalpha}), 
\be\label{expsecond} \h F_1  = \frac{|\nu|}{x}\h R_{\frac{1}{2},\MM}(x) +\h R_{\frac{1}{2},\MM}(x)^2=-\frac{\nu^2}{4x^2}. \ee
Inserting this in the dispersion relation (\ref{disSuper}) and  using $y_M=-\cos(2\pi \sqrt{-x})/\sqrt{-2x}$, 
we get $\h R_{1,\MM}=-\nu^2/4\sqrt 2(-x)^{3/2}$.   When this is inserted in eqn.~(\ref{comres}), the result is in accord with
the supergravity result (\ref{lowestsjt}).   

This discussion applies to the matrix ensemble appropriate for the theory  in which spin structures are weighted with a  factor $(-1)^\zeta$.     For the theory
without $(-1)^\zeta$, one uses a GUE ensemble, and there is no analog of the parameter $\nu$.   In  appendix \ref{punctureapp}, we explain from a bulk point of view why Ramond
punctures are only possible in the theory with $(-1)^\zeta$.   

Now let us consider the ensembles with symmetry group $\O(L+\nu)\times \O(L)$ or $\Sp(L+\nu)\times \Sp(L)$. These are appropriate to a $\sT$-invariant  theory with
$N$ congruent to 0 or 4 mod 8. A similar derivation to the above still applies, with some
modifications.   First of all, a bifundamental of $\O(L+\nu)\times \O(L)$ or $\Sp(L+\nu)\times \Sp(L)$ corresponds, respectively, to an Altland-Zirnbauer ensemble
with $(\upalpha,\upbeta)= (|\nu|,1)$ or $(3+2|\nu|,4)$.  In each case, the $\nu$-dependent shift in $\upalpha$ is $\upalpha\to\upalpha +
2\upbeta |\nu|/\upgamma$. Weighting each Ramond puncture by $\nu$,  the Ramond sector is as before: a disc  with one Ramond puncture
makes the desired contribution $\nu$ to $\Tr\,(-1)^\sF$.

In the NS sector, we now have to use
\be\label{relnres} R_\SJT(x)=\frac{|\nu|}{x}+\gamma R_\MM(x). \ee
A similar calculation to the previous one shows that $\h R_{\frac{1}{2},\MM}=-|\nu|/\gamma x$, as a result of which $R_{\frac{1}{2},\SJT}$ does not depend on $\nu$.   
This is as expected, since in the NS sector it is not possible to have a single Ramond puncture.   For the  same reason, we expect there to be no contribution linear in $|\nu|$
in  $R_{1,\SJT}$.  This depends on a cancellation between various terms in $F_{1,\MM}$ (eqn.~(\ref{firstlineofalpha})) and the identity
$(1-2/\upbeta)-(\upalpha-1)/\upbeta=0$, which holds at $(\upalpha,\upbeta)=(0,1)$ or $(3,4)$.    In order $\nu^2$, we get $F_{1,\MM}=-\nu^2/\gamma^2 x^2$.   This differs
from eqn.~(\ref{expsecond}) by a factor of $(2/\gamma)^2$.     When we insert $F_{1,\MM}$ in the dispersion  relation  to compute $\h  R_{1,\MM}$, we get a factor $\gamma/2$
relative to the previous calculation because for these ensembles, there is an extra factor $2/\gamma$ in  $y_\MM$.    Then in $\h R_{1,\SJT}$, there is another factor of $\gamma/2$
relative to the previous calculation because the 2 on the right hand side of eqn.~(\ref{compar}) is replaced by $\gamma$ in eqn.~(\ref{relnres}).
The upshot is that $\h R_{1,\SJT}$ is  the same as it was for the $\U(L+\nu)\times \U(L)$ ensemble.   The corresponding SJT computation is also the same as before (since it is not
affected by time-reversal symmetry), so again we get agreement between random matrix theory and supergravity.

The analysis in appendix \ref{punctureapp} shows that in the bulk, Ramond punctures are only possible for even $N$.   We expect that a more precise analysis
of the anomaly that was involved will show in the unorientable case that $N$ must be congruent to 0 or 4 mod 8, not to 2 or 6.

\section*{Acknowledgements}

 DS was supported in part by Simons Foundation grant 385600.
Research of EW is partially supported by NSF Grant PHY-1606531.  We are grateful to R.~Dijkgraaf, N.~Seiberg and S.~Shenker for discussions. EW thanks R.~Penner for discussions about hyperbolic geometry and super Riemann surfaces and T.~Voronov for pointing out some references on supergeometry.
   We also thank P. Norbury for discussions of his work.

\appendix
\section{Volumes Of Symplectic Supermanifolds}\label{volsymp}  \subsection{Preliminaries}
 Our goal in this appendix is to obtain a formula in purely bosonic terms for the volume of a symplectic supermanifold -- such as
 the moduli space of super Riemann surfaces.  This will make clear some basic properties of those volumes, which we can compare to random matrix theory.
 It will also enable
 us to make contact with  formulas of Norbury \cite{Norbury2}, who studied the spectral curve $y=\frac{1}{z}\cos 2\pi z$ (where $z=\sqrt{-x}$) from a different starting
 point than that of the present paper.
 
 In the bosonic world, a symplectic manifold is a manifold $M$ of dimension $n=2d$ endowed with a closed two-form $\omega$ that is everywhere nondegenerate.
 Nondegeneracy means that the volume form
 \be\label{woggoApp} \vol_M =\frac{\omega^d}{d!} \ee
 is everywhere nonzero.   The volume of $M$ is $\Vol(M)=\int_M \vol_M$, and is manifestly positive if $M$ is oriented appropriately  (nondegeneracy of $\omega$ implies
 that $M$ is orientable).
An equivalent formula for the volume is
 \be\label{noggo}\Vol(M)=\int_M e^\omega.\ee
 To justify this, one expands $e^\omega$ in powers of $\omega$, and observes that only the term of degree $d$ contributes.
 
 In local coordinates $u^1,\cdots, u^n$, the symplectic form is explicitly $\omega=\frac{1}{2}\sum_{i,j}w_{ij}(u^K)\d u^i \d u^j$, where the coefficient matrix $w_{ij}$ is antisymmetric.
 An alternative formula for the volume form is 
 \be\label{noggoo}\vol_M=\Pf(w) \d u^1\d u^2\cdots \d u^n, \ee
 where $\Pf$ is the Pfaffian.
 
 Now consider a smooth supermanifold $\hM$ of dimension $n|m$, where $n=2d$ is even.   Suppose that $\hM$ can be described by local bosonic coordinates $u^i$,
 $i=1,\cdots, n$ and local fermionic coordinates $\theta^a$, $a=1,\cdots, m$.
The one-forms $\d u^i$ are treated as fermionic variables and the one-forms $\d \theta^a$ are bosonic.\footnote{In  supergeometry, one has to decide whether the exterior
derivative $\d$ commutes or anticommutes with odd constants. The two approaches are equivalent in the sense  that results in one can be reformulated in the other;
for a precise  account, see pp. 62-4 of \cite{DM}.   However, which approach is  more convenient depends on the context.   In gauge theory
on an ordinary manifold with a supergroup as gauge group -- which was the context in our computation of the torsion in section \ref{osp} -- it is more natural to consider
the exterior derivative
 to commute with odd constants.   Symplectic geometry on a  supermanifold is more straightforward if the exterior derivative is considered to anticommute with odd constants,
as we assume here.}   A two-form $\hw$ is homogeneous and quadratic in these
variables.   In local coordinates
\be\label{hurz} \hw =\frac{1}{2} \sum_{i,j}w_{ij} \d u^i \d u^j+ \sum_{i,a}w_{i a }\d u^i \d \theta^a +\frac{1}{2}\sum_{a,b}w_{ab}\d \theta^a\d\theta^b, \ee
where $w_{ij}=-w_{ji}$ but $w_{ab}=+w_{ba}$.   If we write generically $X^I$ for the full set of bosonic and fermionic coordinates $u^i|\theta^a$, and regard
$w_{ij}$, $w_{ia}$, and $w_{ab}$ as blocks of a supermatrix $W$, then we can write simply \be\label{lurz} \hw =\frac{1}{2}\sum_{I,J}W_{IJ}\d X^I \d X^J. \ee
  The closed two-form $\hw$ is said to be nondegenerate if the Berezinian $\Ber\, W$ is everywhere defined and not nilpotent.\footnote{The Berezinian
of a supermatrix $W= \newcommand*{\temp}{\multicolumn{1}{r|}{}}
\left(\begin{array}{cccccc}
A\negthinspace \negthinspace\negthinspace \negthinspace \negthinspace&\temp &B\\
 \cline{1-3}
C\negthinspace \negthinspace\negthinspace\negthinspace \negthinspace&\temp &D\\

\end{array}\right)$ is infinite, or undefined, if $\det D=0$, and it is nilpotent if $\det A=0$. So nondegeneracy of $\hw$ means that the matrices $w_{ij}$ and $w_{ab}$,
which correspond to $A$ and $D$,
both have nonzero determinant.  Note that the square root in eqn.~(\ref{urz}) below does not make sense if $\Ber\,W$ is nilpotent.   Of course, that formula also does not
make sense if $\Ber\,W$ is infinite or undefined. }   In that case, $\hM$ is called a symplectic supermanifold with symplectic form $\hw$.

On a supermanifold, we cannot imitate the definition (\ref{woggoApp}) of the volume form, because a volume form on a supermanifold is not a differential form
(basically because $\d\theta$ regarded as a measure transforms oppositely to $\d\theta$ regarded as a one-form).  There is also no analog of the Pfaffian for a supermatrix.
However, we can use the Berezinian,  and write
\be\label{urz}\vol_\hM = 
 \frac{\sqrt{\Ber\, W}}{(2\pi)^{m/2}}\, \,[\d u^1\d u^2\cdots \d u^n|\d\theta^1\d\theta^2\cdots \d\theta^m]. \ee (The factor of $1/(2\pi)^{m/2}$ will
avoid factors of $2\pi$ in the final results.)  This formula is independent of the local coordinates
used -- except possibly for an overall sign, coming from the choice of sign of $\sqrt{\Ber\, W}$ and the ordering of the factors in $ [\d u^1\d u^2\cdots \d u^n|\d\theta^1\d\theta^2\cdots \d\theta^m]$.
There can be a topological obstruction  to choosing this sign consistently.
When this obstruction vanishes, after picking the sign, one defines the volume of $\hM$ as $\Vol(\hM)=\int_\hM \vol_\hM$.  In contrast to the case
of an ordinary symplectic manifold, this volume has no positivity properties.   When the sign cannot be defined consistently, one may leave
the volume of $\hM$ undefined or say that it is zero.   

We used an arbitrary system of local coordinates in defining $\vol_\hM$.   In general,  we can cover $\hM$ by small open sets $\cO_\alpha$, such that in  each 
$\cO_\alpha$, we pick local bosonic and fermionic coordinates $u^i_\alpha$, $\theta^a_\alpha$.  In an intersection of open sets $\cO_\alpha\cap\cO_\beta$, the local coordinates
$u^i_\alpha|\theta^a_\alpha$ are functions of $u^i_\beta|\theta^i_\beta$.   One says that the supermanifold $\hM$ is ``split'' if the bosonic coordinates $u^i_\alpha$
are functions of $u^j_\beta$ only (and not of $\theta^a_\beta$) while the $\theta^a_\alpha$ are linear in the $\theta^a_\beta$:
\begin{align}\label{polkoApp} u^i_\alpha& = f^i_{\alpha\beta}(u^k_\beta) \cr \theta^a_\alpha&= \sum_b g^a_{b\,\alpha\beta}(u^k_\beta)\theta^b_\beta. \end{align}
In this situation, the $u^i_\alpha$ are local coordinates for a purely bosonic manifold $M$, and the $\theta^a_\alpha$ parametrize the fibers of a vector bundle $V\to M$.
More geometrically, instead of talking about local coordinates, one can say that $\hM$ is the total space of the vector bundle $V\to M$, which has purely fermionic
fibers.\footnote{This is sometimes described by saying that $\hM$ is the total space of $\Pi V\to M$, where the symbol $\Pi$ means that the fibers are fermionic.  We
will just write $V$ rather than $\Pi V$, but we will understand that $V$ is fermionic, that is the fibers of $V\to M$ are parametrized by fermionic coordinates.} 

The fundamental structure theorem of smooth supermanifolds says that every smooth supermanifold can be split.\footnote{The proof proceeds roughly by starting
with any systems of local coordinates $\u^i_\alpha|\theta^a_\alpha$, and then trying to improve the coordinates, by adding terms quadratic and higher
order in the fermionic variables, so as to put the transition functions in the desired form of eqn.~(\ref{polkoApp}).   The possible obstructions that one encounters are cohomology
classes of a smooth manifold with values in a sheaf of smooth functions.  Such cohomology vanishes, so $\hM$ can be split.   The splitting is far from unique, but the
topology of $M$, the cohomology class of $\omega$, and the topological type of the vector bundle $V\to M$ are uniquely determined.  (This is most directly proved
as follows, without  any discussion of coordinates or splitting.  $M$ can be defined as the reduced space of $\hM$, obtained by reducing $\hM$ modulo the odd coordinates.
As such, $M$ is naturally embedded in $\hM$.  $V$ is the normal bundle to $M$ in $\hM$ and $\omega$ is the restriction of $\hw$ to $M$.)}
Thus in analyzing the volume of a smooth symplectic supermanifold $\hM$,
we can assume that $\hM$ is the total space of some $V\to M$.   In this situation, the symplectic form $\hw$ of $\hM$, when restricted to $M$ (by setting $\theta^a=\d \theta^a=0$)
becomes a symplectic form $\omega$ on $M$, and therefore $M$ is an ordinary symplectic manifold.   We will see that the volume of $\hM$ 
depends only on the cohomology class of the symplectic
form $\omega$ of $M$ and the topology of the vector bundle $V\to M$.   

Restricted to $\theta^a=0$, but without setting $\d\theta^a=0$, the symplectic form $\hw$ of $\hM$ takes the form
\be\label{wofo}\hw=\frac{1}{2}\left(\sum_{i,j} w_{ij}(u^k)\d u^i\d u^j+w_{ab}(u^k)\d \theta^a\d\theta^b\right). \ee
Nondegeneracy of $\hw$ means that the matrices $w_{ij}$ and $w_{ab}$ are both everywhere invertible.     In particular, since $w_{ab}$ is symmetric and everywhere
nondegenerate, we can view it as a metric tensor on the vector bundle $V\to M$.    This reduces the structure group of $V$ to 
an orthogonal group $\OO(m)$ or possibly (if $w$ is not positive- or negative-definite) to an indefinite orthogonal group $\OO(m_1,m_2)$ where $m_1+m_2=m$.   We will continue the derivation assuming that
$w_{ab}$ is negative-definite, as is the case\footnote{The negative-definiteness is visible in Omnibus Theorems B and C of \cite{Pennerone}.   The symplectic form $\h\omega$
is described explicitly  (in a coordinate system different from that used in the present paper) in Omnibus Theorem B.   Its fermionic part is negative-definite, and comparison
with Omnibus Theorem C shows that it has been defined so that its bosonic part is the ordinary Weil-Petersson symplectic form  $\omega$  on the bosonic moduli space.   The
negative-definite nature of the  fermionic part  of the symplectic form is also visible in eqn.~(\ref{sympW})  of the present paper, 
where the bosonic part of the symplectic form related  to the
super-Schwarzian was described in a way that is consistent with the standard Weil-Petersson  symplectic form, and  the fermionic part is negative-definite.} for the moduli space of super Riemann surfaces; the generalization does not change much and will be briefly explained at the end.  
 With the nondegenerate
metric $w_{ab}$ at hand, we can restrict ourselves to orthonormal systems of local fermionic coordinates, so we can take henceforth $w_{ab}=-\delta_{ab}$.   

There is actually no problem with the sign of $\sqrt{\Ber\, W}$ if $w_{ab}$ is everywhere invertible.   However, a sign change of one of the $\theta^a$ will
reverse the sign of $[\d u^1\d u^2\cdots \d u^n|\d\theta^1\d\theta^2\cdots \d\theta^m]$ and thus change the sign $\vol_\hM$.    So the overall sign of $\vol_\hM$
is reversed if we reverse the orientation of the bundle $V\to M$.   
Thus the case that the sign of the volume form can be globally defined is the case that  $V$ is orientable, or in other words, that its structure group 
 can be further reduced from $\OO(m)$ to $\SO(m)$.    If $V$ is orientable, then the sign of the volume form depends on a choice of orientation of $V$; this is an inescapable
 choice that must be made to define the volume of a symplectic supermanifold.

\subsection{Computation}\label{volcump}

Now we will make use of the projection $\pi:\hM\to M$ that forgets the odd coordinates to put the symplectic structure in a convenient form (first described in
\cite{Rothstein}).  
The purely bosonic symplectic form   $\omega=\frac{1}{2}\sum_{i,j} w_{ij}(u^k)\d u^i\d u^j$ can be ``pulled back'' to a closed, but degenerate, two-form on $\hM$
that we will call $\pi^*\omega$.    Because the fibers of $\pi$ are contractible (being purely fermionic), the cohomology of $M$ is naturally
isomorphic to the cohomology of $\hM$.   The full symplectic form $\hw$ is therefore equivalent cohomologically to $\pi^*\omega$.   Concretely, this means
that there is a 1-form $\lambda$ such that
\be\label{foggo}\hw=\pi^*\omega+\d\lambda. \ee

On an ordinary symplectic manifold  $M$, adding an exact form to the symplectic form does not change the volume of $M$.   On a symplectic supermanifold,
the same statement holds, as long as the nondegeneracy condition that was needed to define the volume form is maintained.   Therefore, as long as the nondegeneracy
condition is maintained, the volume of $\hM$ will not depend on the one-form $\lambda$.   Because we do need to maintain this nondegeneracy condition, we cannot 
simply set $\lambda=0$.   But we can choose a convenient $\lambda$ to simplify the computation of $\Vol(\hM)$.  

A convenient $\lambda $   may be chosen as follows.  Pick a gauge connection $A$ on $V$, with structure group $\SO(m)$.    Explicitly, $A=\sum_i A_{i\,ab}(u^k)\d u^i$,
where $A_{i\,ab}=-A_{i\,ba}$.  The one-forms $\d\theta^a+A^a_{i\,b}  \d u_i\theta^b$ (where indices are raised and lowered using the metric $w_{ab}=-\delta_{ab}$)
are gauge-covariant, so 
\be\label{gc}\lambda=-\frac{1}{2}\sum_a \theta^a (\d\theta^a+A^a_{i\,b}  \d u^i \theta^b) \ee
is gauge-invariant and globally defined.
Explicitly, with this choice of $\lambda$,
\be\label{noggot} \hw=\pi^*\omega -\frac{1}{2}\sum_a \d \theta^a\d\theta^a +\frac{1}{2}\sum_{a,b} \theta^a\theta^b \partial_i A_{j\,ab}\d u^i \d u^j -\theta^a A_{i\,ab}\d u^i\d\theta^b .\ee

The volume of  a supermanifold with a symplectic structure of this form can be computed rather simply.   It is possible to do this directly from eqn.~(\ref{urz}),
but a particularly transparent approach is as follows.
First we reconsider an ordinary symplectic manifold $M$.   We view the symplectic form $\omega=\frac{1}{2}w_{ij}(u^k)\d u^i\d u^k$ as a function of bosonic
variables $u^k$ and fermionic variables $\d u^k$.   Together these variables parametrize a supermanifold $M'$ which is a fiber bundle over $M$ with fermionic
fibers (parametrized by the $\d u^k$).   Though there is no natural measure on $M$, there is a natural measure on $M'$ (the integration measures for integrating
over the bosonic variable $u^k$ and the fermionic variable $\d u^k$ transform oppositely under a change of coordinates, so their product does not depend on a choice  of
coordinates).  Let us call this measure $\D(u^k,\d u^k)$.
The volume of $M$ is then
\be\label{polo}\Vol(M)=\int\D(u^k,\d u^k)\exp\left(\frac{1}{2}w_{ij}(u^k)\d u^i\d u^j\right). \ee
After thinking through the implications of Berezin integration over the odd variables $\d u^j$    (the usual definition being $\int \D(\d u^j)\cdot 1=0$, $\int\D(\d u^j)\,\d u^j=1$ for each $j$),
the reader can hopefully see that eqn.~(\ref{polo}) is completely equivalent to the definition of the volume in eqn.~(\ref{noggo}).   The same approach holds in the case
of a supermanifold.   If $\hat M$ is a supermanifold with bosonic and fermionic coordinates $X^I$, then one introduces for each $I$ a new variable $\d X^I$
with statistics opposite to those of $X^I$.     Again the $X^I$ and $\d X^I$ parametrize a supermanifold $\h M'$, and this supermanifold has a natural measure
$\D(X^I,\d X^I)$.   Our previous formula for $\Vol(\hM)$ is equivalent to
\be\label{holo}\Vol(\hM)= \int \frac{\D(X^I,\d X^I)}{(2\pi)^m}\exp\left(\frac{1}{2}\sum_{I,J}W_{IJ}(X^K)\d X^I \d X^J\right).\ee
The Gaussian integral over the $\d X^K$ generates the factor $\sqrt{\Ber\,W}$ in the definition of the volume form in eqn.~(\ref{urz}).

The formula (\ref{holo}) is particularly convenient for a symplectic form with the structure in eqn.~(\ref{noggot}), because the Gaussian 
integral over $\d\theta^a$ gives a simple result, after which the integral over the $\theta^a$ is also simple.   We have
explicitly
\begin{align}\label{olo}\Vol(\hM)=\int_{\hM'} \frac{\D(u^i,\d u^i)\D(\theta^a,\d \theta^a)}{(2\pi)^m}\exp& \left(\frac{1}{2}\omega_{ij}(u^k)\d u^i\d u^j -\frac{1}{2}\sum_a \d \theta^a\d\theta^a
\right.\cr&\left. +\frac{1}{2}\sum_{a,b} \theta^a\theta^b \partial_i A_{j\,ab}\d u^i \d u^j -\theta^a A_{i\,ab}\d u^i\d\theta^b\right). \end{align}
The Gaussian integral over the $\d\theta^a$ gives a simple result:
\be\label{wilgo}\Vol(\hM)=\int\frac{\D(u^i,\d u^i)\D(\theta^a)}{(2\pi)^{m/2}}\exp\left(\frac{1}{2}\omega_{ij}(u^k)\d u^i\d u^j +\frac{1}{4}\sum_{a,b}\theta^a\theta^b F_{ij\,ab}\d u^i
\d u^j\right).    \ee
Here $F_{ij\,ab}=\partial_i A_{j\,ab}-\partial_j A_{i\,ab}+[A_i,A_j]_{ab}$ are the components of the curvature two-form $F=\d A+[A,A]$.   The integral over
the $\theta^a$ gives a gauge-invariant polynomial in $F$ of degree $m/2$.    Because of the fermionic nature of the $\theta^a$ and the $\d u^i$, this polynomial is proportional to
what one gets
from the product $F_{i_1i_2\,a_1a_2}F_{i_3i_4\,a_3a_4}\cdots F_{i_{m-1}i_m\,a_{m-1}a_m}$ of $m/2$ copies of $F$ by antisymmetrizing in all indices
$i_1i_2\cdots i_m$ and in all indices $a_1a_2\cdots a_m$.  Taking into account the explicit $1/(2\pi)^{m/2}$ in eqn.~(\ref{wilgo}), this operation builds the Euler class $\chi(V)$,
understood as a function of the $u^a$ and $\d u^a$ that is homogeneous of degree $m$
in the $\d u^a$.   The only other factor in eqn.~(\ref{wilgo}) is $\exp\left(\frac{1}{2}\omega_{ij}\d u^i\d u^j\right)$.  So
\be\label{nilgo}\Vol(\hM)=\int_{\hM'}\D(u^i,\d u^i)\exp\left(\frac{1}{2}\omega_{ij}\d u^i\d u^j\right) \cdot \chi(V). \ee
A more conventional way to write the same formula is
\be\label{gilfo}\Vol(\hM)=\int_M \exp(\omega)\cdot  \chi(V). \ee

Thus we have achieved our goal of expressing the volume of a symplectic supermanifold $\h M$ purely in terms of bosonic geometry.
The result can be  compared to  explicit computations in examples, such as those in \cite{Voronov}.

\subsection{Application}\label{application}

In the main example of the present paper, $\hM$ is the moduli space of super Riemann surfaces, 
and $M$ is the corresponding reduced space, which is the moduli
space of Riemann surfaces with spin structure.  In studying this example, it is useful to be able to give $V$  a complex structure.
For example, in the case of a Riemann surface $\Sigma$ without boundary, $M$ is a complex manifold (the moduli space that parametrizes
a Riemann surface $\Sigma$ with a choice of 
 a square root $K^{1/2}$ of its canonical bundle $K$).   The fiber of the bundle $V$
is $H^1(\Sigma,K^{-1/2})$ (which parametrizes the fermionic directions  in  $\hM$).  So in particular in  this case, $V$ is a holomorphic vector bundle, not just a complex
vector bundle.    If $\Sigma$ has boundaries, then $M$ is no longer a complex manifold so it does not make sense to say that $V$ is holomorphic.  
But for boundaries of NS type, one can still show that $V$ can be given the structure of a complex vector bundle.  The explanation of this is somewhat subtle
and will be given below, along with an explanation that $\chi(V)=0$ if there are boundaries of Ramond type.  Thus  in our main example, we can always assume
that $V$ is a complex vector bundle.   
  A vector bundle with a complex structure acquires a natural orientation, so the complex structure of $V$ gives us a way to define
the sign of $\chi(V)$.  
In fact, the  class that is naturally called $\chi(V)$ if $V$ is viewed as a real vector bundle of some rank  $m$ is equivalently the top Chern class $c_{m/2}(V)$ if $V$ is
viewed as a complex vector bundle of rank $m/2$.  

This gives a succinct way to orient the bundles $V$ for all $\Sigma$ in a uniform fashion, and moreover in a way that behaves well under gluing.  But it is important
to point out that a  choice is involved.  
If $J$ is a complex structure on $V$, then $-J$ is  an equally good complex structure.  For example, if one choice identifies $V$ with  $H^1(\Sigma,K^{-1/2})$,
the other choice will identify $V$ with the dual complex vector bundle, which is $H^0(\Sigma,K^{3/2})$.   
   Reversing the complex structure of $V$  will multiply the orientation of $V$
by $(-1)^{m/2}$, where $m$ is the rank of  $V$ (the odd dimension of the symplectic supermanifold $\h M$).   In our problem of genus $g$ super Riemann surfaces with $s$ boundaries,
$m=4g-4+2s$ so $(-1)^{m/2}=(-1)^s$.     Purely from the point of view of symplectic supergeometry, two formalisms differing by a factor of $(-1)^s$ are equally
natural.   In this paper, we resolve this ambiguity to make the disc partition function positive (as 
one expects from reflection positivity of supergravity).   As a disc has $s=1$, the sign of the  disc amplitude does resolve the ambiguity.  The sign of the measure
of a three-holed sphere  is also sensitive to the same choice.   We only fully fix this sign in appendix \ref{finalcal}, where we fix the meaning  of the symbol $[\d \xi\,\d\psi]$
by using $\int [\d\xi\,\d\psi]\,\xi\psi=+1$ in the derivation of eqn.~(\ref{glos}).   The choice amounted to a choice of orientation of the bundle $V$, for the case of a three-holed
sphere.   Once this choice is made, the measure for all other cases is determined by gluing (eqn.~(\ref{torformt})).   Our choices for the disc and the three-holed sphere
are compatible; this is visible in the fact that the results deduced in appendix \ref{supermirz}, both for the recursion relation and for the input $V_1(b)$ of the recursion
relation, agree with the results of section \ref{superjtnot}.

As explained in footnote \ref{supermeasure} of section \ref{revisited}, we have defined the metric on the  $\osp(1|2)$  Lie algebra
 so that the symplectic form $\hat\omega$ on $\hM$ restricts
on $M$ to the standard Weil-Petersson symplectic form  $\omega$.   On the other hand, the ``triviality'' of the torsion on an orientable two-manifold means that
the measure $\tau$ on $\hM$ that comes from the torsion  is  $\tau=\sqrt{\Ber\,\hat\omega}$.  To  get the natural formula (\ref{gilfo}) with no factors of $2\pi$, we started in eqn.~(\ref{urz})
not with $\sqrt{\Ber\,\hat\omega}$ but with  $\sqrt{\Ber\,\hat\omega}/(2\pi)^{m/2}$, where $m$ is the odd dimension of $\hM$.     In our application to super
Riemann surfaces, $m=-2\chi(\Sigma)$.  So a measure $\mu$ that leads to  the natural
normalization (\ref{gilfo}) of the volumes  is actually not $\tau$ but
\be\label{natnorm} \mu= (2\pi)^\chi\tau.  \ee

The formula (\ref{gilfo}) has a number of interesting corollaries. If $m$ is odd, then  $\chi(V)$ vanishes trivially at least modulo torsion
(as it is supposed to be a polynomial of degree $m/2$ in $F$), so  a symplectic supermanifold with an odd fermionic
dimension has zero volume.    If the rank of the vector bundle $V$ exceeds the dimension of $M$, then the class $\chi(V)$ vanishes
for dimensional reasons, so a supermanifold of dimension $n|m$ with $m>n$  has zero volume.  In general, as $\chi(V)$ is a class of degree $m$,
 the volume $\Vol(\hM)$ is proportional to  $\omega^{(n-m)/2}$.

These statements have interesting implications for our problem of hyperbolic Riemann surfaces with boundary, in which $\hM$ has dimension $6g-6+2s|4g-4+2s$.   In particular, if $g=0$, this is of the form $n|m$ with $n<m$, so the volumes
vanish.   Now let us discuss what happens for $g>0$.  The 
topological  type  of  the vector  bundle $V$ and the integral cohomology  
class $\chi(V)$ are determined  by discrete data, so they are independent
of the $b_i$.  Hence the dependence of $\Vol(\hM)$ on the $b_i$ comes entirely from the dependence of $\omega$ on the $b_i$.
Mirzakhani showed that the cohomology class of $\omega$ is a linear function of the variables $b_i^2$.   (Her analysis was phrased for bosonic
Riemann surfaces, but is not significantly affected by incorporating a spin structure.)      With $\hM$ having dimension $n|m=6g-6+2s|4g-4+2s$, its volume
is a polynomial in $\omega$ of degree $(n-m)/2=g-1$.    So $\Vol(\hM)$ will be a polynomial in the $b_i^2$ of that degree.    We do not have a general
proof of this behavior from the matrix model, but the examples in eqn.~(\ref{examples}) do have the claimed property.\footnote{This property has been proved
by Norbury in unpublished work, using the recursion  relation (\ref{recurVols}), which he discovered independently of the present paper, by extending the topological
recursion constructed in \cite{Norbury2}.}

Now let us explain why $V$ can be given a complex structure in the case of NS boundaries, and why $\chi(V)$ vanishes when there are Ramond boundaries.
A hyperbolic conjugacy class in $\OSp(1|2)$ with spin structure of NS type contains  an  element
\be\label{typical} \newcommand*{\temp}{\multicolumn{1}{r|}{}}
U_b= \left(\begin{array}{cccccc}
e^{b/2} &1  \negthinspace\negthinspace\negthinspace &\temp & 0\\
0 &e^{-b/2}\negthinspace \negthinspace \negthinspace&\temp &0\\
 \cline{1-4}
0 &0\negthinspace \negthinspace \negthinspace&\temp &-1\\
\end{array}\right).\end{equation}
where $b>0$ is a length parameter.   For any positive $b$,  the matrix element of $U_b$ above the diagonal could  be removed by conjugation.   The reason for including
it is that this  ensures that the hyperbolic conjugacy class that  contains  $U_b$ has  a smooth limit for  $b\to 0$.   In that limit, $U_b$ goes over to
\be\label{parabolic} \newcommand*{\temp}{\multicolumn{1}{r|}{}}
U_0= \left(\begin{array}{cccccc}
1 &1  \negthinspace\negthinspace\negthinspace &\temp & 0\\
0 &1\negthinspace \negthinspace \negthinspace&\temp &0\\
 \cline{1-4}
0 &0\negthinspace \negthinspace \negthinspace&\temp &-1\\
\end{array}\right),\end{equation} which is a parabolic element of $\OSp(1|2)$.  
  The  conjugacy class of $\OSp(1|2)$ containing $U_0$ has  the same dimension, namely $2|2$,
 as the hyperbolic  conjugacy class  that  contains $U_b$, and the  conjugacy
class containing $U_b$  goes over smoothly for  $b\to 0$ to the  conjugacy  class containing $U_0$.
$U_0$  is the monodromy of a flat $\OSp(1|2)$ connection  around an NS puncture (not hole). 
So in the NS case, a hole of length $b$ goes over smoothly to a puncture for $b\to 0$.
In the  limit that the holes are replaced  by punctures, the moduli space  of  super Riemann surfaces  becomes a complex manifold,
and  therefore $V$ gets a complex structure.   For questions (like computing the integral cohomology class $\chi(V)$) that are  independent of continuously
variable parameters such as  the lengths,  we can set the length parameters of NS boundaries to 0.
Therefore, in the case of NS boundaries, $V$ can be given a  complex structure, as was claimed earlier.

For holes of Ramond type, a corresponding argument shows instead that the volumes vanish,  as we  found in another way in  section \ref{sec:RBoundaries}.
A hyperbolic conjugacy class with Ramond spin structure contains an element
\be\label{typicalr} \newcommand*{\temp}{\multicolumn{1}{r|}{}}
\t U_b= \left(\begin{array}{cccccc}
e^{b/2} &1  \negthinspace\negthinspace\negthinspace &\temp & 0\\
0 &e^{-b/2}\negthinspace \negthinspace \negthinspace&\temp &0\\
 \cline{1-4}
0 &0\negthinspace \negthinspace \negthinspace&\temp &1\\
\end{array}\right).\ee
for some $b>0$.   Naively  setting $b=0$, we get a parabolic element  of $\OSp(1|2)$:
\be\label{parabolicr} \newcommand*{\temp}{\multicolumn{1}{r|}{}}
\t U_0= \left(\begin{array}{cccccc}
1 &1  \negthinspace\negthinspace\negthinspace &\temp & 0\\
0 &1\negthinspace \negthinspace \negthinspace&\temp &0\\
 \cline{1-4}
0 &0\negthinspace \negthinspace \negthinspace&\temp &1\\
\end{array}\right),\ee
$\t U_0$ is the monodromy of a flat $\OSp(1|2)$  connection around a  Ramond puncture (as opposed to a Ramond boundary).
The limit from $\t U_b$  to $\t U_0$ is not smooth.  The orbit of $\t U_b$ in $\OSp(1|2)$ is  of dimension $2|2$, but the orbit of $\t U_0$  is  of dimension $2|1$,
because $\t U_0$ commutes  with a  certain linear combination of the  odd generators of $\OSp(1|2)$.    Hence, a moduli space of super Riemann surfaces
with a Ramond boundary of length $b$ does  not approach a corresponding moduli space with a Ramond puncture in the limit  $b\to 0$.   If one sets $b=0$,
 one of the odd moduli is lost.    Thus if $V$ parametrizes
the odd moduli  of  a super Riemann surface with a specified Ramond boundary of length $b$  (along with other boundaries), and $V'$ parametrizes
the odd moduli if the boundary in question is replaced by a Ramond puncture, then the rank of $V'$ is 1 less than the rank of $V$.    In the limit $b\to 0$,
one can define  a ``forgetful''
map $V\to V'$ in which one identifies two super Riemann surfaces that are equivalent at $b=0$.    If $V''$ is the kernel of this forgetful map, then $V''$ is a real
vector bundle of rank 1, and there
is an exact sequence $0\to V''\to V\to V'\to 0$.   This implies that  $\chi(V)=\chi(V'')\chi(V')$.  But $\chi(V'')=\chi(V')=0$ (as rational cohomology  classes) since $V'$ and $V''$ are of
odd rank, so $\chi(V)=0$.    Let  us  note that  this argument  also  shows that correlators with Ramond punctures (as opposed  to holes) are essentially  new
quantities.   They are analyzed in section \ref{rpunctures}.

The bundle $V$ can be given a natural complex structure in the case of Ramond punctures, but not in the case of Ramond boundaries.

Two generalizations are worth  mentioning.   First, we assumed that $w_{ab}$ was negative-definite.   More generally, we can decompose $V$ as the direct
sum $V_1\oplus V_2$ of subbundles on which $w_{ab}$ is positive-definite or negative-definite, respectively.   If $V_1$ and $V_2$ have ranks $m_1$ and $m_2$, respectively,
one can  make a similar derivation, leading  to the result that $\chi(V)$ is replaced by  $(-1)^{m_1/2}\chi(V_1)\chi(V_2)=(-1)^{m_1/2}\chi(V)$.
Second, we have implicitly assumed that the smooth supermanifold $\hM$ is real, meaning that the fermions carry a real structure and the vector
bundle $V$ is real.  A generalization that has been called a cs supermanifold is important in superstring theory \cite{DM,Wit}. Here the fermions do not carry any real
structure.    However, in
 the symplectic case, the generalization from real supermanifolds to cs supermanifolds
does not add much, in the following sense.   {\it A priori}, if $\hM$ is a cs supermanifold, the structure group of $V\to M$ is the complex linear group $\GL(m,\C)$.
A general $\GL(m,\C)$ bundle cannot be reduced to a bundle with real structure group $\GL(m,\R)$, so in general a cs supermanifold cannot be given a real structure.
But in the case of a symplectic cs supermanifold, the metric tensor $w_{ab}$ reduces the structure group from $\GL(m,\C)$ to the complex orthogonal group $\OO(m,\C)$.
A further reduction to a maximal compact subgroup is always possible.  A maximal compact subgroup of $\OO(m,\C)$ is the ordinary compact orthogonal group $\OO(m)$.
The $m$-dimensional representation of this group is real, showing that if $\hM$ is a symplectic cs supermanifold, it actually can always be given a real structure.
 
\subsection{Comparison To The Results Of Norbury}\label{comparison}

In  \cite{Norbury2}, Norbury studied the spectral curve $y=\frac{1}{z}\cos (2\pi z)$ (with $z=\sqrt{-x}$) from a different starting point than the one of the present paper.   Seeking to 
generalize the results of Eynard and Orantin for the spectral curve $y=\sin(2\pi z)$ (which they had shown to be related to formulas of Mirzakhani for volumes
of moduli spaces of Riemann surfaces), Norbury was led to introduce the space $M$ that parametrizes a Riemann surface with a choice of spin structure,  along with the
vector bundle that we have called $V$.   He found that the analog for the ``cosine'' curve of the usual moduli space volumes for the ``sine'' curve were the quantities
that appear on the right hand side of eqn.~(\ref{gilfo}).  The sense in which these quantities actually are volumes was unclear, but we have provided an answer: they are
the volumes of the supermanifolds $\hM$ that parametrize super Riemann surfaces.    Norbury's class $\Theta_{g,s}$ is defined as (the sum over spin structures of)
the top Chern class of a  vector bundle $E_{g,s}$ whose fiber is $H^0(\Sigma,K^{3/2})$ (see Definition 2.1 in \cite{Norbury2}).   

The analysis in section \ref{computation} makes it clear that to define the volume of a symplectic supermanifold, one needs to pick an orientation of the vector
bundle $V$ that parametrizes the fermionic moduli.  To compare our results to Norbury's, it is necessary to compare the two orientation conventions, which in fact do
not agree.

The two choices are as follows:

(1) In  this paper, after picking a measure on the Lie superalgebra $\osp(1|2)$, we use the torsion to orient the moduli spaces.   Differently put, the torsion has a sign, and this
sign  is an orientation of the bundle $V$.    Though  it involves more machinery than a purely symplectic approach, the
  torsion has the advantage of making sense on an unorientable two-manifold.
It also has nice gluing properties, in the spirit of topological field theory.   For a two-manifold $Y$, let $\orr_Y$ be the orientation of $V$ determined by the torsion
(in other words, $\orr_Y$ is the orientation of $V$ that matches the sign of the torsion 
measure  $\tau_Y$).   If  $Y$ is built by gluing two-manifolds $Y_1$, $Y_2$ along one or more circles,
then the gluing law for the torsion gives
\be\label{gluor}\orr_Y  = (-1)^{w_\Ra}\orr_{Y_1}\orr_{Y_2}. \ee
Here $w_\Ra$ is the number of circles with Ramond spin structure on which gluing has occurred.   Like the similar factor in eqn.~(\ref{torformt}), this factor
comes from the sign of the torsion of a circle (eqn.~(\ref{waus})).    

(2) For orientable $Y$, Norbury orients $V$ by using the complex structure in which $V=H^0(Y,K^{3/2})$.     As we explained in appendix \ref{application}, a variant of this procedure
would be to use the opposite complex structure on $V$.   This would multiply  the orientation (and hence the volume) for a 
genus $g$ surface with $s$ boundary components by a factor $(-1)^s$.   

To see that these two choices do not agree, it suffices to consider  the example of a genus 1 surface with a  single boundary component.   In eqn.~(\ref{examples}),  we  found
that if we sum over spin structures  with a factor $(-1)^\zeta$, we get $V_1(b)=-1/8$, while at  the end of section \ref{sec:evenodd}, we showed that a similar sum  without the factor
$(-1)^\zeta$ gives $V_1(b)=0$.   The two results together show that with our orientations, the moduli spaces with even and odd spin structures
(and not including  a factor $(-1)^\zeta$ in the odd case) have volumes $-1/16$  and $+1/16$, respectively.

On  the other hand,  Norbury shows
that  in genus 1 with  a single boundary component, and summing over spin structures with no $(-1)^\zeta$, 
$\int_{\M_{1,1}} c_1( E_{1,1})=+1/8$ (see Proposition 2.10, and note that $\int_{\M_{1,1}} \psi_1=1/24$).
Moreover the even and odd spin structures make equal contributions to this integral.\footnote{This was explained to us by  P. Norbury.     Note that  $\int_{\M_{1,1}} c_1( E_{1,1})$
can be a rational number (rather than an  integer) because the moduli space is an orbifold, not a smooth manifold.}  So the integral for either  the even or odd spin  structure is $+1/16$.

Thus  our  results agree with Norbury's for the case of an odd spin structure, but disagree for the case of an even spin structure.     Accordingly, our results will differ
from Norbury's by a factor $(-1)^s (-1)^\zeta$.    The factor $(-1)^s$ is  inessential, in the  sense that it could  be  removed by using the opposite complex structure on $V$.     
But the factor $(-1)^\zeta$ really is essential.

In appendix \ref{punctureapp}, we will show in another way why the two approaches must differ by a factor $(-1)^\zeta$.   

\subsection{Ramond Punctures  And An Anomaly}\label{punctureapp}

In  the  presence of  Ramond punctures  (rather than boundaries), the bundle $V$ has a complex structure, and this gives a natural way to orient it.  
In  particular, the Ramond punctures behave as identical bosons; one can define the volumes without specifying how they should be ordered.
Thus, in approach (2) to orienting the moduli spaces, the volumes with Ramond punctures are manifestly well-defined, as long as one does not include a factor $(-1)^\zeta$.   

However, once Ramond punctures are present, there is no distinction between even and odd spin structures, since the monodromy that occurs when a Ramond
puncture loops around a one-cycle in a two-manifold $Y$ can permute the spin structures of $Y$ in an arbitrary fashion.\footnote{A spin structure in the presence
of Ramond punctures is a spin structure on the complement of the punctures that has a certain type of singularity  at the punctures. (A fermion that is parallel transported
on a  small loop around a Ramond puncture has monodromy $-1$.)   When such a singularity moves around a 1-cycle $\gamma$, the monodromy for a fermion path that has an
odd intersection number with $\gamma$ is reversed in sign.}   Hence it is not possible to define the factor $(-1)^\zeta$ in the presence of Ramond punctures,
and therefore in approach (2), if Ramond  punctures are present, the volumes can only be defined without the factor $(-1)^\zeta$.

If one considers Ramond boundaries rather than Ramond punctures, the situation  is opposite.   The difference between a Ramond boundary and a Ramond puncture,
as explained in appendix    \ref{application}, is that replacing a Ramond puncture  with a very small Ramond boundary adds one fermionic modulus, which is localized near
the boundary.    So as Ramond punctures behave as identical bosons when approach (2) is used  to orient the moduli spaces, Ramond boundaries behave as  identical fermions
in that approach.
This is an anomaly of sorts.  It means that the volume would depend on an ordering of the Ramond boundaries, modulo even permutations.
  This dependence can be canceled by supplying a factor of $(-1)^\zeta$.  To see this, note that  another way to  associate a  fermionic mode to a Ramond boundary
would be to introduce  a Majorana  fermion that propagates  on this boundary.    This is the anomaly that is canceled by a bulk factor $(-1)^\zeta$, as we discussed in  section
\ref{sec:spinButNoT}.

In approach (1) to orienting the moduli spaces, matters are reversed.   The torsion with possible  Ramond boundaries was analyzed in section \ref{osp} and did not depend on
any ordering of the boundaries; the Ramond boundaries  behaved like identical bosons.      So Ramond boundaries make sense in approach (1) to orienting the moduli 
spaces  with  no factor  $(-1)^\zeta$.  Ramond  boundaries become anomalous if one tries to include a factor $(-1)^\zeta$ in the sum over spin structures, because this
causes the Ramond boundaries to behave as fermions.   

What if we  replace Ramond boundaries with Ramond punctures?   This entails  removing a fermionic mode from every Ramond
boundary, so if the Ramond boundaries behave as bosons, the Ramond punctures behave as fermions, and vice-versa.   This means, in approach (1),
that the Ramond punctures are bosonic in the theory with $(-1)^\zeta$ and fermionic in the theory without it.   (It would be possible to show this explicitly  by
extending the torsion computations of section \ref{osp} to include Ramond punctures.)     Hence  in approach (1), in the presence of Ramond punctures,
the volumes can be defined in the theory with $(-1)^\zeta$ but not in the theory without it.

In this discussion, we have treated Ramond punctures and Ramond boundaries in parallel.   However, in the context of the present paper, there is an important difference.
We treat Ramond punctures  (when we do include them)  as an intrinsic part of the theory; their positions  are some  of  the  moduli  that parametrize the moduli  spaces whose volumes we compute.
If the  Ramond punctures behave as fermions, the volumes cannot be defined and the theory is inconsistent.   But we treat Ramond (or NS) boundaries as external probes
of the theory.   If they behave as fermions, these external probes have unexpected properties, but the theory does not become inconsistent.   We could eliminate this asymmetry
by incorporating in the theory D-branes and thus dynamical boundaries.   If a dynamical Ramond boundary  is a fermion, the theory is anomalous or inconsistent.
Hence, which D-branes are possible will depend on whether a factor $(-1)^\zeta$ is present.  This should  not be a surprise, since the same is true
 for related theories such as the 0A and 0B
theories that are briefly described in appendix \ref{app:minimal}.

\section{Behavior Under \texorpdfstring{$N\to N+4$}{N -> N+4}}\label{Euler}

For any of the ten matrix ensembles discussed in section \ref{sec:ensembles}, after picking a suitable matrix potential, one gets a matrix integral that can be expanded in Feynman diagrams.  The
diagrams can be conveniently drawn using the ``double line'' notation of 't Hooft \cite{tHooft:1973alw}, in which a propagator is drawn as a thin ribbon.  The edges of the ribbon
represent the indices of the matrix, and join into ``index loops.''    By gluing  the boundary of  a disc onto each index loop
in such a diagram, one can build a two-manifold.   This is the starting point for the relation between random matrices and two-dimensional gravity.

For the most basic case of a  hermitian matrix with unitary symmetry -- related to GUE statistics -- the  two-manifolds  constructed from Feynman diagrams are orientable. The same is
true for one of the Altland-Zirnbauer ensembles: the bifundamental of $\U(L)\times \U(L)$.  It is undoubtedly not a coincidence that this particular ensemble appears to be related
to the volumes of moduli spaces of orientable super Riemann surfaces.
This fact,  and its analog for other cases,  suggests that Altland-Zirnbauer ensembles are somehow related to random supergeometries just as the Dyson ensembles are related to random geometries.
But it is unclear how to make this precise.  The Feynman diagram expansion of a model based on the bifundamental of $\U(L)\times \U(L)$ was actually studied
some time ago \cite{Morris}; the results were not related to supergeometry in an obvious way.   An alternative idea is that discretization of supergeometry might be described
by a random ensemble of supermatrices; see \cite{Brezin:2018gvn} for recent work on supermatrices.

  For the other eight ensembles, unorientable as well as orientable two-manifolds appear.    These eight ensembles all enter, as we have reviewed in section  \ref{revpairs}, in the random
matrix classification of $\sT$-invariant  models, with or without supersymmetry.    In that context, the eight ensembles are exchanged in pairs under $N\to N+4$.
(For example, for  the case of a supersymmetric matrix model with  $N=1$ or 5 mod 8, $N\to N+4$ exchanges GOE and GSE.)   As we discussed in section \ref{topominus},
from a topological field theory point of view, one expects the exchange $N\to N+4$ to multiply the contribution of a two-manifold of Euler characteristic $\chi$ by $(-1)^\chi$.

One would hope to find the same behavior in the Feynman diagram expansion of a matrix integral.
This appears to be the case, but the literature does not seem to contain a complete proof.   For the basic case of comparing GOE-like and GSE-like ensembles, a proof was given in \cite{MulaseWaldron}.   This proof was surprisingly complicated.
The difficulty of the proof comes from the fact that in this case, as $N\to N+4$ changes the symmetry group, it changes the vertices as well as the propagators
in a Feynman diagram.   
       
\begin{figure}
 \begin{center}
   \includegraphics[width=3in]{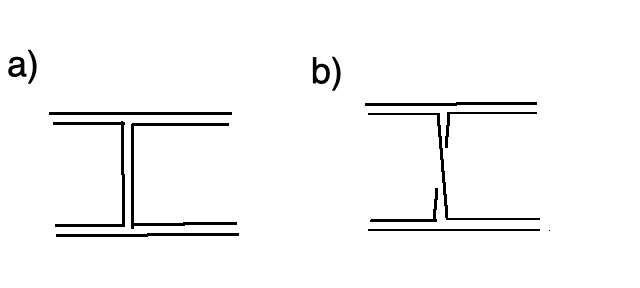}
 \end{center}
\caption{\small Drawn in a) and b) are portions of two Feynman diagrams that differ only by a relative ``twist'' in  one propagator -- the one drawn vertically.
When these pictures are embedded in a more complete Feynman diagram, the twist will change the number of index loops and therefore the Euler
characteristic of the resulting two-manifold by $\pm 1$.  \label{Ribbon}}
\end{figure} 

We will not summarize this proof, and instead will just point out a pair of Altland-Zirnbauer ensembles that differ by $N\to N+4$ and can be compared in a simple
way because the symmetry groups are the same.   We consider the two models in which the  matrix $C$ is a symmetric or antisymmetric second rank tensor of $G=\U(L)$
(corresponding to  supersymmetric models with $N=2,6$ mod 8).
The matrix action is $\Tr\, f(CC^\dagger)$ for a suitable function $f$.   The matrix propagator is, up to a possible constant multiple,
\be\label{donno}\langle C_{ij} C^{\dagger kl}\rangle=\frac{1}{2} \left( \delta^k_i \delta^l_j \pm \delta^k_j\delta^l_i\right), \ee
with a $+$ or $-$ sign depending on whether $C$ is a symmetric or antisymmetric tensor.   The only difference between the Feynman diagram expansions of the two
models comes from this sign.   Now consider two Feynman diagrams that differ only by which term we take in one of the propagators.   In the double line notation,
they differ only by whether an extra ``twist'' is given to the ribbon that corresponds to this propagator (fig.~\ref{Ribbon}).    Adding or removing such a twist changes
the number of index loops by $\pm 1$, so it shifts $\chi$ by $\pm 1$.  If the tensor $C$ is symmetric, adding or removing a twist
does  not affect the sign of the diagram, but if $C$ is antisymmetric,
it does, because of the minus sign in eqn.~(\ref{donno}).   So the Feynman diagram expansion of the antisymmetric tensor model differs from that of the symmetric
tensor model by an extra factor $(-1)^\chi$, as expected.

\section{Super-Schwarzian Path Integrals For The Disk And Trumpet}\label{app:Sch}
As explained for the bosonic case in \cite{Saad:2019lba}, the JT path integral reduces to an integral over the moduli space of bordered (super) Riemann surfaces, together with a path integral over the ``boundary wiggles'' associated to the asymptotically AdS boundary. In this section, we will compute the path integral over the boundary wiggles. There are two cases to consider: the case where the wiggles are at the boundary of the hyperbolic disk, and the case where they are at the ``big end'' of a hyperbolic trumpet whose other end connects to a bordered Riemann surface, see figure \ref{figdisktrumpet}.

In the bosonic case, these boundary wiggles are governed by the Schwarzian theory, and the measure that follows from the JT path integral is the natural symplectic measure with respect to which the Schwarzian theory is one-loop exact \cite{Saad:2019lba}. In the supersymmetric case, the wiggles (together with a fermionic superpartner) are governed by the super-Schwarzian theory \cite{Forste:2017kwy} introduced in \cite{Fu:2016vas}. In principle, for super JT gravity, one could follow the steps in the bosonic case and determine the associated measure by starting from the formulation of JT supergravity as an $\OSp(1|2)$ $BF$ theory. We will assume that the measure that one obtains from this procedure is the natural symplectic one that makes the path integral one-loop exact.

The purely bosonic part of the super-Schwarzian action is the ordinary Schwarzian derivative $\text{Sch}(f(u),u)$, where $u$ is the Euclidean time coordinate of the boundary theory, running from zero to $\beta$. The $u$ coordinate measures rescaled proper time along the wiggly boundary:
\be\label{length}
\mathrm{d}s_{||} = \frac{\mathrm{d}u}{\epsilon},
\ee
where $\epsilon$ is interpreted as a holographic renormalization parameter, see \cite{Maldacena:2016upp}. The total length of the wiggly boundary is therefore $\beta/\epsilon$. The detailed shape of the boundary is specified by the function $f(u)$. This gives the Euclidean ``Poincare time'' of the wiggly boundary as a function of the coordinate $u$. More precisely, the $f$ coordinate is such that the metric of hyperbolic space is
\be
\mathrm{d}s^2 = \frac{\mathrm{d}f^2 + \mathrm{d}z^2}{z^2}.
\ee
Specifying $f(u)$ and (\ref{length}) specifies the boundary curve, up to isometries. There are two special cases that will be important for us. First, we have the classical solution corresponding to a circular boundary for the disk, which is $f = \tan(\pi u/\beta)$ up to $SL(2,\mathbb{R})$ transformations (which becomes $\OSp(1|2)$ in the super-Schwarzian case). For this solution, $\text{Sch}(f(u),u) = 2\pi^2/\beta^2$. Second, we have the classical solution corresponding to a circular boundary for the trumpet, which is $f = \tanh(\frac{bu}{2\beta})$ up to $\U(1)$ transformations, with $\text{Sch}(f(u),u) = -b^2/(2\beta^2)$.

The super-Schwarzian theory \cite{Fu:2016vas} adds to this bosonic piece a fermion $\eta$ that is coupled to $f$. A useful fact is that, with respect to the symplectic measure, the path integral of the super-Schwarzian theory is one-loop exact \cite{Stanford:2017thb}, so we can determine it by analyzing the one-loop path integral around the relevant classical solution. Let's imagine that $\{f = f_0(u),\eta = 0\}$ is a classical solution. The equation of motion implies that $\text{Sch}(f_0(u),u)$ is constant, and we will refer to its constant value as $S$. Then we expand near this solution, by making a small reparametrization $f(u) = f_0(u+\varepsilon(u))$.\footnote{This $\varepsilon(u)$ is unrelated to the holographic renormalization parameter $\epsilon$.} To quadratic order in $\varepsilon$ and the fermion $\eta$, the super-Schwarzian action is \cite{Fu:2016vas}
\be
I = -\gamma\int_0^\beta \mathrm{d}u\left(S -\frac{1}{2} \varepsilon''^2 +S\,\varepsilon'^2 + 2\eta'\eta''- S\,\eta\eta'\right).\label{superSchAc}
\ee
Here $\gamma$ is a constant with dimensions of length that can be absorbed by rescaling $\beta$ (this is equivalent to an overall rescaling of the matrix in the dual matrix integral). We will follow the conventions of \cite{Saad:2019lba} and choose units such that $\gamma = \frac{1}{2}$. Then $\beta$ becomes dimensionless. 

To evaluate the one-loop path integral, it is convenient to expand in Fourier modes
\be\label{superSch}
\varepsilon(u) = \sum_{n}e^{-\frac{2\pi}{\beta}\mathrm{i} n u}\left(\varepsilon_n^{(R)} + \mathrm{i}\varepsilon_{n}^{(I)}\right), \hspace{20pt} \eta(u) = \sum_{m} e^{-\frac{2\pi}{\beta}\mathrm{i}m u}\left(\eta_m^{(R)} + \mathrm{i}\eta_m^{(I)}\right).
\ee
We take $\varepsilon$ to be periodic and $\eta$ to be antiperiodic, so $n\in \mathbb{Z}$ and $m \in \mathbb{Z} + \frac{1}{2}$. In order for the fields to be real, the real parts $\varepsilon^{(R)},\eta^{(R)}$ should be even in $n$ and $m$, and the imaginary parts $\varepsilon^{(I)},\eta^{(I)}$ should be odd. In these variables, the action (\ref{superSch}) is 
\begin{align}
I = -\frac{\beta}{2}S &+ \sum_{n> 0}\frac{\beta}{2}\left(\left(\frac{2\pi n}{\beta}\right)^4 - 2S\left(\frac{2\pi n}{\beta}\right)^2\right) \left((\varepsilon_n^{(R)})^2 + (\varepsilon_n^{(I)})^2\right)\\
&-\sum_{m>0}4\beta\left(\left(\frac{2\pi m}{\beta}\right)^3 - \frac{S}{2}\frac{2\pi m}{\beta}\right)\eta_m^{(R)}\eta_m^{(I)}.
\end{align}
In general, this action has zero-modes. For a generic value of $S$, the only zero mode is the $n = 0$ mode of $\varepsilon$, but for special values of $S$, there can be more. In the super JT theory, these zero modes are interpreted as pure gauge modes. For example, rotations of the wiggly boundary of the trumpet relative to an arbitrary reference coordinate system, or $\OSp(1|2)$ transformations of the wiggly boundary of the disk relative to an arbitrary reference. We should not integrate over these modes. 

An important fact for the one-loop exactness of the super Schwarzian theory is that the quotient of the path integral space by the space of these zero modes is a symplectic supermanifold. The symplectic form is the Kirillov-Kostant-Souriau form for the coadjoint orbits of super Virasoro, and it has the property that with respect to the symplectic form, the Schwarzian action is a Hamiltonian that generates the $\U(1)$ symmetry of translations in $u$. To quadratic order in the fields, the form is the restriction to nonzero modes of
\begin{align}\label{sympW}
\widehat{\omega} &= \frac{\alpha}{2}\int_0^\beta\mathrm{d}u \Big[\mathrm{d}\varepsilon'(u)\mathrm{d}\varepsilon''(u) - 2S\, \mathrm{d}\varepsilon(u)\mathrm{d}\varepsilon'(u) - 4\mathrm{d}\eta'(u)^2 + 2S\mathrm{d}\eta(u)^2\Big].
\end{align}
A quick way to justify this is to check that with respect to this form, the action (\ref{superSchAc}) generates $u$-translations. Concretely, one needs to show that $\mathrm{d}H$ is proportional to the interior product of the $u$-derivative vector field with $\widehat{\omega}$. This is computed by replacing one of the factors of $\mathrm{d}\varepsilon$ and $\mathrm{d}\eta$ by $\varepsilon'(u)$ and $\eta'(u)$.

The constant $\alpha$ that multiplies (\ref{sympW}) is arbitrary, but it is correlated with a similar arbitrary constant in the normalization of the symplectic form on supermoduli space (in the orientable case). This is due to the fact that both symplectic forms are restrictions of the same underlying symplectic form of $BF$ theory. Changing the overall normalization of this form corresponds to the freedom to adjust the coefficient of the topological term in the JT action. We will follow the convention in \cite{Saad:2019lba} and set the constant equal to $\alpha = 1$. The bosonic computation in \cite{Saad:2019lba} implies that the corresponding normalization for the symplectic form on super moduli space is such that the bosonic component is $\sum_{i}\mathrm{d}b_i\,\mathrm{d}\tau_i$ in terms of the Fenchel-Nielsen (length-twist) coordinates. Setting $\alpha = 1$ and inserting the Fourier expansion, one finds
\begin{align}
\widehat{\omega}&=\sum_{n> 0} 2\beta\left(\left(\frac{2\pi n}{\beta}\right)^3 - 2S\frac{2\pi n}{\beta}\right)\mathrm{d}\varepsilon_n^{(R)}\mathrm{d}\varepsilon_n^{(I)}- \sum_{m>0}4\beta\left(\left(\frac{2\pi m}{\beta}\right)^2 - \frac{S}{2}\right)\left((\mathrm{d}\eta_m^{(R)})^2 + (\mathrm{d}\eta_m^{(I)})^2\right).\notag
\end{align}

With these formulas, it is straightforward to evaluate the Gaussian integral
\be
\int \prod_{n>0}\mathcal{D}(\varepsilon_n^{(a)},\mathrm{d}\varepsilon_n^{(a)})\prod_{m>0}\frac{\mathcal{D}(\eta_m^{(a)},\mathrm{d}\eta_m^{(a)})}{(2\pi)^2} \exp\left(\widehat{\omega} - I\right) = e^{\frac{\beta}{2}S}\prod_{n>0}\frac{2\beta}{n}\prod_{m>0}\frac{m}{2\beta}.\label{productnm}
\ee
We are interested in applying this formula to two different cases. The first case is the disk partition function, for which the classical solution has $S = 2\pi^2/\beta^2$. For this case, the action has three bosonic and two fermionic zero modes, corresponding to $n = 0,\pm 1$ and $m = \pm\frac{1}{2}$. These three zero modes are the infinitesimal action of $\OSp(1|2)$ on the classical solution, and we should quotient the path integral space by this group. This means that in (\ref{productnm}), we should take the product over $n = 2,3,4,...$ and $m = \frac{3}{2},\frac{5}{2},\frac{7}{2},...$. Using zeta function regularization or an exponential cutoff, one finds the result
\be
Z^{D}_{\text{SJT}}(\beta) = e^{S_0}\sqrt{\frac{2}{\pi\beta}}e^{\frac{\pi^2}{\beta}}.
\ee

The second case we need is the ``trumpet'' geometry, for which $S = -b^2/2\beta^2$, see \cite{Saad:2019lba}. In this case, the only zero mode is the $n = 0$ mode of $\varepsilon$. Inserting this value of $S$ in (\ref{productnm}) and taking the renormalized product over $n = 1,2,3,...$ and $m = \frac{1}{2},\frac{3}{2},\frac{5}{2},...$, one finds
\be
Z^T_{\text{SJT}}(\beta,b) = \frac{1}{\sqrt{2\pi\beta}}e^{-\frac{b^2}{4\beta}}. \label{nstrumpet}
\ee

It is  also possible to consider the Schwarzian path integral in  the trumpet geometry for the case of a Ramond spin structure on the trumpet.   The difference
is that the fermionic field $\eta$ is now integrally moded and in particular has a zero-mode, which causes the path integral to vanish.   Thus
if $\t T$ is a trumpet with Ramond spin structure, the corresponding Schwarzian path integral is just
\be\label{rtrumpet}
Z^{\t T}_{\text{SJT}}(\beta,b) = 0.
\ee
On the other hand, instead of a trumpet with Ramond spin structure, we can consider a disc with a  Ramond  puncture at the center.   The monodromy around  a Ramond
puncture was described in eqn.~(\ref{parabolicr}).   The important difference between a disc with Ramond puncture and a trumpet with Ramond spin structure is
that a certain linear combination of the fermionic generators of $\osp(1|2)$  is a symmetry of a disc with Ramond puncture, since it commutes with the monodromy.   In the Schwarzian path integral, unbroken
generators of $\osp(1|2)$ are treated as gauge symmetries and the corresponding modes of  the Schwarzian multiplet are omitted.   In the present case, this means
that we omit the zero-mode of the field $\eta$.  The remaining Schwarzian path integral is extremely simple.    The classical action is $S=0$  (we just set $b=0$  in
the formula $S=-b^2/2\beta^2$
for a trumpet, since geometrically a  disc with a puncture is the $b\to 0$   limit of  a trumpet).  Moreover, because of unbroken supersymmetry, the 1-loop path
integral of the fields $\varepsilon,\eta$ is equal to 1  (these fields are both integrally moded, and  their determinants cancel mode by mode).  So the Schwarzian path 
integral for a disc with Ramond puncture is just
\be\label{rpuncture} Z^\Ra_{\text{SJT}}(\beta)=1. \ee

There is no factor of $e^{S_0}$ in  eqn. (\ref{nstrumpet}) or (\ref{rpuncture}), since the Euler characteristic of a trumpet or a once-punctured  disc is 0.

\section{Mirzakhani's  Recursion Relation And Its Superanalog}\label{supermirz}

In this appendix, we will review Mirzakhani's derivation of a recursion relation on volumes of moduli spaces of Riemann surfaces, and generalize this derivation to  super
Riemann surfaces.   The underlying  super McShane identity has also been proved, for the basic  case of a once-punctured torus,  by Y.  Huang, R. Penner, and A. Zeitlin \cite{HPZ}.

\subsection{The Sum Rule}\label{sumrule}

Mirzakhani's sum rule for 
hyperbolic Riemann surfaces with geodesic boundary  generalized an earlier version by McShane for hyperbolic Riemann
surfaces with cusps (or punctures) \cite{McShane}.  There is actually a useful short explanation in section 2 of \cite{Norbury}, where some of the considerations were generalized
to the unorientable case. Here we will consider orientable surfaces only.

We consider a hyperbolic surface $Y$ with a geodesic boundary $\gamma$, of length $b$, on which we will focus.  $Y$ may have  a set $I$ of additional
geodesic boundaries $\gamma_1,\cdots,\gamma_n$, of lengths $B=\{b_1,\cdots, b_n\}$. We assume that $Y$ is not itself a three-holed sphere.  If $Y$ is a three-holed
sphere, then the moduli space of hyperbolic structures on $Y$ with given boundary lengths  is  a point, of volume 1; this will ultimately  be the trivial case with which Mirzakhani's
recursion relation begins.

  Starting at any point $p\in \gamma$, there is  a unique inward going geodesic $\ell_p$  on $Y$
that is orthogonal to $\gamma$.   We decompose $\gamma$ as the disjoint union of three sets $\A$, $\B$, $\CC$, as follows.  A point $p\in\gamma$ is in $\A$ if $\ell_p$ intersects
itself, or returns to $\gamma$, before meeting any  of the other $\gamma_i$; $p$  is in $\B$ if $\ell_p$ reaches one of the other $\gamma_i$ before intersecting itself
or returning to $\gamma$; and $p$ is in $\CC$ if it continues forever in the interior of $Y$  without self-intersection.    In case $\A$ or $\B$, we truncate $\ell_p$ as soon
as it intersects itself or reaches the boundary of $Y$.

A theorem of Birman and Series implies that the set $\CC$ is of measure 0; roughly, it is very difficult for a geodesic in a hyperbolic surface to continue indefinitely  
without self-intersection.  Therefore, if we write $\mu(\A)$ and $\mu(\B)$ for the measures of $\A$ and of $\B$, then the sum of these measures is the total measure of $\gamma$,
namely $b$:
\be\label{worof} b = \mu(\A)+\mu(\B).    \ee

\begin{figure}[t]
\begin{center}
\includegraphics[width=\textwidth]{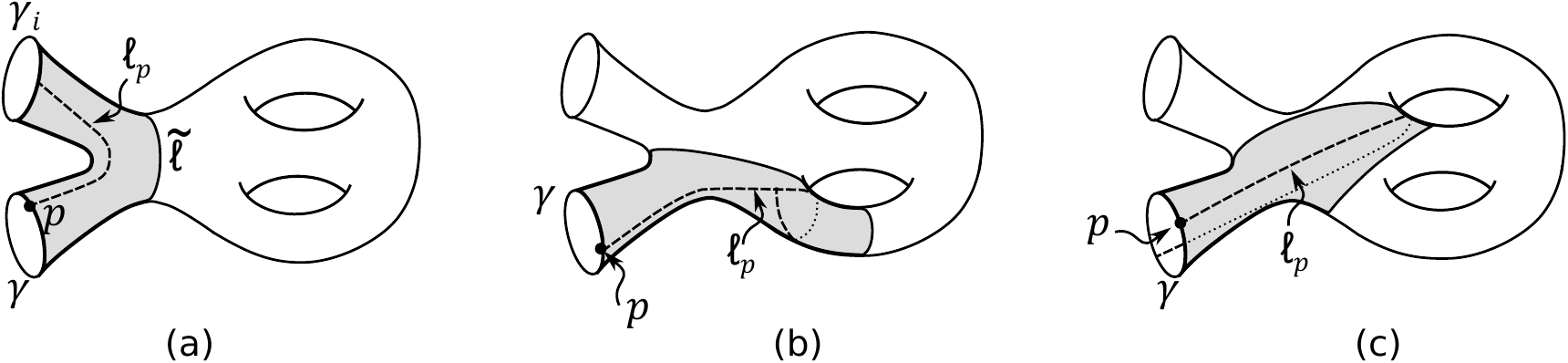}
\caption{{\small We show three-holed spheres associated to the different fates of a geodesic $\ell_p$ starting orthogonally at a point $p$ on boundary $\gamma$. In (a) $p\in \B_i$, and the geodesic makes it to boundary $\gamma_i$. A thickening of $\ell_p, \gamma,\gamma_i$ is topologically the three-holed sphere shown shaded. In (b) and (c), $p\in \A$ and the geodesic either self-intersects or returns to the original boundary. In both cases, a thickening of $\ell_p,\gamma$ determines a topological class of three-holed sphere, shown shaded.}}\label{fig:geodesics}
\end{center}
\end{figure}

The set $\B$ has an obvious decomposition as the union of subsets $\B_i$, where $p\in  \B_i$ if $\ell_p$ reaches $\gamma_i$ first.
Thus $\B=\coprod_{i\in  I} \B_i$, where  $\coprod$ denotes the disjoint union of sets.   However, $\B_i$ has a much more subtle and surprising decomposition,
as follows.   Suppose that $p\in\B_i$, and consider the subset of $Y$   that consists of the union of $\gamma$, $\gamma_i$, and the segment  of  $\ell_p$ that connects
them, see fig.~\ref{fig:geodesics}(a).    Thickening this subset slightly,
we get a subset $\S$ of $Y$ that is topologically a three-holed sphere.  $\S$ has three boundary circles; two of them -- namely $\gamma$ and $\gamma_i$ -- are geodesics.
The third is not a geodesic, but by minimizing its length  within its homotopy class, we do get  a geodesic $\tilde\ell \subset Y$. And then $\gamma$, $\gamma_i$,
and  $\tilde\ell$ are geodesic boundaries of a three-holed sphere $\Lambda\subset Y$.   Because
we have assumed that $Y$ is not itself a three-holed sphere,
$\tilde\ell$  is in the interior  of $Y$, and $\Lambda$ is a  proper subset of $Y$.

Let $\Upsilon_i$ be the set of all three-holed spheres in $Y$ whose  boundary  consists of $\gamma$, $\gamma_i$,  and an internal geodesic   $\tilde\ell\subset Y$.  In general
there are infinitely many choices of $\tilde\ell$, so  $\Upsilon_i$  -- and similarly $\Upsilon$ below -- is a (countably) infinite set.  We  have seen
that every $p\in \B_i$ is naturally associated with some $\Lambda\in \Upsilon_i$.    So we have a  decomposition
\be\label{motto} \B=\coprod_{i\in I,\Lambda\in\Upsilon_i}\B_{i,\Lambda},  \ee where $\B_{i,\Lambda}$ consists of points $p\in \B _i$ such that the three-holed sphere that is obtained from the  construction that was just described is $\Lambda$.
Hence
\be\label{zotto} \mu(\B)=\sum_{i\in  I,\Lambda\in\Upsilon_i} \mu(\B_{i,\Lambda}). \ee

The set $\A$ has a similar decomposition.  If  $p\in \A$, we consider 
the subset of  $Y$ that consists of the union of $\gamma$ with  $\ell_p$.   Thickening this  set slightly gives a subset $\S$ of $Y$ that is
again topologically a three-holed sphere, see fig.~\ref{fig:geodesics}(b) and \ref{fig:geodesics}(c); remember that $\ell_p$ is truncated as soon as it intersects  itself or returns to $\gamma$.    $\S$ has a single geodesic boundary, namely $\gamma$, and two boundary circles that are not geodesics.  By minimizing the
lengths of the nongeodesic boundaries in their homotopy classes, we can find geodesics which together with $\gamma$ bound a three-holed sphere $\Lambda$. It might be the case that one of these two additional geodesics is an external boundary $\gamma_i$, so that $\Lambda \in \Upsilon_i$. Or it might be the case that both are internal boundaries. Then we say $\Lambda \in \Upsilon$, where $\Upsilon$ is the set of all three-holed spheres in $Y$ whose boundary consists of $\gamma$ and two internal geodesics. This leads to
\be\label{otto}\mu(\A)=\sum_{\Lambda\in\Upsilon}\mu(\A_\Lambda) + \sum_{i\in I,\Lambda\in\Upsilon_i}\mu(\A_{i,\Lambda}), \ee
where $\A_\Lambda,\,\A_{i,\Lambda} \subset \A$ are defined by saying that $p\in \A_\Lambda$ or $p\in \A_{i,\Lambda}$ if $\ell_p$ is related to $\Lambda\in \Upsilon$
or $\Lambda\in \Upsilon_i$ in the way just described. Combining (\ref{zotto}) and (\ref{otto}), we can write the sum rule (\ref{worof}) as
\be\label{lotto} b=\sum_{\Lambda\in\Upsilon}\mu(\A_\Lambda) + \sum_{i\in I,\Lambda\in\Upsilon_i}\Big(\mu(\B_{i,\Lambda}) + \mu(\A_{i,\Lambda})\Big). \ee

In this formula, the quantities $\mu(\A_\Lambda)$, $\mu(\A_{i,\Lambda})$,  and $\mu(\B_{i,\Lambda})$ have the nice property that they depend only on $\Lambda$, and not on anything else about
$Y$.  In other words, we can compute them by studying orthogonal geodesics in a three-holed sphere
$\Lambda$, without worrying  about  the rest of $Y$.
 Suppose that $\Lambda_0$ is a three-holed sphere with geodesic boundaries $\gamma,\gamma',\gamma''$ of lengths $b,b',b''$.   Apart from a set of measure
0, analogous to $\CC$ above, we decompose $\gamma$ as the union of disjoint sets $\A_0$, $\B_0'$, and $\B_0''$, where $p\in \A_0$ if $\ell_p$ intersects  itself
or returns  to $\gamma$  before reaching $\gamma'$ or $\gamma''$,  while $p\in \B_0'$ if $\ell_p$ reaches $\gamma'$ first, and $p\in\B_0''$ if $\ell_p$ reaches $\gamma''$ first. 
We set $\TT(b,b',b'')=\mu(\B_0')$, so that by symmetry $\mu(\B_0'')=\TT(b,b'',b')$
.
    Similarly we 
 let $\DD(b,b',b'')=  \mu(\A_0)$.   
We will compute $\TT(b,b',b'')$ in section  \ref{computation}.   This will determine $\DD(b,b',b'')$ because of the sum rule
\be\label{oggo}  b=\DD(b,b',b'')+\TT(b,b',b'')+\TT(b,b'',b').  \ee

Every three-holed  sphere $\Lambda$ is isomorphic to $\Lambda_0$ for some values of $b,b',b''$, so for any  $\Lambda$,
 the  quantities $\mu(\A_\Lambda)$, $\mu(\A_{i,\Lambda})$,   and $\mu(\B_{i,\Lambda})$  in eqn.
(\ref{lotto}), which measure the sets of orthogonal geodesics with a specified behavior  inside $\Lambda$, can be expressed in terms of the functions $\TT(b,b',b'')$ and 
$\DD(b,b',b'')$.   In fact, $\mu(\A_\Lambda)=\DD(b,b',b'')$, while $\mu(\B_{i,\Lambda})=\TT(b,b_i,b')$, $\mu(\A_{i,\Lambda})=\DD(b,b_i,b')$.   So a somewhat more explicit version  of eqn.~(\ref{lotto})  is
\begin{align}\label{koggo} b  &= \sum_{\Lambda\in\Upsilon} \DD(b,b',b'') +\sum_{i\in I,\,\Lambda\in \Upsilon_i} \Big(\TT(b,b_i,b') + \DD(b,b_i,b')\Big)\\
&= \sum_{\Lambda\in\Upsilon} \DD(b,b',b'') +\sum_{i\in I,\,\Lambda\in \Upsilon_i} \Big(b-\TT(b,b',b_i)\Big).\end{align}

\subsection{The Recursion  Relation}\label{recrelation}

As preparation for explaining Maryam Mirzakhani's recursion relation for volumes, we will describe how one might proceed if there were no  need to divide by the mapping
class group.

\begin{figure}[t]
\begin{center}
\includegraphics[width=.9\textwidth]{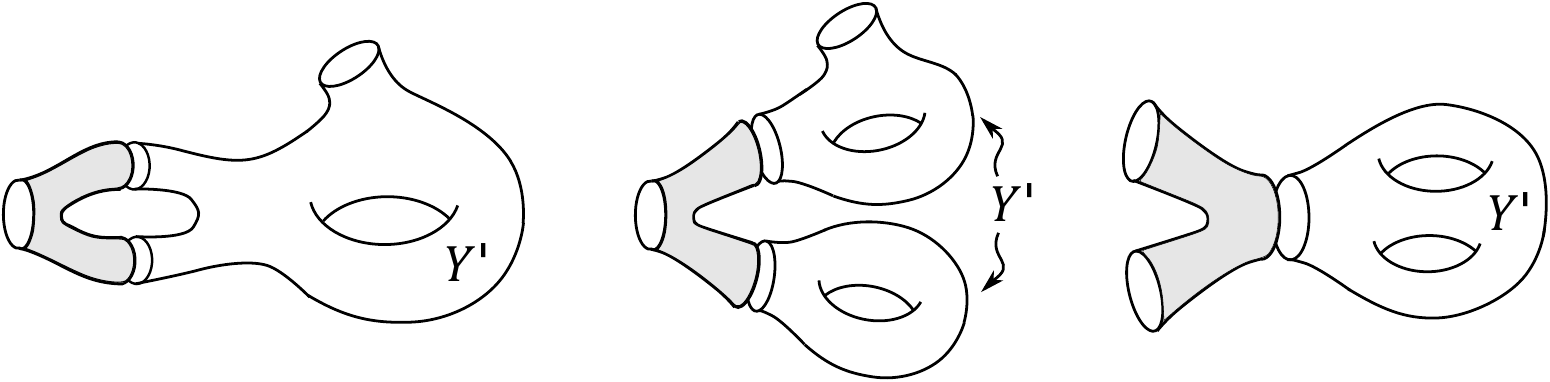}
\caption{{\small We show three ways of building a $g = 2$ surface $Y$  with two boundaries by gluing a three-holed sphere onto another hyperbolic surface $Y'$, possibly disconnected. These three types of gluing correspond respectively to the three types of terms on the RHS of (\ref{recurMirz}).}}\label{fig:gluing}
\end{center}
\end{figure}
Let $Y$ be a hyperbolic surface with boundary lengths $b$ and $b_i,\,i\in I$.   Suppose  that $Y$ can be built by gluing a three-holed sphere $\Lambda$ of  boundary
lengths  $b,b',b''$ onto another hyperbolic  surface  $Y'$ with boundary lengths $b',b'',$ and $b_i,\,i\in  I$, see fig.~\ref{fig:gluing}.  If $Y$  has genus  $g$ with $n+1$ boundaries, then 
 $Y'$ might be, for  example, a surface of genus $g-1$ with
$n+2$ boundaries ($Y'$ can also be disconnected, as shown in the figure).    Looking  back to eqn.~(\ref{zingo}), we see  that the torsions
(or  volume  forms)  of $Y$  and $Y'$ are related  by  $\tau_Y= \d b'\d\varrho'\,\d b'' \d\varrho'' \,\tau_{Y'}$.   If the mapping class group were not relevant, this relation between
the volume forms 
would lead after integration to a relation between the volumes:
\be\label{zofo}V_{g}(b,B)\overset{?}{=}  \frac{1}{2}\int \d b'\d\varrho' \,\d b''\\d\varrho'' \, V_{g-1}(b',b'',B). \ee
Here  $B$ represents the collection $b_1,\cdots,b_n$, and the factor of $1/2$  takes into  account the symmetry under  exchange of $b'$ and $b''$, to avoid double-counting.   We can make this formula  more explicit
by noting that $b',b''$ run over the  half-lines  $[0,\infty)$, while $\varrho', \varrho''$ run over $0\leq \varrho'\leq  b'$, $0\leq \varrho  ''\leq b''$.
Moreover, the volumes do not depend on $\varrho',\varrho''$, so the integration over  those  variables just gives a factor  $b'b''$.   Thus the formula would reduce to
\be\label{roffo}V_{g}(b,B)\overset{?}{=}  \frac{1}{2}\int_0^\infty b'\d b' \,b''\d b'' \, V_{g-1}(b',b'',B). \ee

The analog of this formula for a compact group such as SU(2) is actually correct, as discussed for example in section 2 of \cite{Dijkgraaf:2018vnm}.   However,
for $\SL(2,\R)$, one needs to divide by the mapping class group.   Eqn.~(\ref{zofo}) is wrong because the mapping class group of $Y$ is larger than the mapping
class group of $Y'$, or to  put it differently, because modular invariance would  force  us  in eqn.~(\ref{roffo})  to sum over contributions of infinitely  many  choices
of the three-holed sphere $\Lambda$.   This  sum  will give a divergent factor, so clearly the  formula (\ref{roffo}) is not correct.  (In fact, the volumes $V_g(b_1,\cdots,
b_s)$ are polynomials in $b_1^2,\cdots, b_s^2$, so the integral on the right hand side of eqn.~(\ref{roffo}) will diverge.)

If  one could find a function $f(b,b',b'')$
with the property that its sum over all possible choices of $\Lambda$ is 1, then one could find a correct version of
 eqn.~(\ref{roffo}) by inserting  in the definition $V_g(b,B)=\int_{\M_{g,n+1}}\tau_Y$ 
 a factor of $1=\sum_\Lambda f(b,b',b'')$.   Then, instead of summing over $\Lambda$ and dividing by the mapping class
group of $Y$, it would be equivalent to pick a  particular $\Lambda$ and divide by  the mapping class group of  $Y'$.    The upshot would be a recursion relation rather
like eqn.~(\ref{roffo}), but with an extra factor $f(b,b',b'')$ in the integral on the right hand side.

There is no function $f(b,b,b'')$ that has quite the property  we stated, but as oberved by Mirzakhani, the sum rule of eqn.~(\ref{koggo}) can  be used similarly.
On the right hand side of eqn.~(\ref{koggo}), the summation over $\Lambda$ includes a  sum over  all topological ways of obtaining $Y$ by gluing a three-holed sphere $\Lambda$
onto some other surface $Y'$ (with the constraint that one of the boundaries of $\Lambda$ is a specified boundary of $Y$).
  So a recursion relation derived from this sum rule involves a sum over all topological types of
 gluing.   It  reads\footnote{\label{footnotefactoroftwo} There is a subtlety in terms involving $V_{1}(b)$. Because of the $\mathbb{Z}_2$ symmetry of the torus with one hole, when we glue in such a surface, we should restrict the twist to be between zero and $b/2$ instead of between zero and $b$. In order to  avoid awkward factors of two that result from this, we are simply going to define $V_1(b)$ to be one half of the true moduli space volume. Then we can integrate the twist from zero to $b$ as usual.}
\begin{align}\label{recurMirz}
bV_{g,n+1}(b,B) &= \frac{1}{2}\int_0^\infty b'\mathrm{d}b'\,b''\mathrm{d}b''\,\DD(b,b',b'')\left(V_{g-1}(b',b'',B) + \sum_{\text{stable}}V_{h_1}(b',B_1)V_{h_2}(b'',B_2)\right)\notag\\
&\hspace{20pt} +\sum_{k = 1}^{|B|}\int_0^\infty b'\mathrm{d}b' \,\Big(b-\TT(b,b',b_k)\Big)\,V_g(b',B\setminus b_k).
\end{align}
This is Mirzakhani's recursion relation \cite{mirzakhani2007simple}, which
 was interpreted by Eynard and Orantin \cite{eynard2007weil} 
 in terms of topological recursion for matrix ensembles.   For a certain supersymmetric matrix ensemble, we deduced  in eqn.~
(\ref{recurVols})  a recursion
relation  with  the  same structure.   We will aim to find a superanalog of Mirzakhani's derivation to account for that result.

\subsection{More On Hyperbolic Geometry  And Three-Holed Spheres}\label{uhp}

As  preparation for  computing $\TT(b,b',b'')$, and for  understanding the superanalog of Mirzakhani's derivation, we need some more details on hyperbolic geometry.

Let $z=x+\i y$ with real $x,y$.   The group $\SL(2,\R)$ acts on the upper half-plane $y>0$, leaving fixed the hyperbolic metric
\be\label{monof} \d s^2=\frac{\d z\d\bar z}{\mathrm{Im}^2  z}=\frac{\d x^2+\d y^2}{y^2}.   \ee
We call this half-plane $H$.  The line $x=0$ plus a point at infinity constitute the ``conformal boundary'' of $H$, which we will call $\partial H$.   As is usual in holographic duality, the conformal
boundary is ``at infinity,'' not contained in $H$.   Including the  point at infinity, $\partial H$ is a circle topologically.

\begin{figure}[t]
\begin{center}
\includegraphics[width = .7\textwidth]{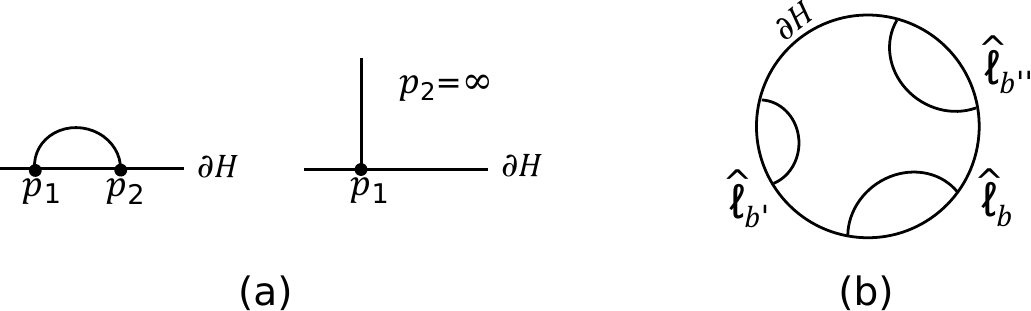}
\caption{{\small In (a) we view $\partial H$ as the real line, and we show a geodesic with virtual endpoints at finite locations in $\partial H$, as well as one with a virtual endpoint at infinity. In (b) we view $\partial H$ as $S^1$, and we show a non-nested configuration of three geodesics.}}\label{fig:UHP}
\end{center}
\end{figure}

$H$ is complete, so a geodesic in $H$ propagates for infinite distance in each direction, but a geodesic in $H$ has at each end a ``virtual endpoint'' in $\partial H$.
Conversely, any two points in $\partial H$ are connected by a unique geodesic in $H$.   These geodesics are easily described in terms of the Euclidean metric
$\d s_E^2=\d x^2+\d y^2$ on $H$.   If $p_1,p_2$ are finite points in $\partial H$, then the geodesic between them is a semicircle, centered at the midpoint of
the interval $[p_1,p_2]$, and with diameter such that it passes through $p_1$ and $p_2$.   If, say, $p_2$ is the point at infinity, then the geodesic between
$p_1$ and $p_2$ is a vertical straight line from $p_1$.  See fig.~\ref{fig:UHP}(a). 

The conformal boundary $\partial H$ can be understood as a copy of $\RP^1$, with homogeneous real coordinates $u,v$.   $\SL(2,\R)$ acts on $\RP^1$ in the natural
way
\be\label{wobblo}\begin{pmatrix} u\cr v \end{pmatrix}\to \begin{pmatrix} a& b \cr c & d\end{pmatrix}\begin{pmatrix} u\cr v\end{pmatrix}, ~~~ad-bc=1.  \ee
The corresponding action on $x=u/v$ is  familiar: 
\be\label{onof} x\to \frac{ax+b}{cx+d}. \ee
The point at $x=\infty$ is, of course, the  point in $\RP^1$ with $(u,v)=(1,0 )$.   

The action of $\SL(2,\R)$ on $H$ is described by the same formula, but with $x$ replaced by a complex variable $z$ (constrained to have $\mathrm{Im}\,z>0$): 
\be\label{wonof} z\to \frac{az+b}{c z+d}. \ee

Now we will discuss some details about three-holed spheres that were not needed in the body of the  paper.
 If $Y$ is a three-holed hyperbolic sphere, with  geodesic boundaries, then its universal cover $\hat Y$ is a region in $H$ bounded by three geodesics.

Let us work this out for the case studied in section \ref{compotwo} of a three-holed sphere with monodromies 
\be\label{montwo} U_0=\vnu_b \begin{pmatrix} e^{b/2}& \kappa \cr 0 & e^{-b/2}\end{pmatrix},~~~~V_0=\vnu_{b'} \begin{pmatrix}e^{-b'/2} & 0 \cr 1 & e^{b'/2}\end{pmatrix} ,\ee
and $W_0=V_0^{-1}U_0^{-1}$.  (We write $b,b',b''$ for what in section \ref{compotwo} were $a,b,c$.)  The length parameters of $U_0$, $V_0$, and $W_0$ are $b,b'$, and a parameter $b''$ that satisfies
\be\label{waffle} 2\cosh \frac{b''}{2}= -2 \cosh\frac{b-b'}{2} -\kappa \ee
or
\be\label{affle}\kappa=-2\cosh\frac{b-b'}{2}-2\cosh\frac{b''}{2}. \ee
Eqn.~(\ref{waffle}) does not determine the sign of $b''$, and we can pick $b''>0$ as a convention.

Let $\ell_b$ be the
the boundary of $Y$ whose monodromy is $U_0$.  Then $\ell_b$  lifts on $H$ to a $U_0$-invariant geodesic $\hat\ell_b$.   For $\hat \ell_b$ 
to be $U_0$-invariant is equivalent to saying
that its virtual endpoints on $\partial H$ are $U_0$-invariant.   A hyperbolic element of $\SL(2,\R)$, such as $U_0$, has precisely two fixed points on $\partial H\cong \RP^1$.
These fixed points correspond to the eigenvectors of the $2\times 2$ matrix $U_0$, since if
\be\label{piffle} U_0\begin{pmatrix} u\cr v\end{pmatrix}=\lambda  \begin{pmatrix} u\cr v\end{pmatrix},\ee
for some $\lambda$, this means that $U_0$ leaves invariant the point in $\RP^1$ that corresponds to $\begin{pmatrix}u\cr v\end{pmatrix}.$  The  endpoints  of $\hat\ell_b$
are thus the points on $\RP^1$ that correspond to eigenvectors of $U_0$.   

The eigenvectors of $U_0$ are $\begin{pmatrix} 1\cr 0\end{pmatrix}$ and $\begin{pmatrix} -\frac{\kappa}{2\sinh \frac{b}{2}}\cr 1\end{pmatrix}$.
So the endpoints of the geodesic $\hat\ell_b$ corresponding to $U_0$ are the points  with $x=\infty$ and $x=-\kappa/(2\sinh \frac{b}{2})$, respectively.

 Similarly, the eigenvectors of $V_0$ are $\begin{pmatrix} 0\cr 1\end{pmatrix} $ and $\begin{pmatrix} -2\sinh\frac{b'}{2}\cr 1\end{pmatrix}$.    So the endpoints of the
 geodesic $\hat\ell_{b'}$ corresponding to $V_0$ are the point  at $x=0$ and the point at $x=-2\sinh\frac{b'}{2}$.
 
 The  endpoints of the third geodesic $\hat\ell_{b''}$ 
similarly correspond to the  eigenvectors of $W_0$, which 
 can be found using the explicit formula (\ref{wurdu}).   The endpoints of $\hat\ell_{b''}$  are at  $x=-(e^{b'/2}+e^{(b\pm b'')/2})$.

At this point, we can verify that the matrices $U_0,V_0,W_0$  really are the monodromies of a three-holed
sphere $Y$  with hyperbolic metric (and not of a flat $\SL(2,\R)$ connection in a different component  of the moduli space).   If such a $Y$ does exist,
then, after we lift to the upper half-plane, the geodesics $\hat\ell_b$, $\hat\ell_{b'}$, and $\hat\ell_{b''}$ are nonintersecting and moreover they are not ``nested'';
the cyclic ordering of the endpoints of the three geodesics on $\partial H=S^1$ must be as shown in fig.~\ref{fig:UHP}(b), up to a possible permutation of the three geodesics.
A sufficient condition is to take $b,b'$ both positive.  (As noted above, there is no loss of generality to assume that also $b''>0$.)  
From the above formulas, the endpoints of $\hat\ell_b$ are more positive than those of $\hat\ell_{b'}$ which in turn are more positive than those of $\hat\ell_{b''}$.
So  the endpoints are arranged appropriately and we do get a hyperbolic metric on a three-holed sphere for any positive boundary lengths $b,b',b''$.

\subsection{Computation}\label{computation}

\begin{figure}
 \begin{center}
   \includegraphics[width=4in]{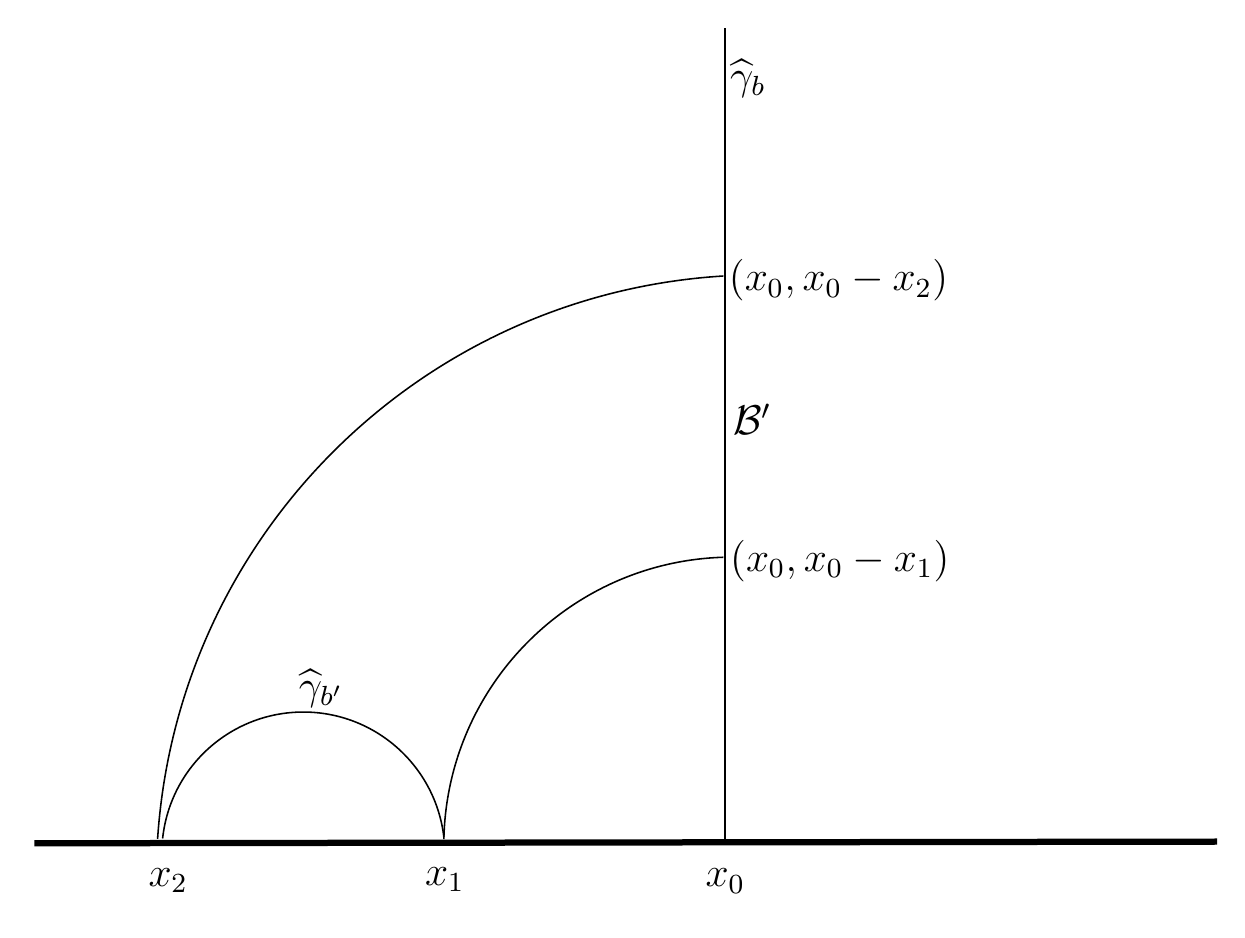}
 \end{center}
\caption{\small Here  the geodesic $\h\gamma_b$ is the vertical straight line $x=x_0$ in $H$ and $\h\gamma_{b'}$ is a semicircle with endpoints at $(x,y)=(x_1,0)$ and $(x_2,0)$.
${\mathcal B}'$ consists of points $p\in \h\gamma_b$ such that the orthogonal geodesic $\ell_p$ to $\h\gamma_b$ at $p$ intersects 
$\h\gamma_{b'}$.
${\mathcal B}'$ is an open interval; the endpoints of its closure are the points $p$ such that $\ell_p$ has an endpoint in common with $\h\gamma_{b'}$.
These are the points  $(x,y)=(x_0,x_0-x_1)$
and  $(x,y)=(x_0,x_0-x_2)$.    The distance between  those two points in the metric on $H$ is $\log  (x_0-x_2)/(x_0-x_1)$. \label{Geodesics}}
\end{figure} 

We will now compute the function $\TT(b,b',b'')$ that was introduced in appendix \ref{sumrule}.  Consider a three-holed sphere $\Lambda$ with geodesic
boundaries $\gamma_b,\gamma_{b'},\gamma_{b''}$ of lengths $b,b',b''$.   The universal cover of $\Lambda$ can be identified as a region in the upper half-plane $H$ bounded
by geodesics  $\h\gamma_b$, $\h\gamma_{b'}$, and $\h\gamma_{b''}$, which are obtained respectively by unwrapping $\gamma_b,\gamma_{b'}, \gamma_{b''}$.

Up to an $\SL(2,\R)$ transformation, we can identify $\h\gamma_b$ as a vertical straight line in $H$ located at, say, $x=x_0$.  In the same description,
$\h\gamma_{b'}$ will be a (Euclidean) semi-circle in  $H$ with endpoints at, say, $x_1, x_2$.    Since $\h\gamma_b$ and $\h\gamma_{b'}$ do not intersect,
we can assume (up to relabeling and reversal of orientation) that the endpoints are arranged with $x_2<x_1<x_0$ (fig.~\ref{Geodesics}).   We want to compute the measure
of the set $\B'\subset \h\gamma_b$ defined by saying that $x\in \B'$ if the orthogonal geodesic $\ell_p$ to $\h\gamma_b$ at $p$ meets $\h\gamma_{b'}$.   From the figure,
it is fairly easy to see that  $\B'$ is an open interval.  The endpoints of the closure of this interval are points $s_1,s_2\in \h\gamma$ with the property that $\ell_{s_1} $ and $\ell_{s_2}$
do not actually intersect $\h\gamma'$, but rather have an endpoint on $\partial H$ that coincides with one of the endpoints of $\h\gamma_{b'}$.

From this description, and the fact that geodesics in $H$ are Euclidean semicircles, it follows that the $(x,y)$ coordinates of the points $s_1,s_2$
are $(x_0,  x_0-x_1) $ and $(x_0,x_0-x_2)$.     $\mu(\B')$ is  simply the distance between these points in the hyperbolic metric of eqn.~(\ref{monof}).  A simple integration gives
 $\mu(\B')= \log (x_0-x_2)/(x_0-x_1)$.  (To express this more invariantly, the ratio $(x_0-x_2)/(x_0-x_1)$ is a cross-ratio  of the four points $x_0,x_1,x_2,\infty$,
 where $\infty$ is the second endpoint of $\hat\gamma_b$.) 
 
 To turn this into  a formula for $\TT(b,b',b'')$, we use the values of $x_0,x_1,x_2$ that were computed in appendix \ref{uhp} for a three-holed sphere with boundary lengths
 $b,b',b''$.    Thus
 $x_0=-\kappa/2\sinh \frac{b}{2}=(\cosh\frac{b''}{2}+\cosh\frac{b-b'}{2})/\sinh\frac{b}{2}$,
 $x_1=0$, $x_2=-2\sinh  \frac{b'}{2}$.  
 Hence
 \be\label{polygo} \TT(b,b ',b'')=\log\frac{x_0-x_2}{x_0-x_1}=\log\frac{\cosh\frac{b''}{2}+\cosh\frac{b+b'}{2}}{\cosh  \frac{b''}{2}+\cosh\frac{b-b'}{2}}, \ee
 which is Mirzakhani's   result. 
 
 In  using this formula, Mirzakhani differentiated  the recursion relation (\ref{recurMirz}) with respect to $b$, so as to replace the functions $\DD$ and $\TT$ with trigonometric
 functions.   We will see that in the supersymmetric case, the integration over the odd moduli automatically gives us trigonometric functions.

  \subsection{Some Supergeometry}\label{superg}
  
 To generalize this computation to the case with $\N=1$ supersymmetry, we will have to be familiar with some aspects of the supersymmetric analog $\hat H$ of the upper
 half-plane.  Some old and new references on this material are \cite{Rabin,Rosly,Baranov:1987cg,Pennerone,Pennertwo}.  
 
 In section \ref{whypin}, we defined $\OSp(1|2)$ as the symmetry group of a copy of $\R^{2|1}$, parametrized by bosonic and fermionic coordinates $u,v|\uptheta$,
 that preserve the symplectic form $\hat\omega=\d u\d v+\frac{1}{2}\d\uptheta^2$.   If we view $u,v|\uptheta$ as homogeneous  coordinates, then they
 parametrize a copy of $\RP^{1|1}$.    This will be the conformal boundary $\partial\h H$ of the supersymmetric analog $\h H$ of the upper half-plane.  
 
 $\RP^{1|1}$ can be parametrized by ordinary coordinates $x|\theta$, defined by $x=u/v$, $\theta=\uptheta/v$, along with a divisor at infinity (which corresponds to the case
 $v=0$).    The Lie algebra $\osp(1|2)$ is represented in these coordinates as follows:\footnote{These vector fields actually generate the ``opposite'' Lie superalgebra
 to the algebra of matrices in eqns. (\ref{antic}) and (\ref{comms}).   The opposite superalgebra  is defined by reversing the signs of commutators and leaving the
 anticommutators unchanged.    (The superalgebra $\osp(1|2)$ and its opposite are equivalent under $(e,f,h,q_1,q_2)\to (-f,-e,h,q_2,q_1).)$}
 \begin{align}\label{liealg}   e & =\partial_x \cr
                                           h& = 2x\partial_x +\theta\partial_\theta \cr
                                            f& = -x^2\partial_x -x\theta\partial_\theta \cr
                                           q_1&= -\partial_\theta +\theta\partial_x \cr
                                            q_2&=x(\partial_\theta-\theta\partial_x). \end{align}
 
 As coordinates on $\partial \h H$, $x$ and  $\theta$  are  real.   To describe $\h H$ and the action on it of $\osp(1|2)$, we simply complexify these formulas.
 Thus we replace the real  even and odd coordinates $x|\theta$ with complex even and odd coordinates $z|\vartheta$, with $z$ constrained by $\mathrm{Im}\,z>0$.      The action of $\osp(1|2)$
 is given by the same formulas as before, with $x|\theta$ replaced by $z|\vartheta$:  
  \begin{align}\label{liex}   e & =\partial_z \cr
                                           h& = 2z\partial_z +\vartheta\partial_\vartheta \cr
                                            f& = -z^2\partial_z -z\vartheta\partial_\vartheta \cr
                                           q_1&= -\partial_\vartheta +\vartheta\partial_z \cr
                                            q_2&=z(\partial_\vartheta-\vartheta\partial_z). \end{align}  
 For our purposes here,\footnote{In superstring theory, a super Riemann surface must be understood in a more subtle way, not simply
 as a real supermanifold.   This is because the holomorphic and antiholomorphic spin structures and odd moduli vary independently (Type II superstrings) or because
 the supersymmetric structure is purely holomorphic (the heterotic string).  See for example section 5 of \cite{Wit}.  Our present problem is more straightforward,
 because as long as we are on an orientable  two-manifold, all supermanifolds we encounter can be understood as real supermanifolds.
 On an unorientable two-manifold, one has to extend $\OSp(1|2)$ in such a way that the odd variables no longer carry a real structure (see section \ref{whypin}),
 and then some of the issues that arise in superstring theory do become relevant.} a super Riemann surface is a real supermanifold of dimension $2|2$.   The complex conjugate coordinates  $\bar z|\bar\vartheta$
 transform under $\osp(1|2)$ in the same way as $z|\theta$; one can simply take the complex conjugate of all formulas in eqn.~(\ref{liex}).  
 
One point about this definition is that the bulk supermanifold $\h H$ is of real dimension $2|2$, while its conformal ``boundary'' is of dimension $1|1$, and thus is
of codimension $1|1$, as opposed to the codimension $1|0$ of an ordinary boundary.   That might come as a surprise, but this behavior is actually typical when  holographic
dualities are described in superspace.   Dimension $2|2$ is appropriate to realize $\OSp(1|2)$ as a  supergroup of  symmetries  in bulk,
and dimension $1|1$  is appropriate  to realize $\OSp(1|2)$ as a supergroup of superconformal symmetries on the boundary.   The relation between $\h H$ and $\partial \h H$
was described from a different perspective in \cite{Pennertwo}. 

Now, recall that in the bosonic case, any two points $p_1$ and $p_2$ in $\partial  H$, say the points with $x=x_1$ or with $x=x_2$, are  the endpoints
of a unique geodesic $\gamma$ in $H$.   Using  the embedding  of  $H$ in $\h H$, we embed $\gamma$ in $\h H$ and declare it to be the geodesic
connecting the points $x_1|0$ and $x_2|0$.  To generalize this to boundary points with nonzero $\theta$,   we just note that
any desired pair $x_1|\theta_1$ and $x_2|\theta_2$  can be reached from $x_1|0$ and $x_2|0$ by an $\OSp(1|2)$ transformation  $g$ (which moreover
is unique up to a symmetry of $\gamma$).   We declare $\tilde \gamma=g(\gamma)$ to be  the geodesic in $\h H$ with endpoints $x_1|\theta_1$  and  $x_2|\theta_2$.  
It is possible to work  out an explicit formula for $\tilde\gamma$, but we will not need it;  our explicit  calculations will only  involve the endpoints of geodesics, and
special geodesics at $\vartheta=\bar\vartheta=0$.


In the bosonic world,  if $\gamma_1$  and   $\gamma_2$ are  geodesics with endpoints  at $x=a_1$  and $x=a_2$, respectively, then we say that $\gamma_1$ and $\gamma_2$
have a common endpoint if $a_1=a_2$.      In the supersymmetric case, if $\gamma_1$ and $\gamma_2$ have respective endpoints $a_1|\theta_1$ and $a_2|\theta_2$,
the naive generalization of this is to require $a_1=a_2,$ $\theta_1=\theta_2$.    This is too strong, however.   It is usually more useful to impose only the single condition
\be\label{onecond} a_1-a_2-\theta_1\theta_2  = 0,\ee
which for most purposes is the closest analog of ``having a common endpoint'' in the bosonic case.

An alternative way to state this condition is as follows.   The vector field $D_\theta=\partial_\theta+\theta\partial_x$  is not $\OSp(1|2)$-invariant,  but it is $\OSp(1|2)$
invariant up to a transformation $D_\theta\to e^{f(x|\theta)}D_\theta$. (More specifically, $D_\theta$  transforms as a superconformal primary of  weight $-1/2$,
a fact that is used in constructing superconformally invariant Lagrangians.)   The orbits generated   by $D_\theta$  are invariant under $D_\theta\to e^f D_\theta$,
so they are superconformally invariant.
Concretely the orbit containing the point $x_1|\theta_1$ is parametrized by an odd variable $\alpha$ as follows:
\be\label{nocord} x|\theta=x_1+\alpha\theta_1|\theta_1+\alpha. \ee
These orbits are said to be the ``leaves'' of an ``unintegrable foliation.''     For our purposes, the relevant point is that the condition
$a_1-a_2-\theta_1\theta_2=0$ is  equivalent to saying that $a_1|\theta_1$ and $a_2|\theta_2$  are contained in the same leaf.  Indeed, if $a_1-a_2-\theta_1\theta_2=0$,
then the leaf through the point $a_1|\theta_1$ contains the point $a_2|\theta_2$, as we see by setting $\alpha=\theta_2-\theta_1$ in eqn.~(\ref{nocord}).

A similar  idea is actually built into the naive statement ``$Y$ is a super Riemann surface with geodesic boundary.''
A super Riemann surface in the sense  of our present discussion has real dimension $2|2$.   Its boundary should have codimension $1|0$, so should be a supermanifold
of dimension $1|2$.   But a geodesic has dimension $1|0$.   The point is that the vector fields on $\h H$ defined by  $D_\vartheta=\partial_\vartheta+\vartheta \partial_z$
and $D_{\bar\vartheta}=\partial_{\bar\vartheta}+\bar\vartheta \partial_{\bar z}$ generate a superconformally invariant  foliation whose leaves are of dimension $0|2$.
Hence a geodesic $\gamma$  (or similarly any generic curve of dimension $1|0$) has a canonical dimension $1|2$ thickening, consisting of all the leaves
that pass through $\gamma$.    It is this thickening that is really the boundary of a super Riemann surface.

In $H$, which is of dimension 2, it is natural for two geodesics $\gamma_1$ and $\gamma_2$, each of which has codimension 1, to intersect.   (Of course, not
all pairs of geodesics  in $H$ do intersect.)    What is a good analog in $\h H$ of the statement that two geodesics in $H$ intersect?
Since geodesics in $\hat H$ have codimension  $1|2$ while $\hat H$ has dimension $2|2$, it is nongeneric for two geodesics in $\h H$ to have a point in common.
The useful notion of ``intersection of geodesics'' in a super Riemann surface is to say that two geodesics $\gamma_1$ and  $\gamma_2$ intersect if the thickening
of $\gamma_1$, in the sense described in the last paragraph, intersects $\gamma_2$.  This condition can be  stated more symmetrically by saying  that the foliation generated  by $D_\vartheta$ and $D_{\bar\vartheta}$ has a leaf that intersects both $\gamma_1 $ and $\gamma_2$ in the naive sense.

Finally, we need to know what it means for two geodesics in $\h H$ to be orthogonal.      If  a  geodesic $\gamma\subset \h H$ actually lies  in $H\subset \h H$,
and $p$ is a  point in $\gamma$, then by   the geodesic $\ell_p$ that is orthogonal to $\gamma$ at $p$, we just mean the geodesic in $H\subset \h H$ that
is orthogonal to $\gamma$ at $p$, in the classical sense.  Any case  can be reduced to this case  by  an $\OSp(1|2)$  transformation,  since  by definition
any geodesic in $\h H$ can be mapped to $H$ by an $\OSp(1|2)$ transformation.                 
  
\subsection{The Supersymmetric Recursion Relation}\label{superrecur}

Now we consider the same setting as in appendix \ref{sumrule}, but in the supersymmetric case.  Thus, $Y$ is a super Riemann surface with a specified geodesic
boundary $\gamma$ of length $b$, and additional geodesic boundaries $\gamma_i$, $i\in I$, of lengths $b_i$.   $\gamma$  has a natural Riemannian measure, since
its lift $\hat \gamma$ to $\h H$ is an orbit of an $\SO(1,1)$ subgroup of $\OSp(1|2)$.      The total measure of $\hat\gamma$ is $b$.

For $p\in \gamma$, let $\ell_p$ be the normal to $\gamma$ at  $p$.   Just as in the bosonic case, we decompose $\gamma$ as a union of disjoint subsets $\A$, $\B$, and $\CC$
according to whether  $\ell_p$ first intersects itself or returns to $\gamma$; first intersects one  of the $\gamma_i$; or does  neither.   The last case corresponds again to a  set of
measure 0, and in the first two cases, just  as before,\footnote{Upon reducing modulo the odd variables (the fermionic coordinates of $Y$ and also the fermionic
moduli), $Y$ reduces to a bosonic Riemann surface $Y_0$ and the same argument as before associates to $\ell_p$ a three-holed sphere $\Lambda_0\subset Y_0$
with geodesic boundaries $\gamma_a$ (one of which is the original $\gamma$).   Picking any map $i:Y_0\to Y$ that reduces to the identity modulo the odd variables, we embed  the 
 $\gamma_a$
in $Y$ as closed loops $i(\gamma_a)$, which reduce modulo the odd variables to the original geodesics $\gamma_a$.   The $i(\gamma_a)$ are in general not geodesics,
but in the homotopy class of each of them 
(and differing from  it only by nilpotent
terms,  proportional to the odd variables) is a unique geodesic.    Indeed, the monodromy of the hyperbolic flat $\OSp(1|2)$ connection around any of the $\gamma_a$
is a hyperbolic element $h_a\in\OSp(1|2)$.   Pass to $\h\H$ by replacing $Y$ with its universal cover.  A hyperbolic element of $\OSp(1|2)$, such as $h_a$,
leaves fixed a unique geodesic in $\h \H$, namely the geodesic  that connects its two fixed points in $\RP^{1|1}$.    This descends to the desired geodesic in $Y$.  The 
three geodesics obtained this way (one of which is $\gamma$) are the boundaries of the desired three-holed sphere $\Lambda\subset Y$.} $\ell_p$ is contained in a distinguished three-holed sphere $\Lambda$ one of whose
geodesic boundaries is $\gamma$.   

Thus after decomposing $\A$ and $\B$ as in (\ref{zotto}) and (\ref{otto}), we have the analog of eqn.~(\ref{lotto}): 
\be\label{wotto} b=\sum_{\Lambda\in \Upsilon}\mu(\A_\Lambda)  +\sum_{i\in I,\Lambda\in \Upsilon_i}\Big(\mu(\B_{i,\Lambda}) + \mu(\A_\Lambda)\Big).  \ee

As before, the measures $\mu(\A_\Lambda)$ and $\mu(\B_{i,\Lambda})$ that appear in these formulas depend only on $\Lambda$ and not on anything else about $Y$.
Thus they come from universal functions.   The main difference is that now a three-holed sphere has two  odd moduli  as well as the three boundary lengths, and
we will have to  take these into account.

Let $\Lambda_0$ be a  three-holed  sphere with geodesic boundaries $\gamma,\gamma',\gamma''$ of lengths $b,b',b''$, and  with odd moduli $\alpha,\beta$.
For fixed $b,b',b''$, the odd variables $\alpha$ and $\beta$ parametrize a supermanifold $\M_{b,b',b''}$ of dimension $0|2$. As before, $\gamma$ is the union of sets $\A_0$, $\B_0'$, and $\B_0''$ (plus a set of measure 0) defined by the behavior of orthogonal geoedesics.  Writing
$\h\DD(b,b',b''|\alpha,\beta)$,  $\h\TT(b,b',b''|\alpha,\beta)$,  and\footnote{$\h\UU$ is obtained from $\h\TT$ by exchanging $b'$ with $b''$ and suitably transforming
$\alpha$ and $\beta$.  We will soon integrate out $\alpha$ and $\beta$, so we avoid writing explicit formulas as this stage.} $\h\UU(b,b',b''|\alpha,\beta)$  for   the measures  of $\A_0$, $\B_0'$, and  $\B_0''$, we get a sum rule
just like eqn.~(\ref{oggo}), with new universal functions:
\be\label{ormby} b=\h\DD(b,b',b''|\alpha,\beta)+\h\TT(b,b',b''|\alpha,\beta)+\h\UU(b,b',b''|\alpha,\beta). \ee

Just as in eqn.~(\ref{recurMirz}), inserting this identity in the definition of the volume as an integral over the supermoduli space leads to a recursion relation that involves
a sum over all ways that $Y$ can be built by  gluing a three-holed sphere $\Lambda$ to some other super Riemann surface $Y'$.
The main difference is that after specifying the boundary lengths $b,b',b''$ of $\Lambda$, there is a  still a moduli space $\M_{b,b',b''}$ of flat
connections on $\Lambda$ over which we have to integrate.  We  will describe in section \ref{finalcal}   the integration measure $ \mu$ for this integral.   Including some more trivial factors that are explained in a moment, the analog of eqn.~(\ref{recurMirz})  is then 
\begin{align}\label{recursuper}
bV_{g}(b,B) &=\frac{1}{2} \int_0^\infty b'\mathrm{d}b'\,b''\mathrm{d}b''\,\int_{\M_{b,b',b''}}\hspace{-18pt}\d\mu\,\h\DD(b,b',b''|\alpha,\beta)\left(V_{g-1}(b',b'',B) + \sum_{\text{stable}}V_{h_1}(b',B_1)V_{h_2}(b'',B_2)\right)\notag\\
&\hspace{20pt} +2\sum_{k = 1}^{|B|}\int_0^\infty b'\mathrm{d}b' \,\int_{\M_{b,b_k,b'}}\d\mu\,\Big(b-\h\UU(b,b_k,b'|\alpha,\beta)\Big)\,V_g(b',B\setminus b_k).
\end{align}

 In the sum over spin structures, we may or may not include a factor $(-1)^\zeta$.
However, as explained in section \ref{superjtnot}, the more interesting case is that we do include such a factor (otherwise the volumes are expected to all vanish).  We
have written the formula for this case.  
With a factor $(-1)^\zeta$ included, if 
any of the circles on which we glue $\Lambda$ to $Y'$ has a spin structure of Ramond type, then the sum over the spin structure ``orthogonal'' to that circle
 vanishes, as we learned in section
\ref{bulk}.  So we assume that all of those circles have NS spin structure.  Similarly, we assume the same for  all external boundaries.
 A factor of 2 in the  term proportional to $b-\h\UU$ has the following origin.   We define the volumes to include a sum over spin structures, which includes a sum over spin structures
 ``orthogonal'' to the boundary components (that is, the spin bundle is trivialized on $\partial Y$, and spin structures are considered equivalent only if there is an
 equivalence between them that respects this trivialization).   In the gluing that leads to the $b-\h\UU$ term in eqn.~(\ref{recursuper}), but not in the gluing that leads
 to the $\h\DD$ term, the spin structure sum on the left has an extra factor of 2 compared to the spin structure sum on the right.   We compensate for this with an explicit
 factor of 2 multiplying the $b-\h\UU$ term.

To  get a formula for the volumes in purely bosonic terms, we should integrate  over $\M_{b,b',b''}$.    We define
\begin{align}\label{integrated} \int_{\M_{b,b',b''}} \d \mu \, \h\DD(b,b',b''|\alpha,\beta)&=\cD(b,b',b'') \cr
                                              \int_{\M_{b,b',b''}} \d \mu \, \h\TT(b,b',b''|\alpha,\beta)&=\cT(b,b',b'').  \end{align}
Once we  integrate over the odd variables, $\h\TT$ and $\h\UU$ differ only by exchange of $b'$ and $b''$,  so 
\be\label{flipped}\int_{\M_{b,b',b''}} \d\mu\,\h\UU(b,b',b''|\alpha,\beta)=\cT(b,b'',b') \ee
The supersymmetric recursion relation then reduces to
\begin{align}\label{recurexp}
bV_{g}(b,B) &=\frac{1}{2} \int_0^\infty b'\mathrm{d}b'\,b''\mathrm{d}b''\,\cD(b,b',b'')\left(V_{g-1}(b',b'',B) + \sum_{\text{stable}}V_{h_1}(b',B_1)V_{h_2}(b'',B_2)\right)\notag\\
&\hspace{20pt} -2\sum_{k = 1}^{|B|}\int_0^\infty b'\mathrm{d}b' \,\cT(b,b',b_k)\,V_g(b',B\setminus b_k).
\end{align}
In writing this equation, we used that the moduli space $\M_{b,b',b''}$ has zero volume, like all of the genus 0 moduli spaces, so the integral over this moduli space of a constant function vanishes:
\be\label{lipped}\int_{\M_{b,b',b''}} \d\mu\cdot b =0. \ee
This fact has another useful implication. When we integrate the sum rule (\ref{wotto}) over $\M_{b,b',b''}$, we learn that
\be\label{simplerel}0= \cD(b,b',b'')+\cT(b,b',b'')+\cT(b,b'',b'). \ee
Therefore, to make eqn.~(\ref{recurexp}) explicit, it suffices to compute $\cT(b,b',b'')$.

\subsection{Final Calculations}\label{finalcal}

To compute the function $\h     \TT$, we consider a three-holed sphere $\Lambda$ with the monodromies of eqn.~(\ref{twom}), namely 
\begin{equation}\label{nwom}\newcommand*{\temp}{\multicolumn{1}{r|}{}}
U_0=\vnu_b\left(\begin{array}{cccccc}
e^{b/2} &\kappa    \negthinspace\negthinspace\negthinspace &\temp & 0\\
0 &e^{-b/2}\negthinspace \negthinspace \negthinspace&\temp &0\\
 \cline{1-4}
0 &0\negthinspace \negthinspace \negthinspace&\temp &\vnu_b\\

\end{array}\right)\exp(\xi q_1), ~~~~
V_0=\vnu_{b'}\left(\begin{array}{cccccc}
e^{-b'/2} &0 \negthinspace\negthinspace\negthinspace &\temp & 0\\
1 &e^{b'/2}\negthinspace \negthinspace \negthinspace&\temp &0\\
 \cline{1-4}
0 &0\negthinspace \negthinspace \negthinspace&\temp &\vnu_{b'}\\

\end{array}\right)\exp(\psi q_2),
\end{equation}
(again we replaced $a,b,c$ with $b,b',b''$) and $W_0=V_0^{-1}U_0^{-1}$.   The parameter $\kappa$ can be expressed in terms of the third boundary
parameter $b''$ as  in eqn.~(\ref{noflox}):
\be\label{loflox}\kappa =- 2\cosh\frac{b''}{2}-2\cosh\frac{b-b'}{2} +\psi\xi(e^{b/2}\vnu_{b'}+e^{b'/2}\vnu_b). \ee

Let $\gamma_b$, $\gamma_{b'}$, and $\gamma_{b''}$ be geodesic boundaries of $\Lambda$ with respective monodromies $U_0$, $V_0$, and $W_0$.  In the universal
cover of $\Lambda$,  these geodesics  lift to ``unwrapped'' geodesics $\hat\gamma_b$, $\hat\gamma_{b'}$, and $\hat\gamma_{b''}$.  The endpoints of these
unwrapped  geodesics can be found, by the same logic as in the bosonic case, from  the eigenvectors of the monodromies acting on $u,v|\uptheta$.   To be more precise,
each monodromy has two bosonic eigenvectors, which are  the endpoints of the corresponding  unwrapped geodesic.
(The monodromy also has a fermionic eigenvector, but it will not play a role.)   

For example, one bosonic eigenvector of $U_0$ is $\begin{pmatrix} u \cr v\cr  \uptheta\end{pmatrix}= \begin{pmatrix}  1\cr 0 \cr 0\end{pmatrix}$.   This corresponds to an endpoint
of $\hat\gamma_b$ at infinity.   The second bosonic eigenvector is $\begin{pmatrix} -\kappa/(2\sinh\tfrac{b}{2}) \cr 1\cr \xi (1-\updelta_b e^{-b/2})^{-1} \end{pmatrix}$.
So the second endpoint of $\hat\gamma_b$ is
\be\label{secondendpt} x_0|\theta_0 = -\kappa/(2\sinh \tfrac{b}{2})|\xi(1-\vnu_b e^{-b/2})^{-1}. \ee

Similarly, the  bosonic eigenvectors of $V_0$ are $\begin{pmatrix} 0 \cr 1\cr 0\end{pmatrix}$  and  $\begin{pmatrix}-2\sinh  b'/2 \cr 1\cr \psi(e^{b'/2}+\vnu_{b'}  ) \end{pmatrix}$, so $\hat\gamma_{b'}$ has endpoints 
\begin{align}\label{bpfirst} x_1|\theta_1 & =0|0 \cr x_2|\theta_2 &= -2\sinh (\tfrac{b'}{2})| \psi(e^{b'/2}+\vnu_{b'}).\end{align}

Before making use of these formulas to compute the function $\h\TT$, it is convenient to first make a supersymmetry transformation $x|\theta\to x+\theta_0\theta|\theta-\theta_0$,
where $\theta_0$ was defined in eqn.~(\ref{secondendpt}).  After this transformation, the endpoints of $\gamma_b$ are at infinity and  at
\be\label{firstprime}x_0'|\theta_0' = -\kappa/(2\sinh \tfrac{b}{2})|0 . \ee
Accordingly, in this description, $\hat\gamma_b$ is simply a vertical straight line in $H\subset \h H$, and all of the orthogonals to $\h\gamma_b$ are likewise contained
in $H$.   Thus, we do not need to consider more general geodesics in $\h H$.

The supersymmetry transformation $x|\theta\to x+\theta_0\theta|\theta-\theta_0$ maps the endpoints of $\h\gamma_{b'}$ to
\begin{align}\label{bppfirst}  x_1'|\theta_1'& = x_1+\theta_0\theta_1|\theta_1-\theta_0\cr
 x_2'|\theta_2' &= x_2+\theta_0\theta_2|\theta_2-\theta_0. \end{align}

The function $\h\TT$ is the length of the segment of $\B'\subset \hat\gamma_b$ consisting of points $p$ such that the orthogonal geodesic $\ell_p$ to
$\hat\gamma_b$ at $p$ intersects $\hat\gamma_{b'}$.   The endpoints of $\B'$ are the points  $p\in\hat\gamma_b$ such that  $\ell_p$
and $\gamma_{b'}$  satisfy the  condition (\ref{onecond}), which says that $\ell_p$ and $\hat\gamma_{b'}$ just barely fail to intersect.
   This  happens if the endpoints of $\ell_p$ (which are at $\theta=0$, since  $\ell_p$ is
contained in $H$) are  $x_1'|0$ or $x_2'|0$.   Since $\ell_p$ is a semicircle   in  $H$ centered at $x=x_0$, the same reasoning as in the bosonic case tells us that the endpoints of 
$\B'$ are at  the  points in $H$ with  $(x,y)= (x_0',x_0'-x_1')$ and $(x,y)=(x_0',x_0'-x_2')$.     Hence rather as in the bosonic case,
the length  of $\B'$ is $\h\TT= \log (x_0'-x_2')/(x_0'-x_1')$.    In view of eqn.~(\ref{bppfirst}), this is equivalent to
$\h\TT=\log\left((x_0-x_2-\theta_0\theta_2)/(x_0-x_1-\theta_0\theta_1)\right)$.   Here  in fact, $(x_0-x_2-\theta_0\theta_2)/(x_0-x_1-\theta_0\theta_1)$
is a superconformal cross ratio of the points  $x_0|\theta_0$, $x_1|\theta_1$,  $x_2|\theta_2$, and a fourth  point at infinity.   So  the result is a sort of minimal
supersymmetric  generalization of the bosonic formula.  

Using the above formulas for the endpoints and eqn.~(\ref{loflox}) for $\kappa$, we get
\begin{align}\label{almostfinal}\h\TT(b,b',b''|\xi,\psi)=&\log\left(\frac{-\kappa +4\sinh\frac{b}{2}\sinh\frac{b'}{2} -\xi\psi  (e^{b/2}+\vnu_{b})(e^{b'/2}+\vnu_{b'})}{-\kappa}\right) \cr=& \log\frac{  \cosh\frac{b''}{2}+\cosh\frac{b+b'}{2}-\frac{\xi\psi}{2}\left(e^{(b+b')/2}+\vnu_b\vnu_{b'}\right)  }{   \cosh\frac{b''}{2}+\cosh\frac{b-b'}{2}+\frac{\xi\psi}{2}\left(e^{b/2}\vnu_{b'}+e^{b'/2}\vnu_b\right)    }.\end{align}

To reduce this to the purely  bosonic function $\cT(b,b',b'')$ that appears in the final form (\ref{recurexp}) of the recursion relation, we have to integrate this expression
over the odd moduli  $\xi,\psi$.   The torsion measure can be read off from\footnote{The overall sign of this measure depends on some conventions.   For example,
in the symplectic approach of appendix \ref{volsymp}, the sign depends on a choice of orientation of the fermionic bundle $V$.
Analogous issues arise in the torsion.  Reversing this sign would multiply  the volumes
by $(-1)^\chi$, which for an orientable surface is the same as $(-1)^n$, where $n$ is the number of boundary components.}  eqn.~(\ref{torformt}) (where we substitute $a_t,b_t,c_t|\xi_t,\psi_t\to b,b',b''|\xi,\psi$):
\be\label{wendo}\tau =\frac{\vnu_b\vnu_{b'}}{4} e^{-(b+b')/4}(e^{b''/4}-\vnu_{b''} e^{-b''/4})[\d\xi\,\d\psi]. \ee
But according to eqn.~(\ref{natnorm}), if we want  to  define volumes that satisfy the particularly  natural normalization of eqn.~(\ref{gilfo}), the measure we must use is not
$\tau$ but
\be\label{newmeasure}\mu=(2\pi)^\chi\tau=\frac{\tau}{2\pi},\ee
where we use the fact that $\chi=-1$ for a three-holed sphere.
So
\begin{align}\label{glos}\cT(b,b',b'')&=\int_{\M_{b,b',b''}}\d\mu\,\h\TT(b,b',b''|\xi,\psi) \cr 
&=-\frac{\vnu_b\vnu_{b'}}{16\pi}\frac{(e^{(b+b')/4}+\vnu_b\vnu_{b'}e^{-(b+b')/4})(e^{b''/4}-\vnu_{b''} e^{-b''/4})}{ \cosh\frac{b''}{2}+\cosh\frac{b+b'}{2}} \cr&~~
~-\frac{\vnu_b \vnu_{b'}}{16 \pi} \frac{(e^{(b-b')/4}\vnu_{b'}+e^{-(b-b')/4}\vnu_b)(e^{b''/4}-\vnu_{b''} e^{-b''/4})}{ \cosh\frac{b''}{2}+\cosh\frac{b-b'}{2}}.  \end{align}

This formula is valid for any spin structures on the boundary of $\Lambda$.  For the application to the recursion relation (\ref{recurexp}), we want the spin structures to be all of NS type, so we take
$\vnu_b=\vnu_{b'}=\vnu_{b''}=-1$.   After a few uses of the identity $2\cosh A \cosh B=\cosh(A+B)+\cosh(A-B)$, one finds
\be\label{finalt} \cT(b,b',b'')=\frac{1}{16\pi}\left(\frac{1}{\cosh\frac{b-b'+b''}{4}}+\frac{1}{\cosh\frac{b-b'-b''}{4}}-\frac{1}{\cosh\frac{b+b'+b''}{4}}-\frac{1}{\cosh\frac{b+b'-b''}{4}}\right). \ee
From eqn.~(\ref{simplerel}), we then have also
\be\label{finald} \cD(b,b',b'')=-\cT(b,b',b'')-\cT(b,b'',b')=-\frac{1}{8\pi}\left( \frac{1}{\cosh\frac{b'+b''-b}{4}} -\frac{1}{\cosh\frac{b'+b''+b}{4}}   \right).\ee
When these  results for $\cD$ and $\cT$ are inserted in the recursion relation (\ref{recurexp}), one finds perfect agreement with the recursion relation (\ref{recurVols}) derived
from random matrix theory.

It is interesting to observe that the start of the topological recursion with $V_1(b)=-1/8$ can actually be computed similarly.   We can build a torus with a single
boundary of length $b$ by starting with a one-holed sphere of boundaries $b,b',b'$  and gluing together the two boundaries of length  $b'$ to make a single circle $C$. Repeating the derivation
that leads to eqn.~(\ref{recurexp}), assuming that spin structures are weighted with $(-1)^\zeta$, we get 
\be\label{specialcase} b V_1(b)=\int_0^\infty b' \d b'  \,\cD(b,b',b'). \ee
Here we should divide by 2 because (due to the $\mathbb{Z}_2$ symmetry of the torus with one hole) we define $V_{1}(b)$ to be one-half of the true moduli space volume, as explained in footnote \ref{footnotefactoroftwo}.   But we also get  a factor  of  2 from the sum over spin structures.    {\it A priori}, the spin structure on  $C$  could be of NS
or R type.   However,  in the theory with the factor $(-1)^\zeta$, contributions  in which the spin structure is of R type
vanish because of the usual cancellation in the sum over the spin structure ``orthogonal'' to $C$.
If the spin structure on $C$ is of NS type; then the sum  over orthogonal spin structures gives a factor of 2, compensating for the factor
of $1/2$ from the symmetry and leading to (\ref{specialcase}).   One can evaluate the integral in (\ref{specialcase})  using eqn.~(\ref{finald}), and one recovers $V_1(b)=-1/8$.

If  instead one considers the theory without the factor $(-1)^\zeta$, one can recover  the expectation that $V_1(b)=0$.
For this, we have to allow both types of spin structure on the circle $C$.   The volume will then satisfy
\be\label{specialcaseR} b \t V_1(b)=\int_0^\infty b' \d b'  \,\left(\cD(b,b',b')- \t\cD(b,b',b')\right), \ee
where $\t\cD(b,b',b'')$ is  a function similar to $\cD(b,b',b'')$, but for a three-holed sphere with spin structures of types NS, R, R.
The minus sign has the same origin  as the factor $(-1)^{w_\Ra} $ in eqn.~(\ref{torformt}); it comes from the factor $-\vnu$ in the formula (\ref{waus}) for the torsion of a circle.
We have, as in eqn.~(\ref{simplerel}), $\t\cD(b,b',b'')=-\t \cT(b,b',b'')-\t\cT(b,b'',b') $, where now $\t \cT(b,b',b'')$ is obtained by setting $\vnu_b=-1$, $\vnu_{b'}=\vnu_{b''}=1$
in eqn.~(\ref{glos}).    We get 
\be\label{modt} \t\cT(b,b',b'')=\frac{1}{16\pi}\left(-\frac{1}{\cosh\frac{b-b'+b''}{4}}+\frac{1}{\cosh\frac{b-b'-b''}{4}}-\frac{1}{\cosh\frac{b+b'+b''}{4}}+\frac{1}{\cosh\frac{b+b'-b''}{4}}\right), \ee
leading to $\t\cD(b,b',b'')=\cD(b,b',b'')$, and therefore to the expected $\t V_1(b)=0$. Using the equality of $\cD$ and $\t\cD$ and the vanishing of volumes in genus zero, a recursion similar to (\ref{recurexp}) implies that all volumes vanish in the theory without $(-1)^\zeta$.

\section{Comments On Nonperturbative Effects}\label{app:nonperturbative}
In the main text of the paper, we focused on aspects of JT gravity and of matrix integrals that are perturbative in $e^{-S_0}$. In this appendix, 
we make a few comments on nonperturbative effects. For the most part, we will consider an Atland-Zirnbauer-type matrix integral, with measure of the form
\be
\prod_{i<j}|\lambda_i-\lambda_j|^\upbeta \prod_i \lambda_i^{\frac{\upalpha-1}{2}}e^{-L V(\lambda_i)}\d\lambda_i, \hspace{20pt} \lambda_i>0,
\ee
and with $\upalpha = 0,1,2$ and $\upbeta = 2$. These three cases are dual to JT supergravity 
theories without a crosscap divergence. We will also comment on a fourth case with this property in section \ref{sec:fourthcase}.

\subsection{The Bessel Curve}

A useful starting point is the solvable ``Bessel'' spectral curve $y^2 = -1/x$, or equivalently $\rho_0 = 1/(\pi\sqrt{x})$. This spectral curve describes the small $x$ region of any Altland-Zirnbauer ensemble, including the ones dual to JT gravity. It plays the role of the ``Airy'' curve $y^2 = -x$ in the standard Dyson ensembles. A matrix integral with spectral curve $y^2 = -1/x$ can be obtained as a double-scaled limit of a conventional matrix integral with $V(\lambda) = \lambda$. For this potential, the orthogonal polynomials are the $\widetilde{\upalpha}$-Laguerre polynomials, with $\widetilde\upalpha = (\upalpha-1)/2$. Using asymptotics for these functions, Nagao and Slevin \cite{nagao1993nonuniversal} computed the exact density of eigenvalues in the double-scaled limit: 
\begin{align}
\langle \rho(x)\rangle &= e^{2S_0}\left(J_{\widetilde{\upalpha}+1}(\xi)^2 - J_{\widetilde{\upalpha}}(\xi)J_{\widetilde{\upalpha}+2}(\xi) + 2 \frac{J_{\widetilde{\upalpha}}(\xi)J_{\widetilde{\upalpha}+1}(\xi)}{\xi}\right)\hspace{20pt} \xi = 2e^{S_0}\sqrt{x}.\label{NagaoSlevin}
\end{align}
This function has an asymptotic expansion for large $e^{S_0}$. The expansion contains two types of terms: odd powers of $e^{-S_0}$, and both even and odd powers of $e^{-S_0}$ multiplied by a nonperturbative term oscillating with fixed frequency. The leading terms of both types are
\be
\langle \rho(x)\rangle \approx \frac{e^{S_0}}{\pi\sqrt{x}}+\dots + \frac{\sin\left(- \frac{\pi}{2} \upalpha+4 e^{S_0}\sqrt{x} \right)}{4\pi x} +\dots.\label{leadingTerms}
\ee
The terms that form the first class of dots can be obtained from the recursive treatment of the loop equations that we descibed in section \ref{sec:loopEquations}. The oscillating terms are nonperturbative in $e^{-S_0}$ and are not obtained this way.

When $\upalpha \in\{0,2\}$, we argued in section \ref{sec:higherOrdersZero} that after double-scaling, the only nonzero terms in the genus expansion were the disk $R_0(x)$, the cylinder $R_{0}(x_1,x_2)$, and the crosscap $R_{\frac{1}{2}}(x)$. Indeed, for $\upalpha \in \{0,2\}$, the dots in (\ref{leadingTerms}) vanish and the two terms are the exact answer. We can use this to compute the resolvent $R(x) = \int_0^\infty \mathrm{d}x' \frac{\rho(x')}{x-x'}$. It is natural to define the resolvent with a branch cut along the positive real axis, where the eigenvalue density is nonzero. Then
\be
\langle R(x)\rangle = -\frac{e^{S_0}}{\sqrt{-x}} \pm\frac{1}{4x} \mp \frac{\exp\left(-4e^{S_0}\sqrt{-x}\right)}{4x}.
\ee
where the top sign is for $\upalpha = 0$ and the bottom sign is for $\upalpha = 2$. The first term is $e^{S_0}y(x)$. The second term is the crosscap contribution $R_{\frac{1}{2}}(x)$, and the final term is a nonperturbative correction. The truncation of the perturbative series is precisely consistent with the expectations of section \ref{sec:higherOrdersZero}.

When $\upalpha = 1$, the Bessel functions in (\ref{NagaoSlevin}) have a nontrivial asymptotic expansion, which implies an expansion of the resolvent that we will find useful below. To compute this, it is convenient to take a slight detour and first compute the expectation value of $Z(\beta) = \int_0^\infty \mathrm{d}x \rho(x')e^{-\beta x'}$. This can be done by inserting an integral representation
\be
J_\nu(\xi) = \frac{(\frac{1}{2}\xi)^\nu}{2\pi\mathrm{i}}\int_{\epsilon + \mathrm{i}\mathbb{R}}\exp\left(t - \frac{\xi^2}{4t}\right)\frac{\mathrm{d}t}{t^{\nu+1}}
\ee
of each of the Bessel functions in (\ref{NagaoSlevin}), with integration parameters $t_1,t_2$ for the two factors of $J$. Then $\langle Z(\beta)\rangle$ beomces an integral over $x,t_1,t_2$. The $x$ and $t_1$ integrals can be done easily. After a change of variables, the final $t_2$ integral leads to
\be
\langle Z(\beta)\rangle = \int_0^a\frac{\mathrm{d}u}{2\pi}\sqrt{\frac{a-u}{u}}e^{-u}, \hspace{20pt} a = \frac{4 e^{2S_0}}{\beta}.
\ee
To find the $e^{-S_0}$ expansion of this formula, we expand the square root in a power series in $u/a$, and integrate from zero to infinity. At each order in $1/a$, this gives a simple expression involving gamma functions. Using $R(x) = -\int_0^\infty \mathrm{d}\beta e^{\beta x}Z(\beta)$, and integrating the expansion of $\langle Z(\beta)\rangle$ term-by-term, one finds a result that can be written (for $g\ge 1$) as
\be
R_{g}(x) = \frac{(2g)!}{2^{8g}(2g-1)}{2g\choose g}^2\frac{1}{(-x)^{g+\frac{1}{2}}}.\label{RBessel}
\ee

\subsection{More General Spectral Curves}
Still with $\upbeta = 2$ and $\upalpha = 0,1,2$, we would like to generalize (\ref{leadingTerms}) beyond the curve $y^2 = -1/x$. It is convenient to start out by considering the the operator $\psi(x) = e^{-LV(x)/2}\det(x-H)$. For the Bessel curve $y^2 = -1/x$, the expectation value of $\psi(x)$ is
\be
\langle \psi(x)\rangle = (\text{const.})\frac{J_{\widetilde{\upalpha}}(\xi)}{\xi^{\widetilde{\upalpha}}} \approx (\text{const.})\left[\frac{\cos\left(2 e^{S_0}\sqrt{x} -\frac{\pi}{4}\upalpha\right)}{x^{\frac{\upalpha}{4}}} + \dots\right]\label{leadingA}
\ee  
where $\xi$ is defined in (\ref{NagaoSlevin}), and the final expression is the leading asymptotics for large $e^{S_0}$ and positive $x$.\footnote{For a conventional finite $L$ matrix integral of type $(\upalpha,2)$, the expectation value of $\det(x-H)$ is equal to the order $L$ monic polynomial from the set of polynomials orthogonal with respect to the measure $x^{(\upalpha-1)/2}e^{-LV(x)}$ on the positive real axis. For the potential $V(x) = x$, these polynomials are the $\widetilde{\upalpha}$-Laguerre polynomials. Double-scaling to obtain $y^2 = -1/x$ gives the Bessel function quoted.} For a general spectral curve, we can't hope to find an exact expression, but we would like to generalize the leading asymptotics. To do so, one can write
\be
\det(x-H) = \exp\left(\int^x dx'R(x')\right)
\ee
in terms of the resolvent $R$, and then expand in correlation functions of the resolvent using
\be
\langle e^{X}\rangle = \exp\left(\langle X\rangle + \frac{1}{2}\langle X^2\rangle_c + \dots\right).
\ee
In this expansion, terms of order $e^{S_0}$ and order one come from only three places: $R_0(x)$, $R_{\frac{1}{2}}(x)$, and $R_0(x_1,x_2)$. So to compute the expectation value of the determinant to this order, we exponentiate these terms. This leads to an approximate expression for $\langle \psi(x)\rangle$ in terms of the function $\Psi(x)$, defined by
\be
\Psi(x) = \exp\left[\text{Disk}(x) + \text{Xcap}(x) + \frac{1}{2}\text{Cyl}(x,x)\right] = (\text{const.})\frac{\exp(\text{Disk}(x))}{(-x)^{\frac{\upalpha}{4}}}.\label{twoBrances}
\ee
Here, the disk, crosscap and cylinder functions are defined for $x<0$ by
\begin{align}
\text{Disk}(x) &= \int_0^{x} \mathrm{d}x'\left(e^{S_0}R_0(x') - \frac{L V'(x')}{2}\right) = e^{S_0}\int_0^{x} \mathrm{d}x'\, y(x')\\
\text{Xcap}(x) &= \int^{x}\mathrm{d}x'\, R_{\frac{1}{2}}(x) = -\frac{\upalpha{-}1}{4}\,\log (-x) + (\text{const.})\\
\text{Cyl}(x_1,x_2) &= \int_{-\infty}^{x_1}\mathrm{d}x_1'\int_{-\infty}^{x_2}\mathrm{d}x_2'\;R_{0}(x_1',x_2') = -\log(\sqrt{-x_1} + \sqrt{-x_2}) + (\text{const.}).
\end{align}

The main complication is that that $\langle \psi(x)\rangle$ is entire, while the function $\Psi(x)$ is multi-valued. As explained in \cite{Maldacena:2004sn}, one has to sum over branch choices for $\Psi(x)$ with the right coefficients. These can be determined by matching to the formula (\ref{leadingA}) for small $x$. A shortcut for doing this is to start with (\ref{leadingA}) and simply replace the factor $\pm 2\mathrm{i} e^{S_0}\sqrt{x}$ by its generalization $\text{Disk}(x) = \pm \mathrm{i}\int_0^x \mathrm{d}x' \rho_0(x')$. This leads to the following formula for the leading asymptotics for large $e^{S_0}$ and positive $x$:
\be
\langle \psi(x)\rangle \approx (\text{const}.)\frac{\cos\left(-\frac{\pi}{4}\upalpha + \pi e^{S_0}\int_0^x\mathrm{d}x \rho_0(x)\right)}{x^{\frac{\upalpha}{4}}}+\dots
\ee
This simple replacement is possible because the crosscap and cylinder terms $R_{\frac{1}{2}}(x)$ and $R_0(x_1,x_2)$ are completely universal for double-scaled $(\upalpha,2)$ theories. So, to this order, the only dependence on the spectral curve comes from the disk amplitude, which is easy to isolate as the term in the exponential multiplying $e^{S_0}$.

So far we have just discussed the determinant operator, but the logic for the density of eigenvalues is similar. One can write $\langle \rho(x)\rangle$ in terms of correlation functions of determinants, so we expect the same universality to apply. Starting with the oscillating term in (\ref{leadingTerms}) and making the replacement of $\pm 2\mathrm{i} e^{S_0}\sqrt{x}$ by $\text{Disk}(x) = \pm \mathrm{i}\int_0^x \mathrm{d}x' \rho_0(x')$, one finds
\begin{align}
\langle \rho(x)\rangle \approx e^{S_0}\rho_0(x)+\dots + \frac{\sin\left(-\frac{\pi}{2}\upalpha + 2\pi e^{S_0}\int_0^x \mathrm{d}x' \rho_0(x')\right)}{4\pi x} +\dots.\label{toDerive}
\end{align}
We emphasize that this formula is only supposed to apply for double-scaled $(\upalpha,2)$ theories. 

If we further specialize to the cases $\upalpha \in\{0,2\}$, both sets of dots should vanish in perturbation theory. We don't want to claim that (\ref{toDerive}) is the exact answer though, because there could be ``instanton'' corrections related to nontrivial critical points of the effective potential of the matrix integral. However, unlike the oscillating term in (\ref{toDerive}), these instanton effects (if present) are expected to be exponentially small, of order $\exp(-\# e^{S_0})$.

\subsection{A Case Without The Oscillating Term}\label{sec:fourthcase}
There is one more JT supergravity theory without a crosscap divergence: the case without time-reversal symmetry, and with equal weighting for even and odd spin structures in the bulk. It makes sense to ask about nonperturbative effects for this model. The random matrix ensemble for the supercharge $Q$ is a standard Dyson $\upbeta = 2$ ensemble, double-scaled in such a way that both endpoints of the cut move to $\pm \infty$. As we saw in section \ref{sec:evenodd}, all perturbative correlators vanish except $R_0(x)$ and $R_{0}(x_1,x_2)$. 

We expect the oscillating contribution in (\ref{toDerive}) to be absent for this case. The reason is that for a standard $\upbeta = 2$ ensemble, the oscillating corrections to the density of eigenvalues fade as we move farther from the endpoints of the cut, and in a limit where the endpoints have been scaled away, the oscillating correction should vanish. Note, however, that in $\langle \rho(x)\rangle$ there could still be $\exp(-\# e^{S_0})$ instanton effects that arise from nontrivial critical points of the effective potential.

\subsection{Comments On Volumes}
The formula (\ref{RBessel}) can be used to determine the large $b$ asymptotics of volumes $V_g(b)$ for the case that corresponds to a matrix integral with $(\upalpha,\upbeta) = (1,2)$. This is the case discussed in section \ref{sec:diffEvenOdd}: orientable surfaces and a bulk topological field theory $(-1)^\zeta$.

The logic is as follows. First, since the spectral curve for JT gravity reduces for small $x$ to a multiple of the Bessel curve $y^2 = -1/x$, the small $x$ asymptotics of the resolvent are determined by the formula (\ref{RBessel}):
\begin{align}
R_{g\;\text{SJT}}(x)&= 2\cdot \sqrt{2}^{2g-1}\hspace{-3pt}\cdot R_{g\;\text{Bessel}}(x)+\dots \\&= \frac{\sqrt{2}}{2^{7g}}\frac{(2g)!}{2g-1}{2g \choose g}^2\frac{1}{(-x)^{g+\frac{1}{2}}}+\dots\label{smallx}
\end{align}
where the dots are subleading by powers of $x$ as $x\rightarrow 0$. In the first line, the first factor of two is to account for the difference between $R_{g\,\text{SJT}}$ and $R_{g,\text{M}}$ as in (\ref{ncu}). The second factor $\sqrt{2}^{2g-1}$ is to account for the fact that for small $x$, the relevant spectral curve (\ref{specCurvebeta2}) limits to $y^2 = -1/(2x)$ instead of $y^2 = -1/x$. To convert we need to rescale $y$ by $\sqrt{2}$, which introduces the factor shown. In the second line we inserted (\ref{RBessel}) for $R_{g\,\text{Bessel}}(x)$.

Now, $R_{g,\text{SJT}}(x)$ is an integral transform of $V_g(b)$, and in order to produce the small $x$ limit (\ref{smallx}), we need the leading term in $V_g(b)$ as $b\rightarrow \infty$ to be 
\be
V_{g}(b)= -\frac{2^{2-7g}g}{2g-1}{2g\choose g}^2 b^{2g-2}+\dots.\label{largeb}
\ee
where the dots are subleading by powers of $1/b^2$ as $b\rightarrow\infty$. We emphasize that this formula is valid for arbitrary $g$, but it only gives the leading term in powers of $b$.

In the bosonic case, an analogous formula for the leading coefficient in powers of $b$ is also known exactly. Apart from this formula, and low-genus results, little is known exactly. However, there are good large genus approximate formulas, based on Zograf's \cite{zograf2008large} conjectures for the asymptotics of intersection numbers. These imply asymptotic formulas for the bosonic volumes $V_{g}(b_1,\dots,b_n)$ for large $g$, with the restriction that all lengths $b_i$ be much smaller than $g$. Zograf's formulas were motivated by comparison to explicit low-genus computations, but aspects of his conjectures were later established rigorously \cite{mirzakhani2013growth,mirzakhani2011towards}. For the special case of one boundary, a formula was proposed in \cite{Saad:2019lba} for $V_g(b)$ in a somewhat larger domain: large $g$ and arbitrary $b$. This formula was inspired by a matrix integral calculation and has not been established rigorously.

For the volumes of super moduli space $V_g(b)$, we used a method similar to that of \cite{Saad:2019lba}, together with (\ref{toDerive}), to generate the following conjecture for large $g$ and arbitrary $b$:
\be
V_{g}(b) \approx -\frac{\Gamma(2g-1)}{2^{g}\pi^{3-2g} \mathrm{i}}\oint \frac{\mathrm{d}z}{bz}\frac{\sinh(b z)}{\left(\sin(2\pi z)\right)^{2g-1}},\hspace{20pt} g\gg 1.\label{volConj}
\ee
The contour integral around the origin reduces to a residue, which is an even polynomial in $b$. The method that led us to (\ref{volConj}) is not rigorous, and we prefer not to try to present a partial derivation. But we feel that it is worth writing the formula, because agreement with low genus exact results (computed by topological recursion) seems promising. As a function of $b$, the maximum percentage error for $g = 12$ is around 2.1\%. One can also try to extrapolate the error in $1/g$, and we found the best results by holding $b g$ fixed in the extrapolation. Using a five-term extrapolation in $1/g$ from $g = 7,\dots,12$, the maximum over $b$ of the extrapolated error is around 0.015\%.

As another piece of evidence, one can check that the leading term in (\ref{volConj}) for large $b$ is proportional to $b^{2g-2}$, and the coefficient agrees with the exact result (\ref{largeb}) for large $g$. This is encouraging, especially since the large $b$ region is where the largest percentage error is for $g = 1,\dots,12$. Finally, as an opposite limit one can take $g$ to be much larger than $b$. This is analogous to the limit where the Zograf conjecture applies in the ordinary bosonic case. In this limit, saddle points at $z = \pm\frac{1}{4}$ dominate the integral in (\ref{volConj}), and we find
\be
V_{g}(b)\approx -\frac{\Gamma(2g-\frac{3}{2})}{2^{g-\frac{5}{2}}\pi^{\frac{7}{2}-2g}}\frac{\sinh(\frac{b}{4})}{b}\hspace{20pt} g\gg b,\hspace{10pt} g\gg 1.
\ee
We don't have a known result to compare to in this case, but we note that the dependence on $b$ is similar to the bosonic case, although with $\sinh(b/4)$ instead of the bosonic $\sinh(b/2)$.

\section{Comments On The Minimal String}\label{app:minimal}
A matrix integral is described by a discrete choice of symmetry class, together with a choice of spectral curve. In the main text of this paper, we encountered all ten of the standard symmetry classes of matrix integral, but we only studied two essentially different spectral curves: one for the bosonic models and one for the supersymmetric ones.

To access different spectral curves, one would have to replace JT gravity or supergravity with a different bulk theory. A class of such bulk theories has been studied extensively in the literature: the ``minimal string'' theories. These are 2d worldsheet gravity theories that consist of a $(2,p)$ minimal model, together with Liouville theory and the reparametrization ghosts. There are both bosonic and supersymmetric versions of these theories. We will briefly comment on both cases.

In the bosonic case, the connection was made in \cite{Harris:1990kc,Brezin:1990dk} between the $\upbeta = 1,4$ ensembles and unoriented versions of the bulk minimal string theories, with the $\upbeta = 4$ case distinguished by including a $(-1)^{n_c}$ factor. In the context of the present paper, there is one interesting detail. Up to arbitrary choices of normalization, the leading density of eigenvalues for the bosonic $(2,p)$ minimal string is \cite{Moore:1991ir,Seiberg:2003nm}
\be
\rho_0(x) = \frac{1}{4\pi^2}\sinh\left(\frac{p}{2}\text{arccosh}\left(1 + \frac{8\pi^2}{p^2}x\right)\right)\label{minimalString}
\ee
where $p$ is an odd integer. In the limit $p\rightarrow \infty$, this reduces to the spectral curve of JT gravity \cite{Saad:2019lba}. However, for fixed $p$, the large $x$ behavior is dramatically modified. Indeed, substituting (\ref{minimalString}) and $y = \pm \mathrm{i}\pi \rho_0$ into the formula for the ``crosscap'' contribution (\ref{R12}), we find a finite integral. As $p\rightarrow\infty$, one has
\be
R_{\frac{1}{2}}(x) = \frac{\log(p)}{\sqrt{-x}} + \text{(finite)}
\ee
so crosscap divergence in the JT theory is regulated in the finite $p$ minimal string.

Now we turn to the supersymmetric case. For orientable surfaces, two versions of the supersymmetric minimal string have been defined. These are referred to as 0A and 0B, and they differ in the weighting of the sum over spin structures. The 0B theory involves a sum with uniform weighting, and the 0A theory includes a factor of $(-1)^\zeta$.\footnote{In general for type 0 string theories (unlike in type II) the left-moving and right-moving worldsheet fermions experience the same spin structure, and the GSO projection depends only on the total (left + right) fermion number; see section 10.6 of \cite{Polchinski:1998rr}. The projection simply requires that NS sector states must be bosonic, and R sector states must be fermionic in type 0A and bosonic in type 0B. This projection is equivalent to summing over all spin structures, with equal weighting for type 0B, and with weighting $(-1)^\zeta$ for type 0A. Note that the GSO projection that defines type II string theories can also be understood as a particular sum over spin structures. However, one has to define separate spin structures for the left-moving and right-moving worldsheet fermions. Corresponding super Riemann surfaces (and their moduli spaces) can be defined, with no relationship between holomorphic and antiholomorphic odd coordinates. Such ``cs supermanifolds'' are essentially different from the real supermanifolds that arise in JT supergravity, and in 0A and 0B theory,
on an orientable  manifold. (Some of the subtleties familiar in superstring theory
 do become relevant in the unorientable case, because $\pin^-$ structures in two dimensions are not real.)}

These two versions of the minimal string were related to matrix integrals in \cite{Klebanov:2003wg,Seiberg:2003nm,Takayanagi,Douglas}, building in part on earlier work
\cite{Morris,Johnson1,Johnson2,Johnson3,Crnkovic}. The type 0B theory (with an appropriate sign of the super-Liouville cosmological constant $\mu$) was related to a double-scaled unitary matrix integral with no endpoints in the eigenvalue distribution. In the double-scaled limit, the distinction between unitary and hermitian matrix integrals disappears, so this is the same type of matrix ensemble that we found for the supercharge $Q$ in the case without $(-1)^\zeta$. The type 0A theory was related to a double-scaled complex matrix integral. For square complex matrices, this is the same thing as an $(\upalpha,\upbeta) = (1,2)$ Altland-Zirnbauer ensemble, which is the symmetry class that we found in the case with $(-1)^\zeta$. 
For rectangular matrices, this is the generalization that we studied in section \ref{rpunctures}.
 
So the symmetry classes that we identified in this paper for the supersymmetric case without time-reversal symmetry are consistent with the previous results. We expect that large $p$ limits of these minimal string theories (with the appropriate sign of $\mu$) will coincide with JT supergravity.

The supersymmetric minimal string has a time-reversal symmetry, and one can gauge this symmetry, leading to a theory that includes a sum over unorientable as well as orientable surfaces. In this situation, each of the 0A and 0B theories have a four-fold bifurcation, for a total of eight different theories. These differ by weighting the pin$^-$ sum by a factor  $\exp(-\i\pi \eta N'/2)$, where $N'$ is an integer mod eight. The cases with odd $N'$ generalize the 0A theory, and the cases with even $N'$ generalize 0B. We expect these to be related to matrix integrals of the classes described in table \ref{table1}.

\bibliography{references}
\bibliographystyle{utphys}

\end{document}